\documentclass[twocolumn,amsmath,nofootinbib,superscriptaddress]{revtex4}

\usepackage{url}
\usepackage{amssymb}
\usepackage{amsfonts}
\usepackage{amsbsy}
\usepackage{graphicx}
\usepackage{hyperref}
\usepackage{array}
\usepackage{multirow}
\usepackage{epsfig}
\usepackage{graphics}
\usepackage{amssymb}
\usepackage{amsmath}
\usepackage{comment}
\usepackage{enumerate}
\usepackage{subcaption}
\usepackage{natbib}
\usepackage{xcolor}
\usepackage{fancyhdr}
\def\spose#1{\hbox to 0pt{#1\hss}}
\def\simlt{\mathrel{\spose{\lower 3pt\hbox{$\mathchar"218$}}
   \raise 2.0pt\hbox{$\mathchar"13C$}}}
\def\simgt{\mathrel{\spose{\lower 3pt\hbox{$\mathchar"218$}}
     \raise 2.0pt\hbox{$\mathchar"13E$}}}
 \def\simpropto{\mathrel{\spose{\lower 3pt\hbox{$\mathchar"218$}}
     \raise 2.0pt\hbox{$\propto$}}}

\def\beq#1{\begin{equation}\label{#1}}
\def\eeq{\end{equation}}
\def\beqa#1{\begin{eqnarray}\label{#1}}
\def\eeqa{\end{eqnarray}}

\usepackage{float}
\usepackage{placeins}
\usepackage{mathtools}

\DeclarePairedDelimiter\floor{\lfloor}{\rfloor}

\fancypagestyle{plain}{
\fancyhf{}
\fancyhead[R]{\fontfamily{cmss}\selectfont
  \fcolorbox{black}{white}{\parbox{1.9in}{\raggedleft\small
    \color{black}DISTRIBUTION STATEMENT A.\\Approved for public release.}}}
}

\parskip=\medskipamount

\begin{document}
\title{Sequence Aggregation Rules for Anomaly Detection in Computer Network Traffic}
\author{Benjamin J. Radford}
\email[]{benjamin.radford@gmail.com}
\affiliation{KeyW Corporation}
\thanks{This research was developed with funding from the Defense \mbox{Advanced} Research Projects Agency (DARPA). The views, opinions and/or findings expressed are those of the authors and should not be interpreted as representing the official views or policies of the Department of Defense or the U.S. Government.}
\author{Bartley D. Richardson}
\affiliation{KeyW Corporation}
\author{Shawn E. Davis}
\affiliation{KeyW Corporation}

\begin{abstract}
We evaluate methods for applying unsupervised anomaly detection to cybersecurity applications on computer network traffic data, or flow. We borrow from the natural language processing literature and conceptualize flow as a sort of ``language'' spoken between machines. Five sequence aggregation rules are evaluated for their efficacy in flagging multiple attack types in a labeled flow dataset, CICIDS2017. For sequence modeling, we rely on long short-term memory (LSTM) recurrent neural networks (RNN). Additionally, a simple frequency-based model is described and its performance with respect to attack detection is compared to the LSTM models. We conclude that the frequency-based model tends to perform as well as or better than the LSTM models for the tasks at hand, with a few notable exceptions. 
\end{abstract}

\maketitle

\thispagestyle{plain}

\date{\today}
\vspace{10mm}


\section{Introduction}

Vulnerabilities lurking undiscovered on computer networks threaten to be the vectors of the next major data breach, service disruption, or illicit appropriation of hardware. Malicious actors exploit these previously unknown, little-known, or unpatched vulnerabilities to misuse network resources, sometimes to costly or disastrous effect. When the set of possible vulnerabilities is well-defined, detection is a matter of maintaining software that can scan systems and networks for the signatures of actors exploiting those vulnerabilities. However, sophisticated actors will sometimes rely on novel, or 0day, vulnerabilities that are typically resistant to signature-based detection. Unsupervised learning, a subset of machine learning that relies on data without annotations or labels, is well-suited for the task of detecting the exploitation of obscure or unknown vulnerabilities.

Cybersecurity encompasses a broad set of challenges including, but not limited to, intrusion detection, incident response, malware analysis, and attribution. Here we focus on the detection of malicious network activities, broadly defined. Historically, intrusion detection software, antivirus software, and similar tools have relied on heuristic-based rules and malware signatures to perform detection. Matching techniques like these are effective when similar attacks are known to security software developers a priori. Detecting malicious events on a network that are previously unknown or for which detection rules do not yet exist requires a different approach. Here, we demonstrate the use of unsupervised machine learning to identify anomalies on a network by modeling the network's ``normal'' behavior and scoring deviations from this baseline. 

We proceed by first providing a very brief background of machine learning for cybersecurity applications.\footnote{For a more thorough review on this topic, readers may be interested in \cite{buczak:guven:2016}.} We then discuss CICIDS2017, a publicly available dataset for cybersecurity and intrusion detection research. Close attention is paid to data processing and feature generation. The methodology section introduces the models to be used and evaluation criteria. We present the results of our analysis and conclude with a short discussion of open questions and directions for future research.
\section{Background}

Machine learning for cybersecurity is a growing area of research. Researchers from Pacific Northwest National Laboratory and Western Washington University have used recurrent neural networks (RNN) to analyze authentication log data \citep{tuor:etal:2017a, tuor:etal:2017b, brown:etal:2018}. These works not only demonstrate the capacity for anomaly detection via RNNs in system log data but also introduce an attention mechanism to the model that provides context for the flagged anomalies. These works follow earlier an application of long short-term memory (LSTM) models for intrusion detection given system call logs captured at the host level \citep{kim:etal:2016}. Machine learning has also found its way into both commercial and open source security software solutions. Mirsky et al. introduce Kitsune, a network intrusion detection system (NIDS) based on an ensemble of artificial neural network (ANN) autoencoders \citep{mirsky:etal:2018}. Autoencoders are models that learn an identify function to map data points to themselves often via an intermediate step that compresses the data before it is reconstructed. Kitsune is open source software and is available online \citep{mirsky:2018}. 

Recent work has demonstrated that Long Short-Term Memory (LSTM) Recurrent Neural Networks (RNN) can be applied to the problem of anomaly detection in computer network flow data \citep{radford:etal:2017}. We follow this previous work here and evaluate four new aggregation rules, in addition to the previously introduced IP-dyad-hour, as well as a simpler model for anomaly detection. 

\section{Data}


We rely on a public dataset for intrusion detection (IDS) tasks from the University of New Brunswick's (UNB) Canadian Institute for Cyberseurity (CIC) \citep{unb:2018}. The dataset, CICIDS2017, comprises 3.1 million flow records \citep{sharafaldin:etal:2018}. This dataset covers five days in 2017 and includes labeled attack types shown in Table~\ref{tab:attacktypes}. The documentation that accompanies CICIDS2017 lists additional attack types that are not present in the flow data. We believe these to be undifferentiated subcategories of those attacks listed here.\footnote{These include \emph{Infiltration -- Dropbox download} and \emph{Infiltration -- Cool disk -- MAC} among others.} Benign traffic was artificially generated to simulate user behaviors while attacks were performed manually. After dropping incomplete records, approximately 2.9 million flows remain. UNB CIC also provides full packet capture for CICIDS2017 though those data are not utilized for this work. 

\begin{center}
\begin{table}
\caption{Attack Types in CICIDS2017}\label{tab:attacktypes}
\begin{tabular}{| l | r | r |}
\toprule
Attack & Flow Count & Proportion \\ 
\hline 
Benign & 2358036 & 8.3e-1 \\
DoS Hulk & 231073 & 8.2e-2 \\
Port Scan & 158930 & 5.6e-2 \\
DDoS & 41835 & 1.5e-2 \\
DoS GoldenEye & 10293 & 3.6e-3 \\
FTPPatator & 7938 & 2.8e-3 \\
SSHPatator & 5897 & 2.1e-3 \\
DoS Slow Loris & 5796 & 2.0e-3 \\
DoS Slow HTTP Test & 5499 & 1.9e-3 \\
Botnet & 1966 &  7.0e-4 \\
Web Attack: Brute Force & 1507 & 5.3e-4 \\
Web Attack: XSS & 652 & 2.3e-4 \\
Infiltration & 36 & 1.3e-5 \\
Web Attack: SQL Injection & 21 & 7.0e-6 \\
Heartbleed & 11 & 4.0e-6 \\
\hline
\end{tabular}
\end{table}
\end{center}

Atheoretical detection of malicious actions or events in cybersecurity-relevant data is difficult to validate due to the observed heterogeneity in attack style and the ingenuity of malicious actors. CICIDS2017 provides us the opportunity only to demonstrate that anomaly detection techniques can or cannot identify events of the specific types and characteristics as those present in the data. Caution should therefore be taken when extrapolating the results presented here to unsupervised anomaly detection in computer network data more generally. Theorizing about a network's vulnerabilities and potential adversaries (i.e. having a threat model) will likely produce better results than fully unsupervised machine learning techniques.\footnote{Grok your data generating processes.}

In keeping with the this paper's goal of a generalizable anomaly detection approach to cybersecurity, the models presented herein are entirely unsupervised. The ``ground truth'' labels provided by CIC are used only for post-facto evaluation and are ignored prior (including during model selection, tuning, and training). However, we recognize that ``unsupervised'' is something of a misnomer. While no information whatsoever about the attack types is used in model training, the modeling process starts long before the training phase. Feature selection, unit aggregation, and other data pre-processing steps necessarily influence what characteristics of network traffic are learned and what characteristics are not. To the extent that attacks are anomalous with respect to only certain features or levels of aggregation, the data pre-processing phase encodes a prior estimate of those features that will best reveal malicious activities. Attacks vary in purpose and implementation and therefore resist a singular feature set for identification. Attacks that manifest in unusual byte count patterns may not manifest in unusual port sequences and vice versa. 

Incomplete records with respect to required fields are dropped from the data.\footnote{The authors recognize that multiple imputation may be the preferred method for addressing incomplete records in some fields but opt for row-wise deletion in keeping with how we anticipate a deployed cybersecurity solution would behave.} We also omit all 1253 records that do not include at least one internal IP address. Internal IP addresses are listed in the CICIDS2017 documentation.

\subsection{Feature Set}

We evaluate two feature sets. The feature sets are taken directly from \citep{radford:etal:2017}. We adopt these features for several reasons including demonstrated past performance, their applicability to many flavors of network flow data, and to avoid feature over-engineering. Complex feature engineering could run afoul of our goal to eschew heuristic-based attack detection.\footnote{If a network owner understands the specifics of their threat model, more nuanced feature engineering is likely to result in better performance.} We refer to the two feature sets as \emph{protobytes} and \emph{ports}. Protobyte sequences are generated by concatenating a time-ordered series of protobyte tokens; a protobyte token is defined as \texttt{protocol:}$\floor{log_2(bytes)}$. An example protobyte sequence would look like: \texttt{TCP:10|TCP:12|UDP:04}. The second feature set, port sequences, is generated by first determining the likely service port for each record in the flow data. Flow data contain two ports per record, one per device. Of those, one port is typically a ``service port,'' a port reserved for a known application or service. For simplicity, we rely on heuristic rules to select the likely service port for each record. In particular, we set \texttt{service port} to be the result of $min(min(\text{\texttt{src port}}, 10000), min(\text{\texttt{dst port}}, 10000))$. An example port sequence would look like \texttt{80|443|80}.

\subsection{Aggregation Rules}

For both feature sets, we evaluate five sequence aggregation rules. We refer to these by their units of analysis: \emph{source}, \emph{destination}, \emph{dyad}, \emph{internal}, and \emph{external}. For each aggregation rule, we first order the data by time and then group by the unit of analysis per hour. Source and destination aggregation concatenate tokens in time-order after grouping by \texttt{(source IP, date, hour)} and \texttt{(destination IP, date, hour)}, respectively. A dyad-hour is defined as the tuple \texttt{(source IP, destination IP, date, hour)}. Finally, internal and external are defined by the tuples \texttt{(internal IP, date, hour)} and \texttt{(external IP, date, hour)}, respectively. Each aggregation rule results in a unique set of sequences -- sequences are only guaranteed to be contiguous with respect to the chosen aggregation rule. 
\section{Methodology}

Our approach diverges from \cite{radford:etal:2017} in our method of outlier, or anomaly, scoring. Previous works assigned an \emph{attack/benign} label to every dyad-hour unit. Each dyad-hour unit's anomaly score was essentially the minimum predicted probability for any token in the given dyad-hour. In the current paper, we score every individual sequence token conditional on the learned model and on the 10 tokens that immediately precede the predicted token. A target token's outlier score is the predicted probability of the correct token given the model parameters and previously-observed tokens within the same sequence. A token here is defined as a single element of a sequence; tokens correspond to individual rows of flow data.

\subsection{Sequence Modeling}

For sequence modeling, we follow previous work \citep{radford:etal:2017}. Sequences of length 10 are produced for each dyad via a rolling window. Sequences are left-padded with zero-values that are masked during training. For each length 10 sequence, the value of the subsequent ($11^{th}$) token, the target, is predicted. 

We train all models on the first day of data, July 3, 2017. A 10-token sliding window is utilized to generate sequences. All sequences are left zero-padded and the sliding window is moved one token at a time. Because the first day contains no attack data, this training method corresponds to the \emph{clean baseline} method described in \cite{radford:etal:2017}. That previous work demonstrated that it is not necessary to train models on clean network traffic data to perform successful anomaly detection. We elect to train on the first day of data both for simplicity and with the understanding that future applications of this work will likely require that models be trained in advance of deployment rather than on an ongoing basis. The training sets comprise approximately 450,000 sequences (with 50,000 reserved for validation) with slight variations depending on aggregation rule.

\begin{center}
\begin{figure}
\includegraphics[width=0.5\columnwidth]{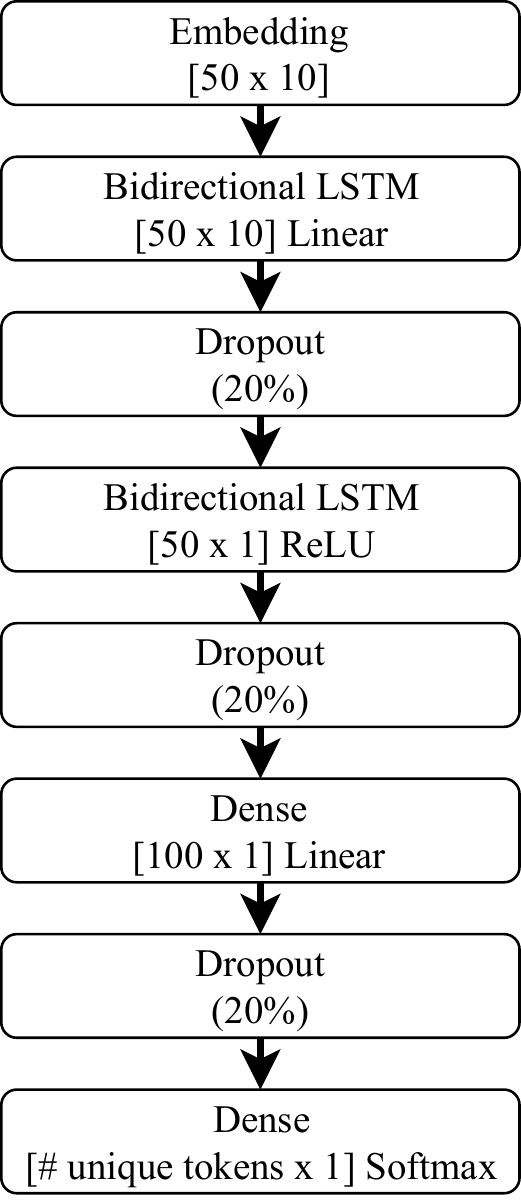}
\caption{LSTM Architecture. Layer output dimensions indicated by [$m \times n$].}\label{fig:lstmarchitecture}
\end{figure}
\end{center}

\subsection{Long Short-Term Memory Model}

Long short-term memory (LSTM) recurrent neural networks (RNN) are utilized here. The chosen model architecture is summarized in Figure~\ref{fig:lstmarchitecture}. We begin with an embedding layer that projects the sequence token values into a dense 50-dimensional vector. The embedded sequences are then passed through two bidirectional LSTM layers. Bidirectional LSTMs read the input sequence both forwards and backwards and then concatenate the output vectors. The second LSTM layer introduces non-linearity via the use of rectified linear unit activation. The output of the second LSTM is passed to a fully-connected dense layer and finally to an output fully-connected dense layer with softmax activation. Because the target is multiclass, models are trained to minimize multiclass cross-entropy. We optimize via TensorFlow's Lazy Adam optimizer; due to the sparsity of input vectors, we do not need to update all weights at every step, only those that are active on the forward pass.\footnote{\url{https://www.tensorflow.org/api_docs/python/tf/contrib/opt/LazyAdamOptimizer}} Class weights are provided during training due to class imbalance.

\subsection{Evaluation}

After a model is trained, it is used to predict the out-of-sample tokens for the remaining sequences (July 4 -- July 7). For the protobytes sequences, all out-of-sample tokens are predicted and assigned an outlier score. For the port sequences, the sliding window is shifted by 3 tokens at every step. Therefore, only one-third of all out-of-sample port sequence tokens is assigned an outlier score. This is done to alleviate memory constraints encountered when operating on the full set of port sequence tokens. The remaining two-thirds of out-of-sample port sequence tokens are omitted. Outlier scores are the negative of the predicted probability assigned to the correct value of the target token.

Every model is run three times. For each replication of a model, the training data are resampled with replacement. We then average over the results of each set of models to produce a bootstrapped estimate of the mean area under (AUC) the receiver operating characteristic curve (ROC). A grand total of 30 models are trained: 2 feature sets $\times$ 5 aggregation rules $\times$ 3 resampled training sets. 

\subsection{Baseline Comparison}

In addition to LSTM models, we also evaluate a simple frequency-based method of outlier detection. For both sets of tokens, protobytes and ports, the training set frequencies of those tokens are calculated. Outlier scores are assigned to test set tokens according to $score(token_j) = -\frac{1}{n}\sum_{i=1}^{n} I_{i = token_j}$, where $n$ is the number of examples in the training set and $I_{i=token_j}$ is an indicator that takes a value 1 if $i=token_j$ and 0 otherwise.

\begin{table*}
  [ht] \caption{Protobyte Sequence AUC} \label{tab:protobytes}
  \begin{tabular}{rcccccc} 
  \toprule
	\multicolumn{1}{c}{} & \multicolumn{5}{c}{\underline{LSTM}} & \multicolumn{1}{c}{\underline{Frequency}}\\   
	\multicolumn{1}{c}{} & Source & Destination & Dyad & Internal & External & \\ \hline
	All Attacks & 0.44 & 0.27 & 0.24 & 0.37 & 0.46 & \textbf{0.80} \\
	Botnet & 0.47 & 0.35 & 0.30 & \textbf{0.61} & 0.54 & 0.57 \\
	DoS GoldenEye & 0.63 & 0.63 & 0.64 & 0.61 & 0.58 & \textbf{0.65} \\
	DoS Hulk & 0.50 & 0.25 & 0.24 & 0.46 & 0.34 & \textbf{0.66} \\
	DoS Slow HTTP Test & 0.32 & 0.40 & 0.39 & 0.34 & 0.38 & \textbf{0.59} \\
	DoS Slow Loris & 0.52 & 0.61 & \textbf{0.62} & 0.47 & 0.52 & 0.53 \\
	FTPPatator & 0.46 & 0.62 & 0.73 & 0.47 & \textbf{0.69} & 0.64 \\
	Heartbleed & \textbf{1.00} & 0.99 & 0.99 & \textbf{1.00} & \textbf{1.00} &\textbf{1.00} \\
	Infiltration & \textbf{0.93} & 0.86 & 0.86 & 0.91 & 0.82 & 0.73 \\
	Port Scan & 0.32 & 0.29 & 0.23 & 0.25 & 0.63 & \textbf{0.96} \\
	SSHPatator & 0.48 & \textbf{0.60} & 0.57 & 0.51 & 0.47 & 0.51 \\
	Web Attack Brute Force & 0.18 & 0.36 & 0.25 & 0.20 & 0.34 & \textbf{0.51} \\
	Web Attack SQL Injection & 0.66 & \textbf{0.81} & 0.72 & 0.57 & 0.68 & 0.56 \\
	Web Attack XSS & 0.08 & 0.31 & 0.10 & 0.09 & 0.24 & \textbf{0.48} \\
	\hline \\ 
	\multicolumn{7}{r}{\footnotesize Highest AUC per row in bold.}
 \end{tabular}
\end{table*}

\begin{table*}
  [ht] \caption{Port Sequence AUC} \label{tab:ports}
  \begin{tabular}{rcccccc} 
  \toprule
	\multicolumn{1}{c}{} & \multicolumn{5}{c}{\underline{LSTM}} & \multicolumn{1}{c}{\underline{Frequency}}\\   
	\multicolumn{1}{c}{} & Source & Destination & Dyad & Internal & External & \\ \hline
	All Attacks & 0.49 & 0.68 & 0.70 & 0.55 & 0.71 & \textbf{0.87} \\
	Botnet & 0.88 & 0.90 & 0.82 & 0.91 & 0.86 & \textbf{0.93} \\
	DoS GoldenEye & 0.22 & 0.47 & 0.48 & 0.32 & 0.51 & \textbf{0.71} \\
	DoS Hulk & 0.17 & 0.46 & 0.48 & 0.29 & 0.51 & \textbf{0.74} \\
	DoS Slow HTTP Test & 0.25 & 0.50 & 0.48 & 0.34 & 0.52 & \textbf{0.71} \\
	DoS Slow Loris & 0.24 & 0.51 & 0.48 & 0.34 & 0.52 & \textbf{0.71} \\
	FTPPatator & 0.82 & 0.85 & 0.89 & 0.79 & 0.88 & \textbf{0.91} \\
	Heartbleed & 0.96 & \textbf{1.00} & 0.99 & 0.95 & 0.98 & 0.96 \\
	Infiltration & \textbf{0.98} & 0.95 & 0.95 & \textbf{0.98} & 0.95 & 0.96 \\
	Port Scan & 0.97 & \textbf{0.98} & \textbf{0.98} & 0.97 & \textbf{0.98} & \textbf{0.98} \\
	SSHPatator & 0.69 & 0.80 & 0.84 & 0.65 & 0.84 & \textbf{0.87} \\
	Web Attack Brute Force & 0.21 & 0.60 & 0.48 & 0.32 & 0.54 & \textbf{0.71} \\
	Web Attack SQL Injection & 0.21 & \textbf{0.81} & 0.48 & 0.31 & 0.63 & 0.71 \\
	Web Attack XSS & 0.21 & 0.65 & 0.48 & 0.31 & 0.56 & \textbf{0.71} \\
	\hline \\ 
	\multicolumn{7}{r}{\footnotesize Highest AUC per row in bold.}
 \end{tabular}
\end{table*}

\section{Results}

Tables~\ref{tab:protobytes} and ~\ref{tab:ports} report the average AUC value for every protobyte and port sequence model, respectively. For every attack type, the best AUC score is shown in bold. The ROC plots themselves are included in the Appendix, Tables~\ref{tab:frequencya} -- \ref{tab:portsb}. The average curve is bold and the three bootstrap curves are thin and dashed. All four curves are presented to provide the reader with a visual representation of the variance in model performance with respect to repeated sampling.\footnote{In Table~\ref{tab:frequencya} the three columns correspond to performance in the protobyte-based frequency model, the port sequence model, and a combination of the two called PC1. PC1 is the projection of frequency-based protobyte anomaly scores and frequency-based port anomaly scores onto their first principal component. This column is included for interested readers but is not discussed further.} 

We measure the effectiveness of a model with AUC. The AUC can be interpreted as the probability that a model assigns a higher anomaly score to an attack token than to a randomly chosen non-attack token. An AUC of 1 corresponds to perfect separation of attacks from non-attacks while an AUC of 0.5 indicates that a model is working no better than chance. Negative AUC values indicate that the model is performing worse than chance; a possibility in situations where attack labels are unavailable (i.e. for unsupervised tasks) and therefore modelers are unable to determine whether outlier scores should be inverted.\footnote{Note that the computation of AUC requires labeled data points and is therefore typically found in the context of supervised modeling. Here, we are using fully unsupervised models but have the benefit of a labeled data set with which to evaluate performance post facto.} ``All Attacks'' indicates that all attack types have been collapsed to a single category. For all other individual attack types, the attack type of interest is labeled ``attack'' and all other attack types are combined with benign data to form the ``non-attack'' category.

Overall, there is a substantial amount of variation in attack detection performance across aggregation rules and attack types. The simple frequency-based models tend to outperform the LSTM models. In all but two cases, the frequency-based model performs at least comparably to, if not better than, the best of the corresponding LSTM models. The exceptions are in the LSTMs' ability to detect SQL injection attacks and infiltration. Given protobyte sequences, the LSTMs outperform the frequency model for both attack types across all aggregation rules. 

All aggregation rules and both feature sets are able to flag the use of Heartbleed; AUC scores for detection of this particular attack fall between 0.95 and 1.0. Unsurprisingly, port scans are faithfully detected under all aggregation rules when analyzing port sequences and are not detected when analyzing protobyte sequences. Curiously, while the protobyte LSTMs are unable to catch the port scan, the frequency-based model detects it with an AUC of 0.96. Future work should examine what protobyte tokens are flagged by the frequency-based model with respect to the port scan and determine why the LSTM models assign these tokens relatively high predicted probabilities.

The port frequency-based model scores an AUC of 0.71 or above for all attack types and averages 0.87. The highest average AUC achieved by an LSTM model is 0.71, the score for port sequences aggregated by external IP address. The average AUC may not be the best metric for practical applications as there should be no expectation that any single feature set or any single aggregation rule should produce the best attack detector for all attacks. Instead, clever methods for ensemble anomaly detection are needed to leverage the relative strengths of a variety of models and feature sets. 
\section{Conclusion}

That the chosen aggregation rule makes, on the whole, minimal difference in attack detection is intriguing. Perhaps event sequences are less important in detecting the given attacks than is the relative frequency of the target token itself. In other words, perhaps aggregation rules are entirely unnecessary and a probability model based on token frequency would perform equally well. This would seem an unsatisfactory cybersecurity solution, however, for all but the most homogeneous of networks with respect to flow generation. 

Alternatively, perhaps the LSTM models are best-suited for capturing attack onsets rather than all flows associated with attacks. Once an attack has begun, perhaps the transitions between attack tokens are easily modeled by the LSTM. This could be the case, for instance, if the model learns a simple autoregressive process in which high probability is assigned to the most recently seen value in the sequence. The first token of an attack sequence may then appear anomalous but subsequent tokens would appear normal, given that the model has already seen previous attack tokens. In future work, we intend to examine this possibility by repeating our analyses but re-coding attack tokens that are immediately preceded (within a sequence) by a same-valued attack token. 

Another possibility is that sequences of flow-like data simple do not encode much information. Flow is already an aggregation of underlying time-ordered data. One flow can represent dozens or hundreds of packets, each with their own unique attributes. By aggregating to the flow level, the variability of data within that flow is lost and only summary statistics remain. In a natural language analogy, this would be like taken a series of sentences and replacing them with metadata entries that encode the number of words and characters per sentence. The language-like structure of communication between machines (proxied by IP address) has been lost in favor of a summary that omits all of the content and much of the context of what is said. 

Future research should apply sequence modeling and deep learning techniques directly to the packet-level data. It is possible that, in doing so, flow aggregation is made unnecessary and that determinations about data retention and importance could be made at the packet level or lower. This may come with high computation and storage costs and so researchers should take care to balance their solutions against real-world constraints that may not be apparent in lab settings. 

While the results presented above may not initially appear encouraging with respect to the applicability of certain deep learning techniques to flow-like network data, we caution against undo pessimism. We make no claim to have identified the ideal, or even good, feature sets, aggregation rules, model architectures, or hyperparameter settings. Furthermore, CICIDS2017 is built on largely simulated traffic and so the generalizability of results on these data is open for debate, especially with respect to networks that do not resemble the simulated network in terms of size, roles, hardware, software, and threat models. Instead, we present these results in the hope that they encourage further research into machine learning approaches to cybersecurity and as a reminder that complex models are not always the best models. 

\bibliographystyle{IEEEtran}
\bibliography{master}

\begin{thebibliography}{10}
\providecommand{\url}[1]{#1}
\csname url@samestyle\endcsname
\providecommand{\newblock}{\relax}
\providecommand{\bibinfo}[2]{#2}
\providecommand{\BIBentrySTDinterwordspacing}{\spaceskip=0pt\relax}
\providecommand{\BIBentryALTinterwordstretchfactor}{4}
\providecommand{\BIBentryALTinterwordspacing}{\spaceskip=\fontdimen2\font plus
\BIBentryALTinterwordstretchfactor\fontdimen3\font minus
  \fontdimen4\font\relax}
\providecommand{\BIBforeignlanguage}[2]{{%
\expandafter\ifx\csname l@#1\endcsname\relax
\typeout{** WARNING: IEEEtran.bst: No hyphenation pattern has been}%
\typeout{** loaded for the language `#1'. Using the pattern for}%
\typeout{** the default language instead.}%
\else
\language=\csname l@#1\endcsname
\fi
#2}}
\providecommand{\BIBdecl}{\relax}
\BIBdecl

\bibitem{buczak:guven:2016}
\BIBentryALTinterwordspacing
A.~L. Buczak and E.~Guven, ``A survey of data mining and machine learning
  methods for cyber security intrusion detection,'' \emph{IEEE Communications
  Surveys \& Tutorials}, vol.~18, no.~2, 2016. [Online]. Available:
  \url{https://ieeexplore.ieee.org/document/7307098/?part=1}
\BIBentrySTDinterwordspacing

\bibitem{tuor:etal:2017a}
A.~Tuor, R.~Baerwolf, N.~Knowles, B.~Hutchinson, N.~Nichols, and R.~Jasper,
  ``Recurrent neural network language modles for open vocabulary event-level
  cyber anomaly detection,'' \emph{Proceedings of AAAI-2018 Artificial
  Intelligence in Cyber Security Workshop}, 2017.

\bibitem{tuor:etal:2017b}
A.~Tuor, S.~Kaplan, B.~Hutchinson, N.~Nichols, and S.~Robinson, ``Deep learning
  for unsupervised insider threat detection in structured cybersecurity data
  streams,'' \emph{Proceedings of AI for Cyber Security Workshop at AAAI 2017},
  2017.

\bibitem{brown:etal:2018}
A.~Brown, A.~Tuor, B.~Hutchinson, and N.~Nichols, ``Recurrent neural network
  attention mechanisms for interpretable system log anomaly detection,''
  \emph{arXiv:1803.04967}, 2018.

\bibitem{kim:etal:2016}
G.~Kim, H.~Yi, J.~Lee, Y.~Paek, and S.~Yoon, ``Lstm-based system-call language
  modeling and robust ensemble method for designing host-based intrusion
  detection systems,'' \emph{arXiv:1611.01726}, 2016.

\bibitem{mirsky:etal:2018}
Y.~Mirsky, T.~Doitshman, Y.~Elovici, and A.~Shabtai, ``Kitsune: An ensemble of
  autoencoders for online network intrusion detection,'' \emph{Network and
  Distributed Systems Security Symposium}, 2018.

\bibitem{mirsky:2018}
\BIBentryALTinterwordspacing
Y.~Mirsky. (2018) Kitnet-py. [Online]. Available:
  \url{https://github.com/ymirsky/KitNET-py}
\BIBentrySTDinterwordspacing

\bibitem{radford:etal:2017}
B.~J. Radford, L.~M. Apolonio, A.~J. Trias, and J.~A. Simpson, ``Network
  traffic anomaly detection using recurrent neural networks,''
  \emph{Proceedings of the 2017 National Symposium on Sensor Data and Fusion},
  2017.

\bibitem{unb:2018}
\BIBentryALTinterwordspacing
(2018) Intrusion detection evaluation dataset (cicids2017). Canadian Institute
  for Cybersecurity, University of New Brunswick. [Online]. Available:
  \url{http://www.unb.ca/cic/datasets/ids-2017.html}
\BIBentrySTDinterwordspacing

\bibitem{sharafaldin:etal:2018}
I.~Sharafaldin, A.~H. Lashkari, and A.~A. Ghorbani, ``Toward generating a new
  intrusion detection dataset and intrusion traffic characterization,''
  \emph{4th International Conference on Information Systems Security and
  Privacy (ICISSP)}, 2018.

\end{thebibliography}

\clearpage

\appendix
\section{Appendix}
\begin{table*}
  [ht] \caption{Frequency Model A} \label{tab:frequencya}
  \begin{tabular}{ccc} 
  \hline 
  Protobyte Sequences & Port Sequences & PC1\\
      \hline 
      \includegraphics[width=.19\textwidth]{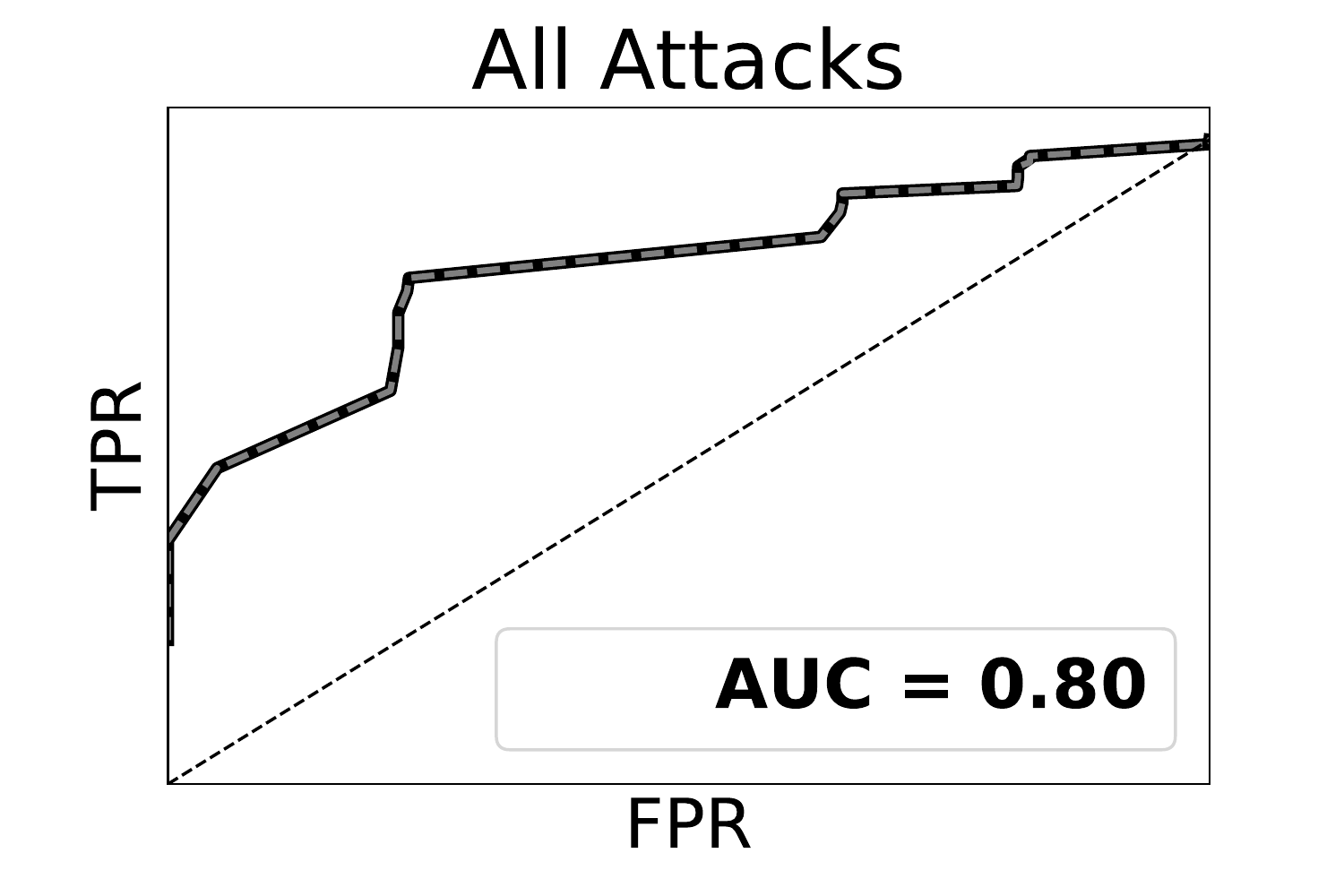} &
      \includegraphics[width=.19\textwidth]{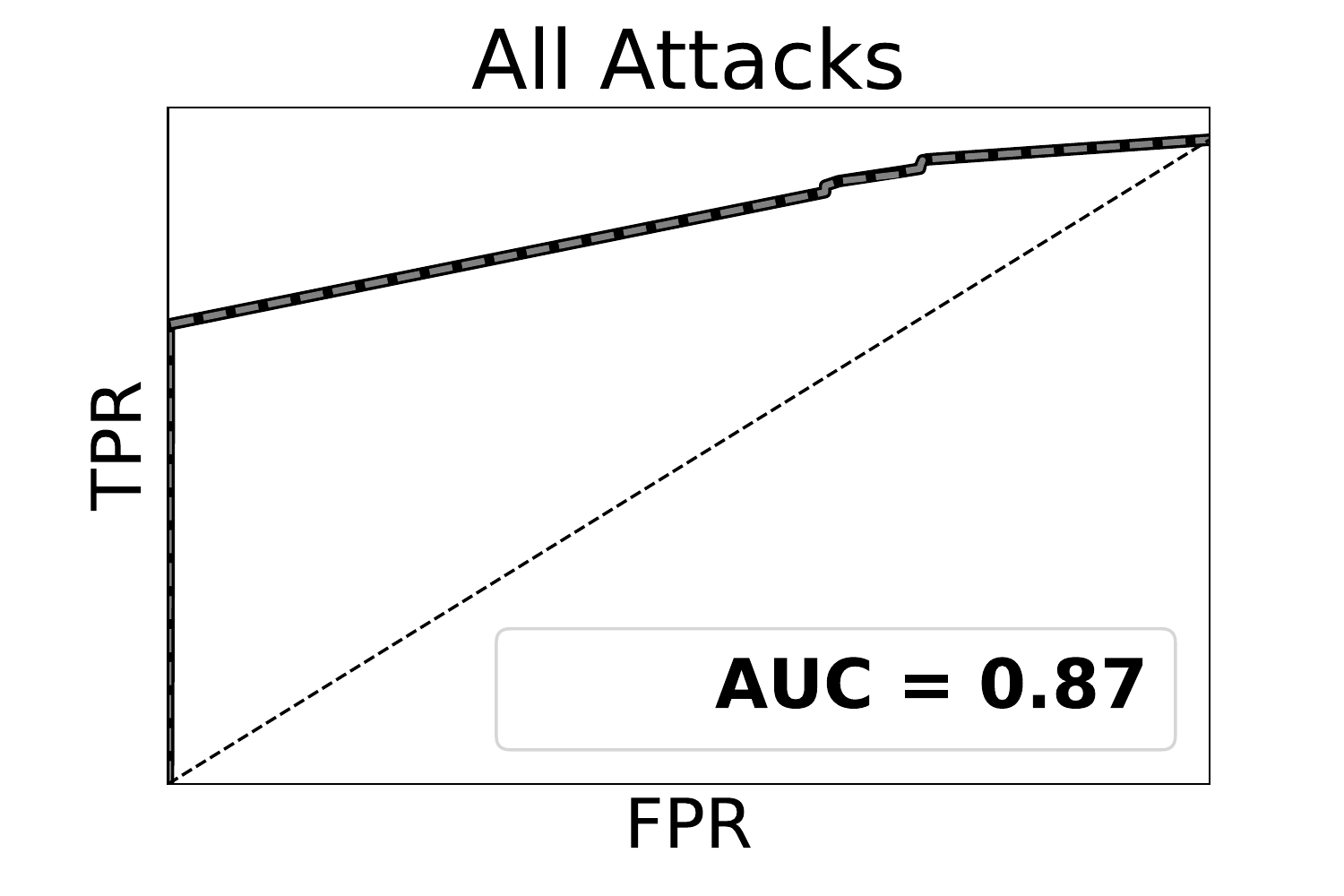} &
      \includegraphics[width=.19\textwidth]{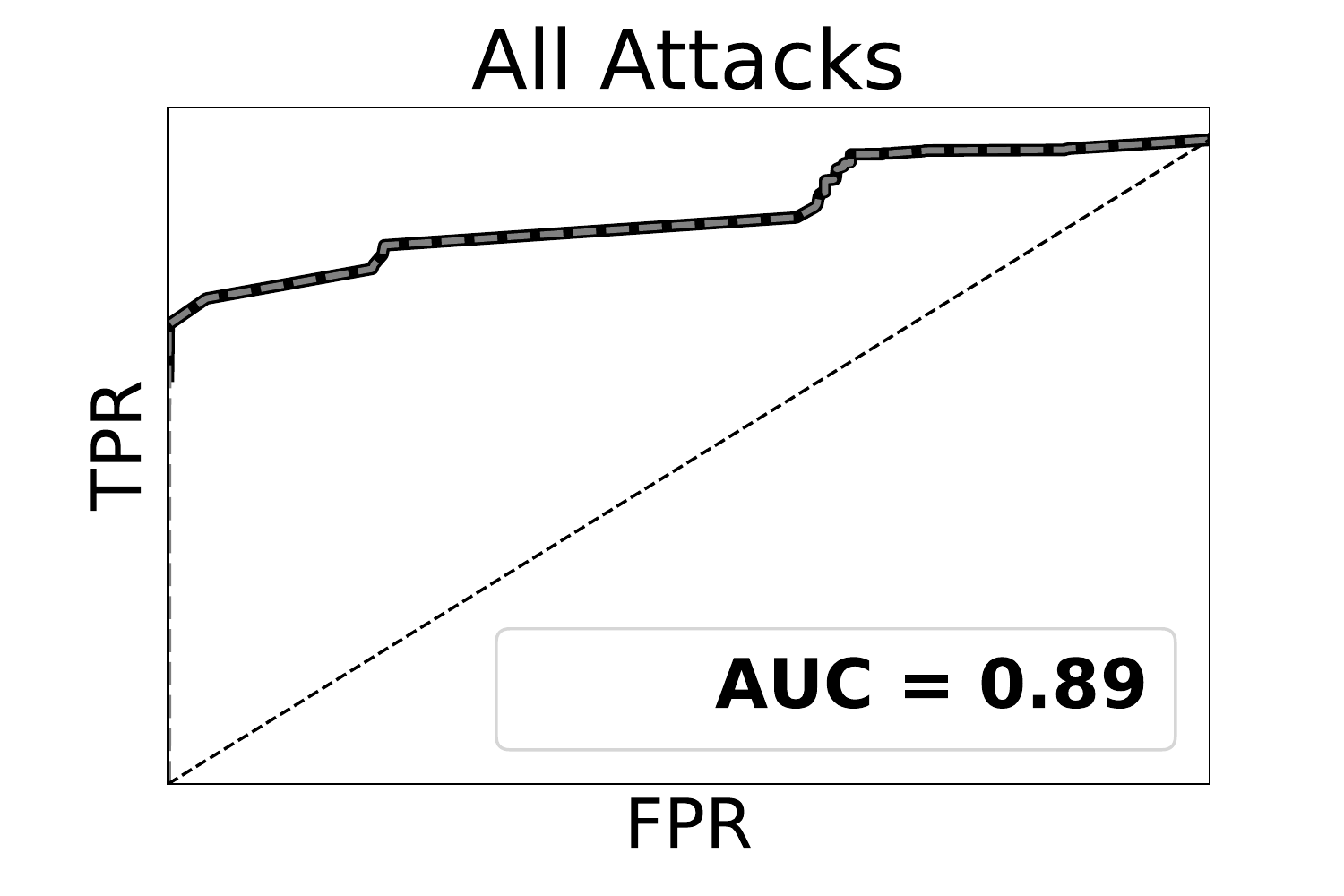} 
      \\
            \includegraphics[width=.19\textwidth]{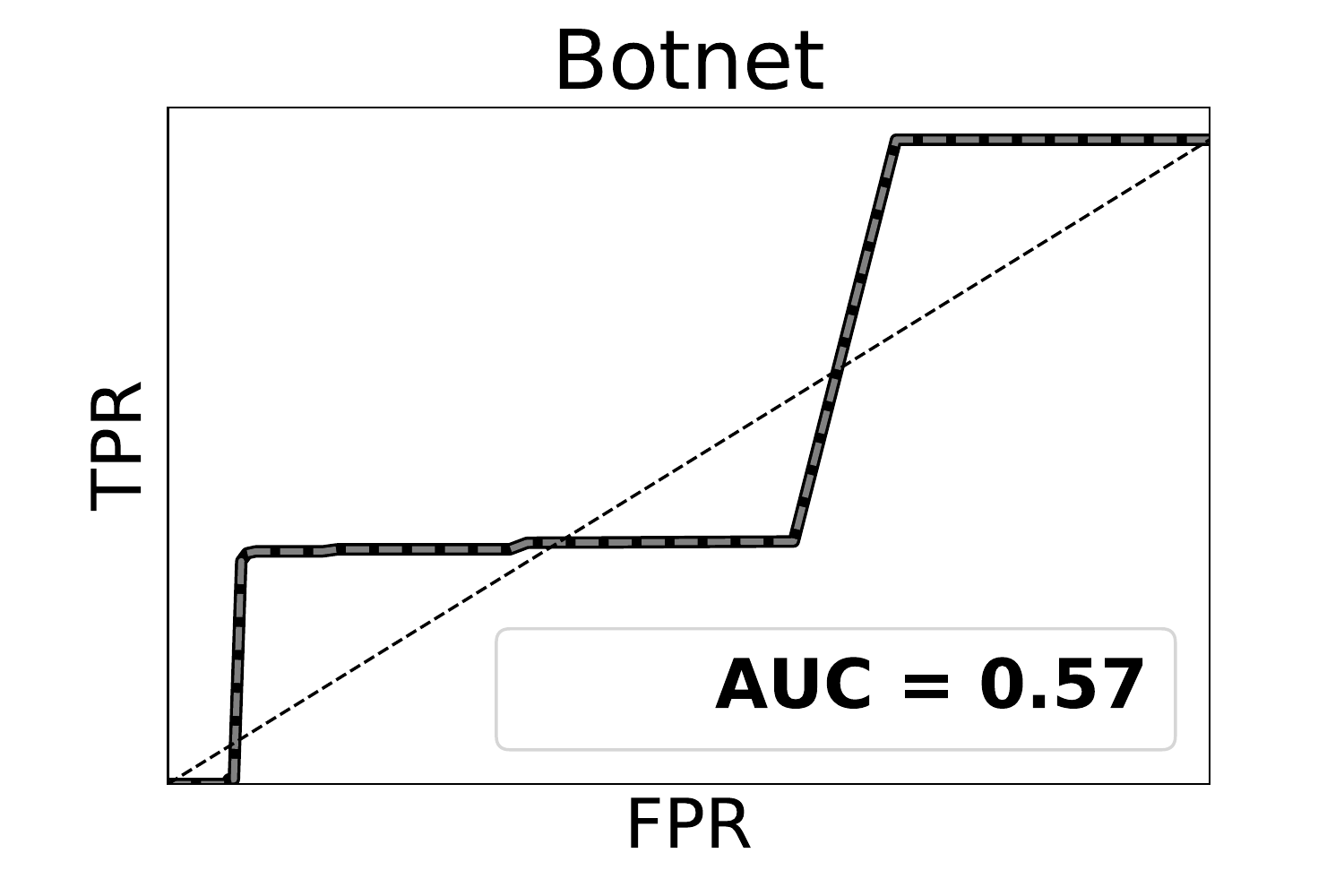} &
      \includegraphics[width=.19\textwidth]{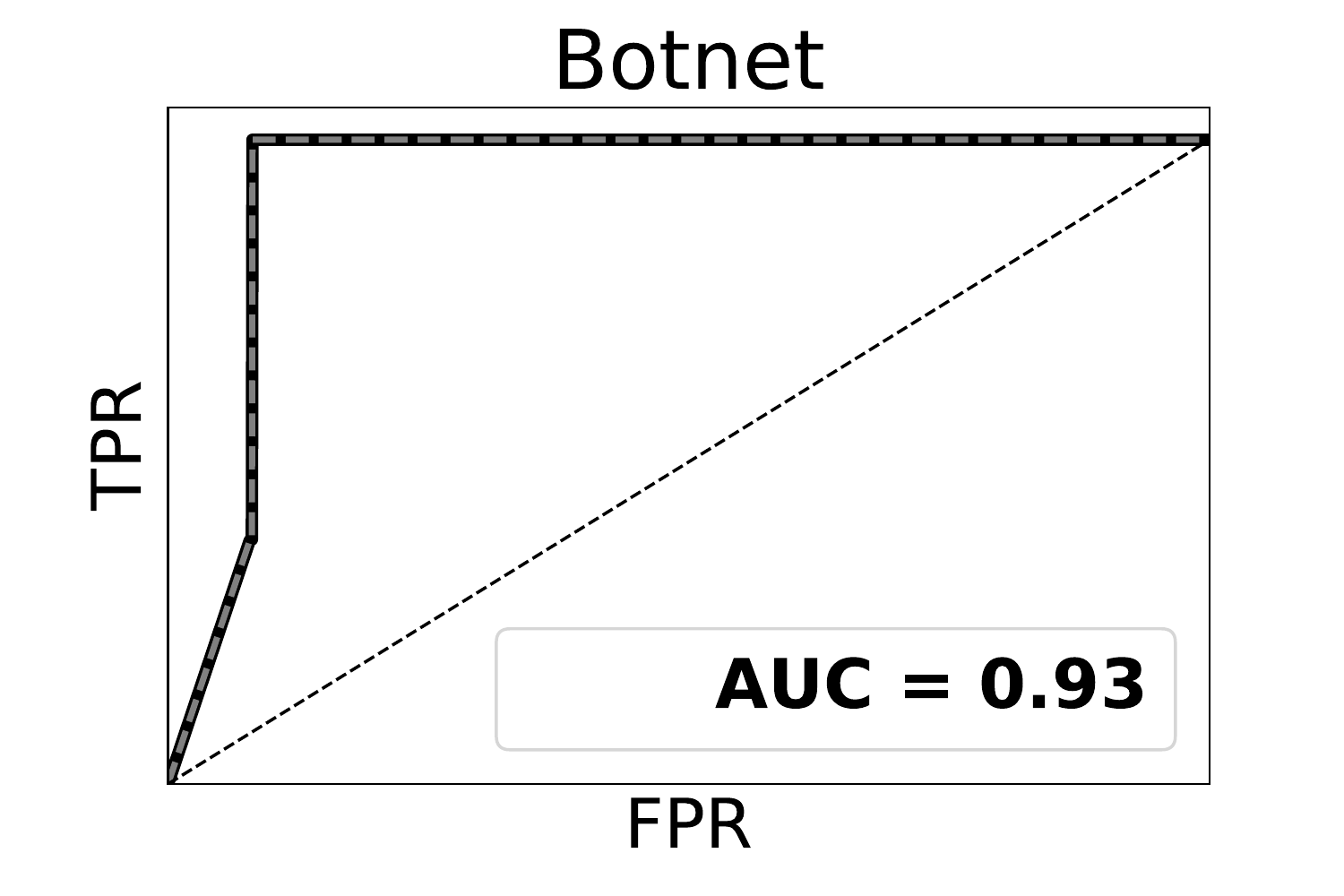} &
      \includegraphics[width=.19\textwidth]{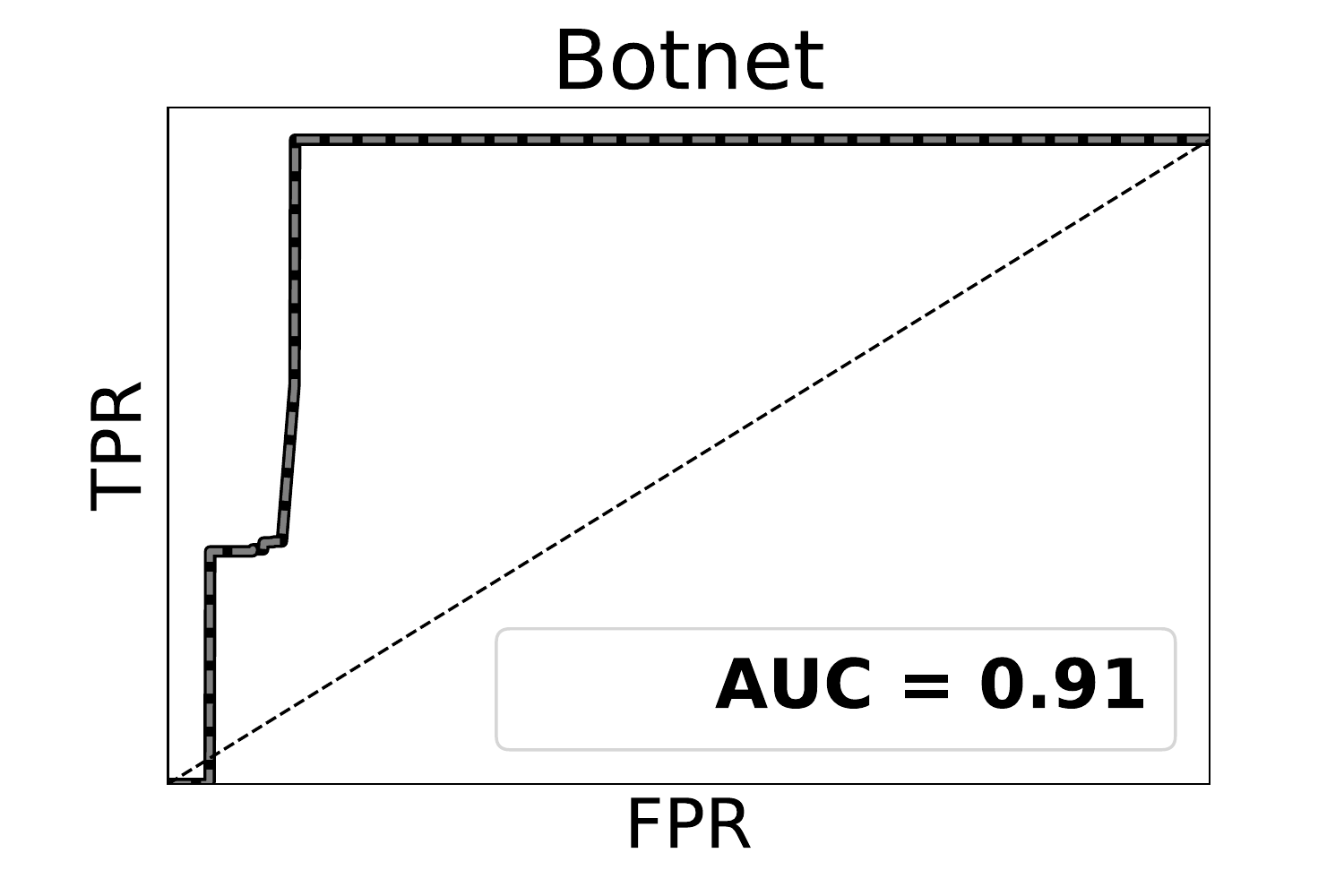} 
      \\
            \includegraphics[width=.19\textwidth]{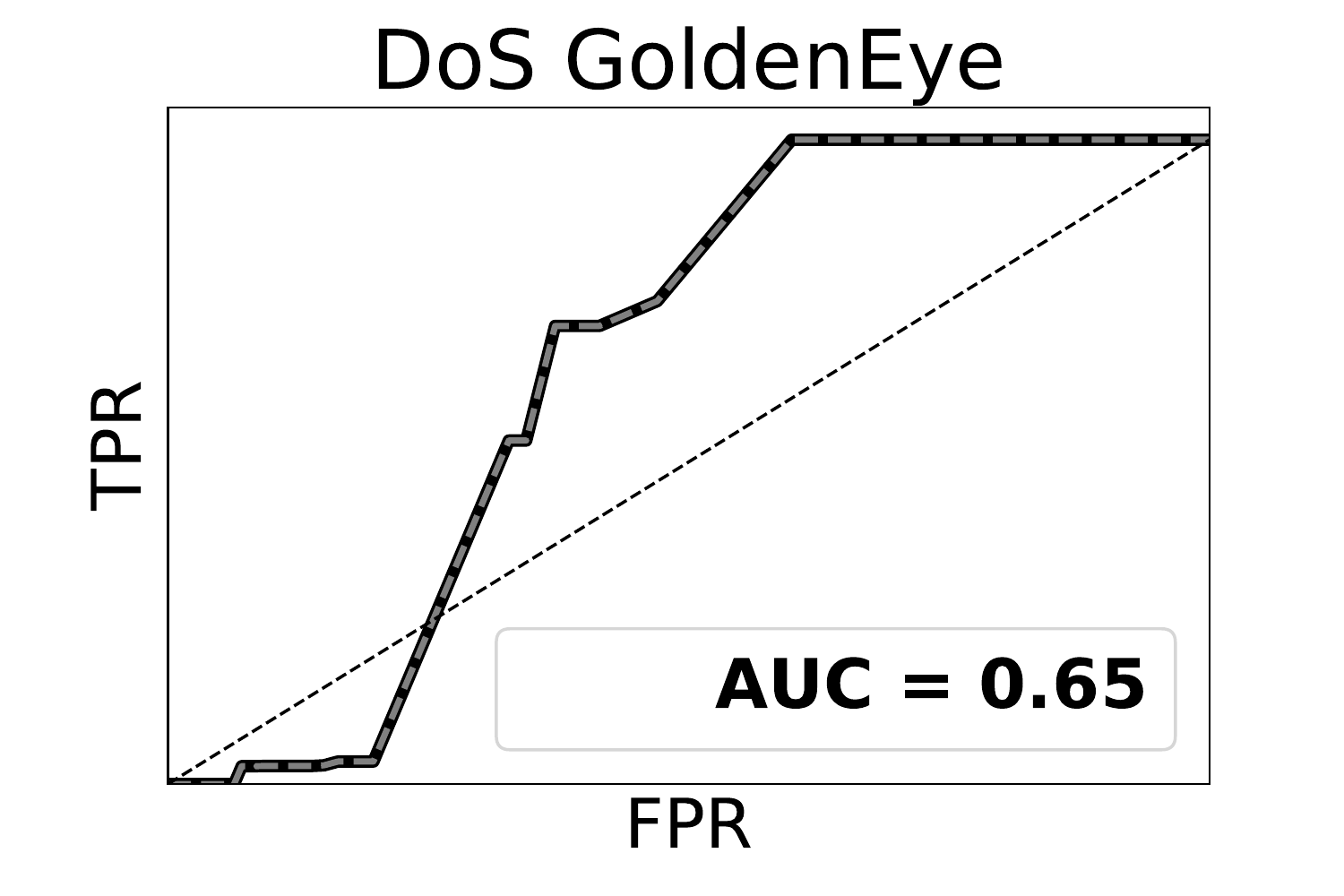} &
      \includegraphics[width=.19\textwidth]{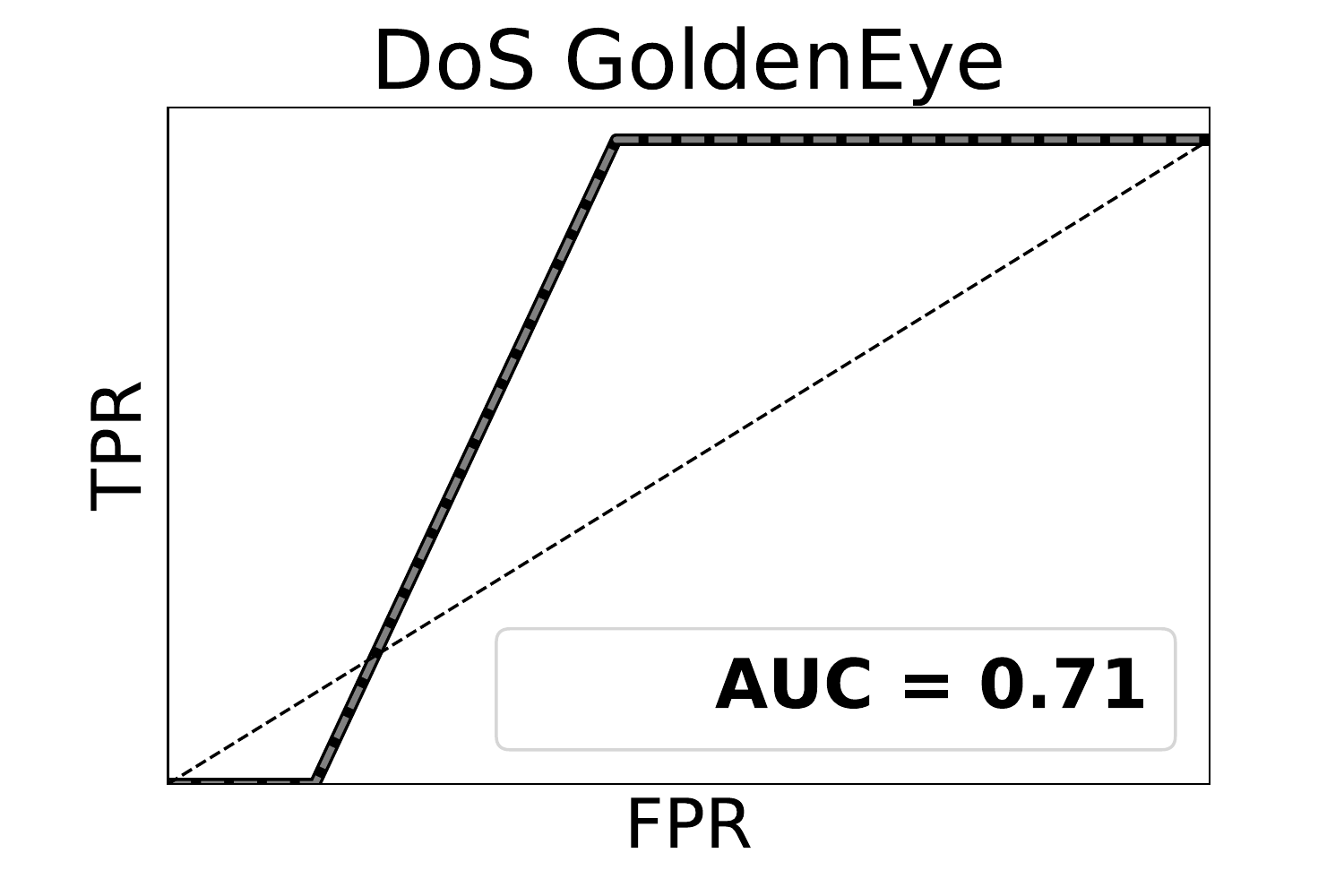} &
      \includegraphics[width=.19\textwidth]{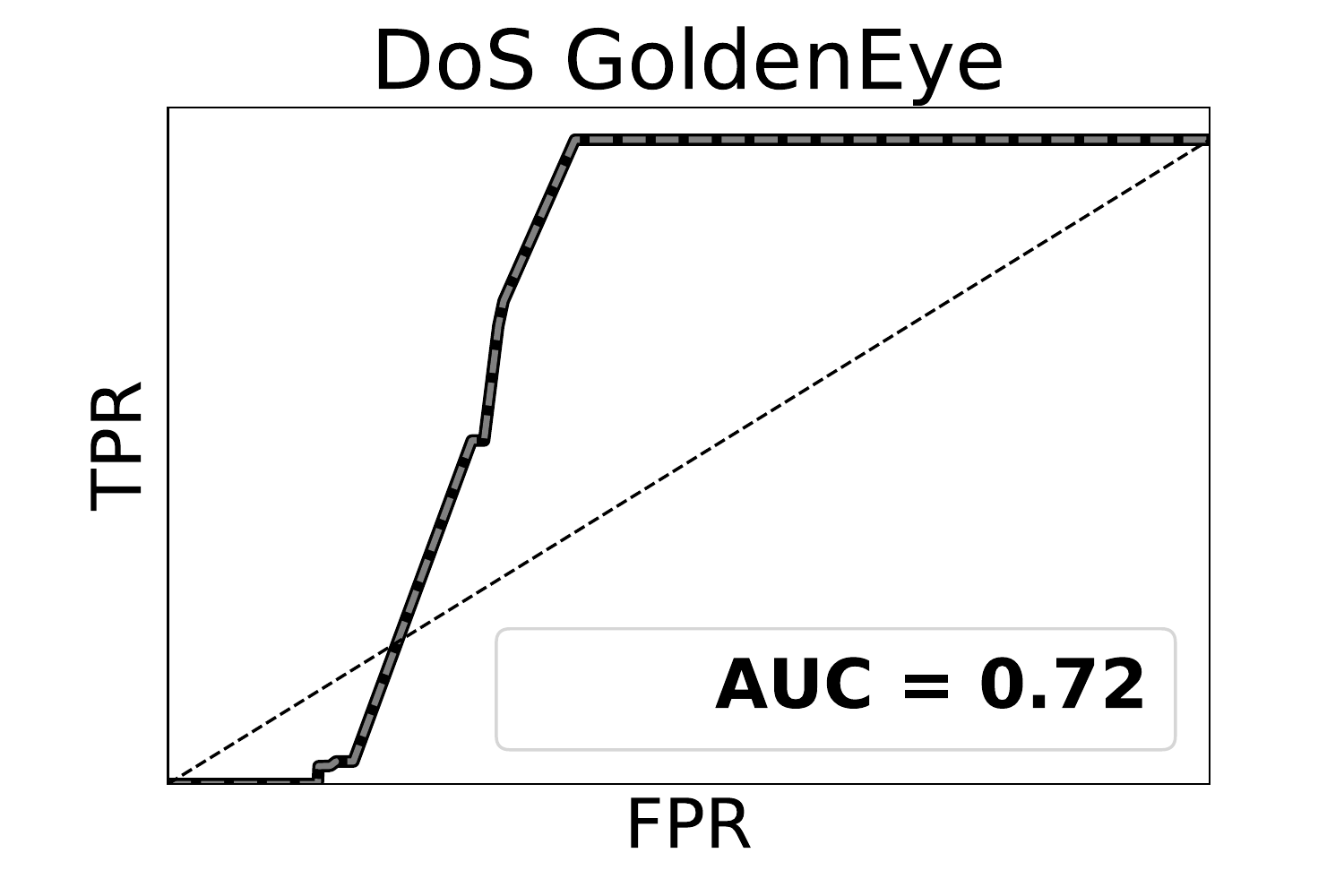} 
      \\
      \includegraphics[width=.19\textwidth]{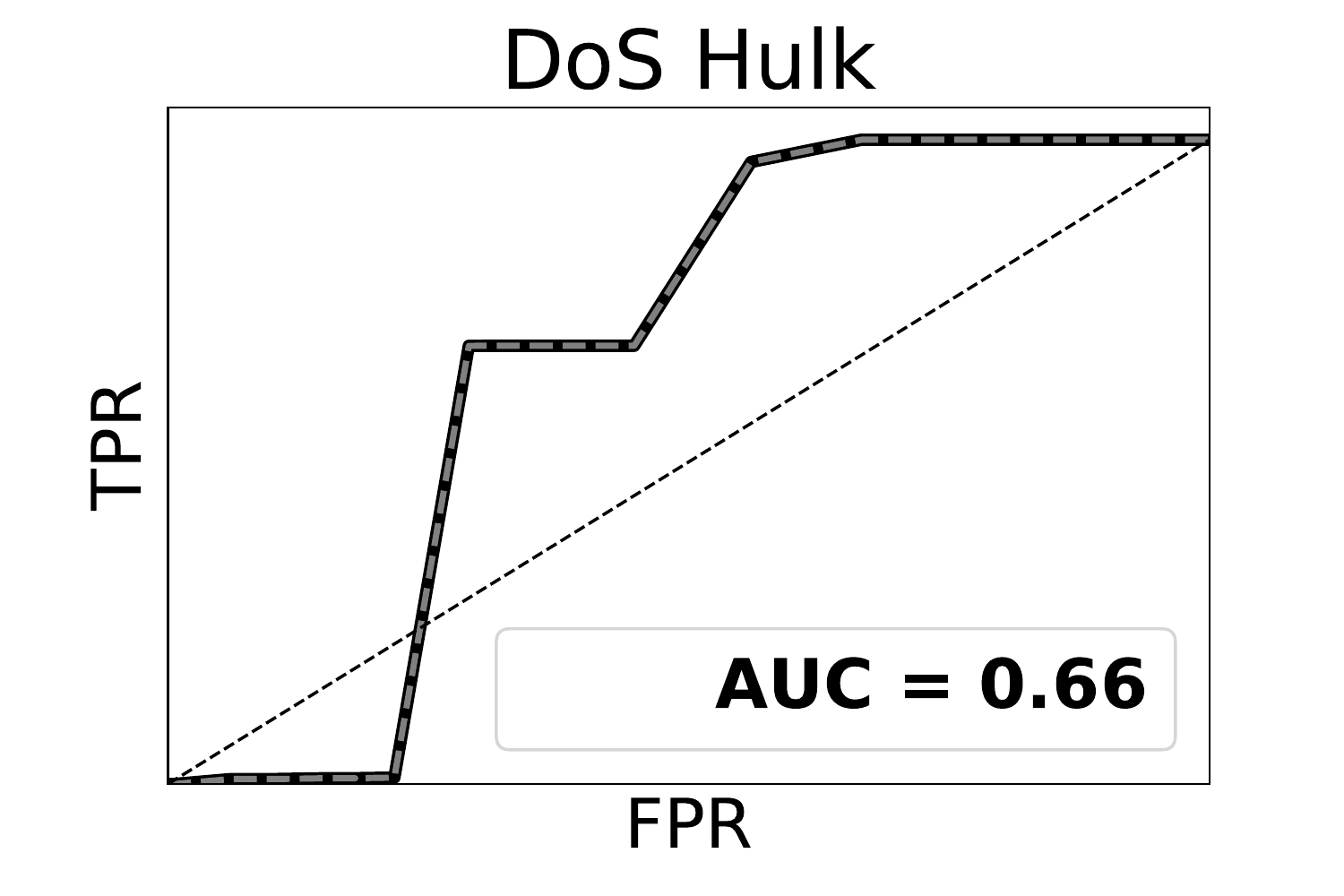} &
      \includegraphics[width=.19\textwidth]{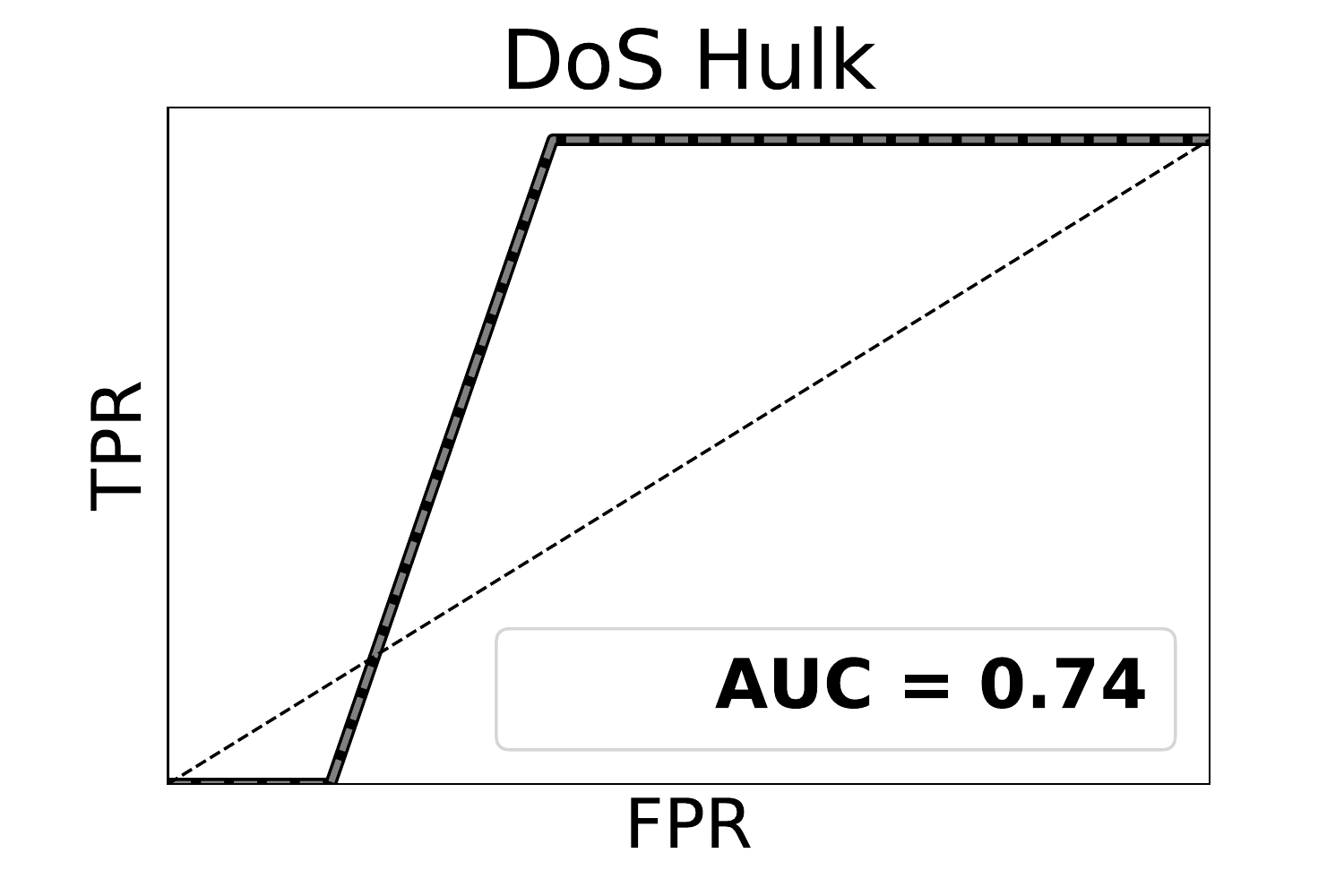} &
      \includegraphics[width=.19\textwidth]{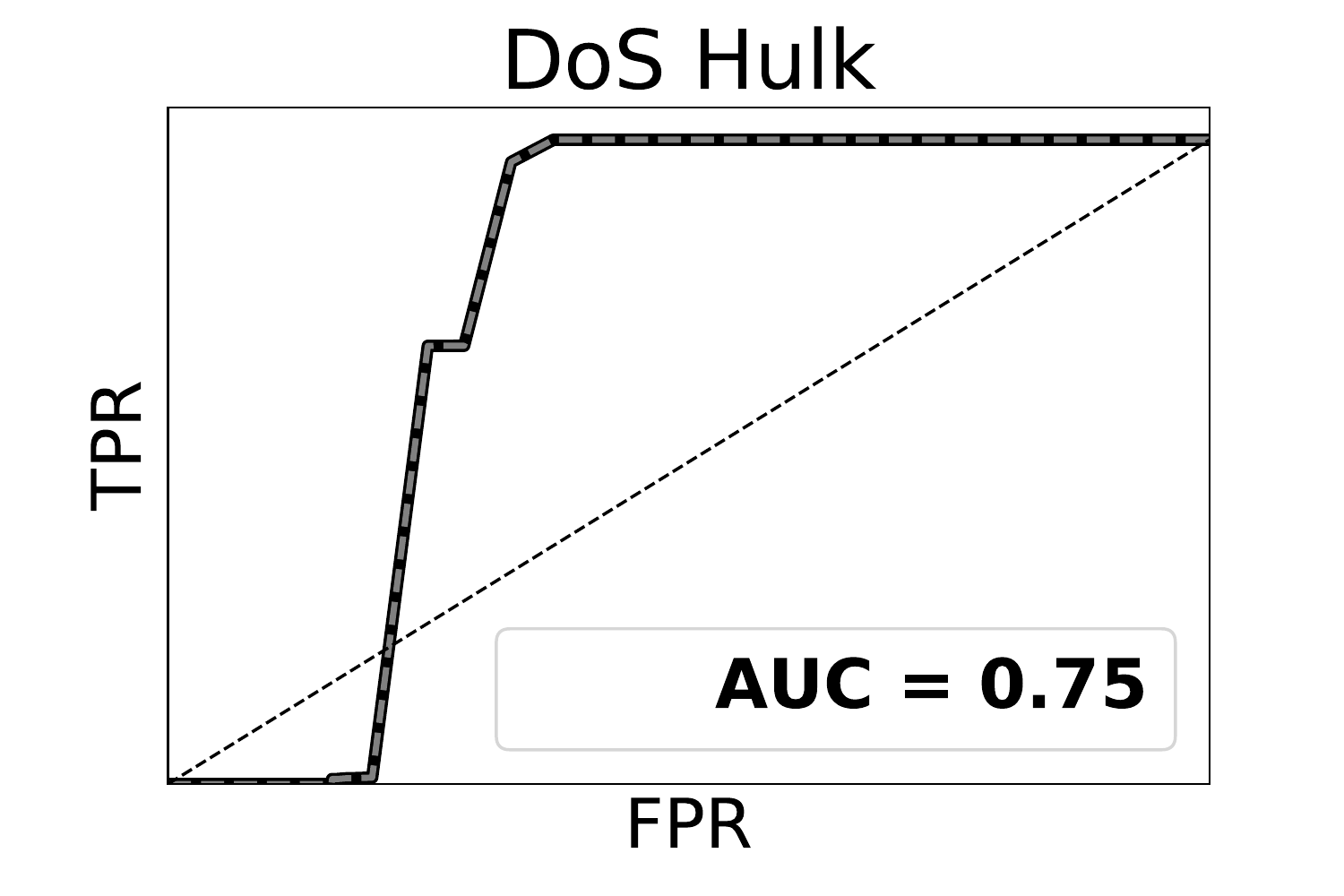} 
      \\
            \includegraphics[width=.19\textwidth]{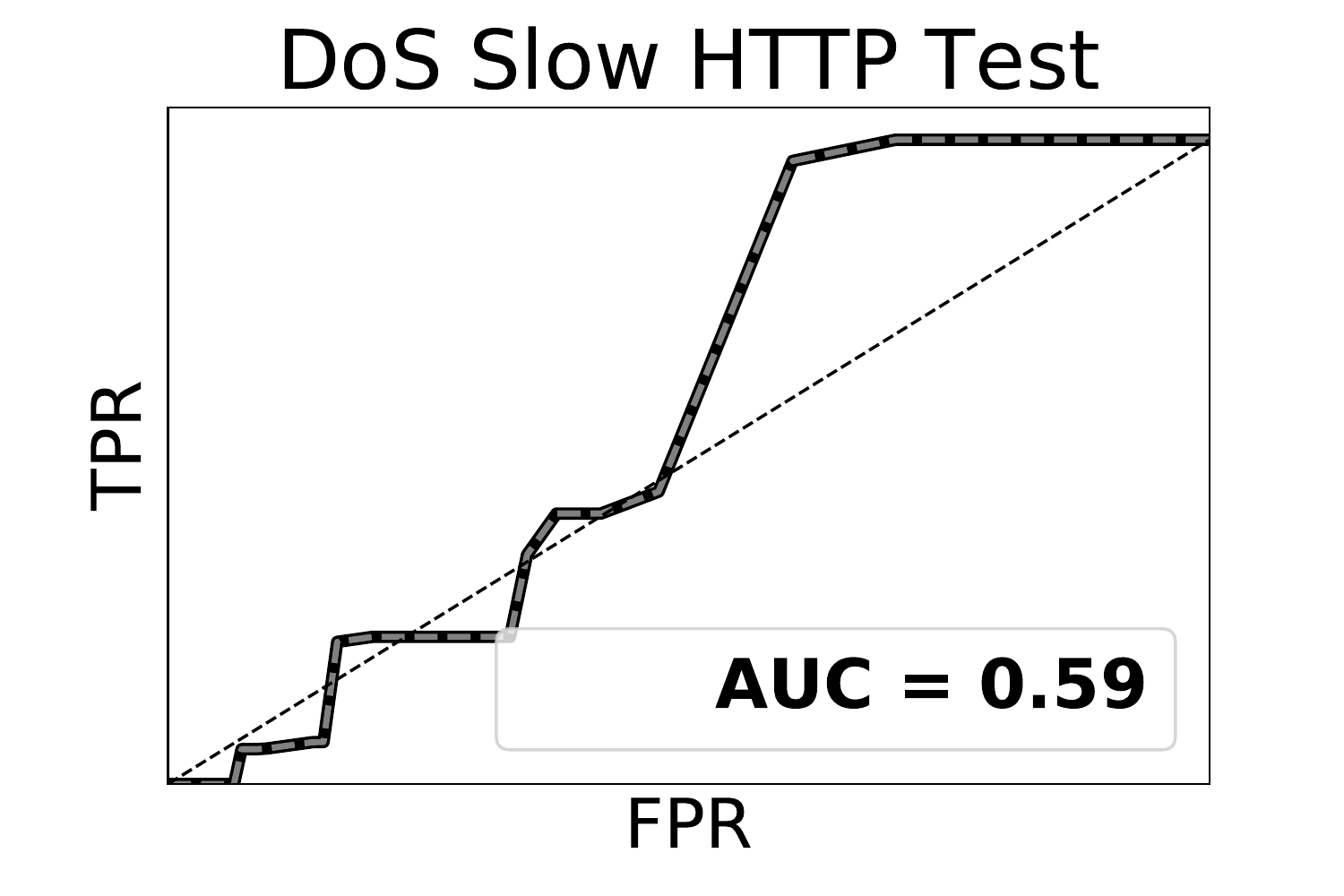} &
      \includegraphics[width=.19\textwidth]{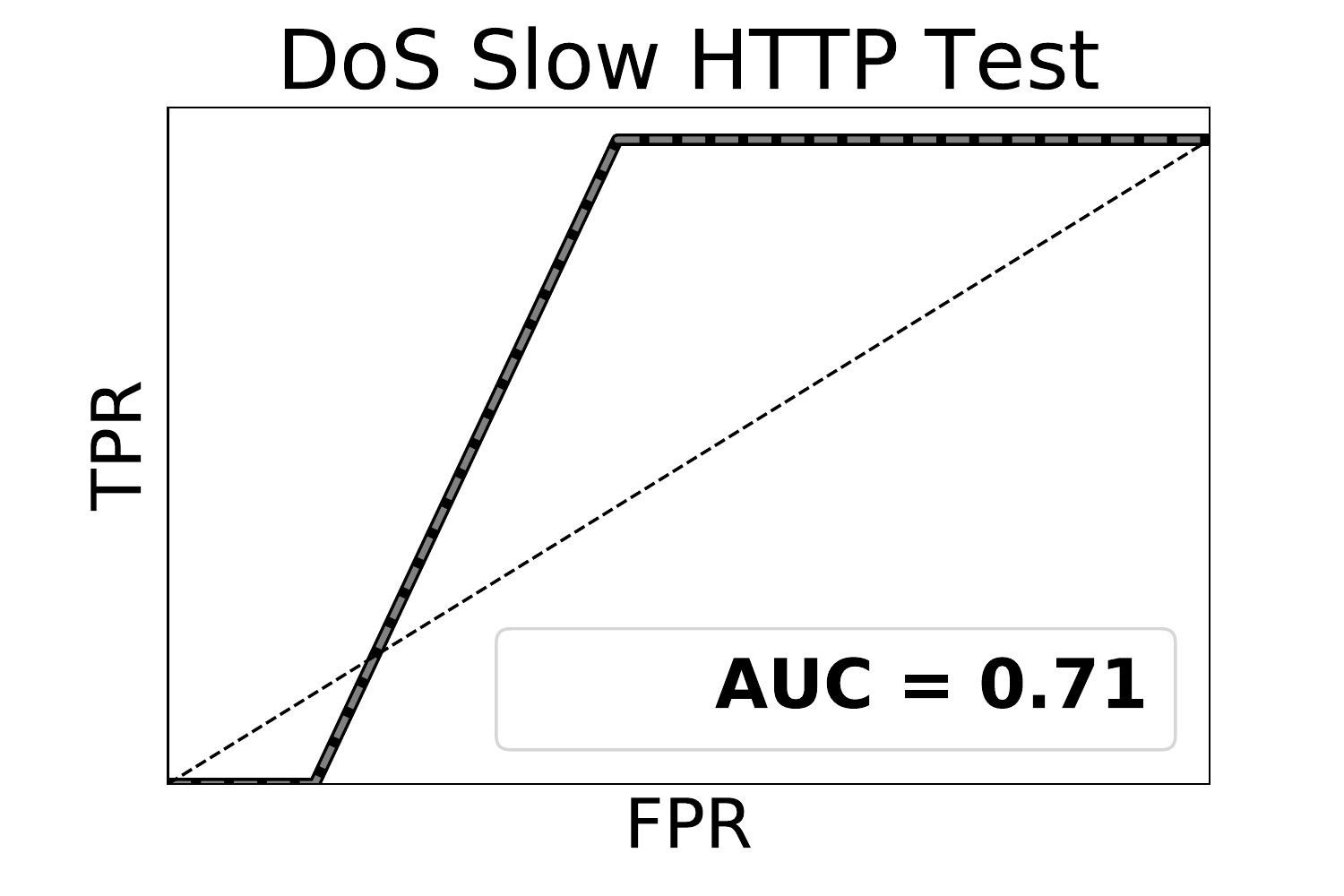} &
      \includegraphics[width=.19\textwidth]{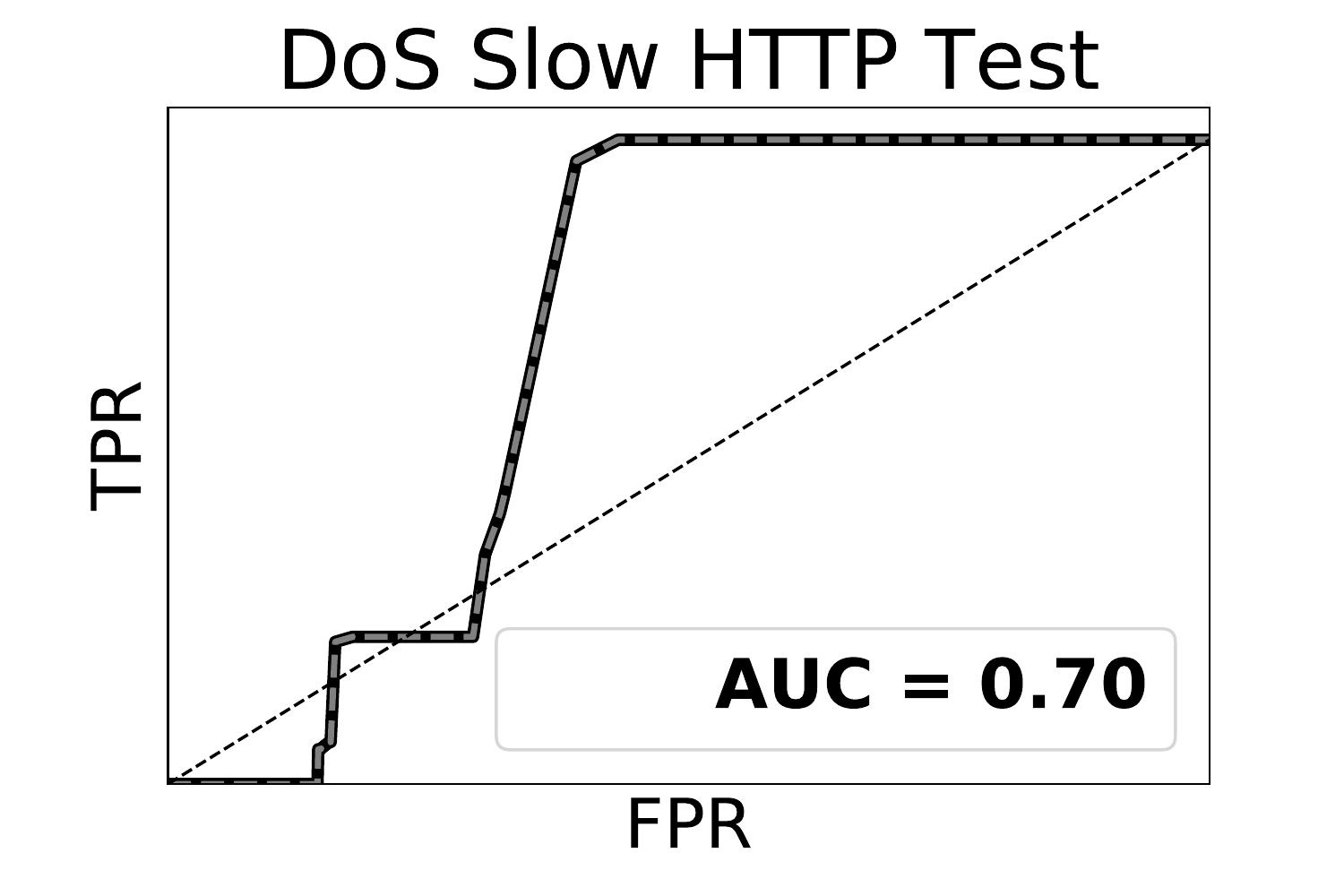} 
      \\
            \includegraphics[width=.19\textwidth]{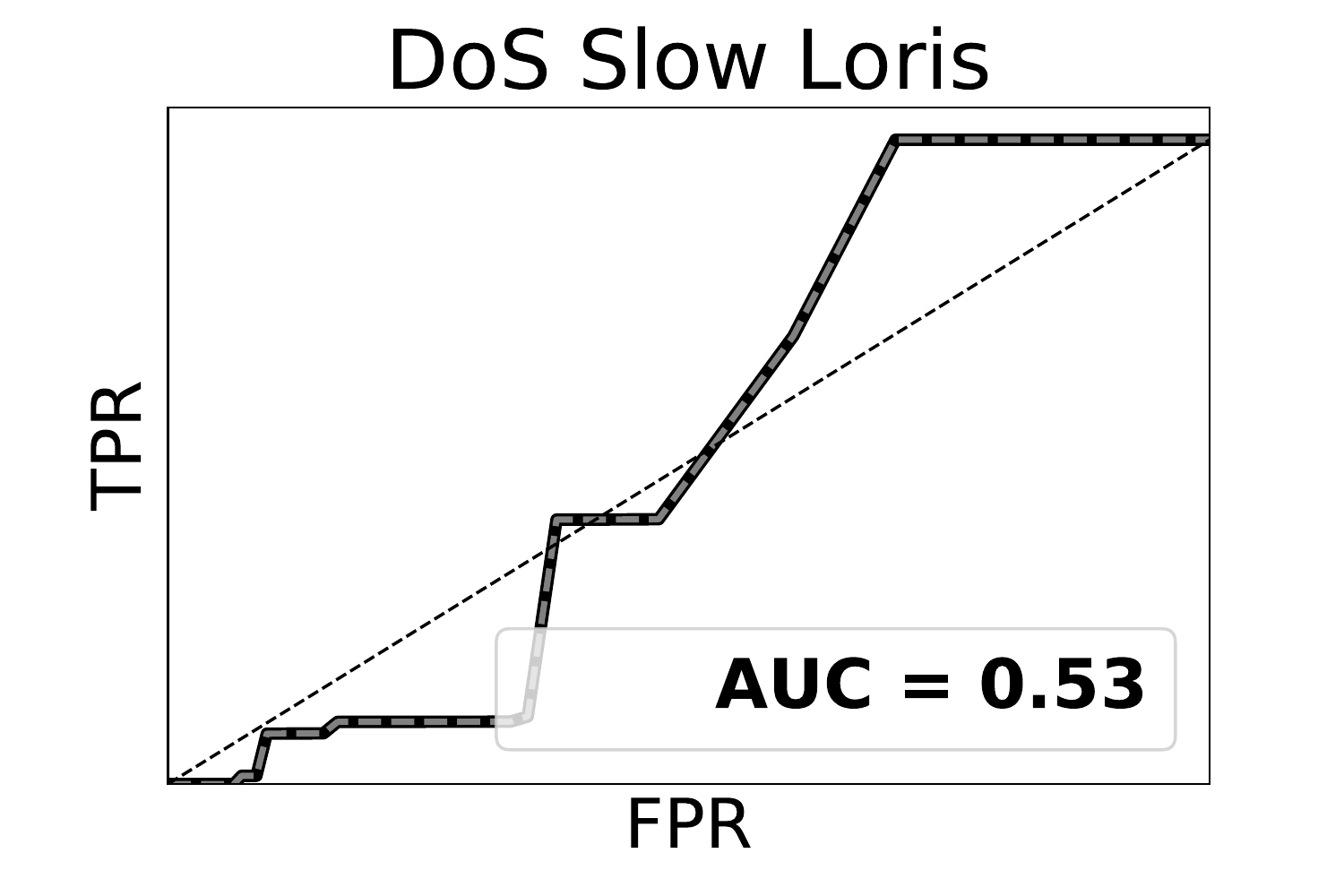} &
      \includegraphics[width=.19\textwidth]{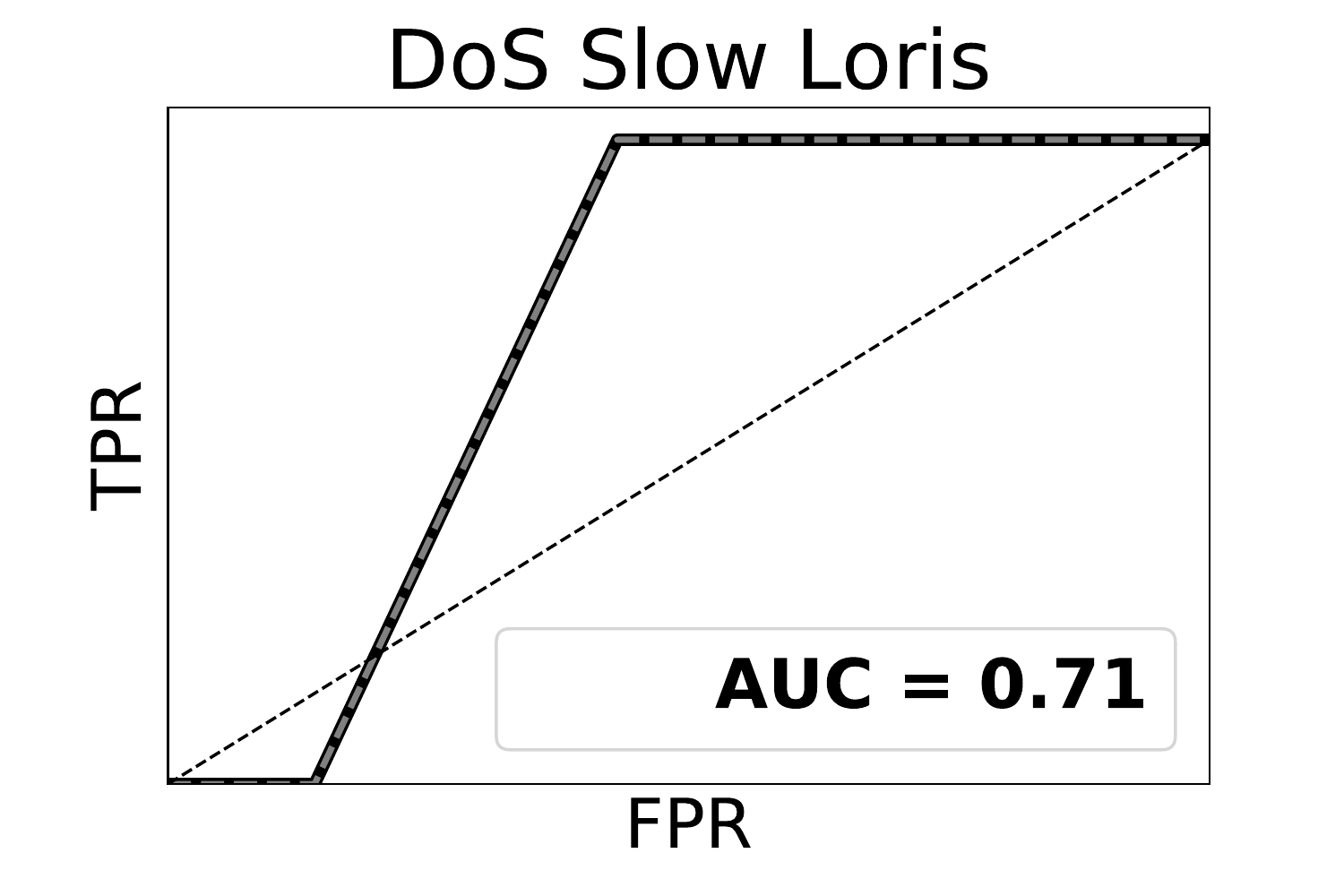} &
      \includegraphics[width=.19\textwidth]{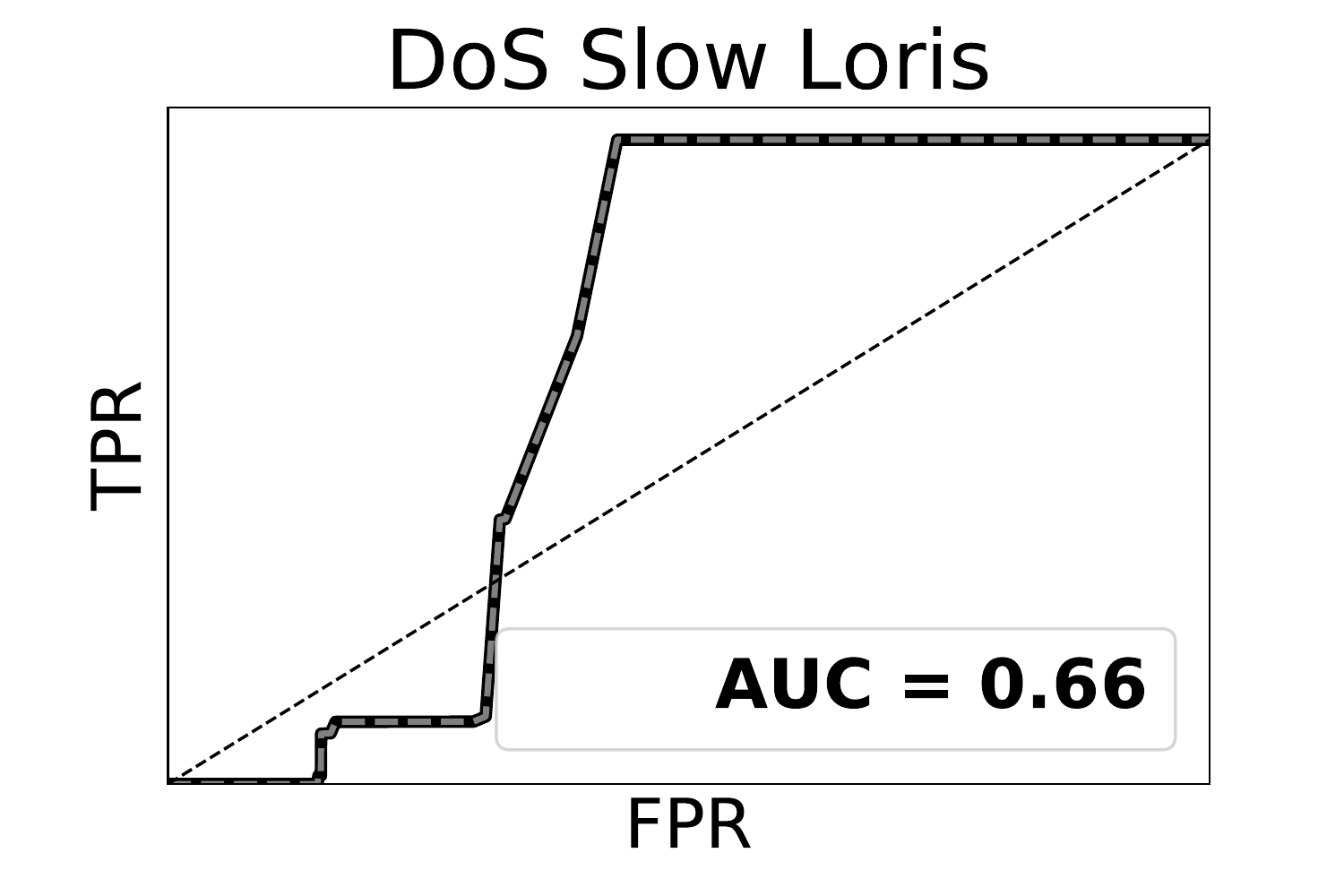} 
      \\
      \includegraphics[width=.19\textwidth]{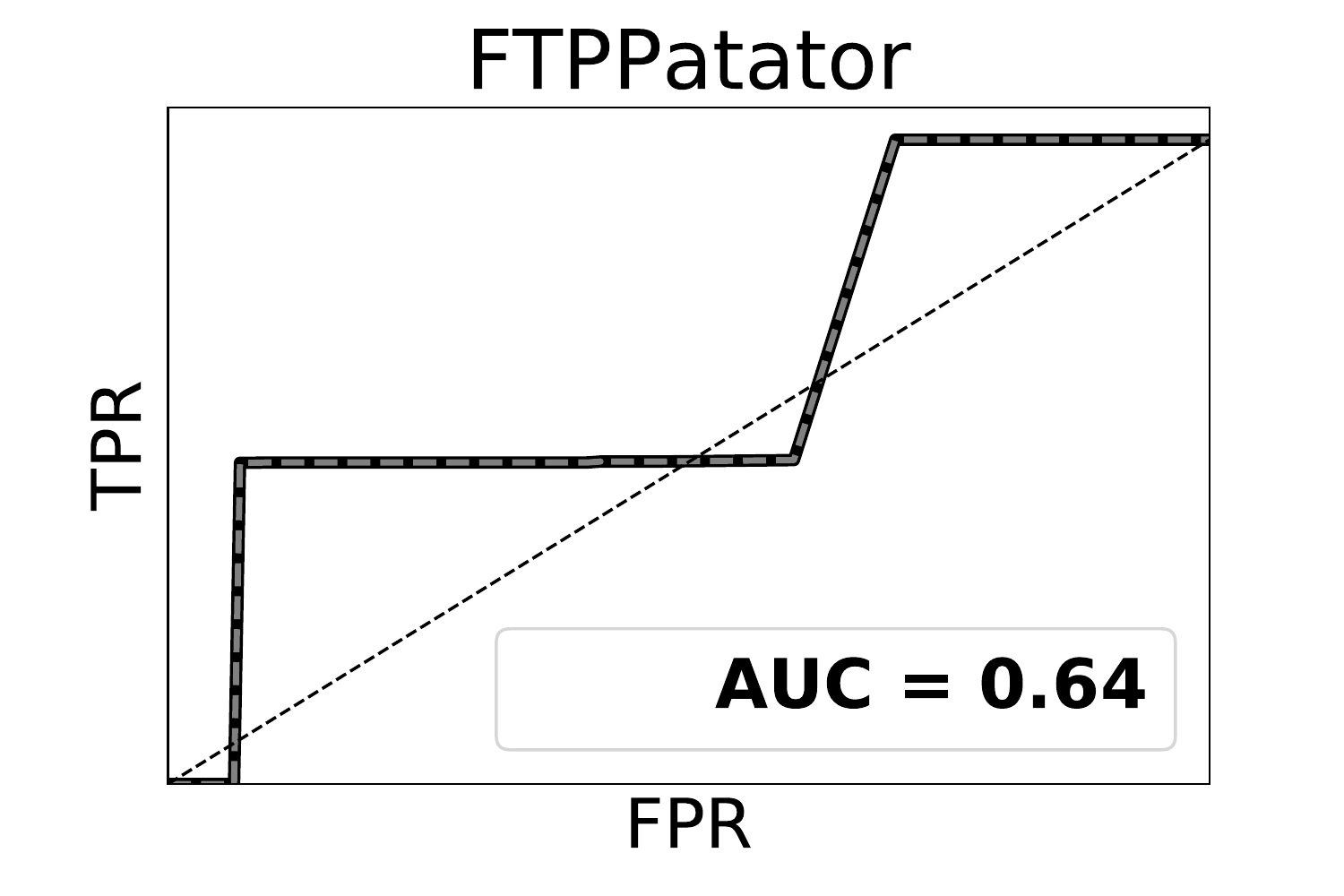} &
      \includegraphics[width=.19\textwidth]{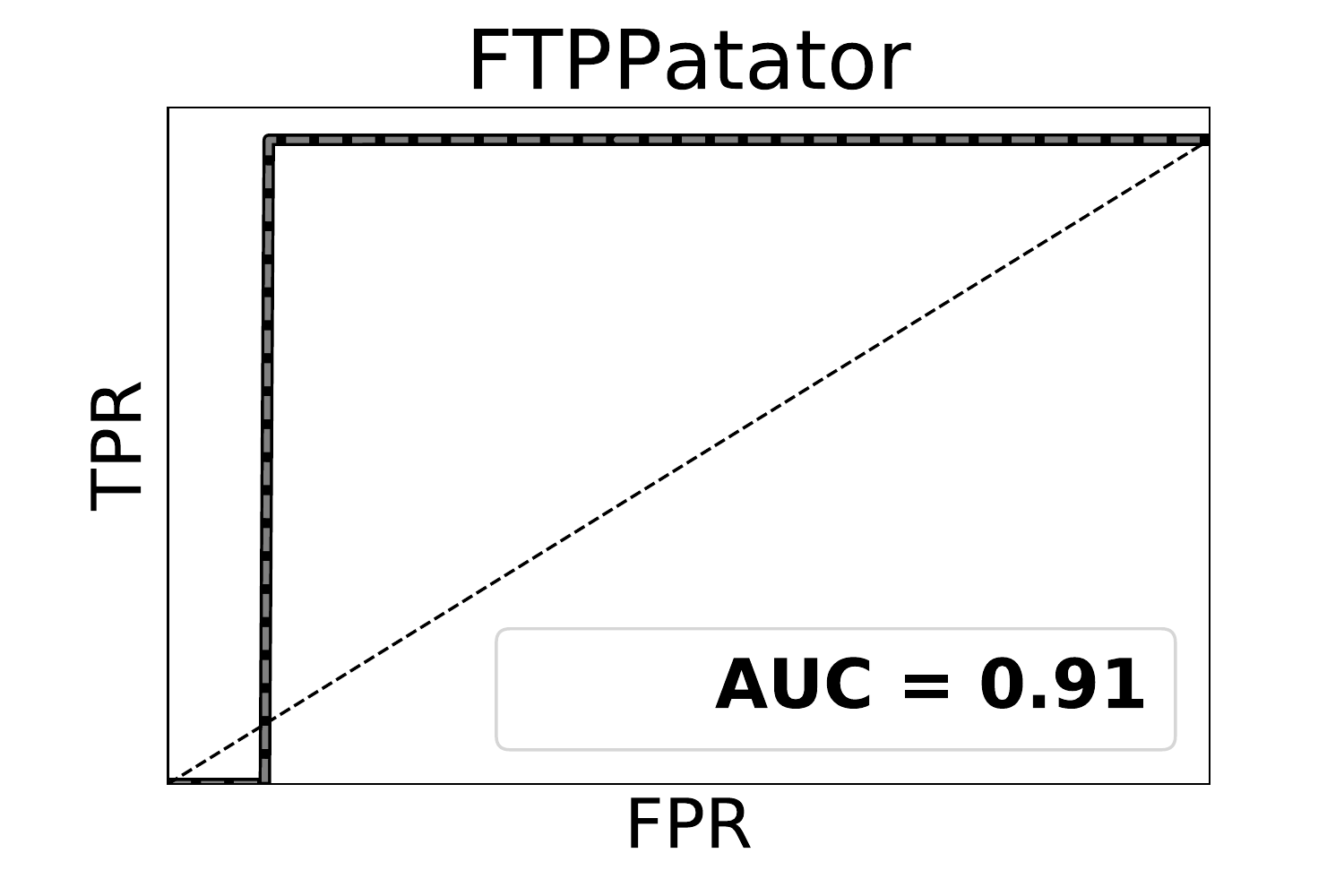} &
      \includegraphics[width=.19\textwidth]{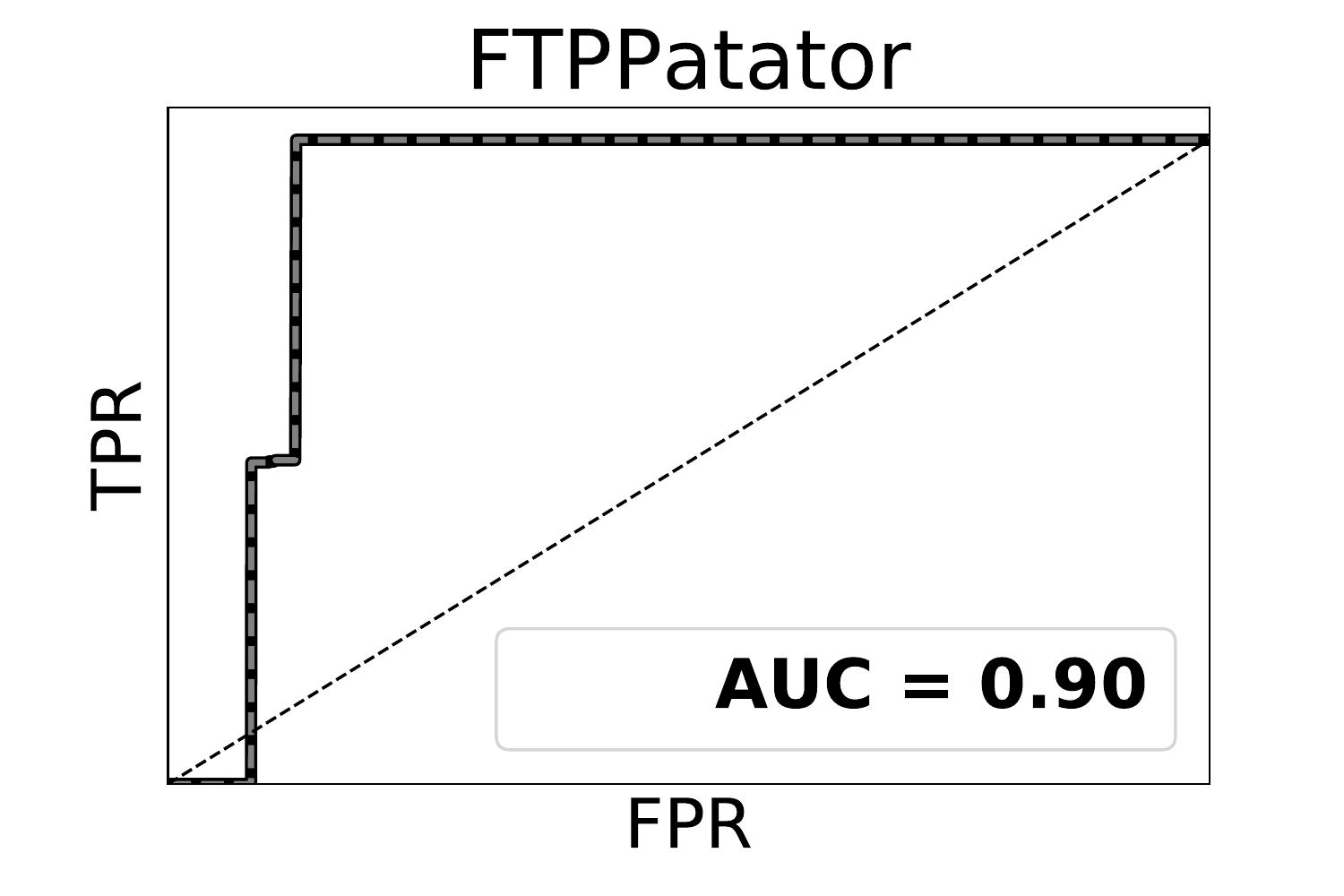} 
      \\
  \end{tabular}
\end{table*}      
      
 \begin{table*}
  [ht] \caption{Frequency Model B} \label{tab:frequencyb}
  \begin{tabular}{ccc} 
  \hline 
  Protobyte Sequences & Port Sequences & PC1 \\
      \hline      
      \includegraphics[width=.19\textwidth]{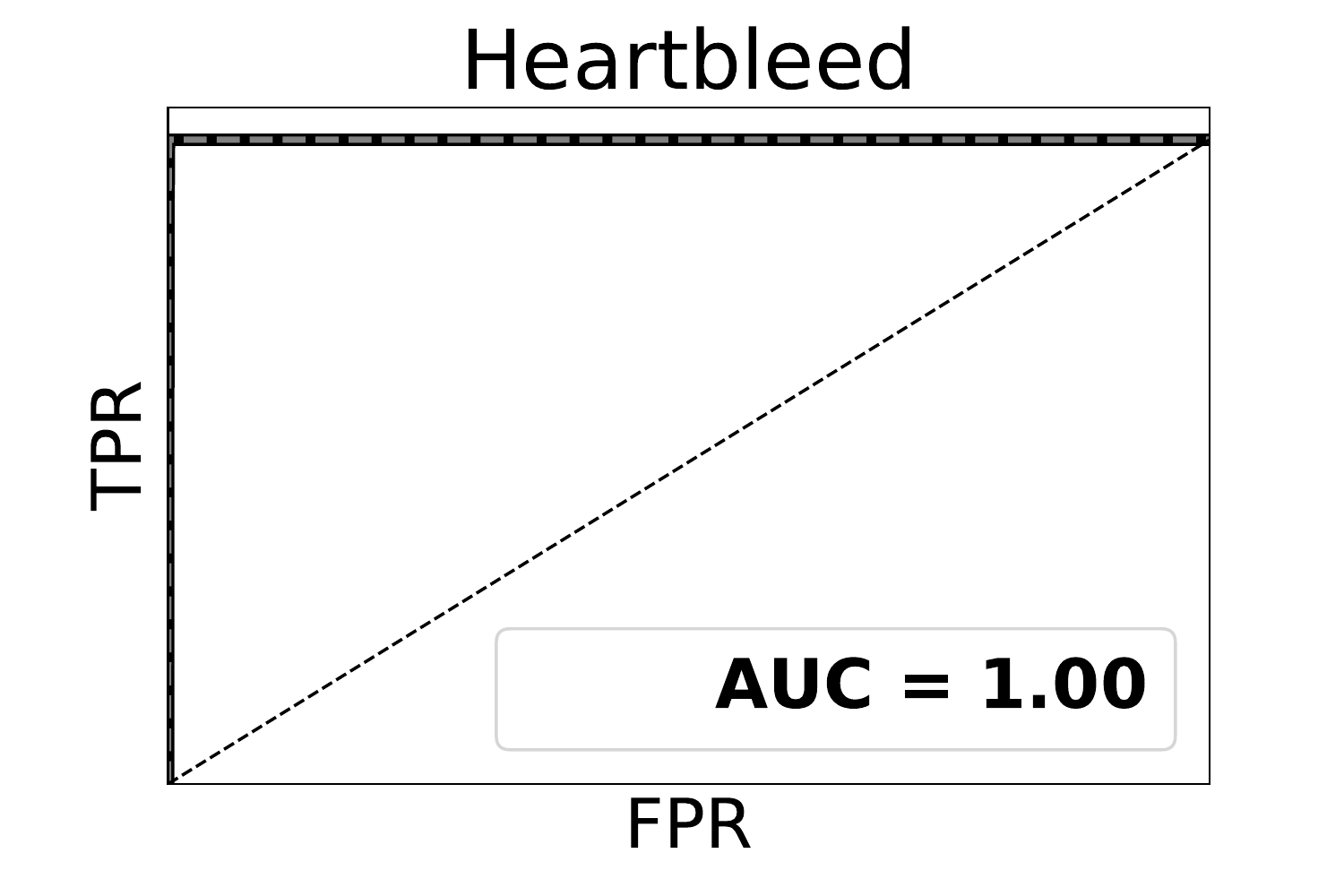} &
      \includegraphics[width=.19\textwidth]{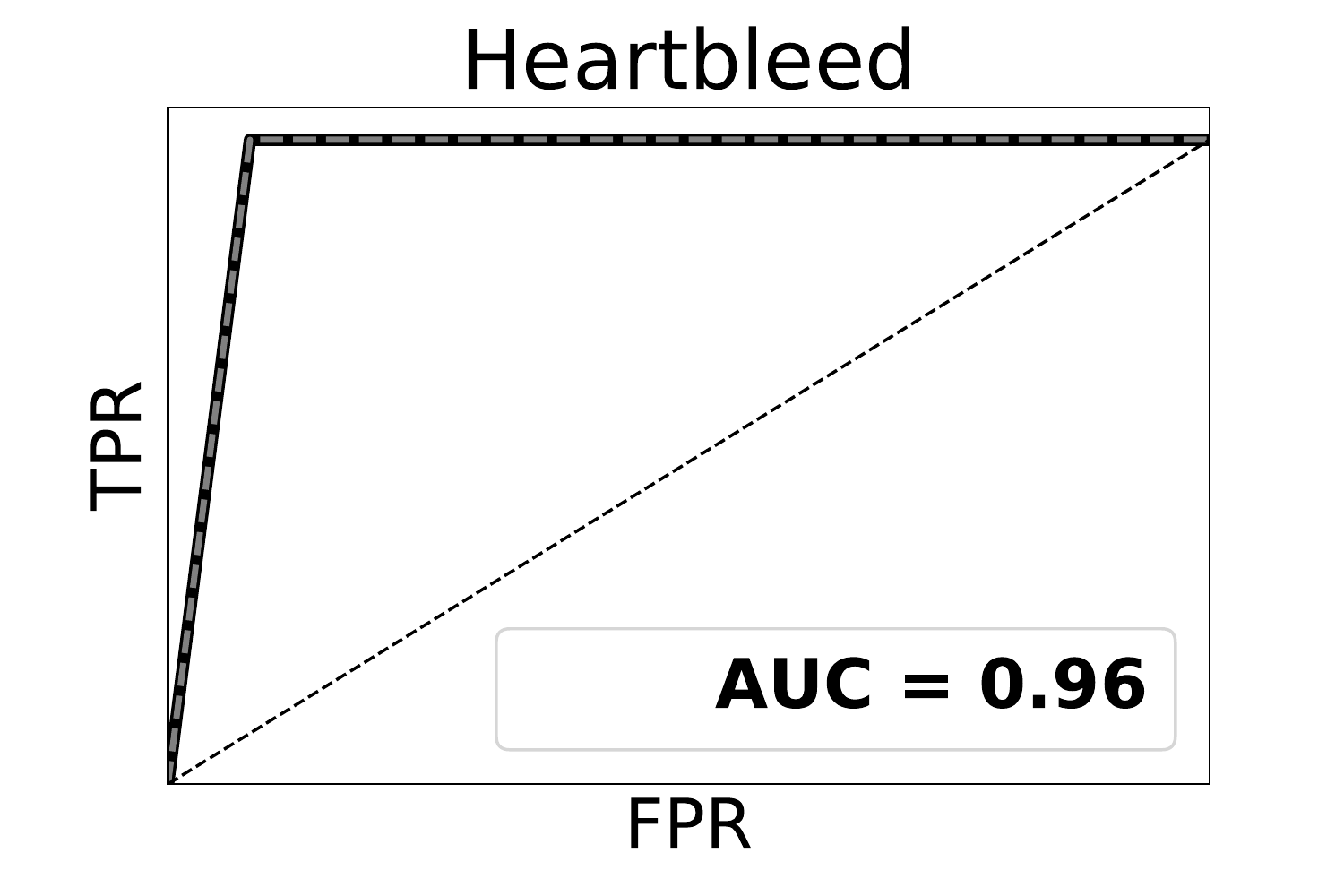} &
      \includegraphics[width=.19\textwidth]{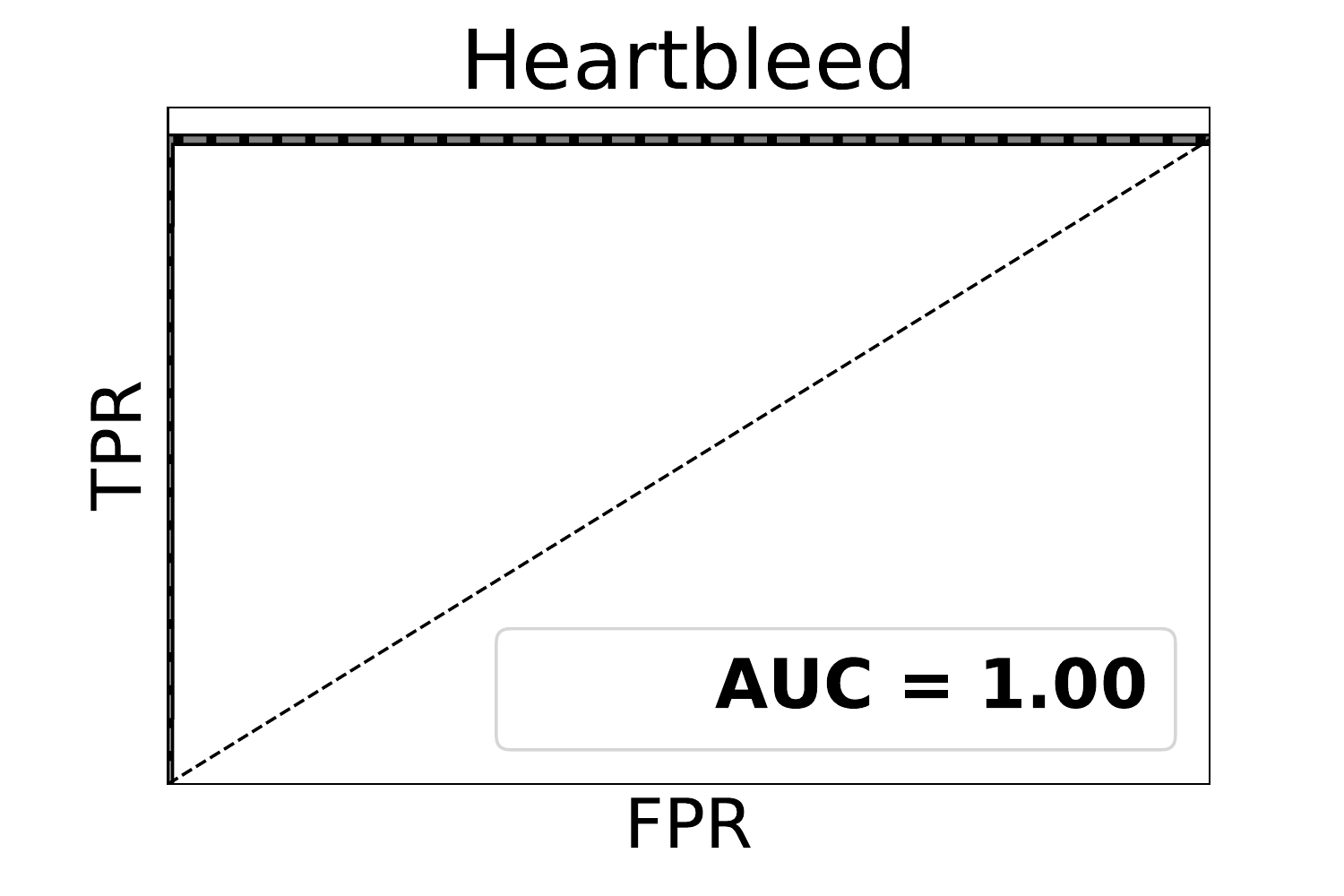}
      \\
      \includegraphics[width=.19\textwidth]{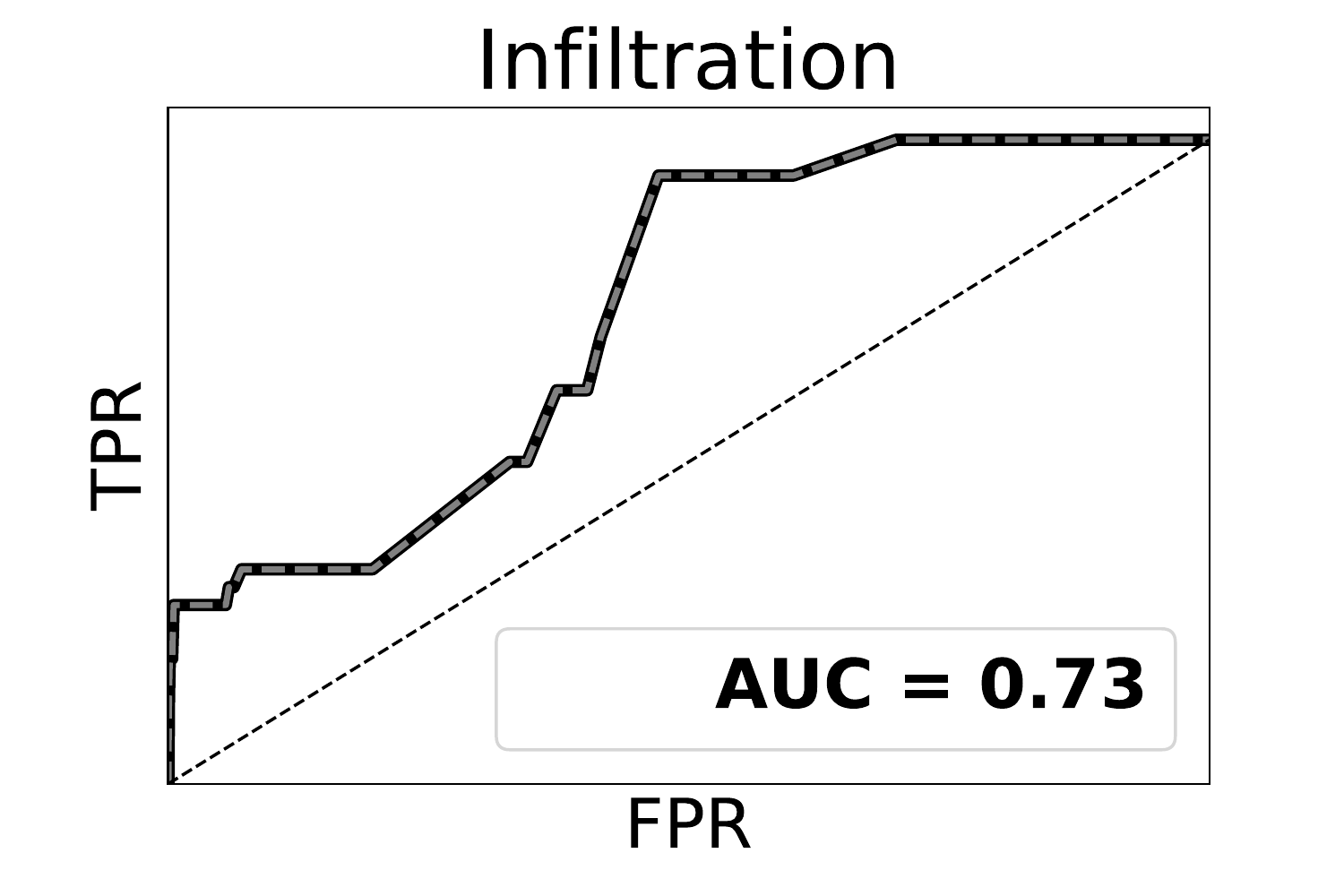} &
      \includegraphics[width=.19\textwidth]{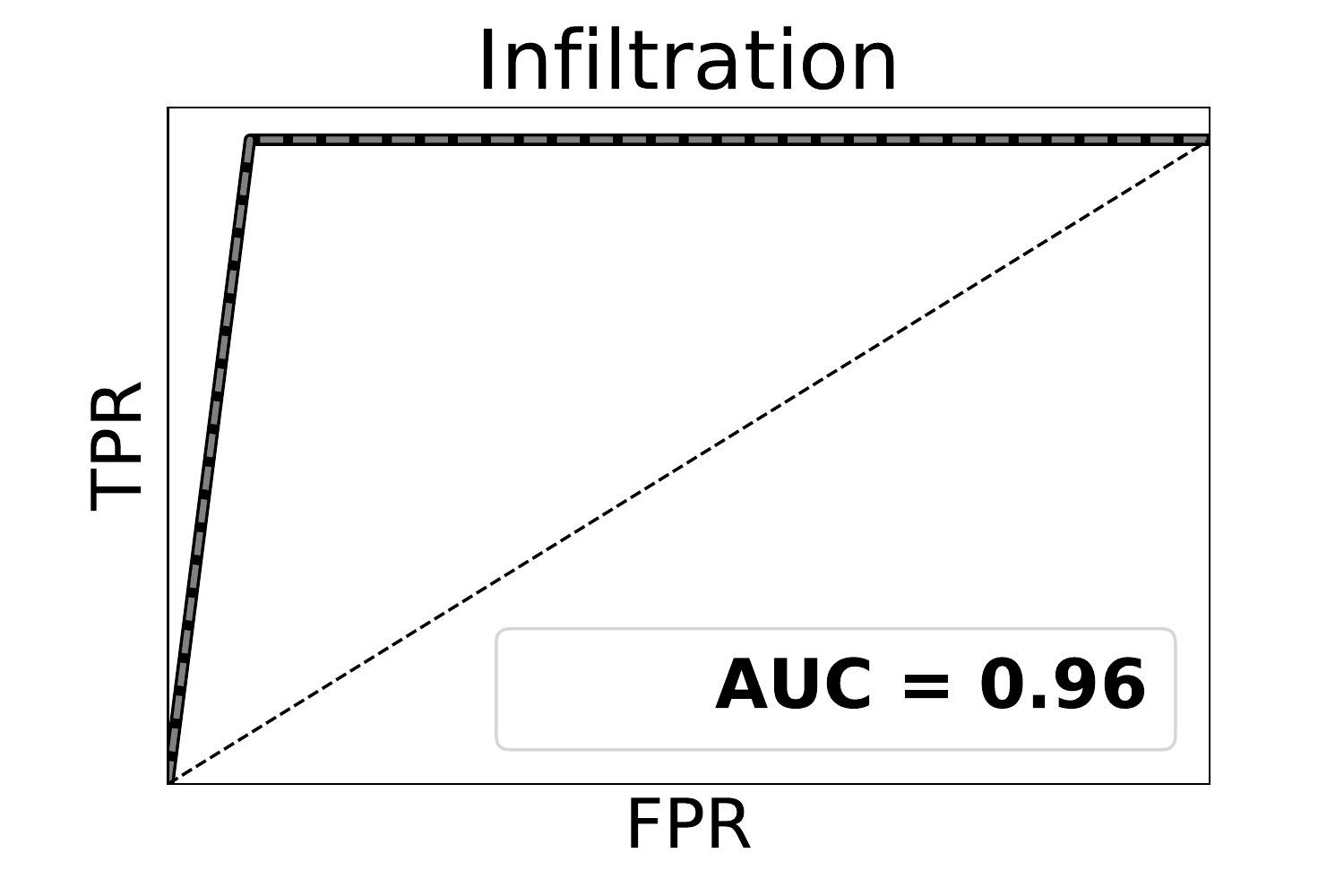} &
      \includegraphics[width=.19\textwidth]{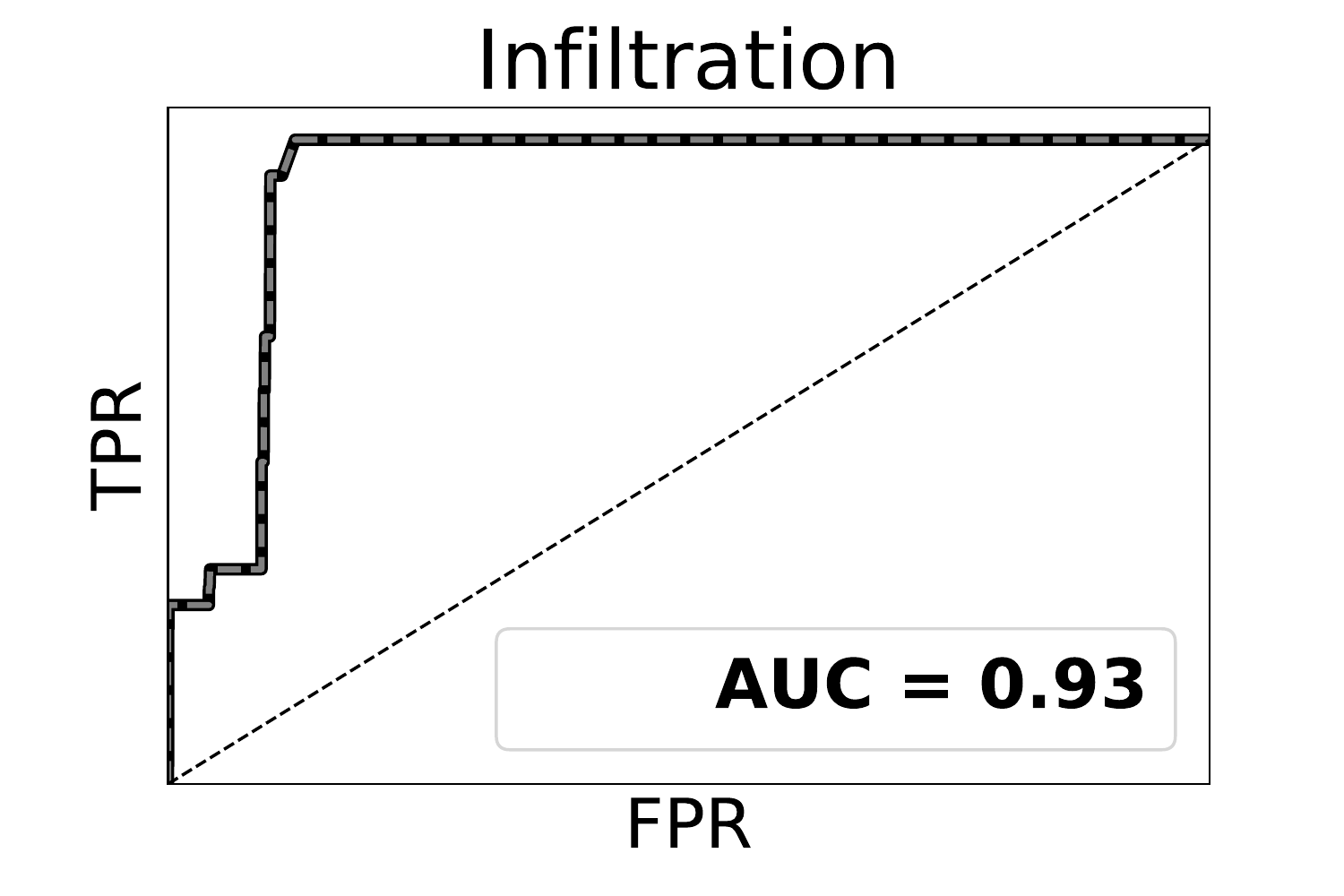} 
      \\
      \includegraphics[width=.19\textwidth]{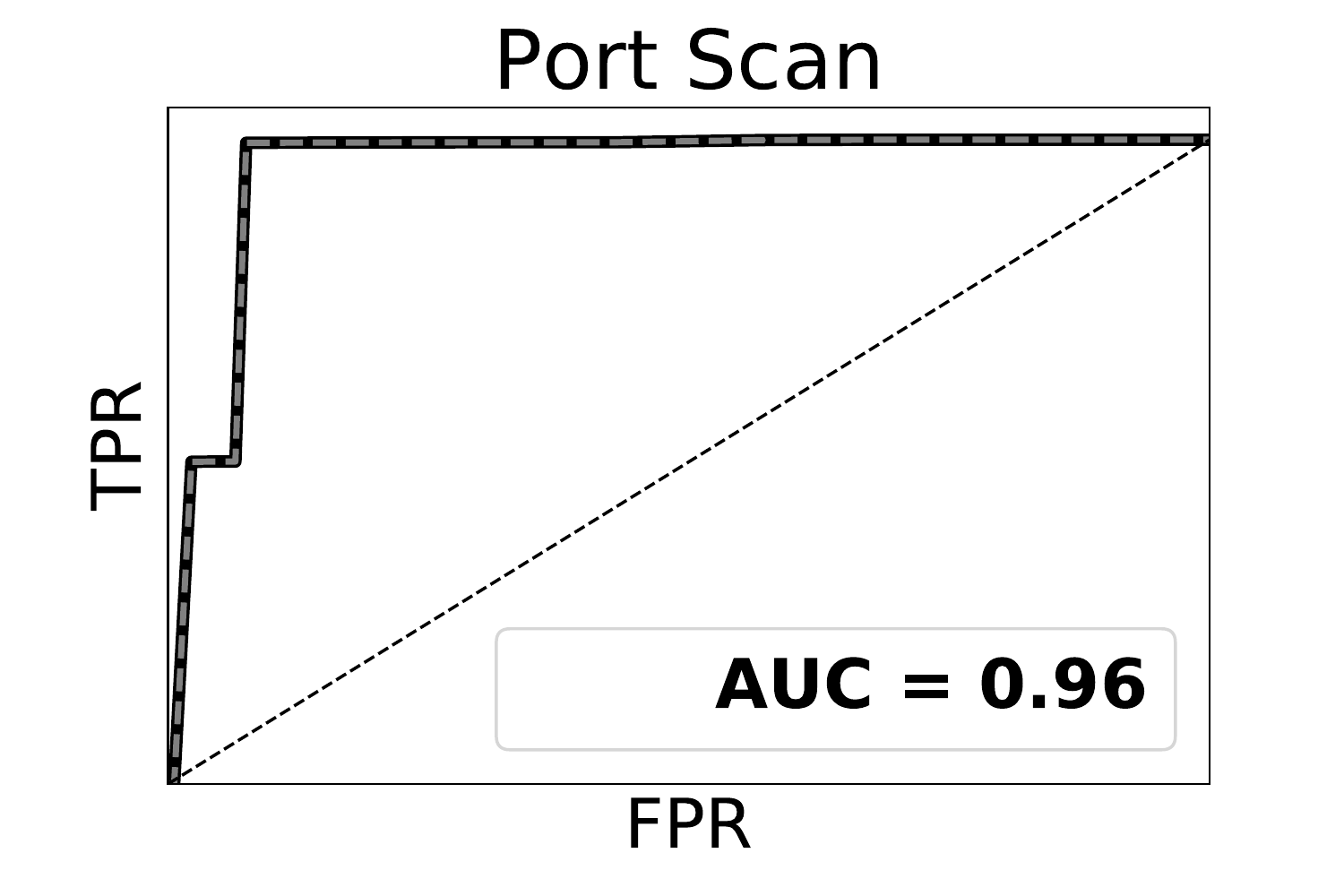} &
      \includegraphics[width=.19\textwidth]{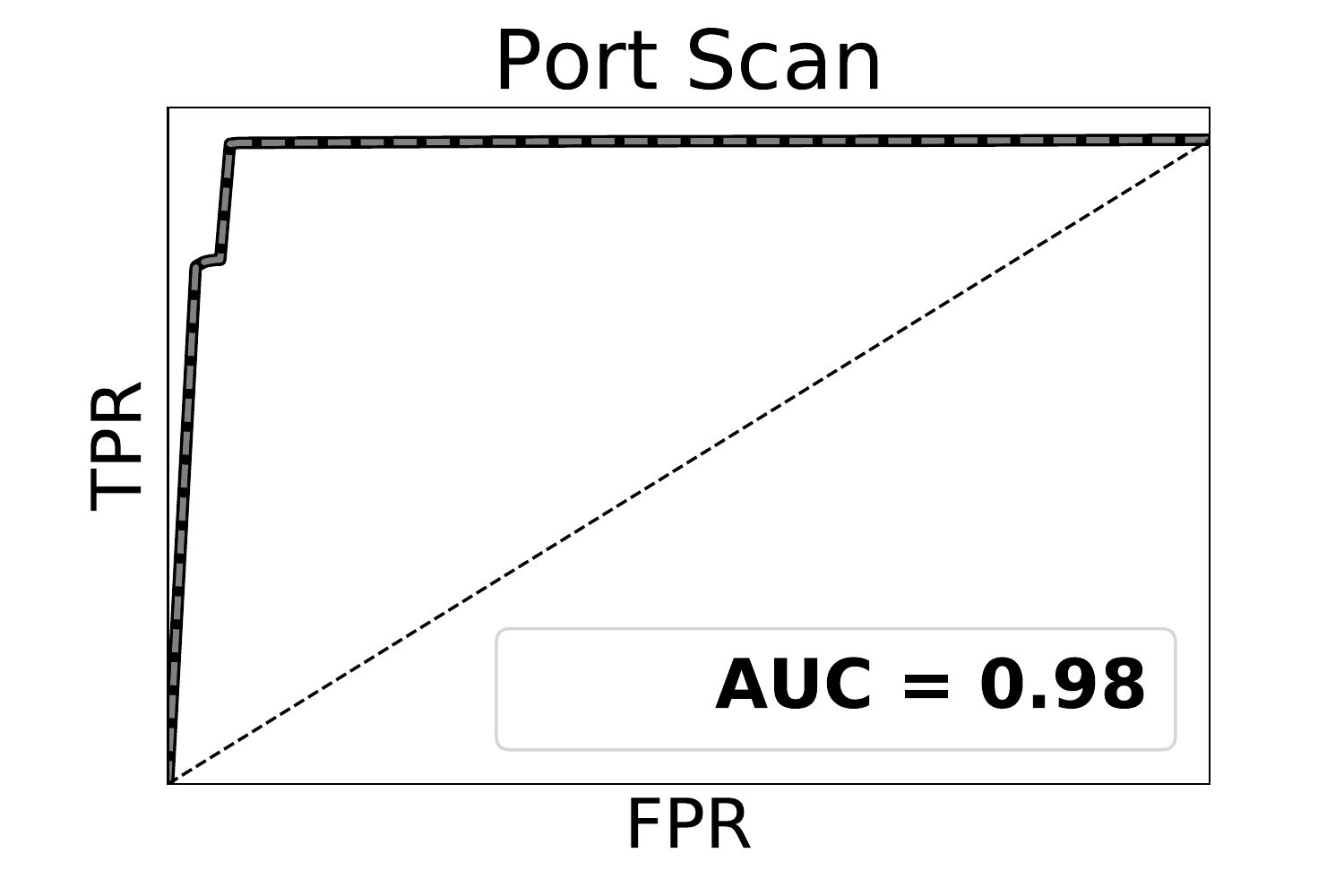} &
      \includegraphics[width=.19\textwidth]{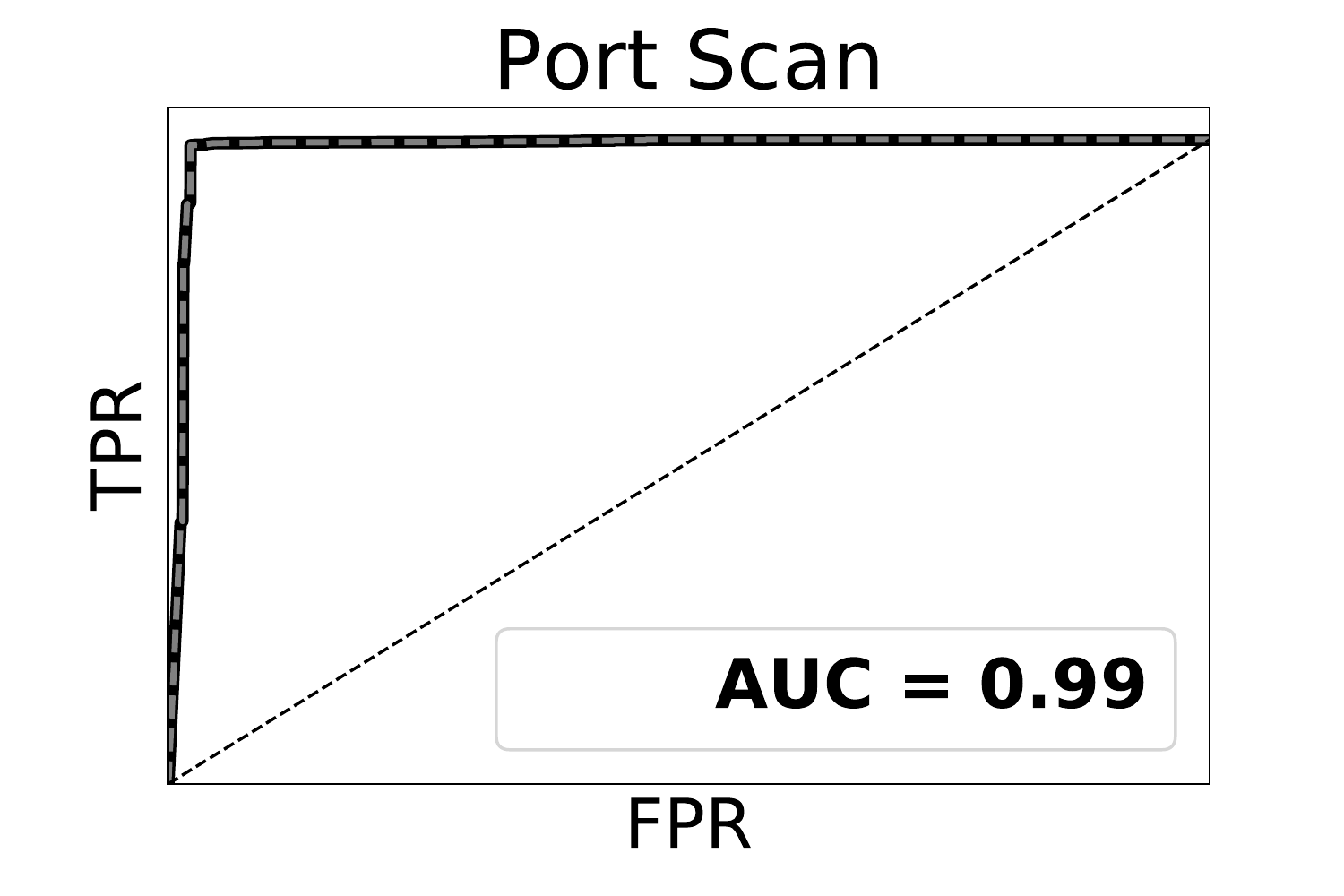} 
      \\
      \includegraphics[width=.19\textwidth]{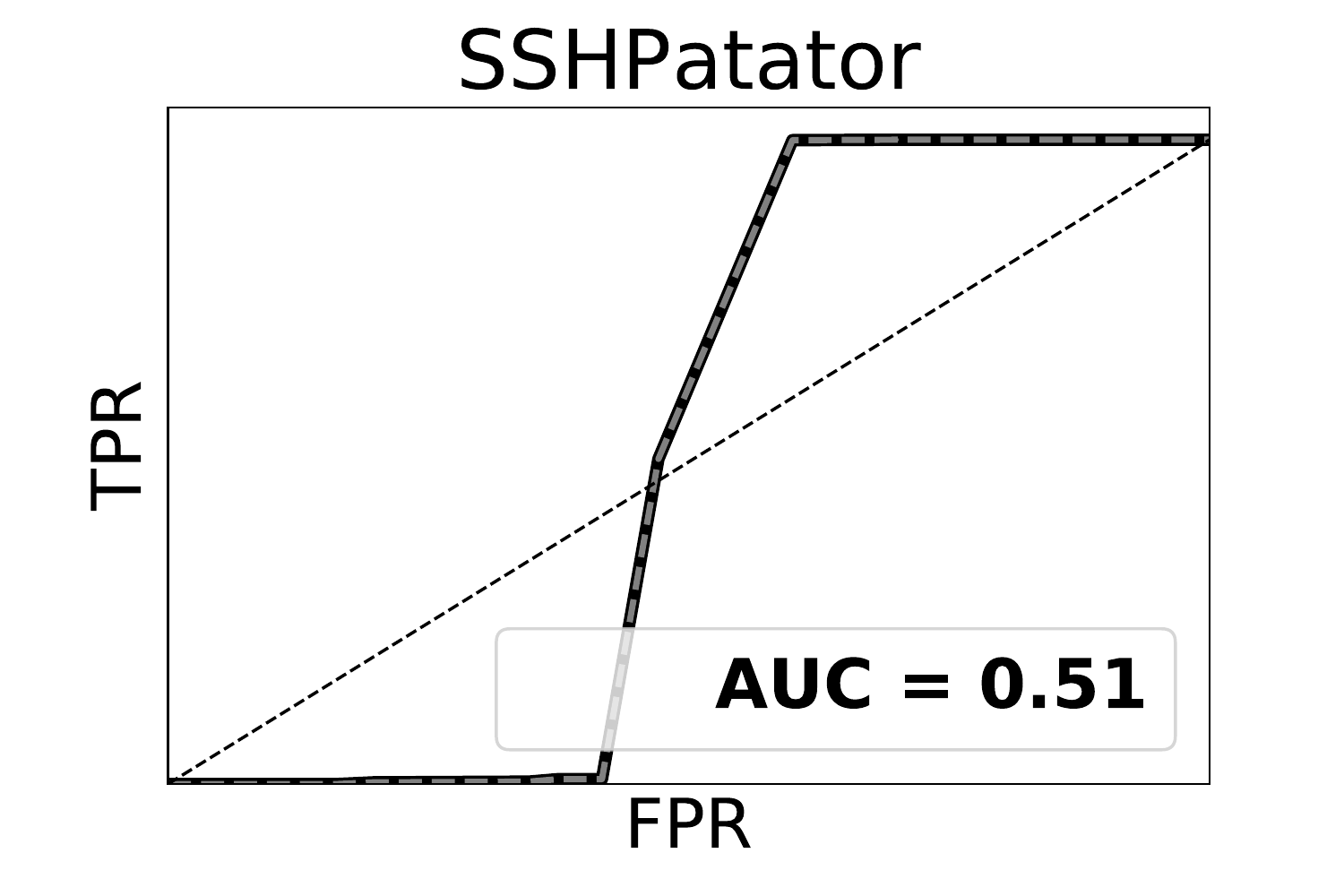} &
      \includegraphics[width=.19\textwidth]{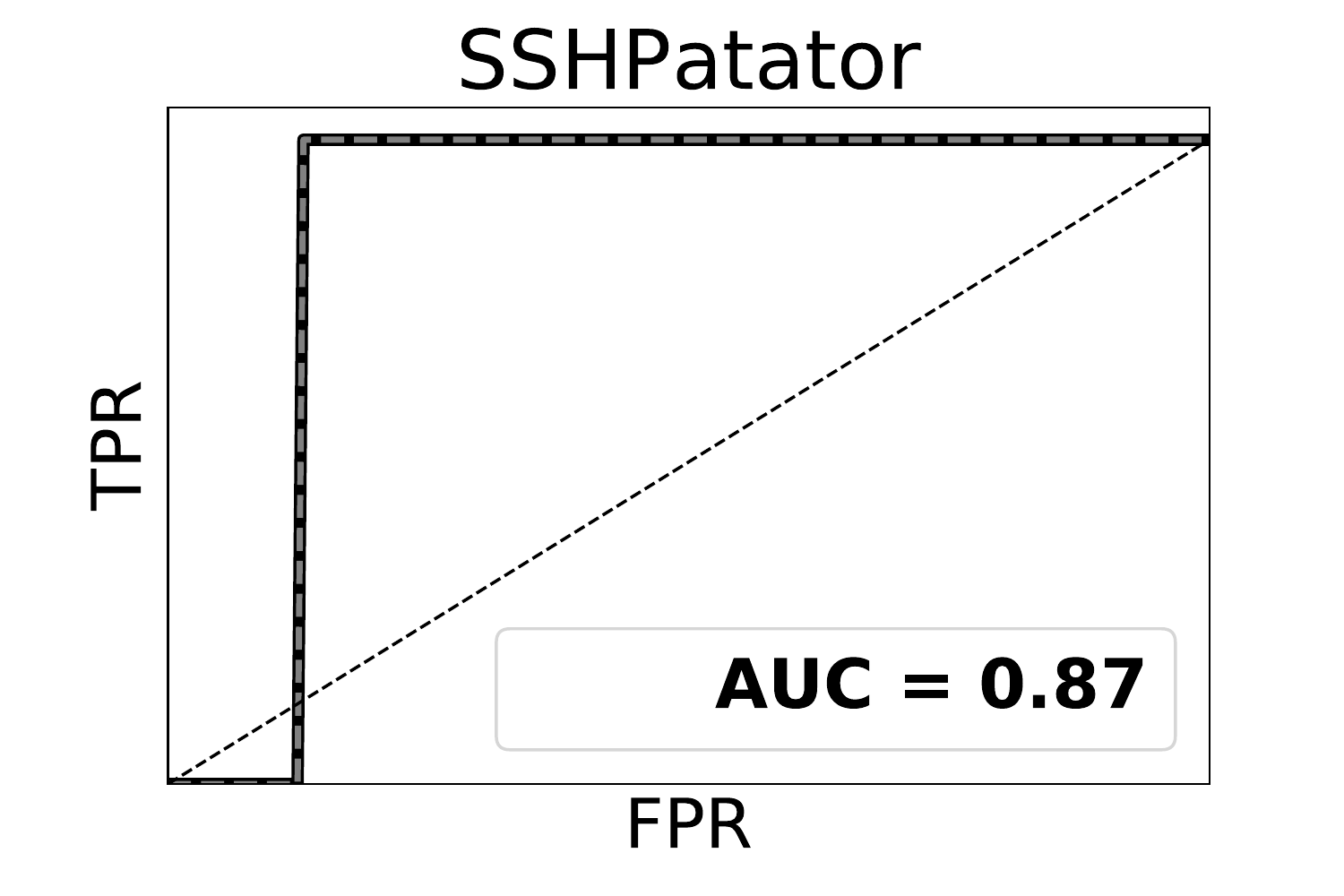} &
      \includegraphics[width=.19\textwidth]{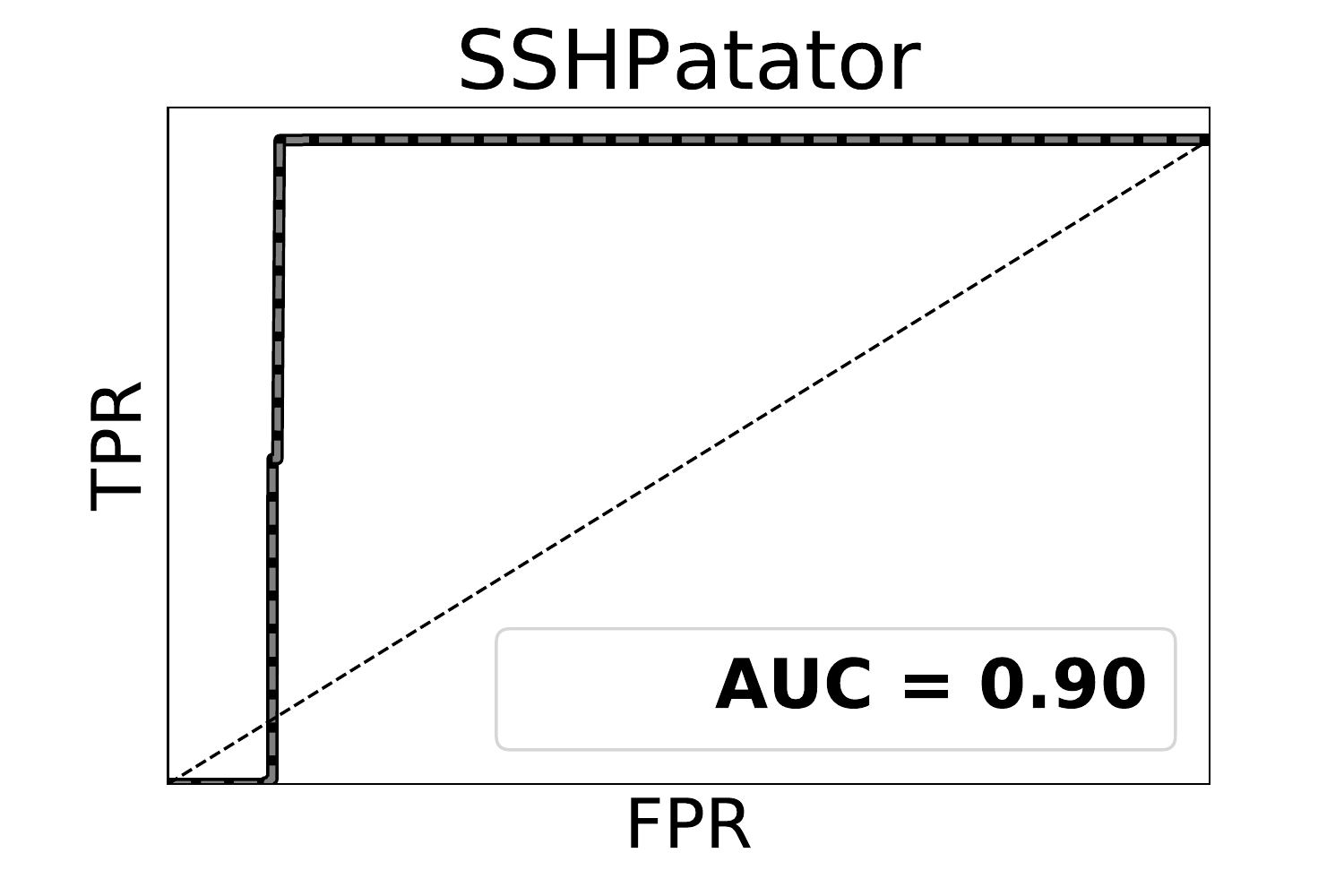} 
      \\
      \includegraphics[width=.19\textwidth]{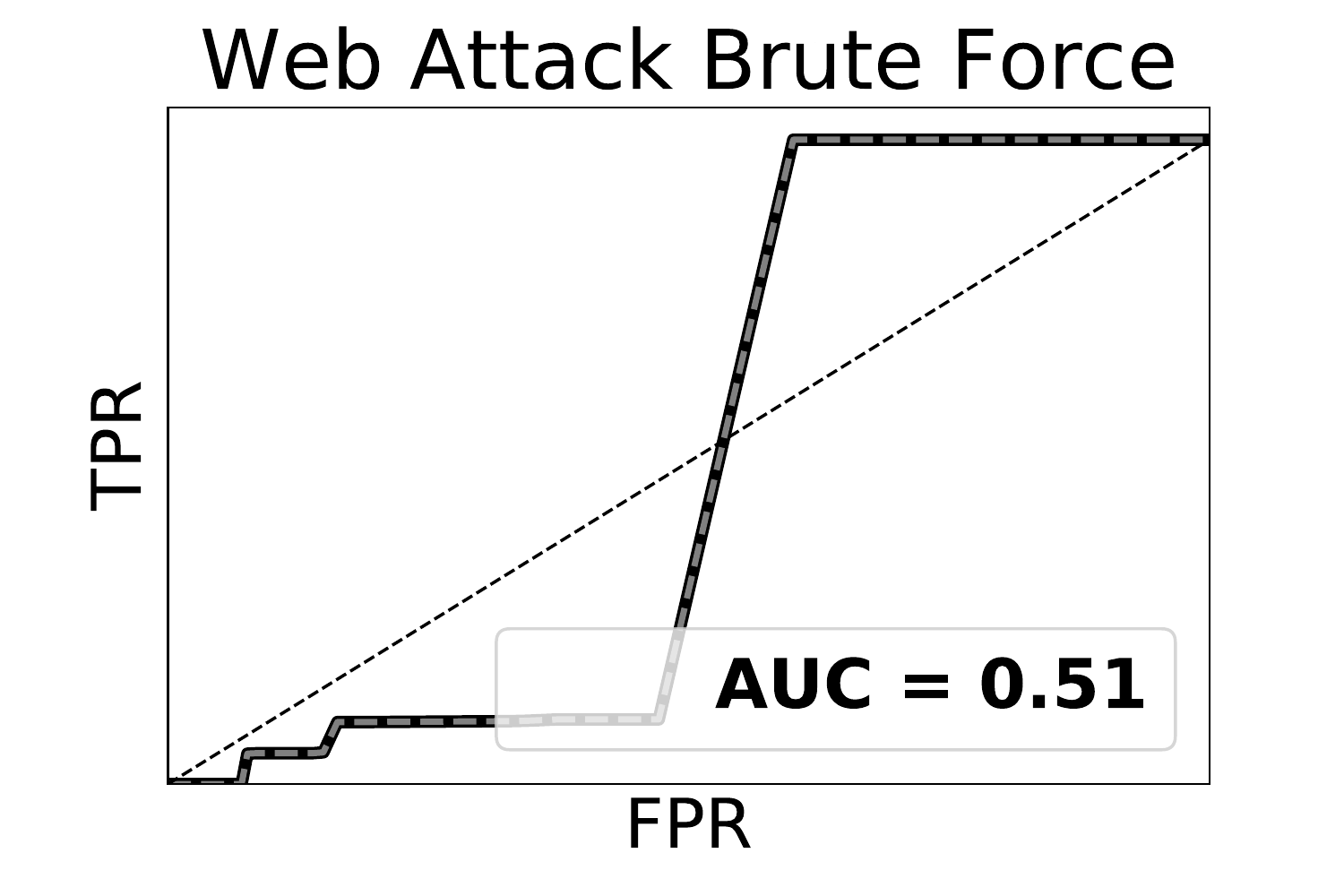} &
      \includegraphics[width=.19\textwidth]{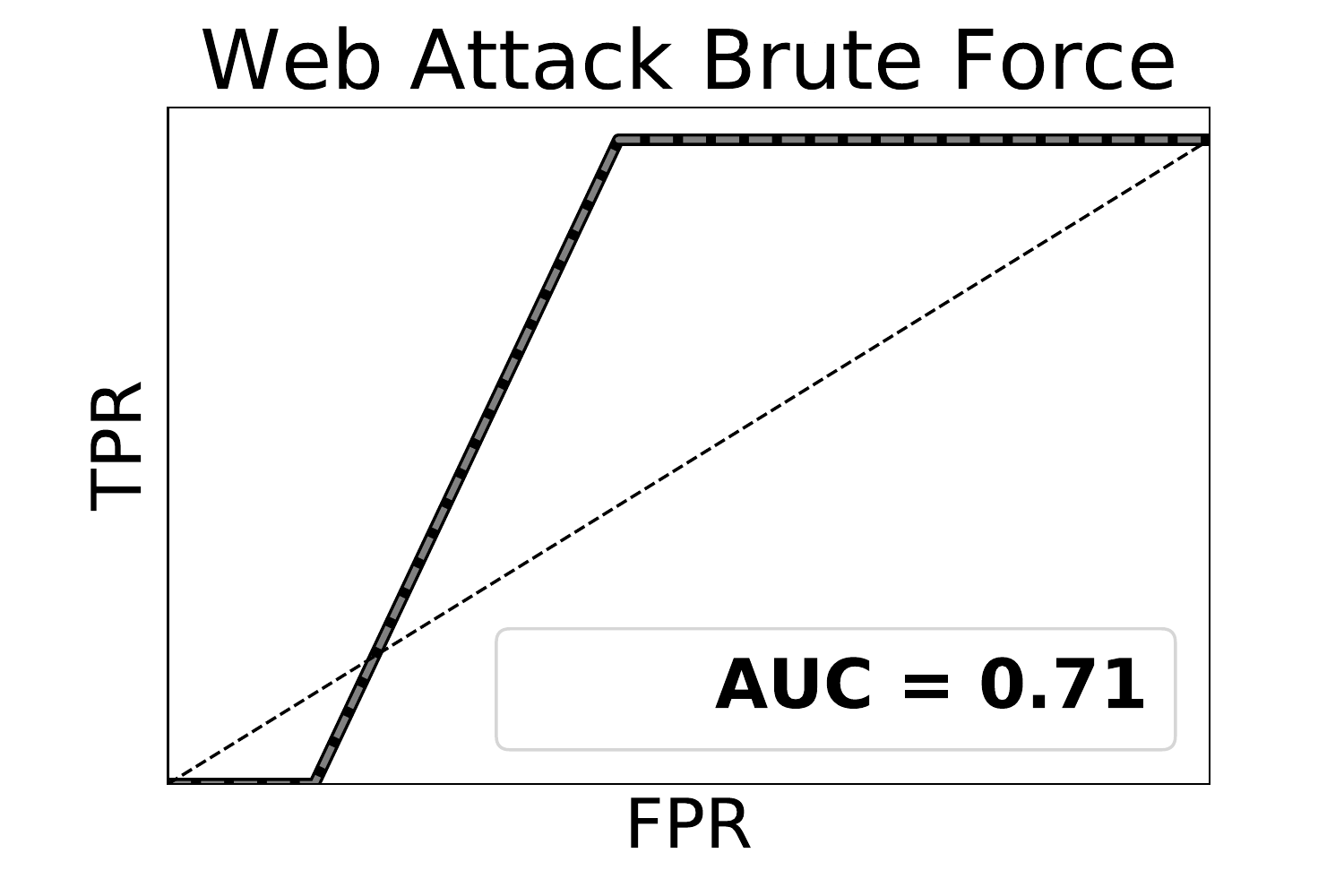} &
      \includegraphics[width=.19\textwidth]{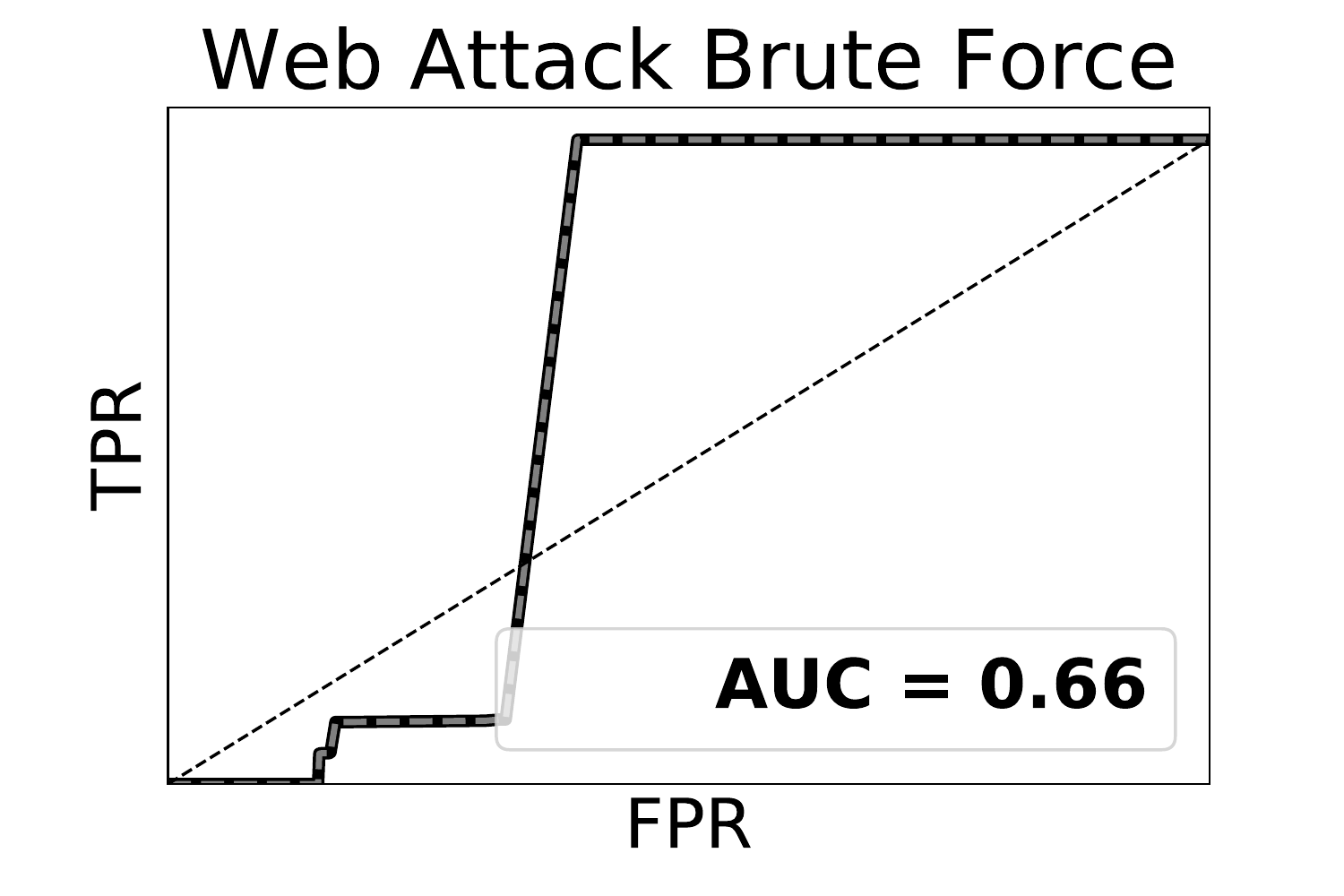} 
      \\
      \includegraphics[width=.19\textwidth]{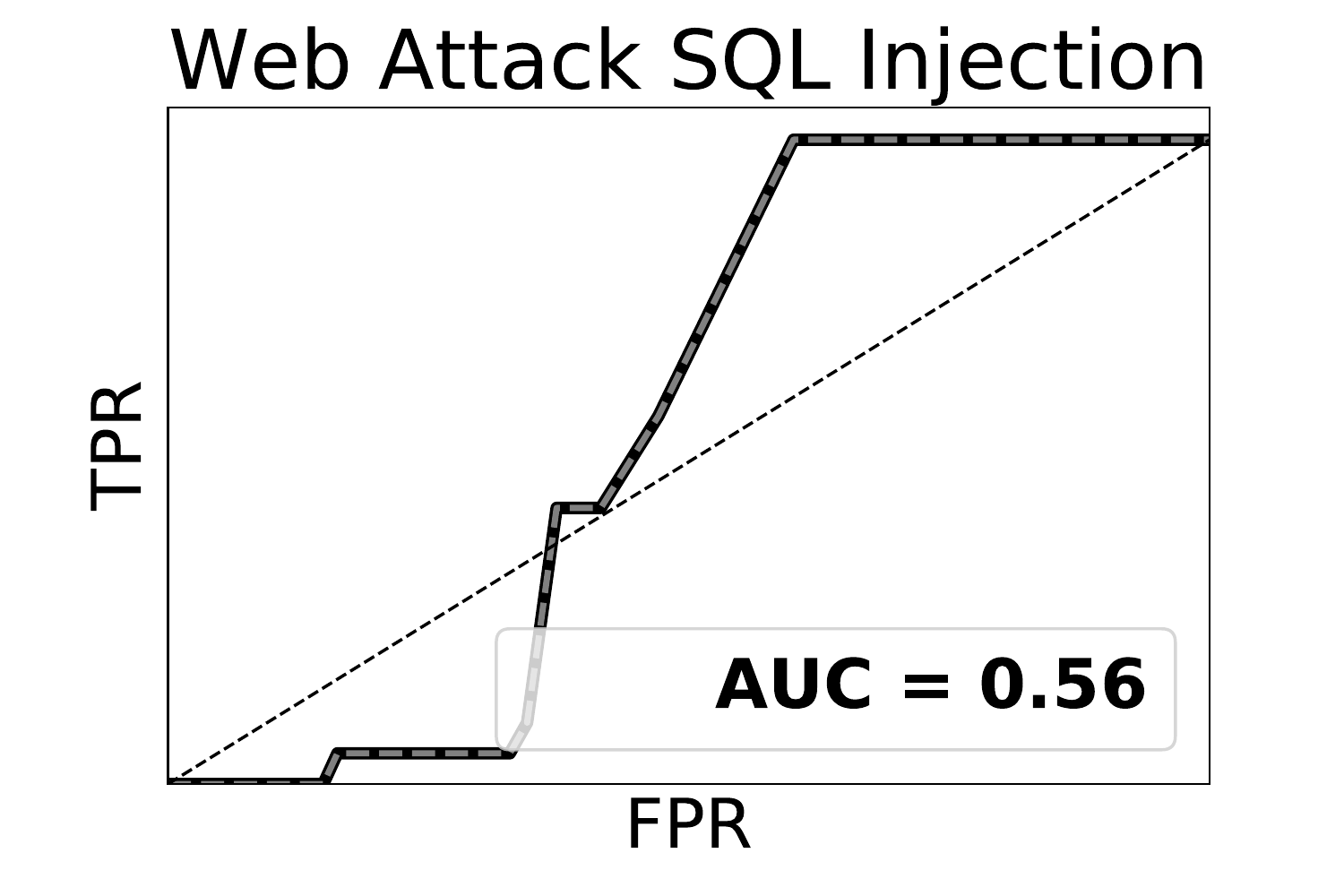} &
      \includegraphics[width=.19\textwidth]{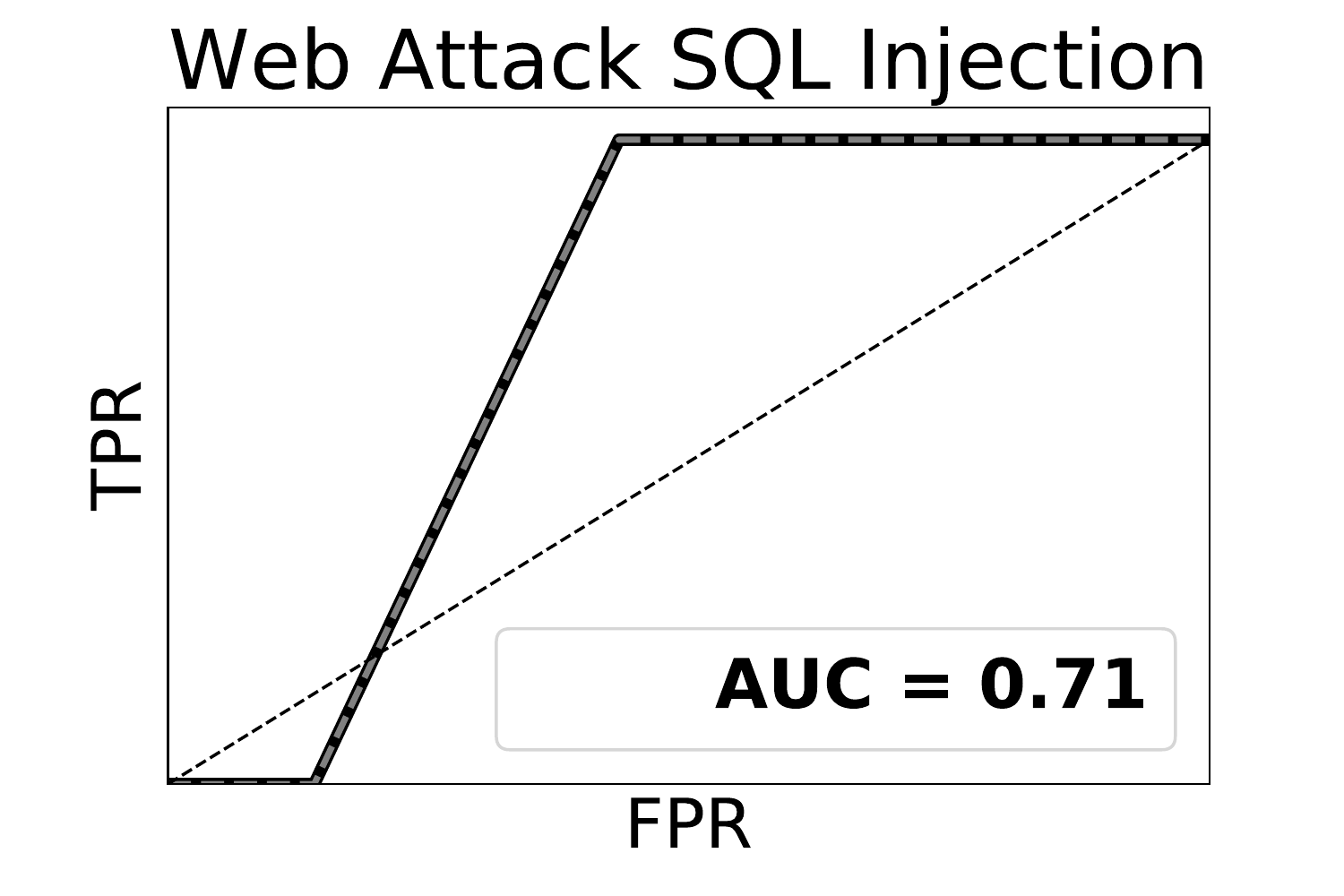} &
      \includegraphics[width=.19\textwidth]{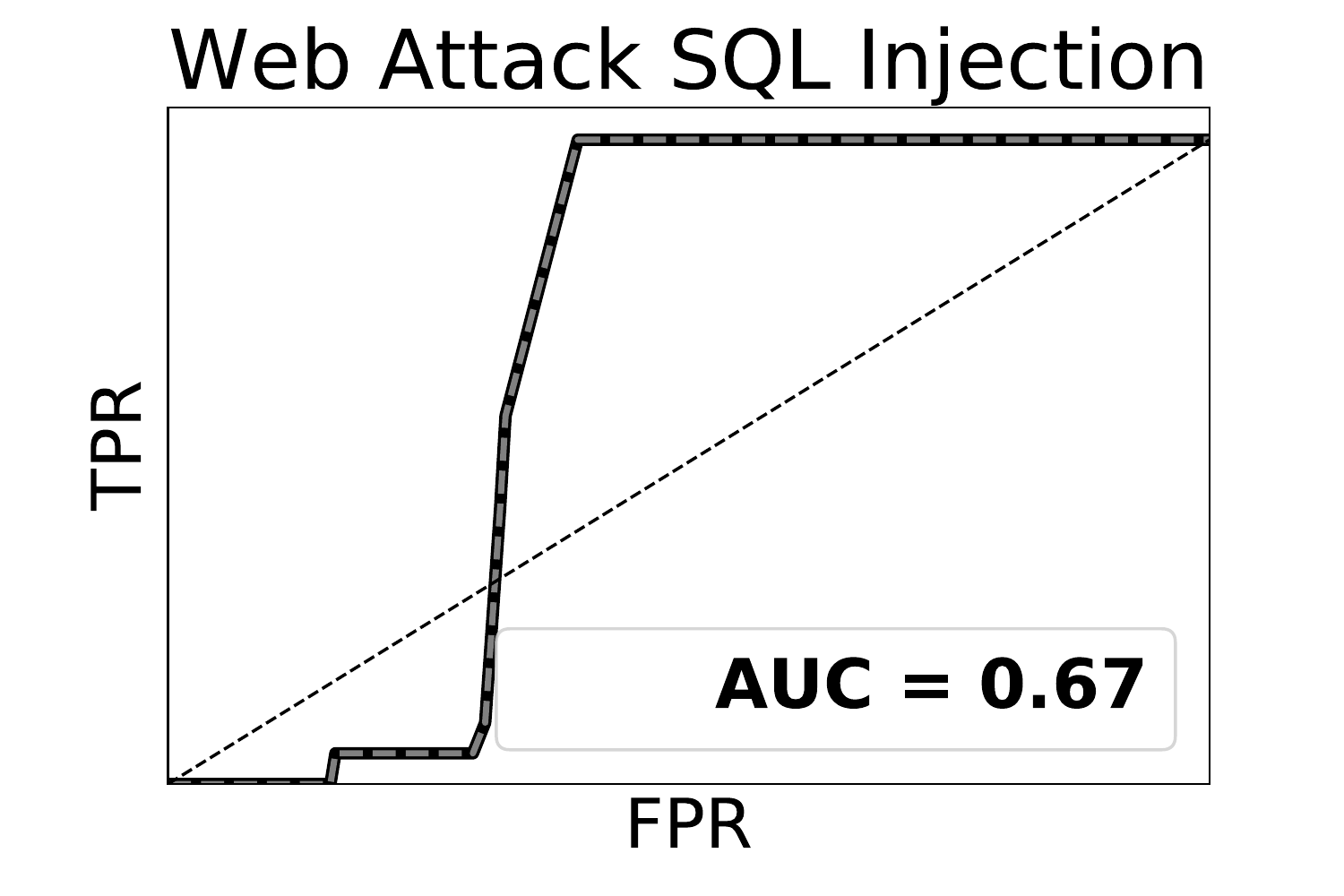} 
      \\
      \includegraphics[width=.19\textwidth]{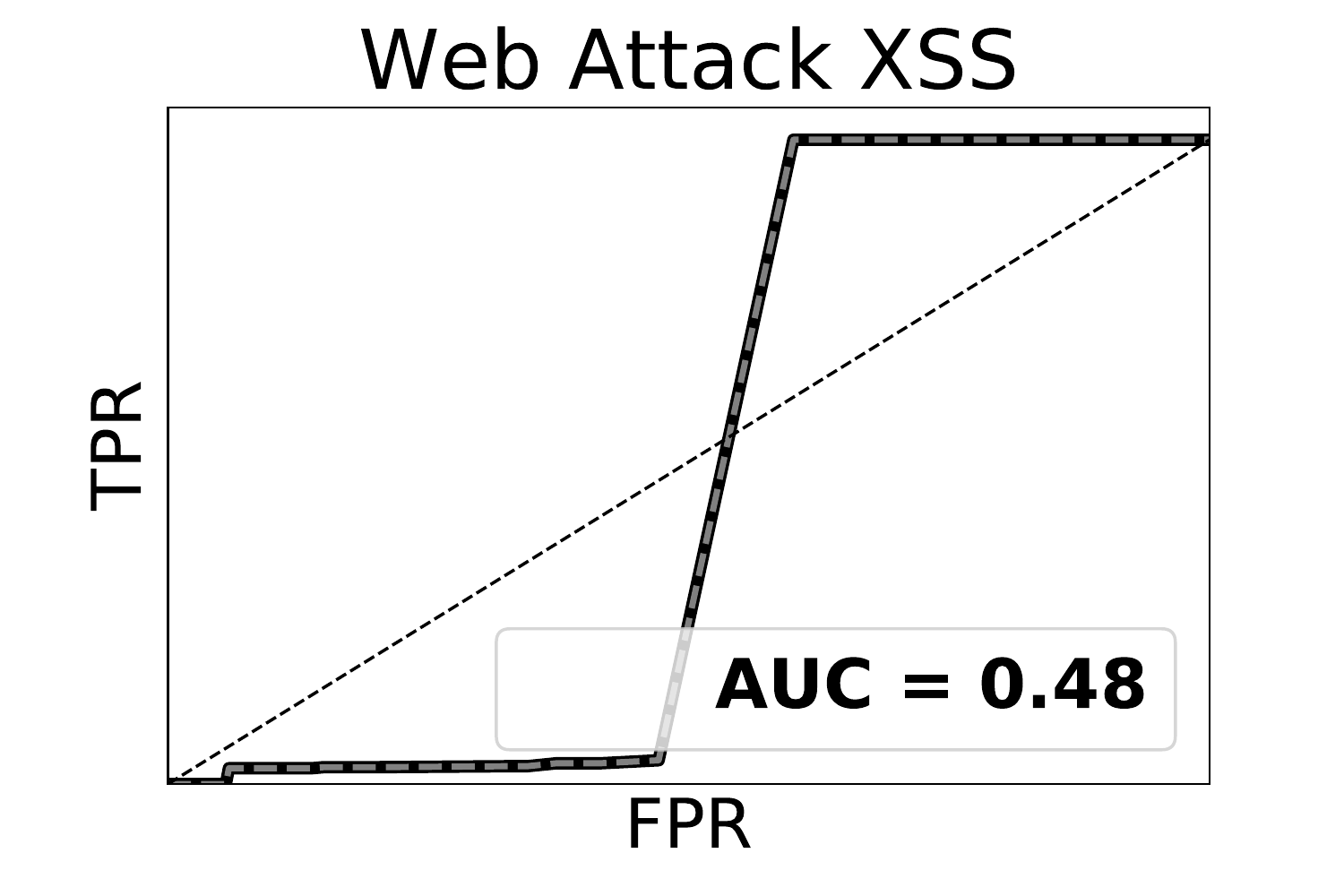} &
      \includegraphics[width=.19\textwidth]{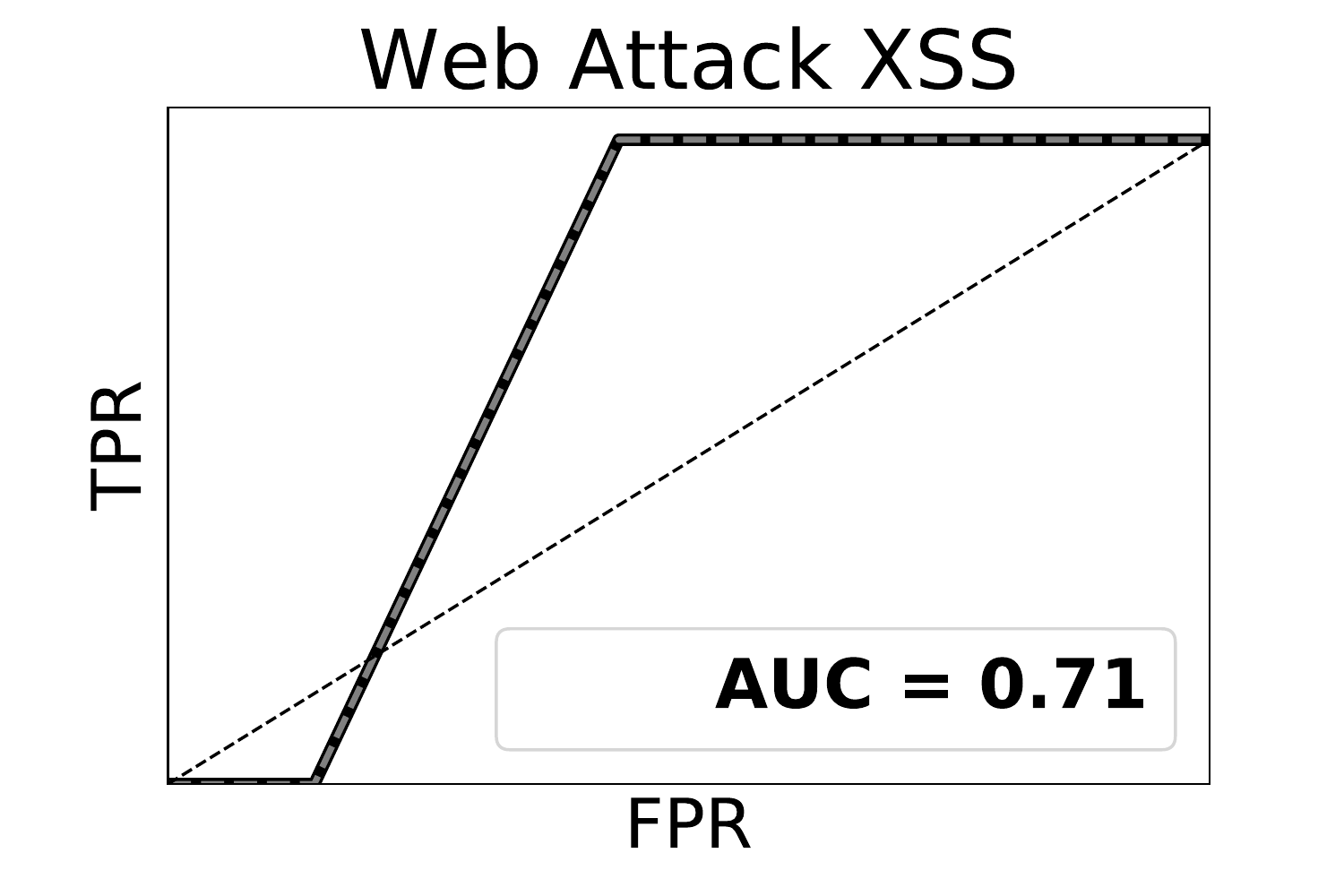} &
      \includegraphics[width=.19\textwidth]{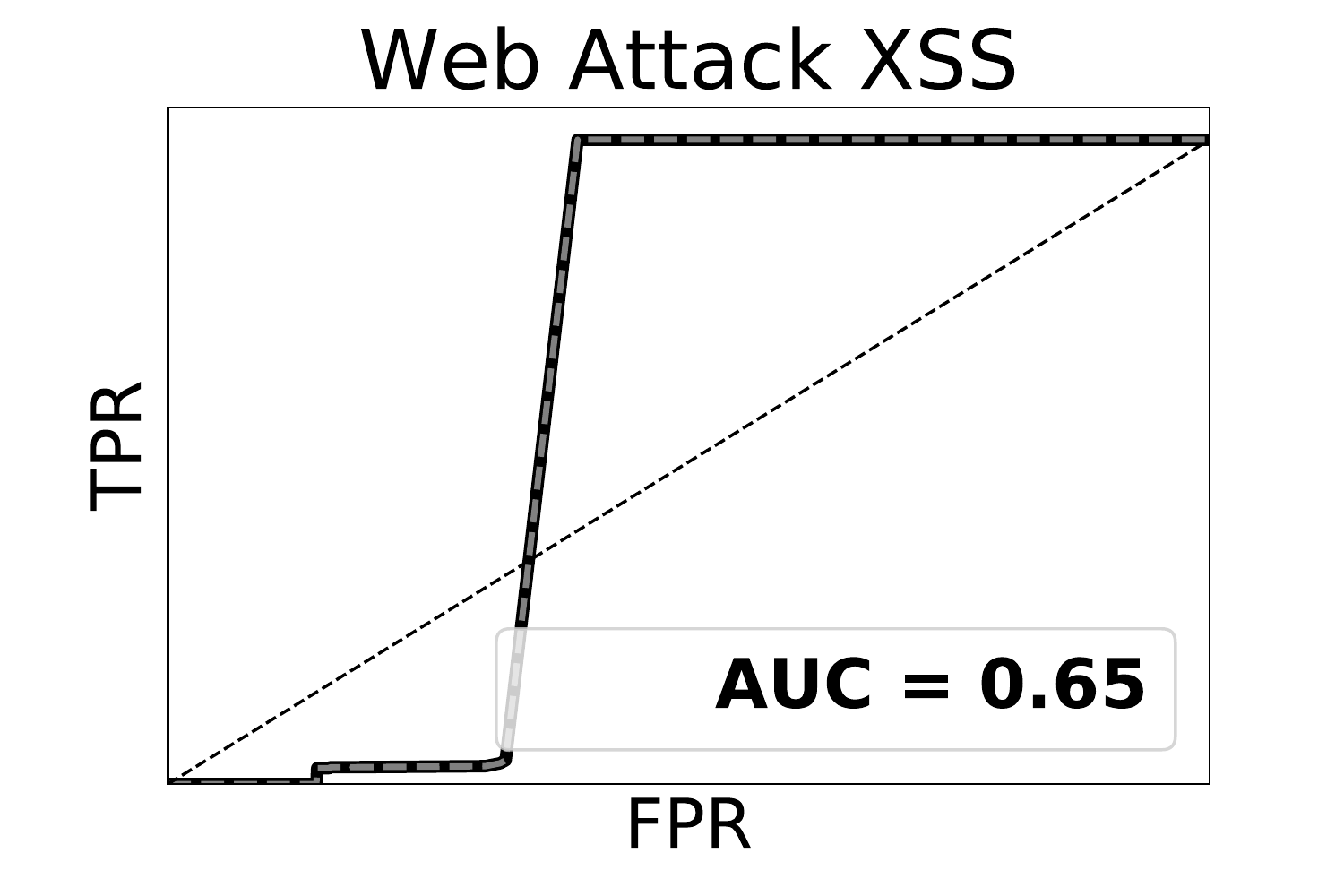} 
      \\
  \end{tabular}
\end{table*}

\begin{table*}
  [ht] \caption{Protobyte Sequences A} \label{tab:protobytesa}
  \begin{tabular}{ccccc} 
  \hline 
  Source & Destination & Dyad & Internal & External \\
      \hline 
      \includegraphics[width=.19\textwidth]{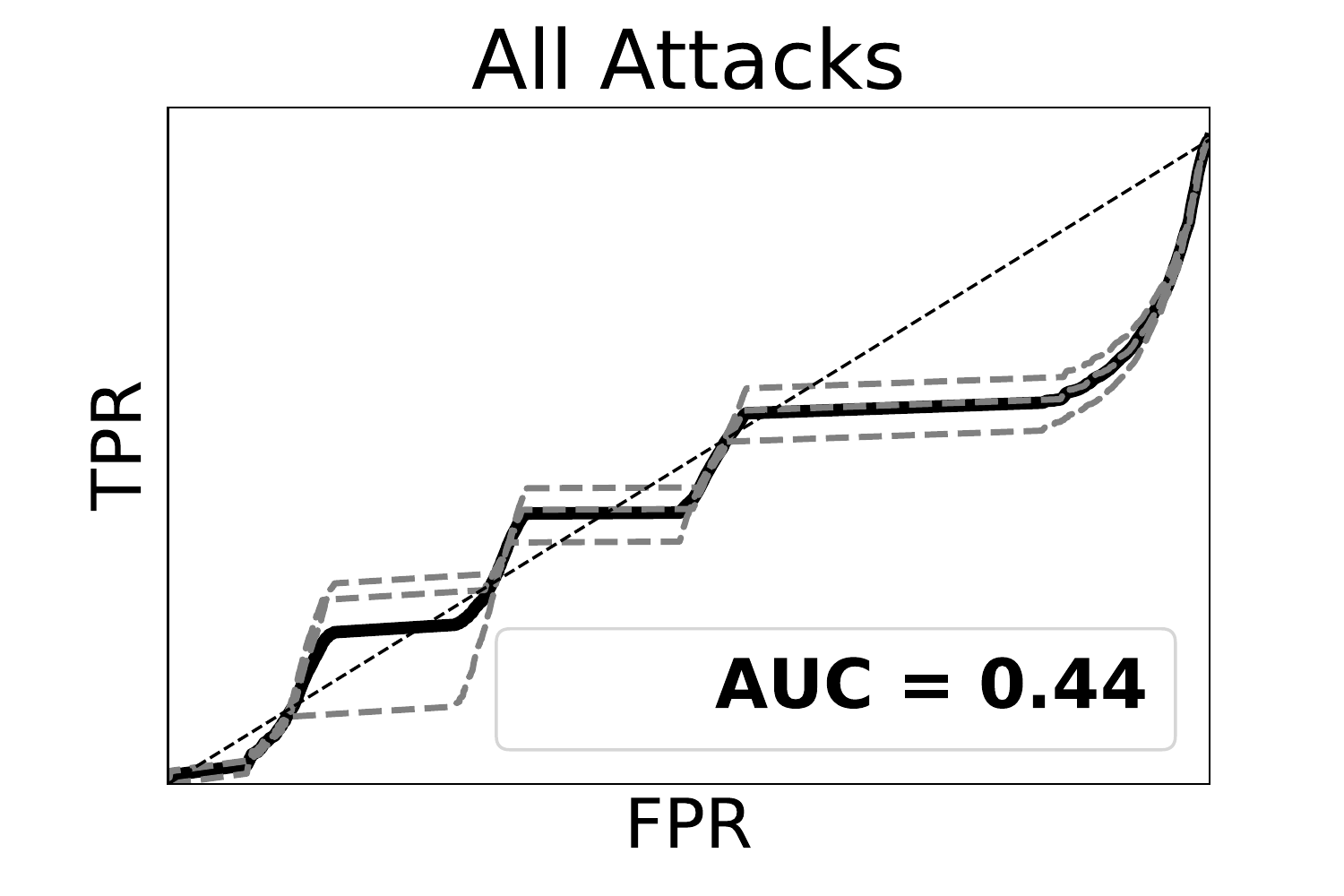} &
      \includegraphics[width=.19\textwidth]{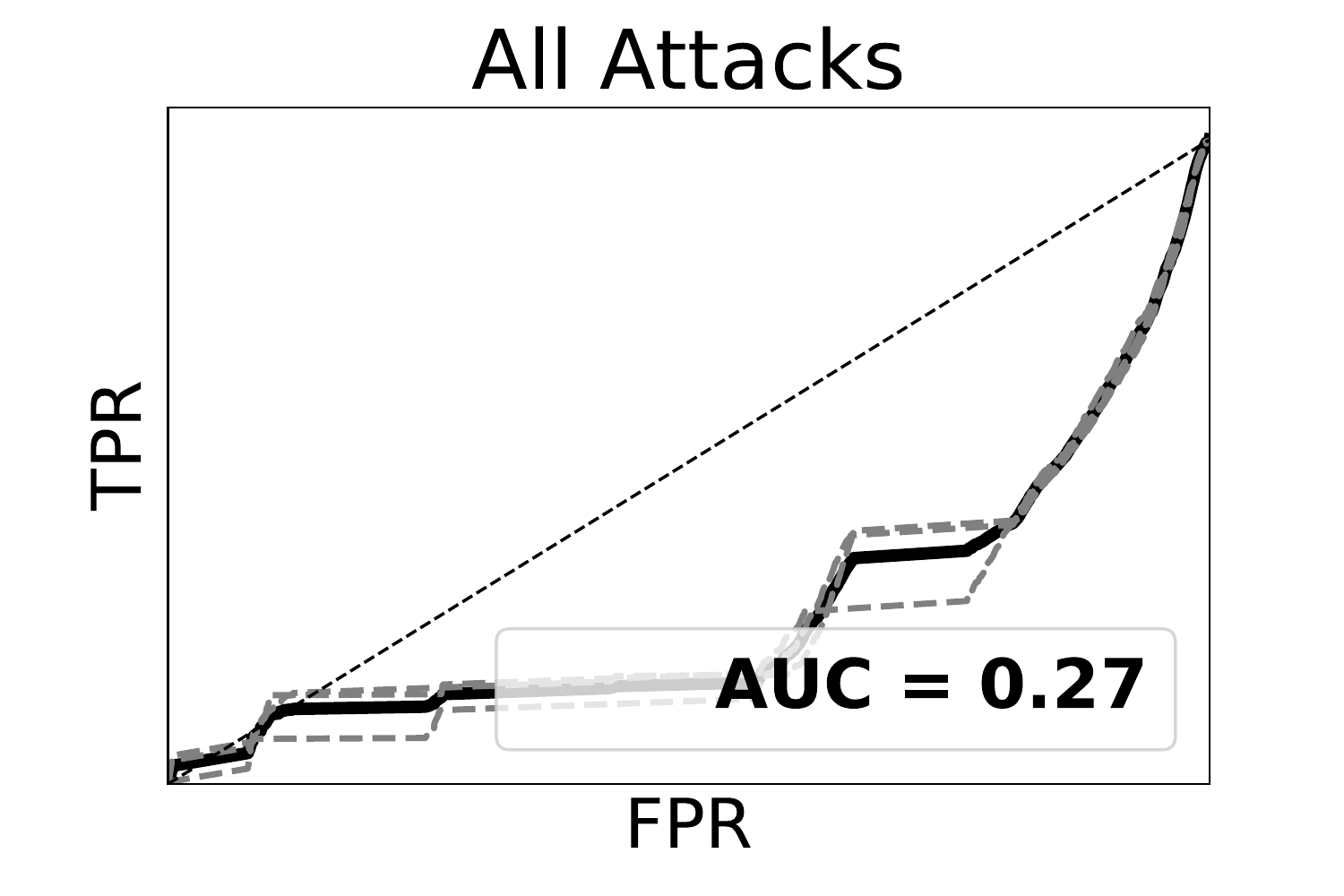} &
      \includegraphics[width=.19\textwidth]{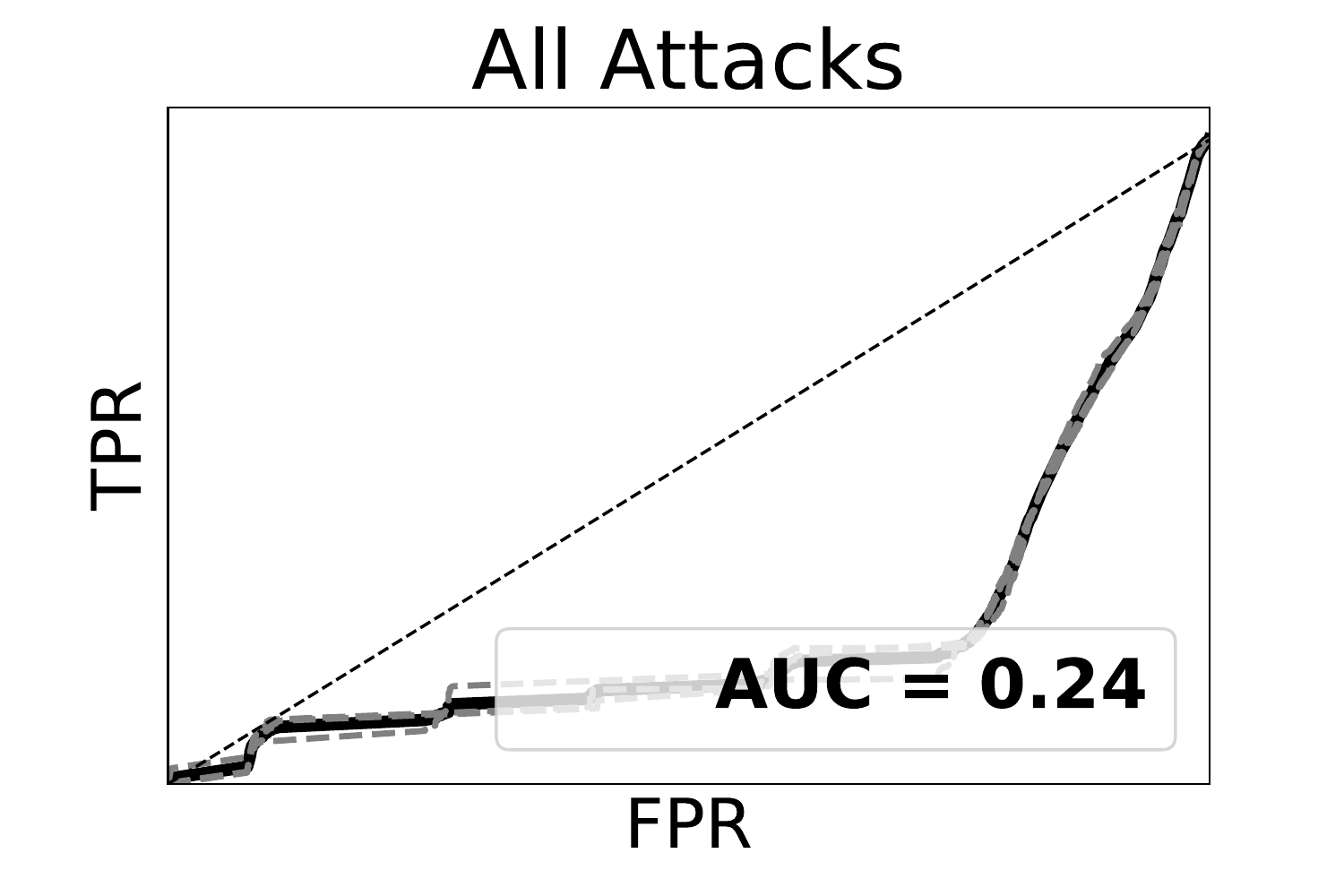} &
      \includegraphics[width=.19\textwidth]{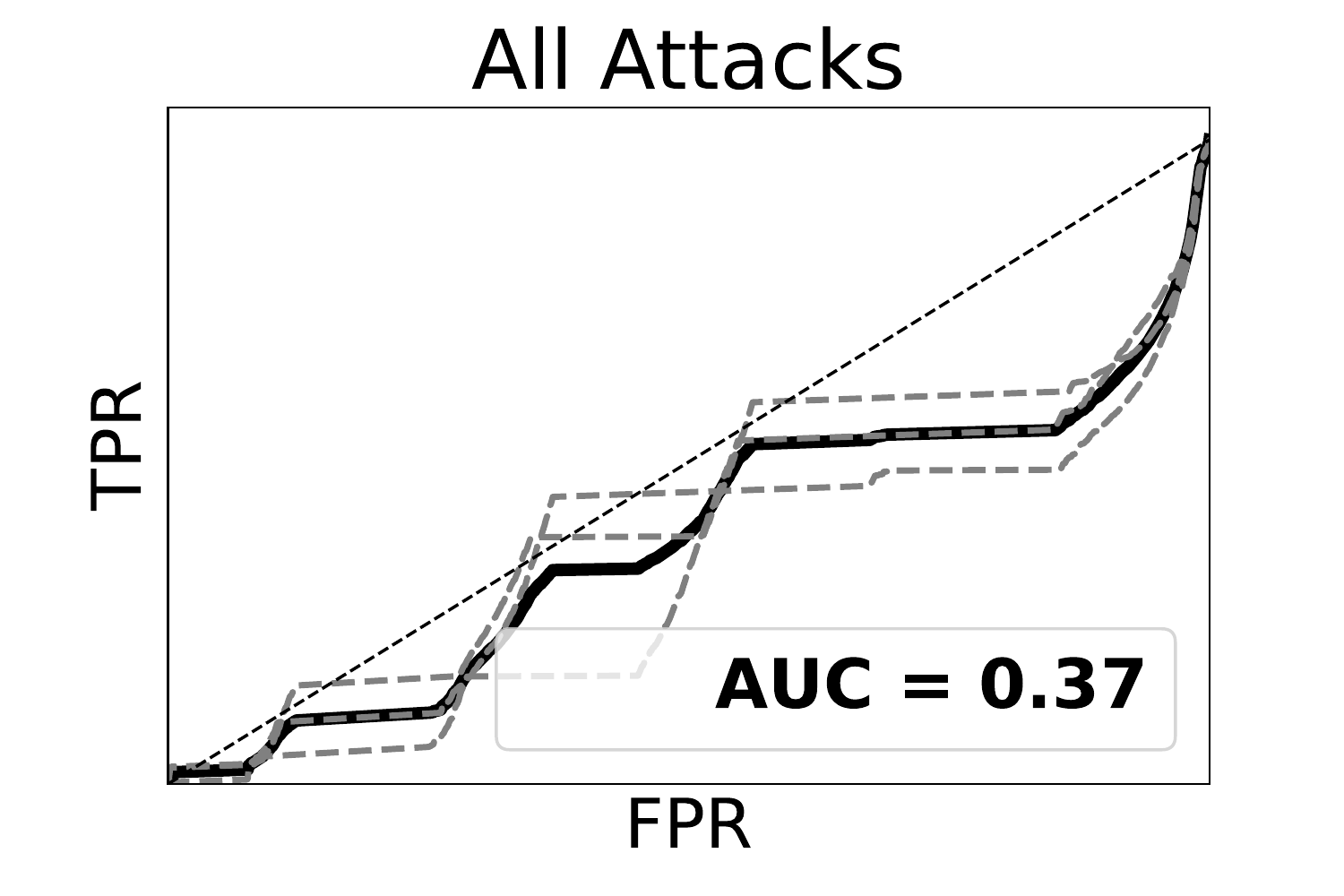} &
      \includegraphics[width=.19\textwidth]{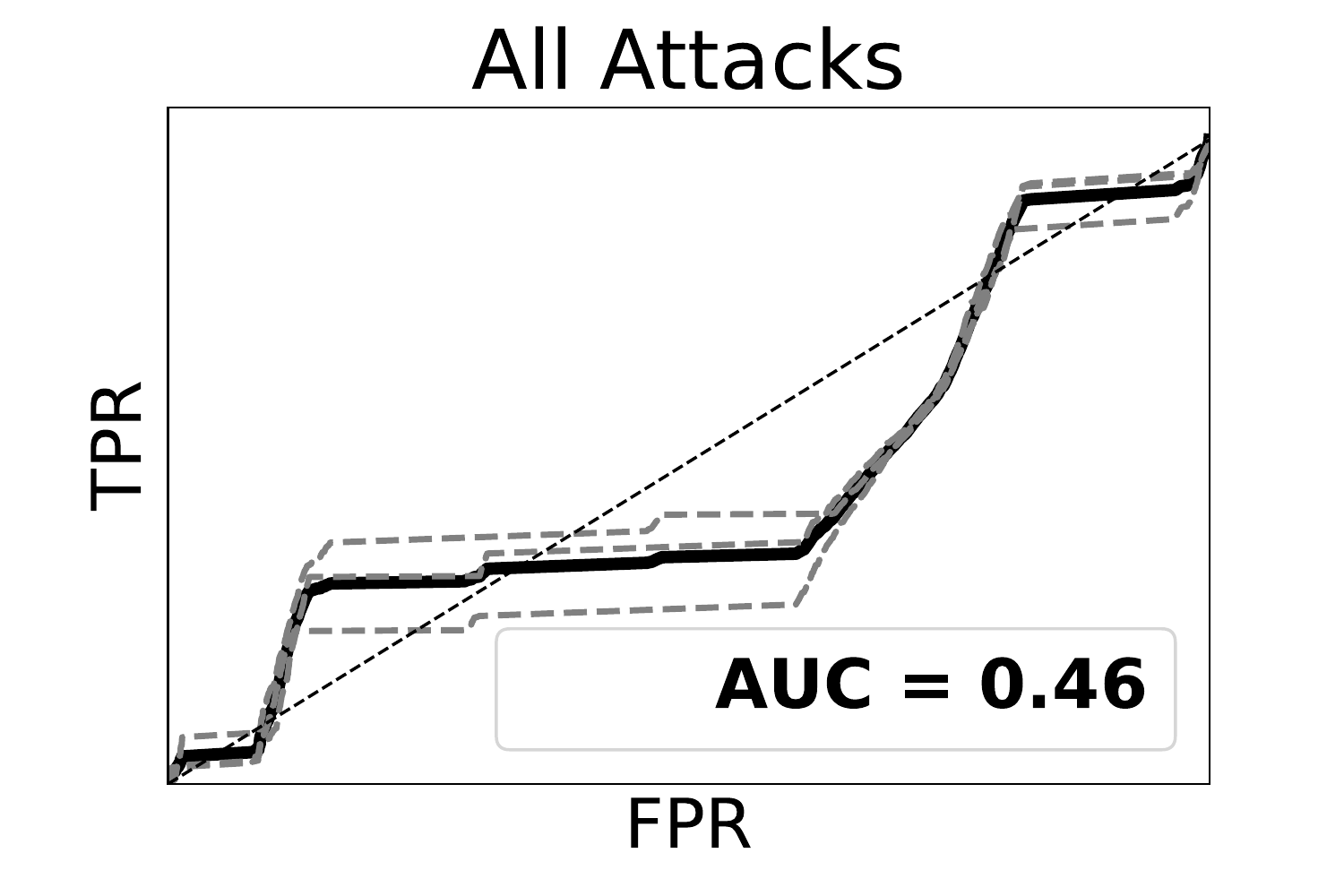}
      \\
            \includegraphics[width=.19\textwidth]{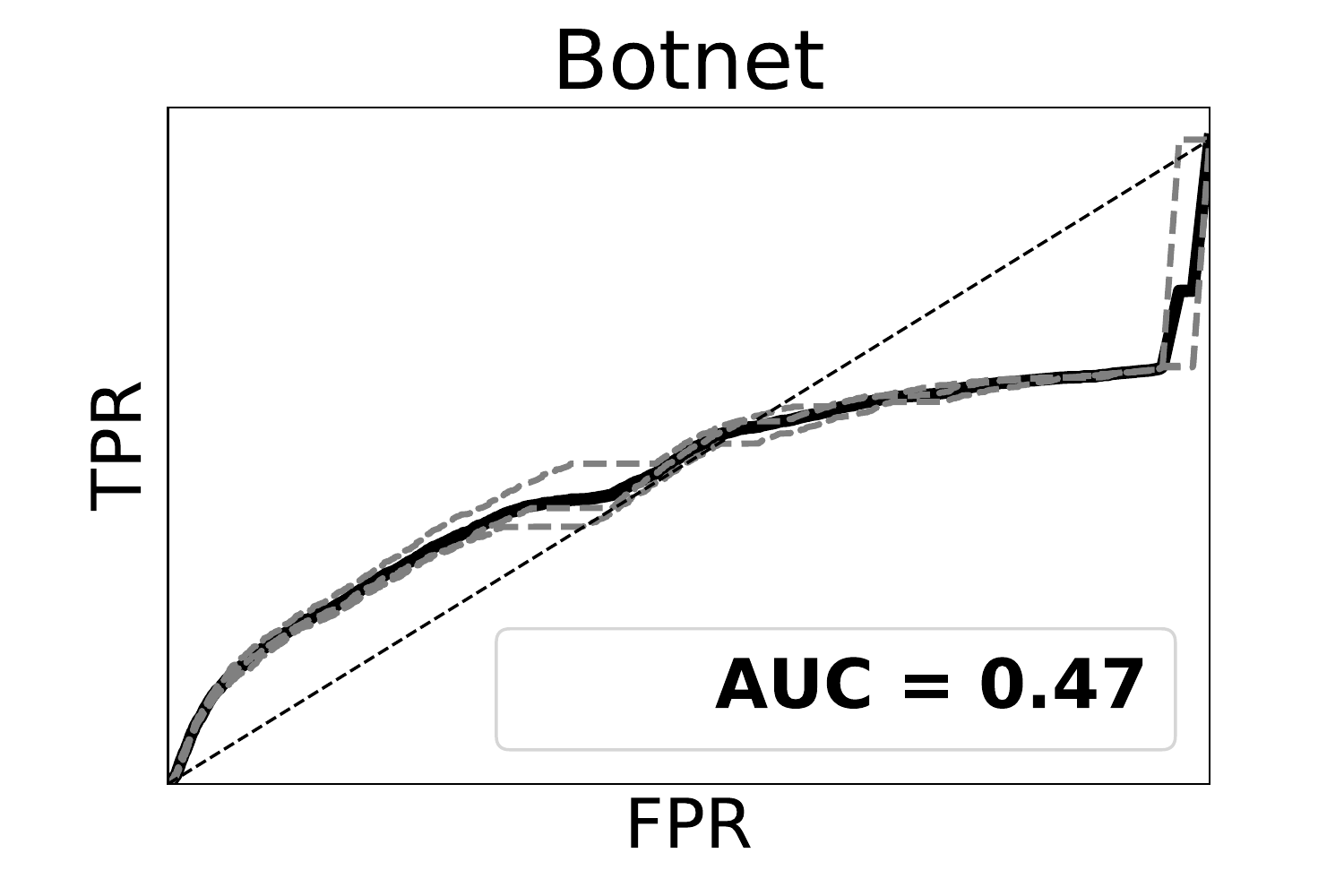} &
      \includegraphics[width=.19\textwidth]{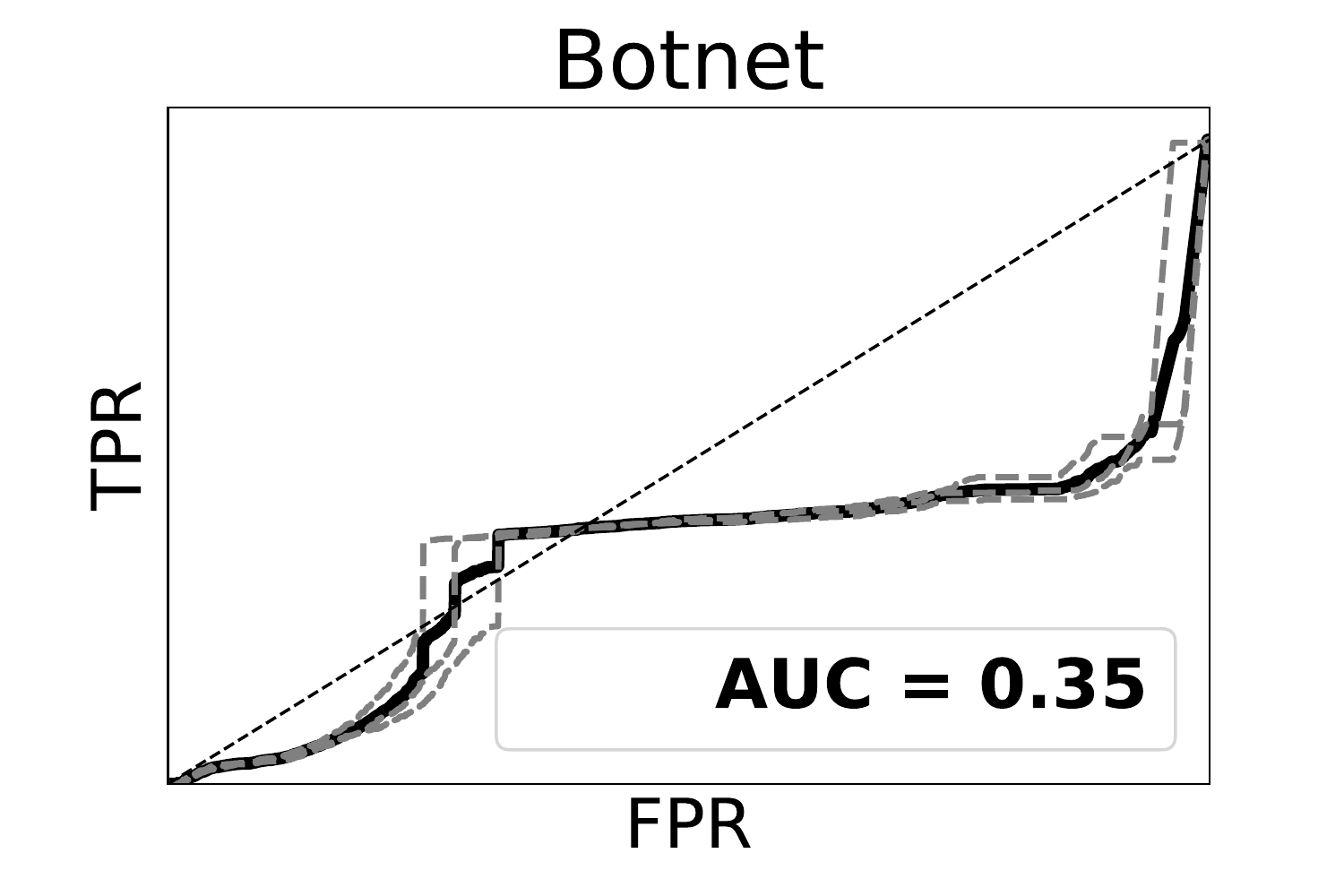} &
      \includegraphics[width=.19\textwidth]{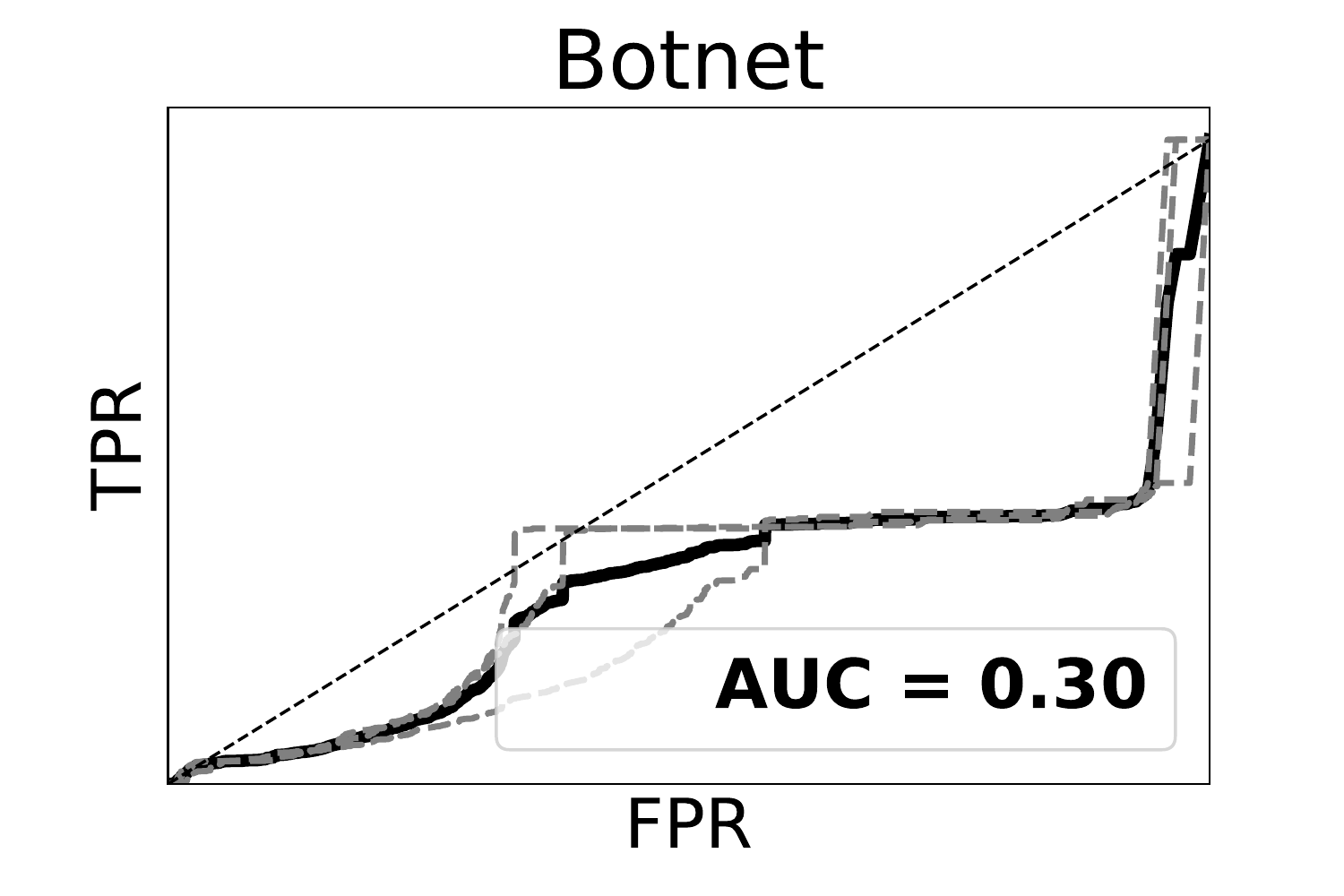} &
      \includegraphics[width=.19\textwidth]{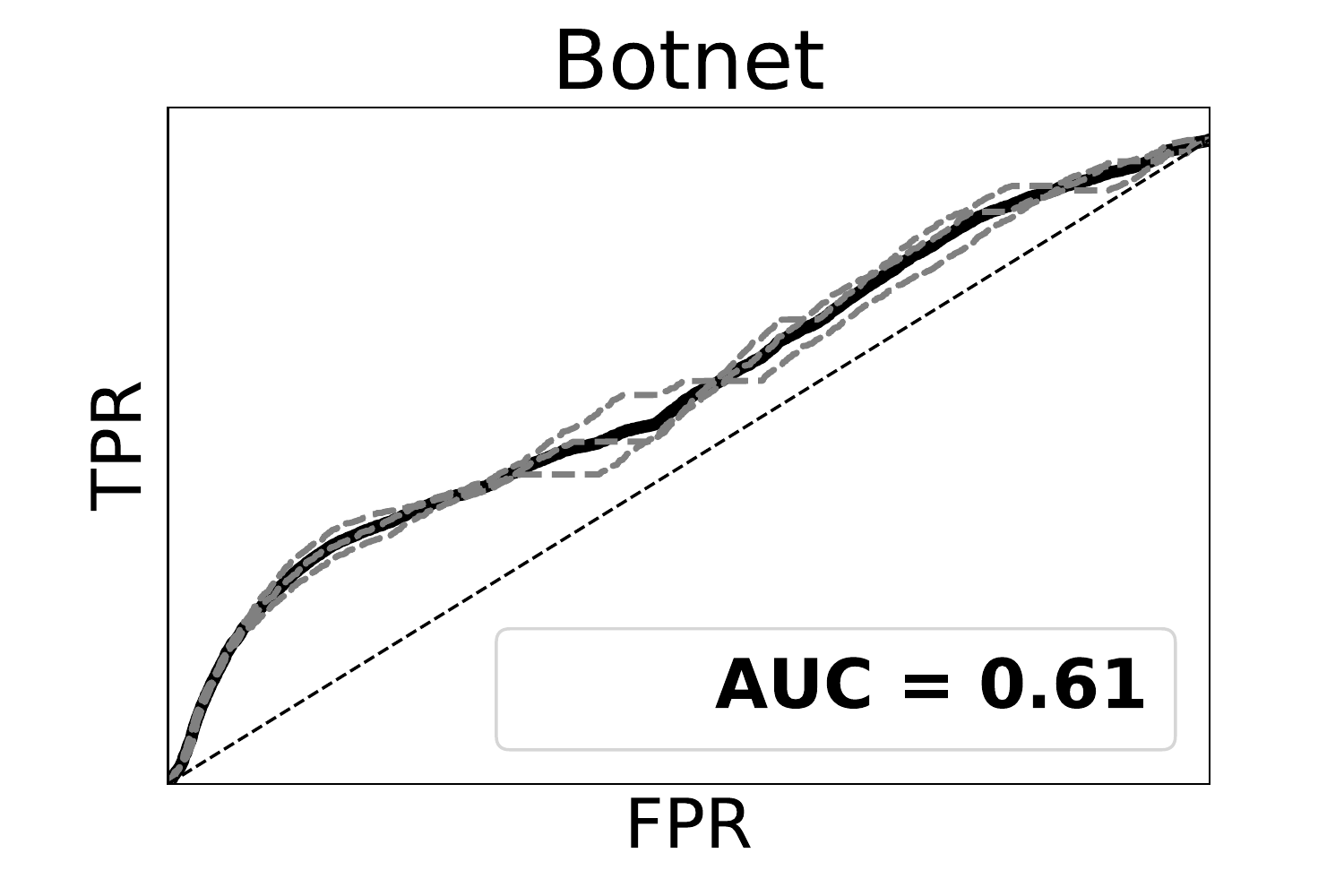} &
      \includegraphics[width=.19\textwidth]{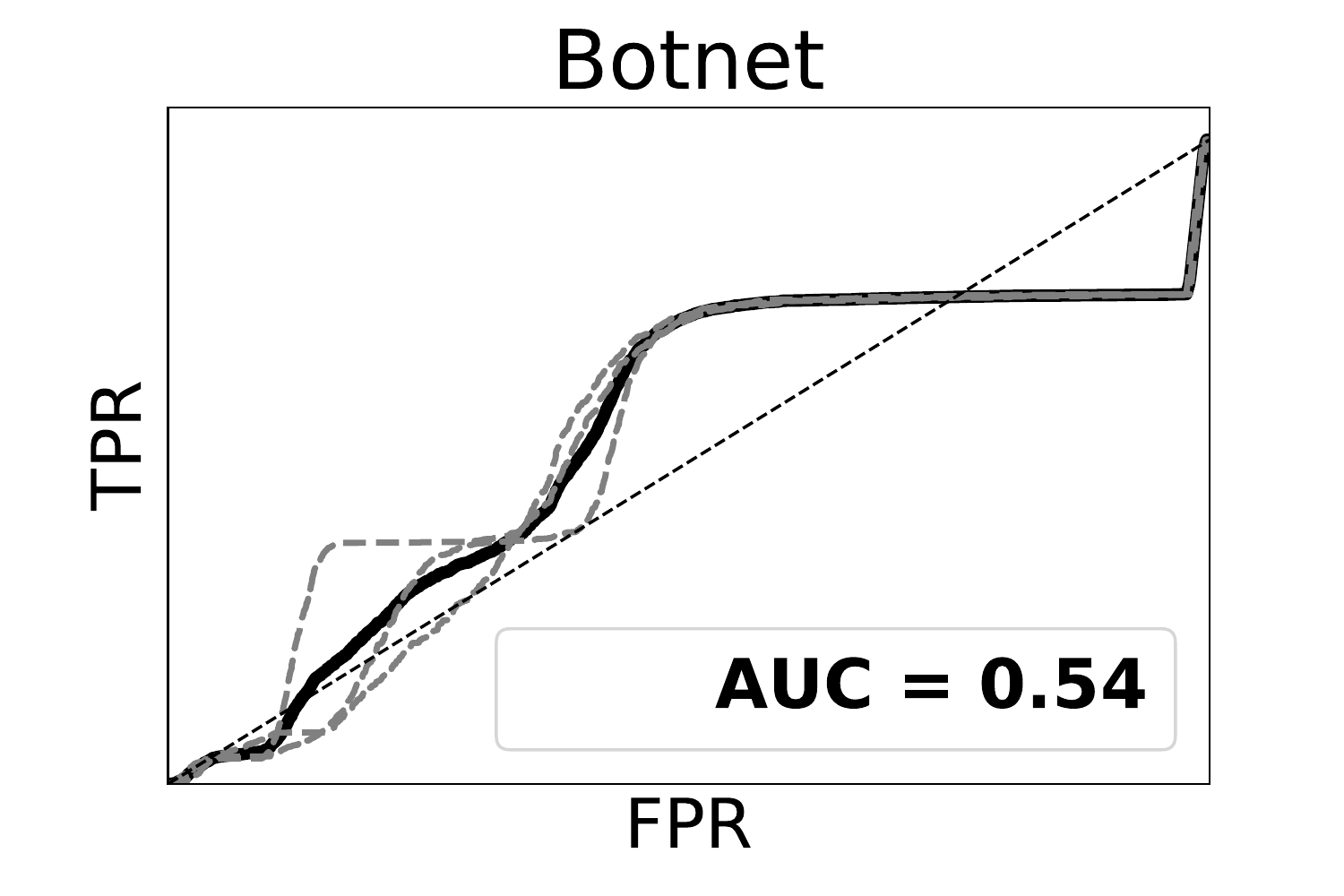}
      \\
            \includegraphics[width=.19\textwidth]{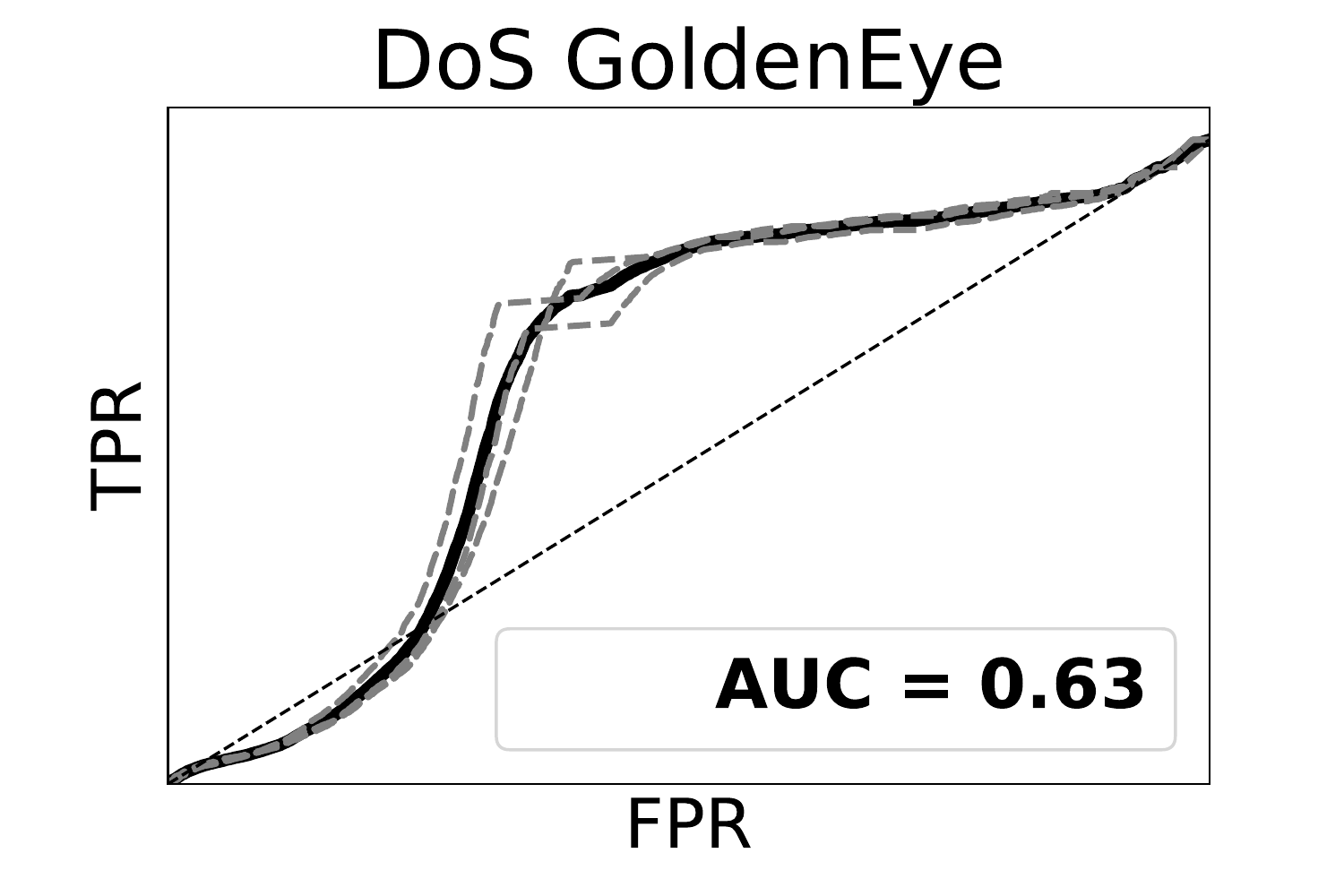} &
      \includegraphics[width=.19\textwidth]{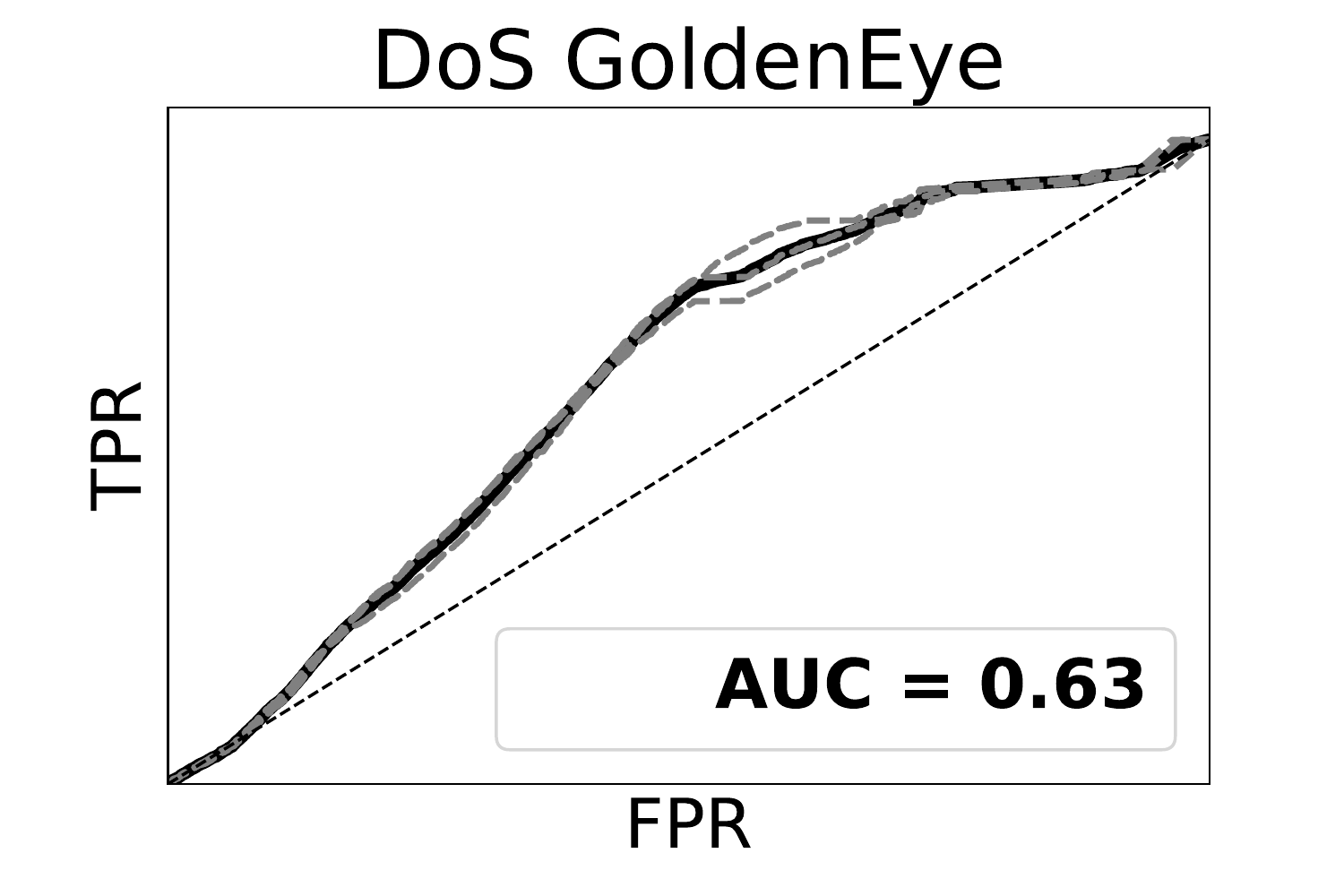} &
      \includegraphics[width=.19\textwidth]{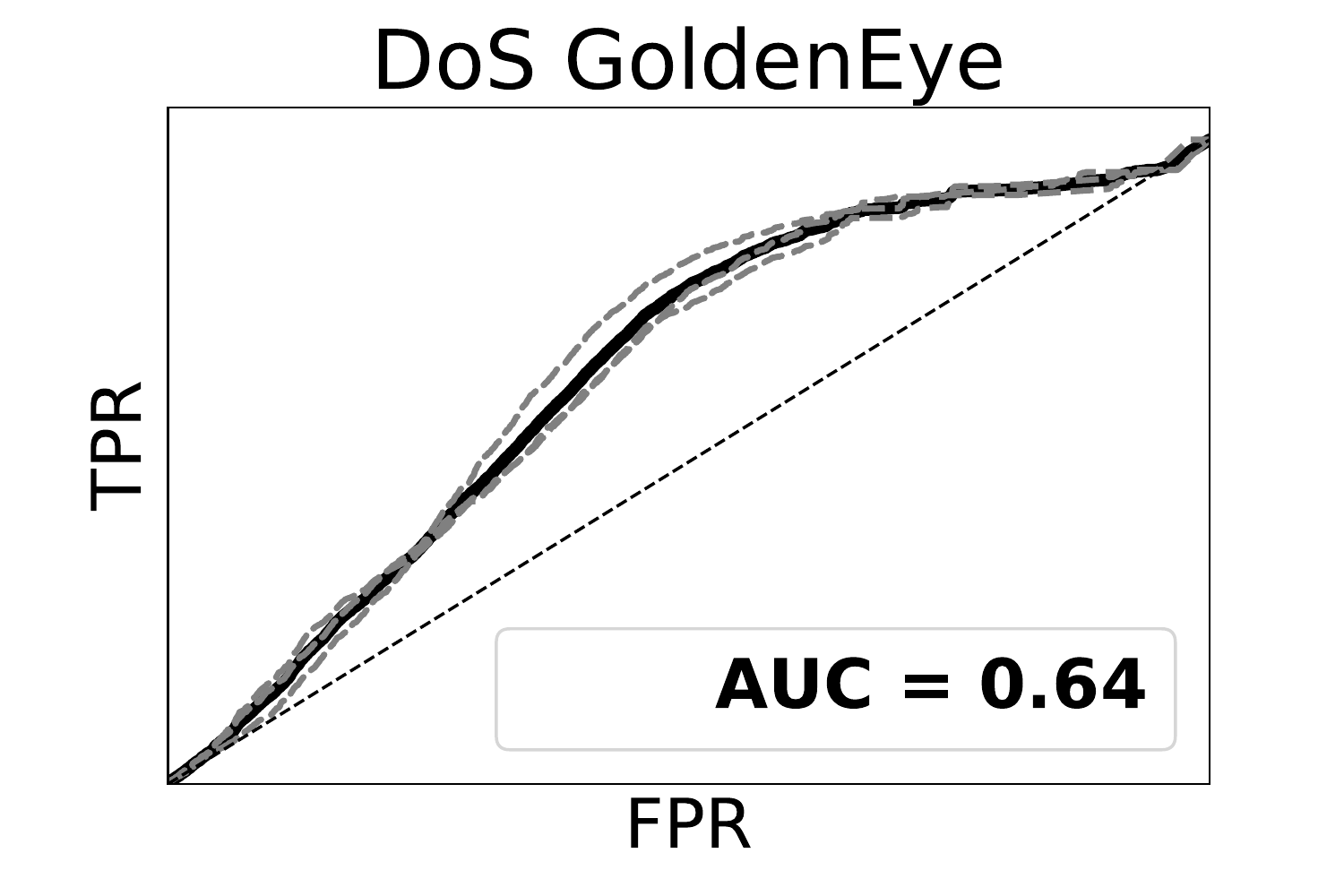} &
      \includegraphics[width=.19\textwidth]{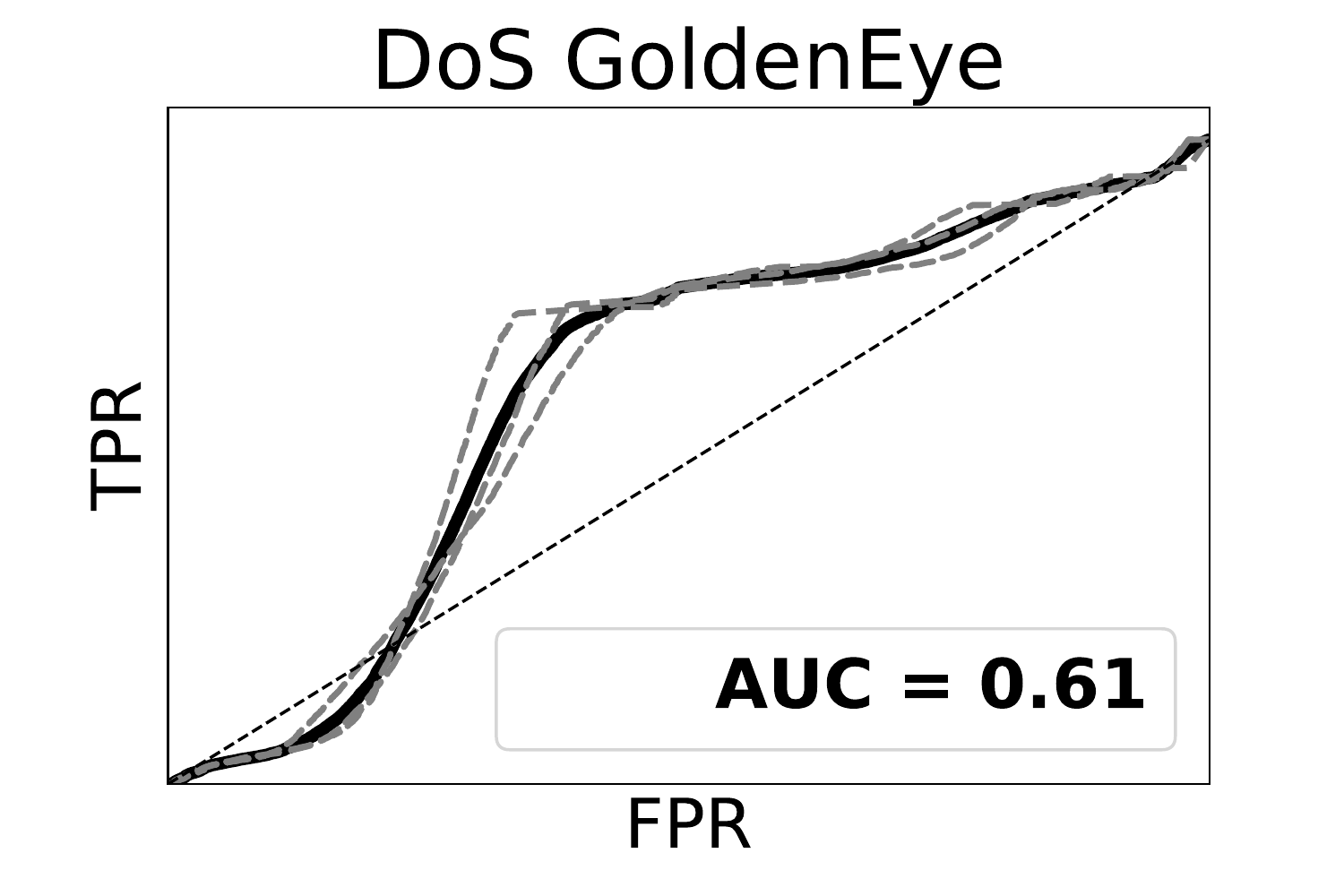} &
      \includegraphics[width=.19\textwidth]{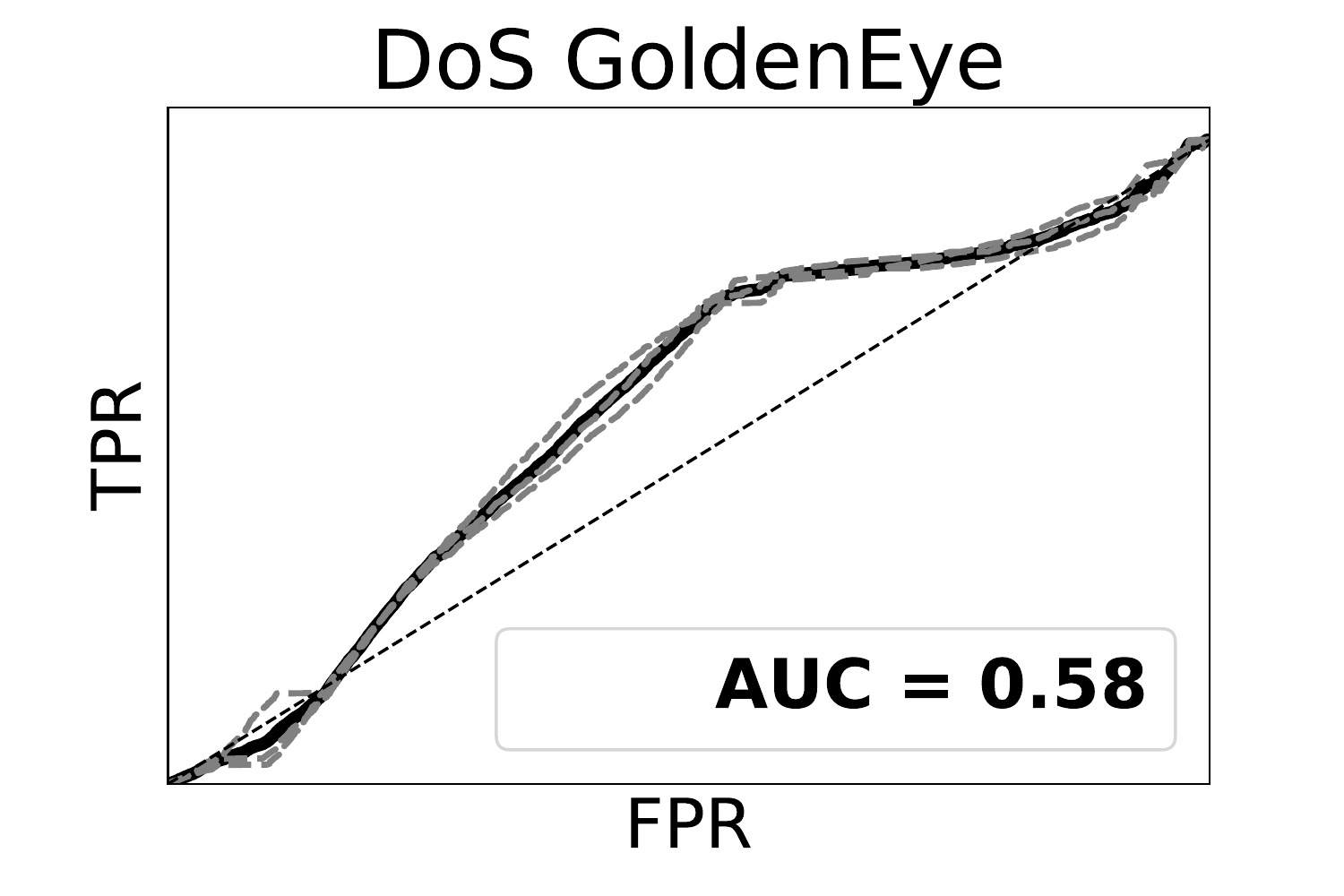}
      \\
      \includegraphics[width=.19\textwidth]{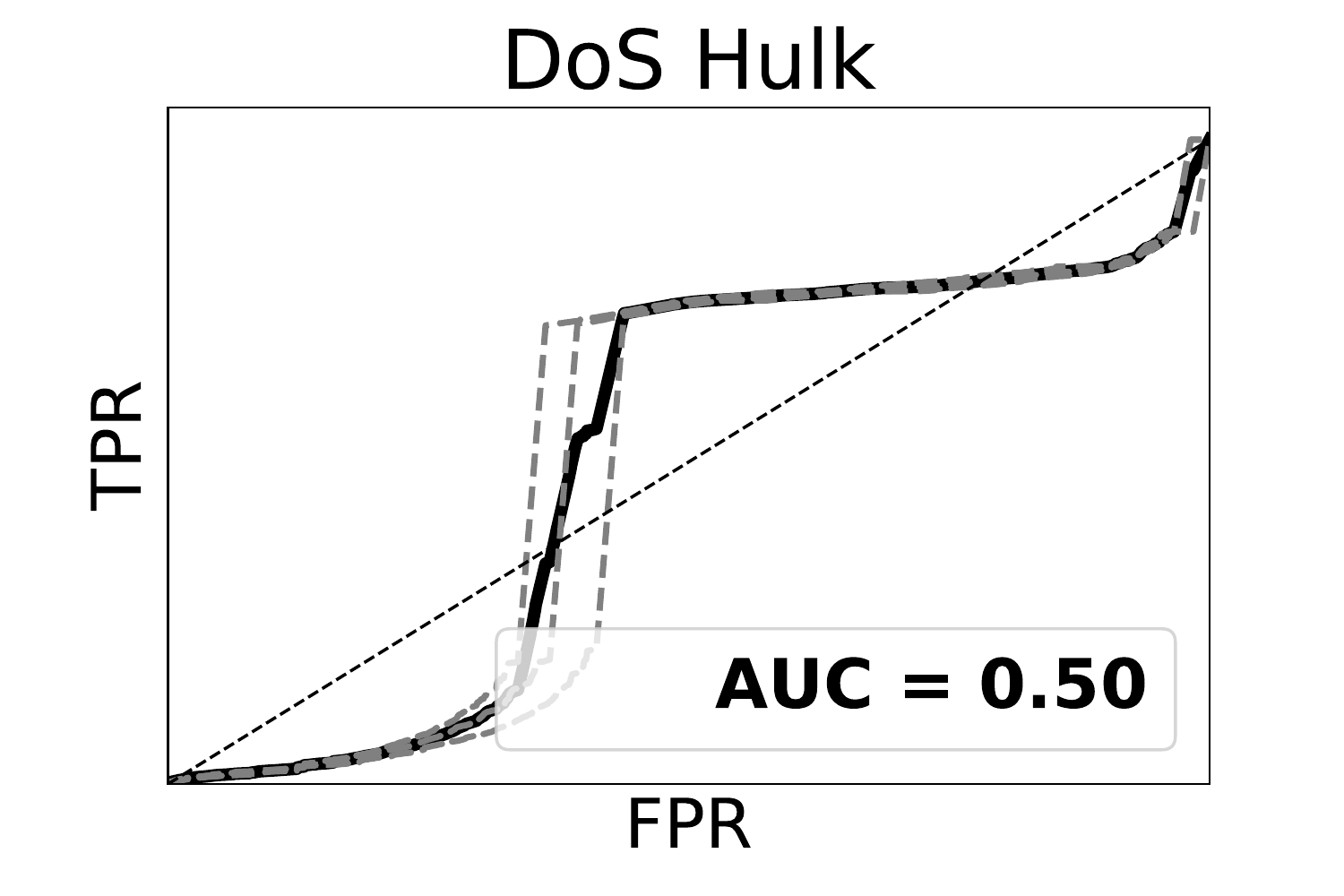} &
      \includegraphics[width=.19\textwidth]{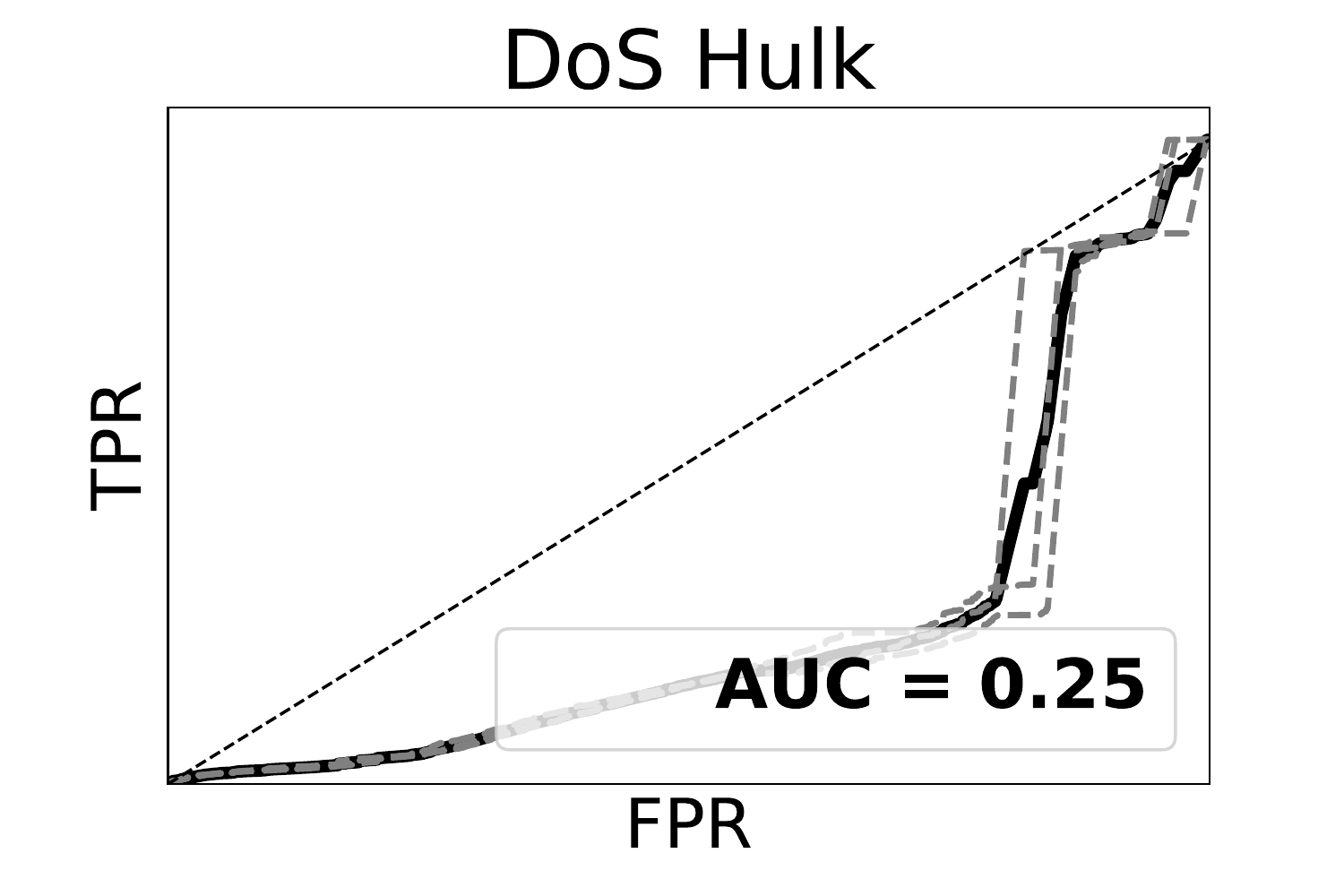} &
      \includegraphics[width=.19\textwidth]{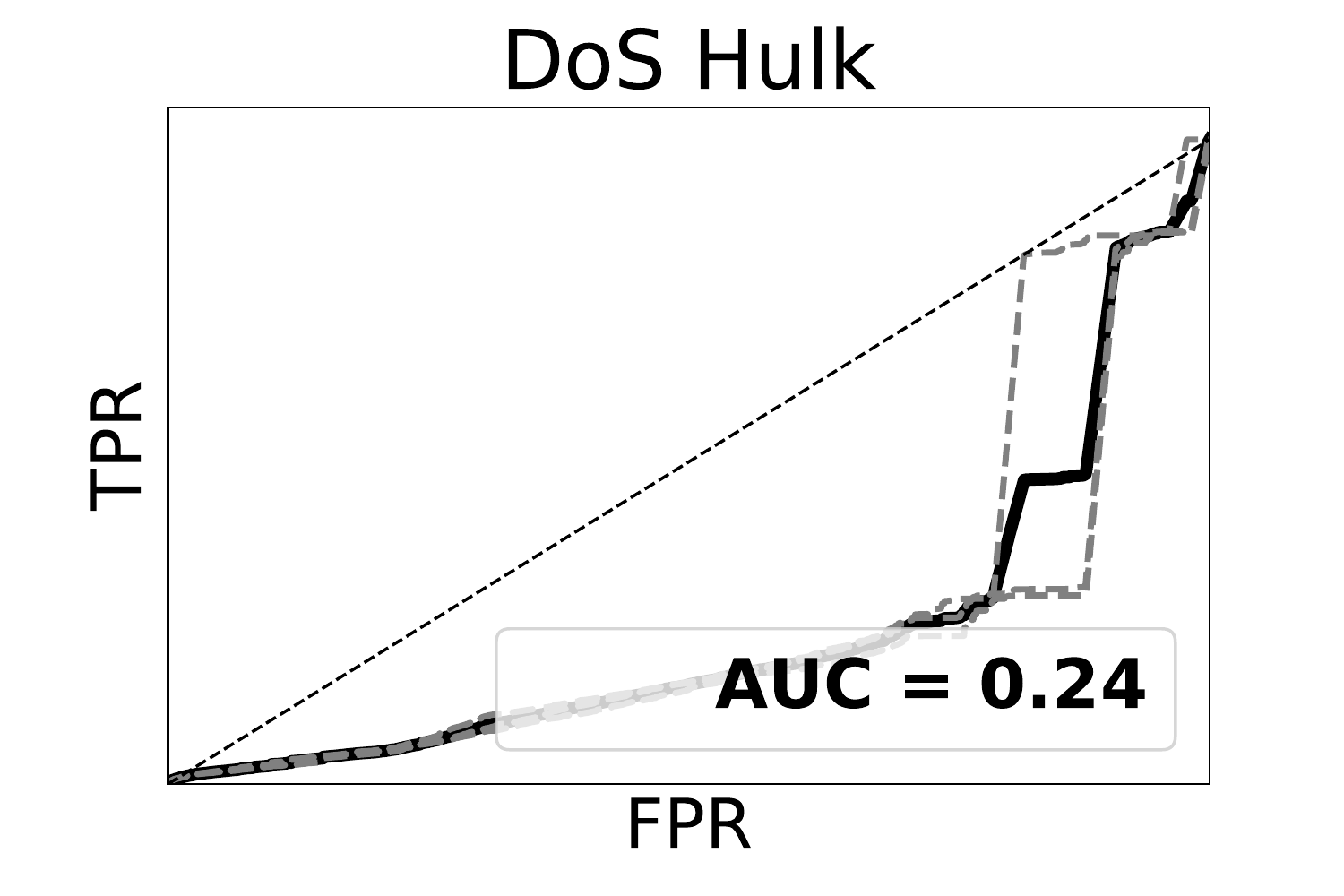} &
      \includegraphics[width=.19\textwidth]{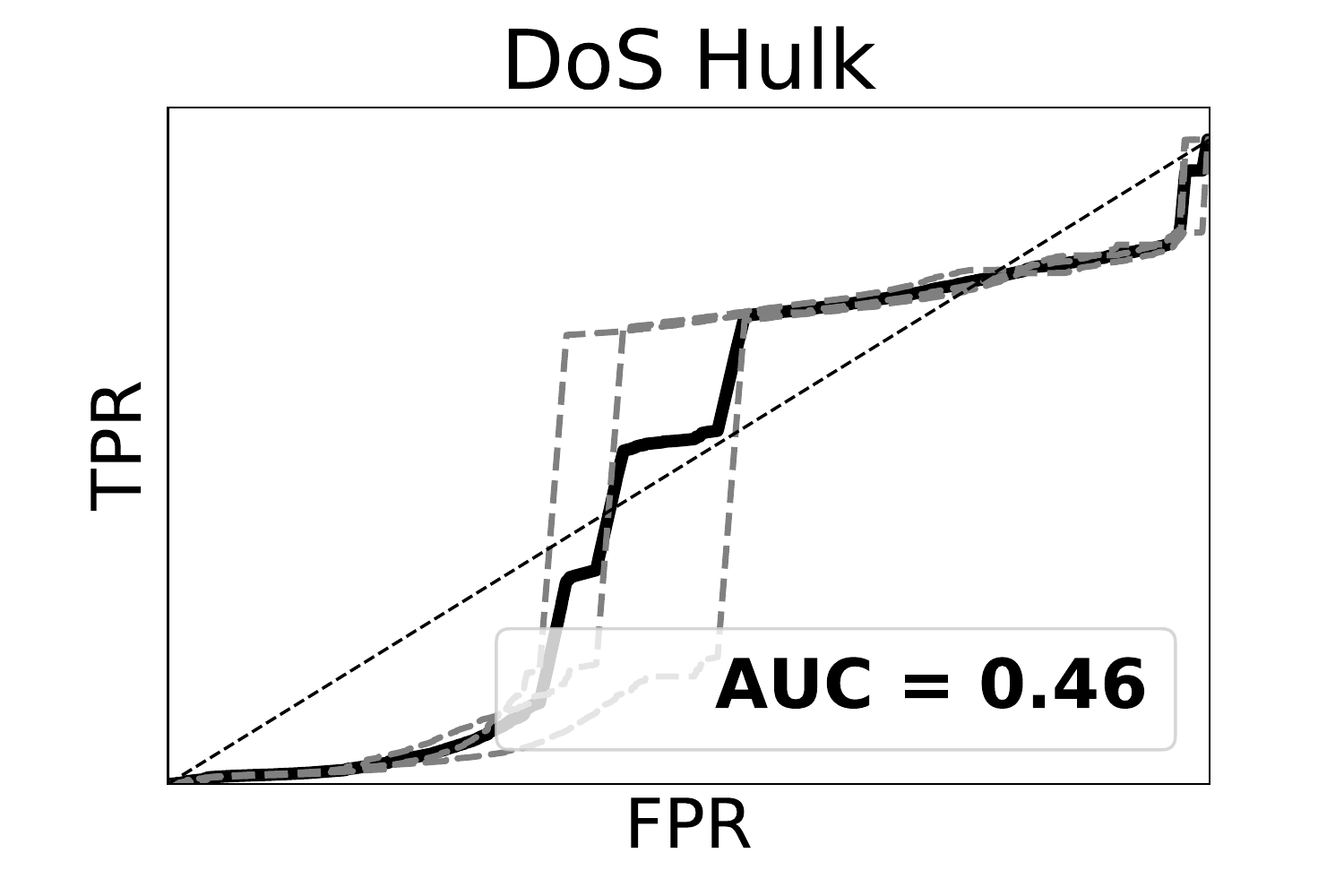} &
      \includegraphics[width=.19\textwidth]{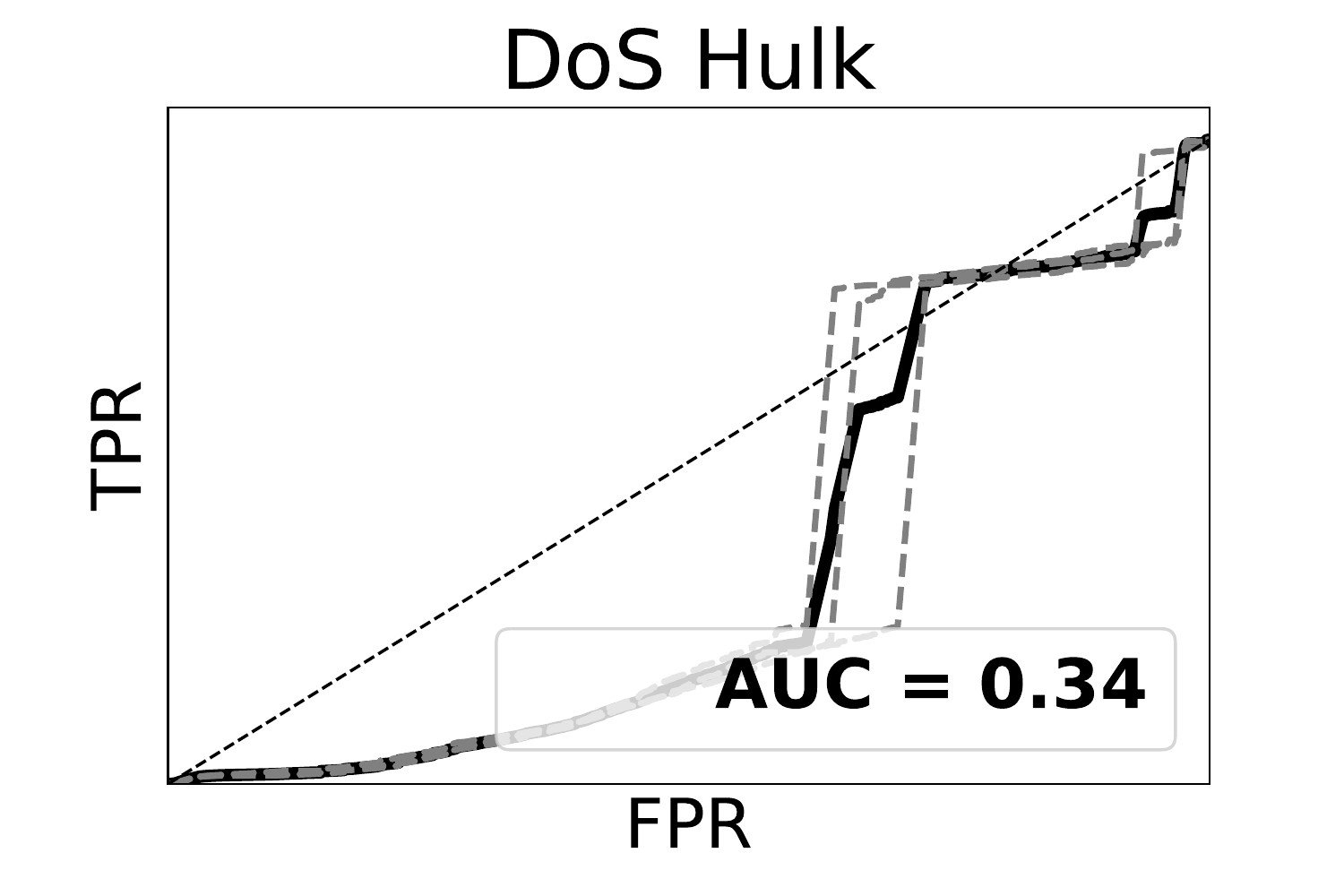}
      \\
            \includegraphics[width=.19\textwidth]{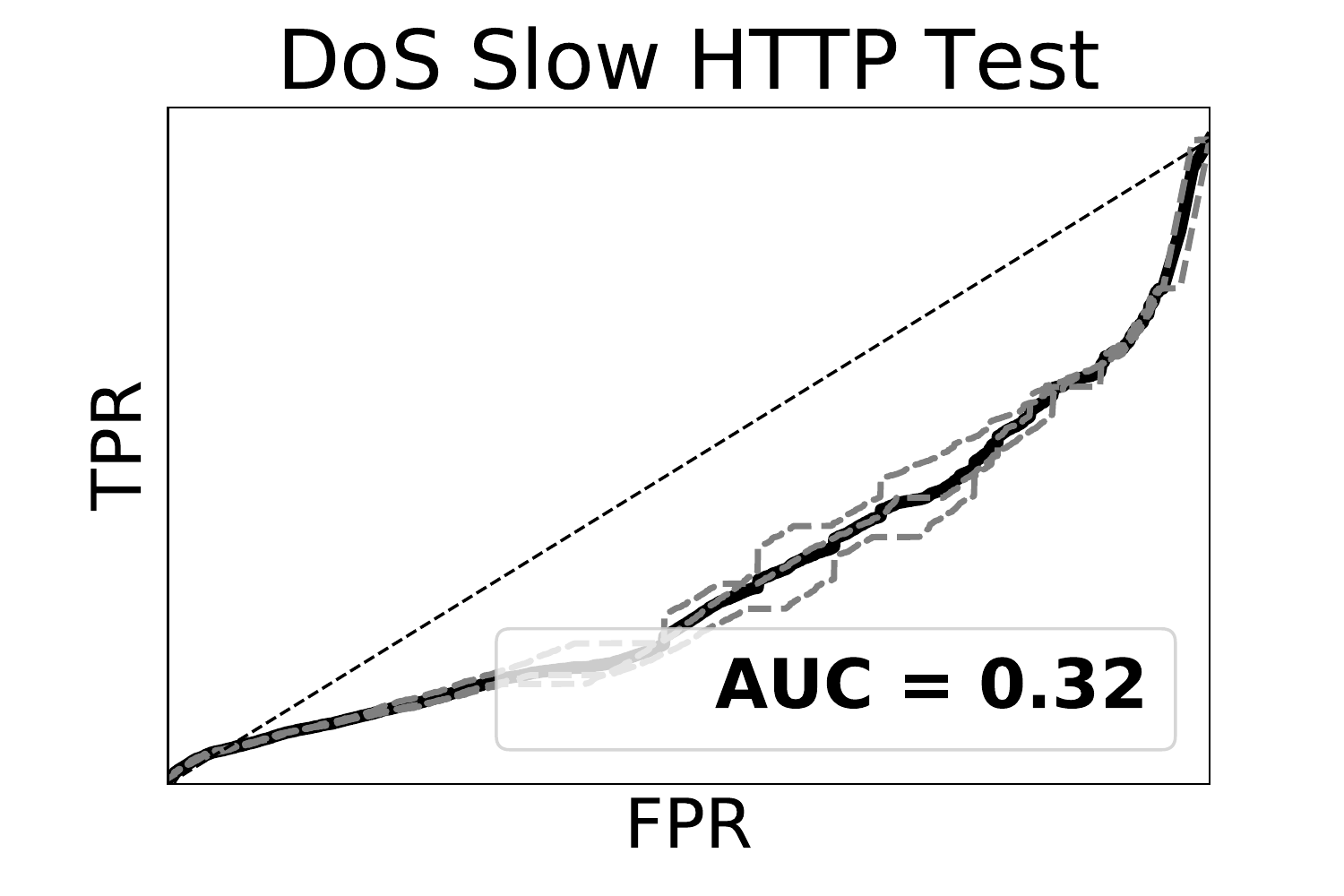} &
      \includegraphics[width=.19\textwidth]{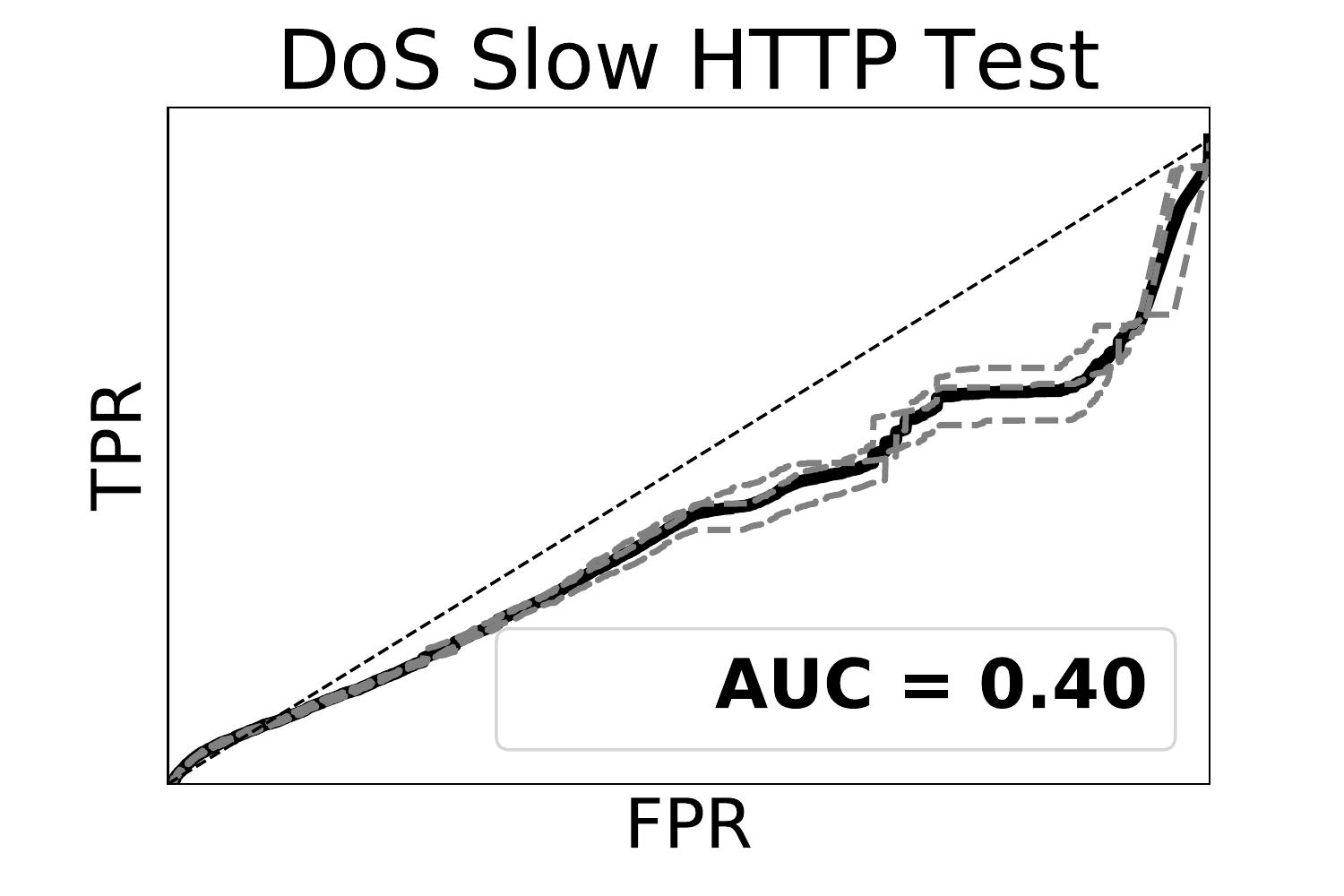} &
      \includegraphics[width=.19\textwidth]{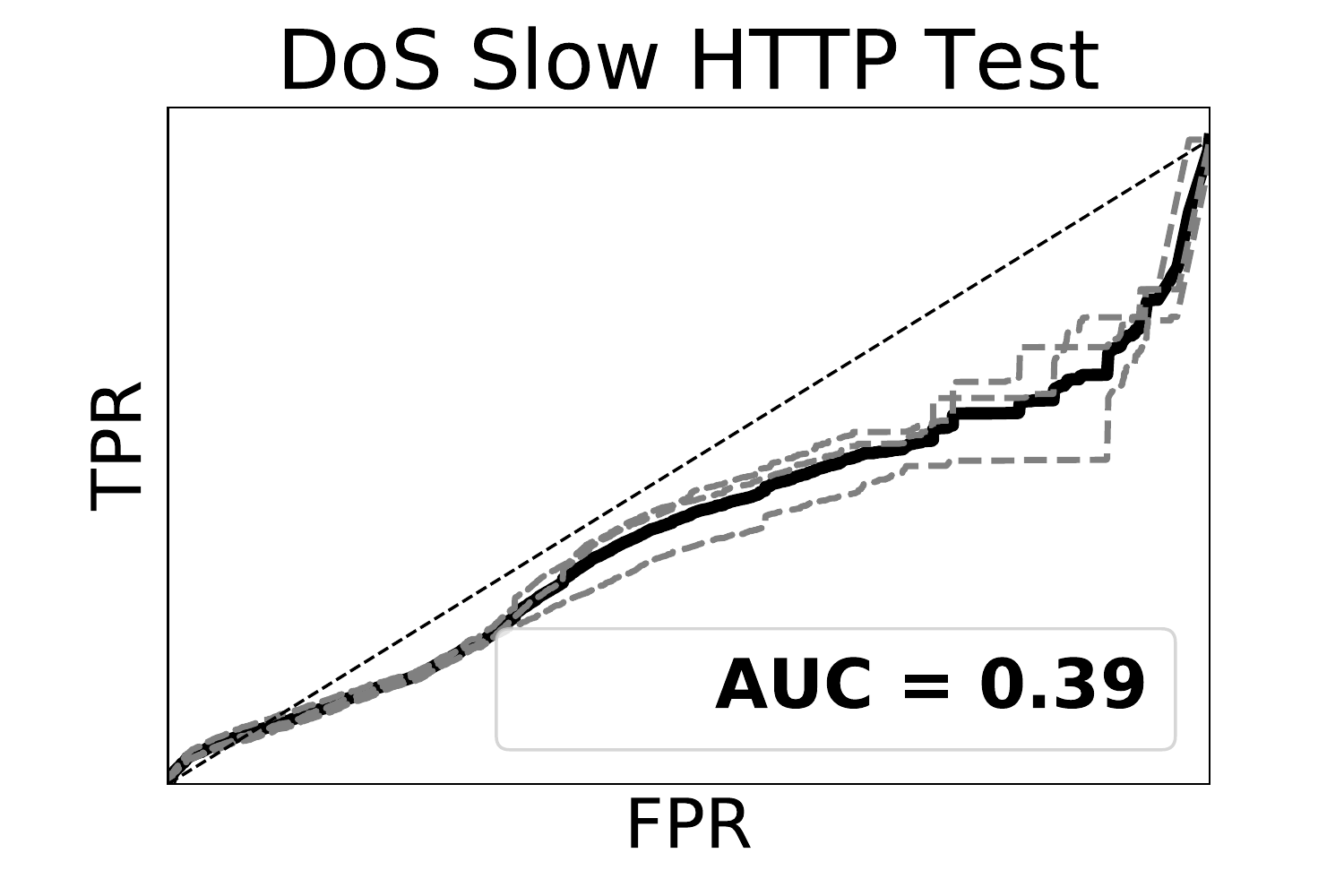} &
      \includegraphics[width=.19\textwidth]{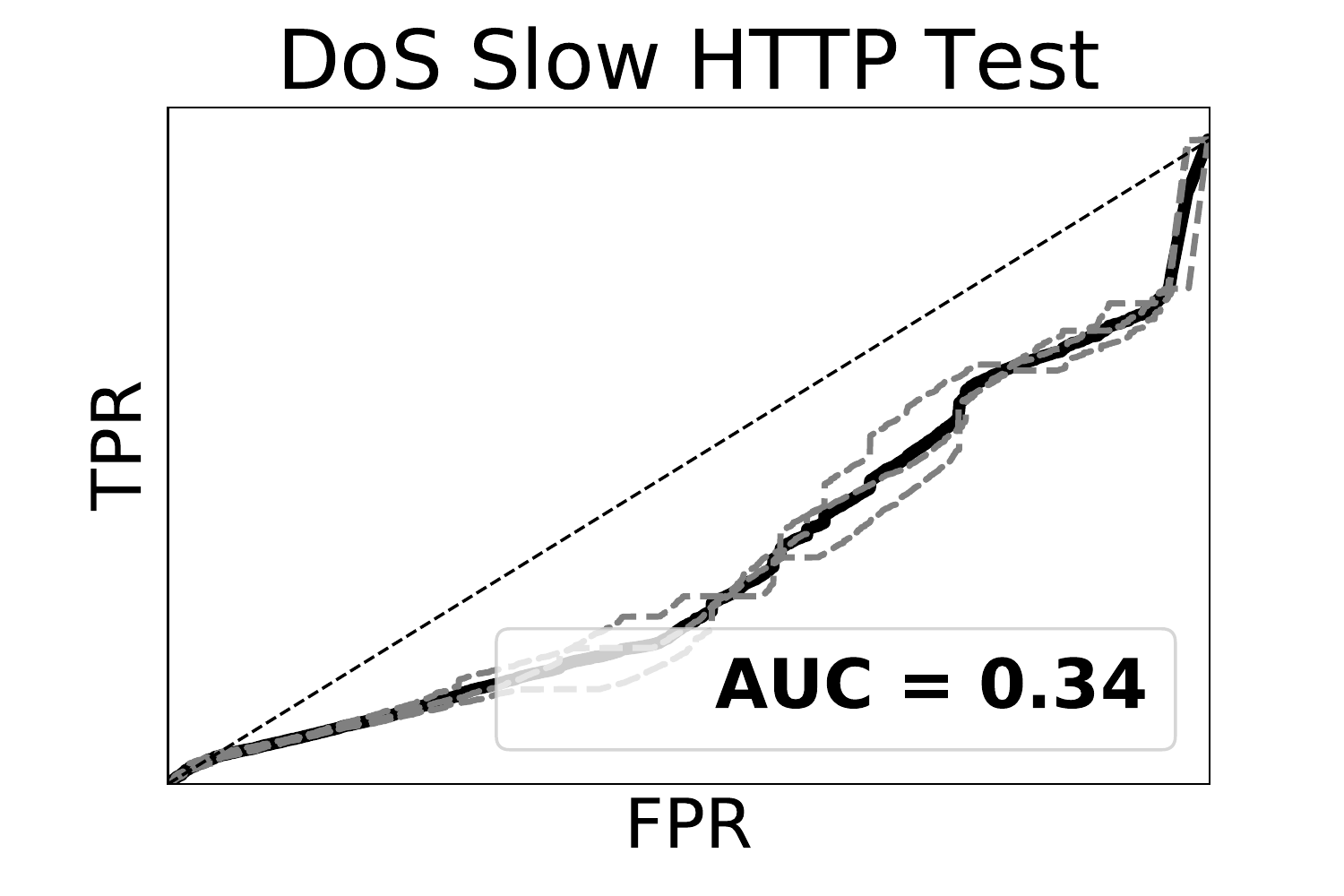} &
      \includegraphics[width=.19\textwidth]{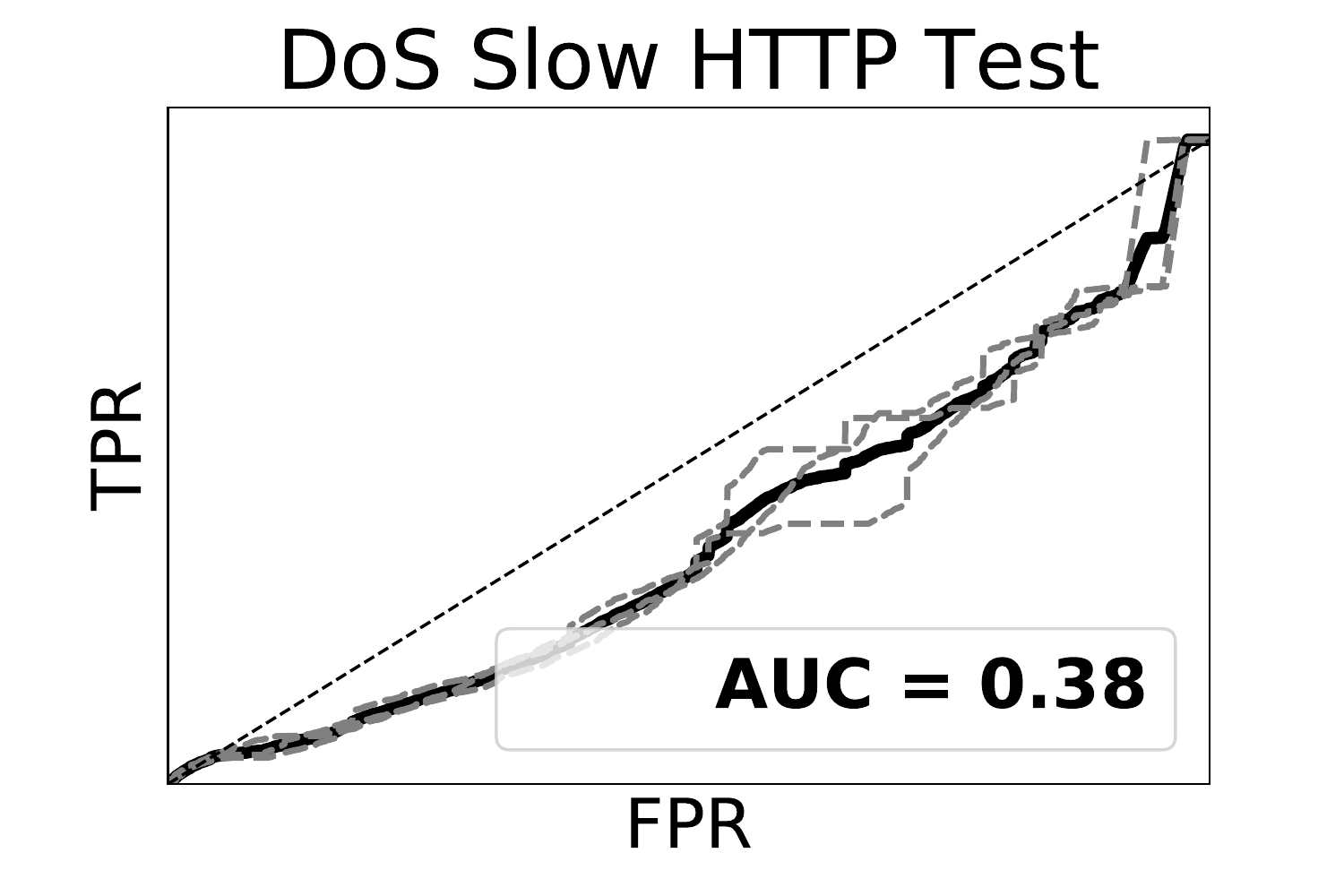}
      \\
            \includegraphics[width=.19\textwidth]{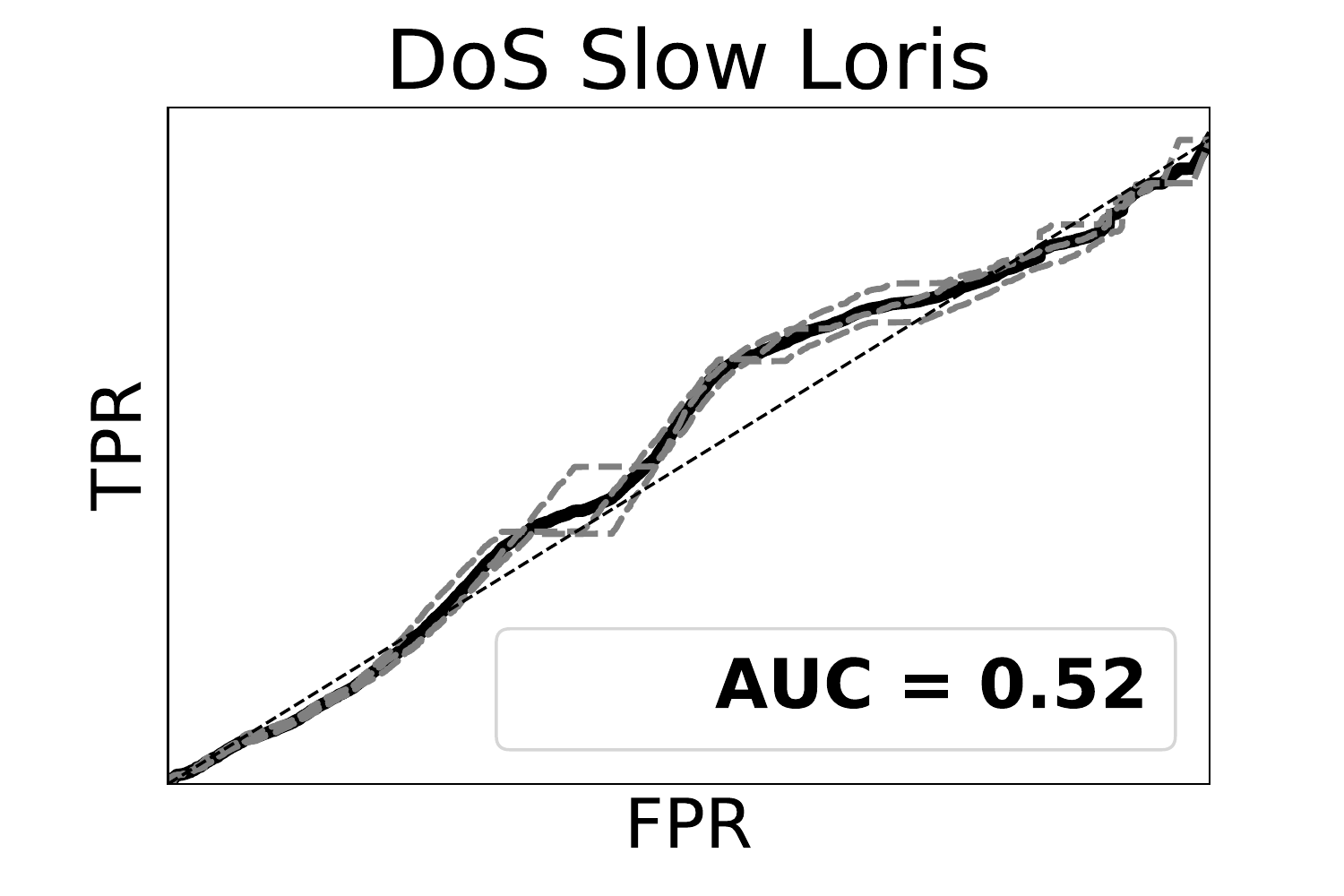} &
      \includegraphics[width=.19\textwidth]{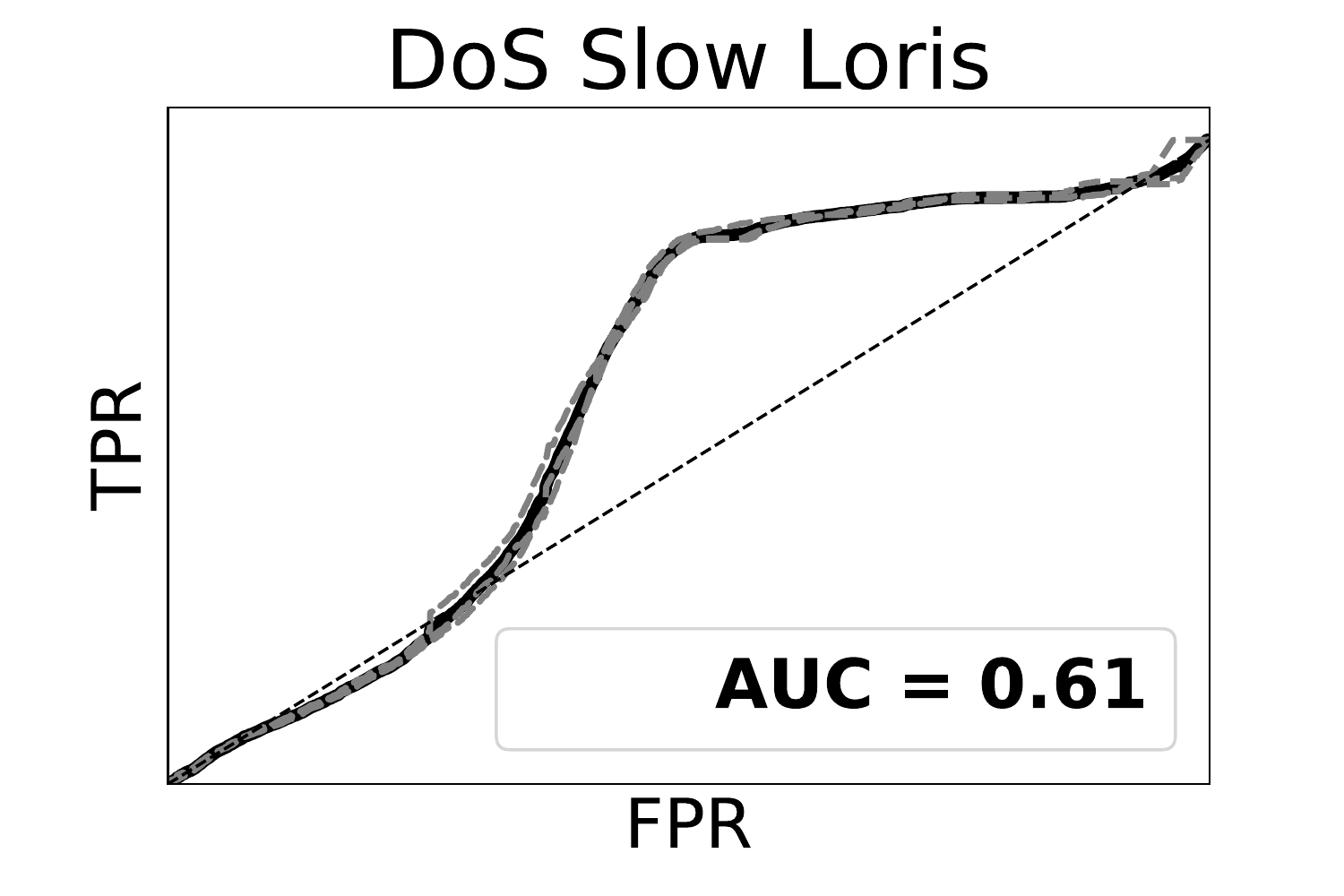} &
      \includegraphics[width=.19\textwidth]{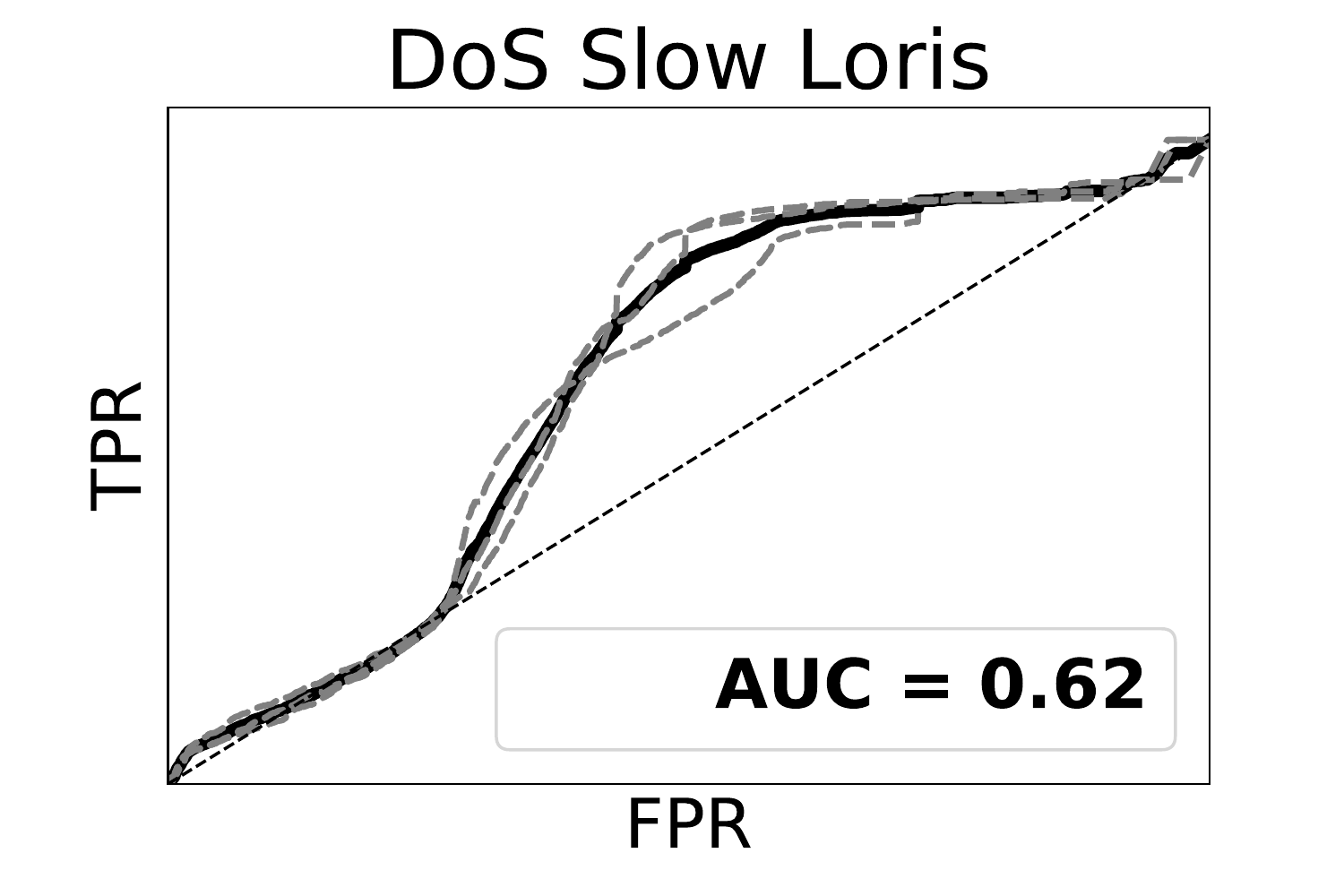} &
      \includegraphics[width=.19\textwidth]{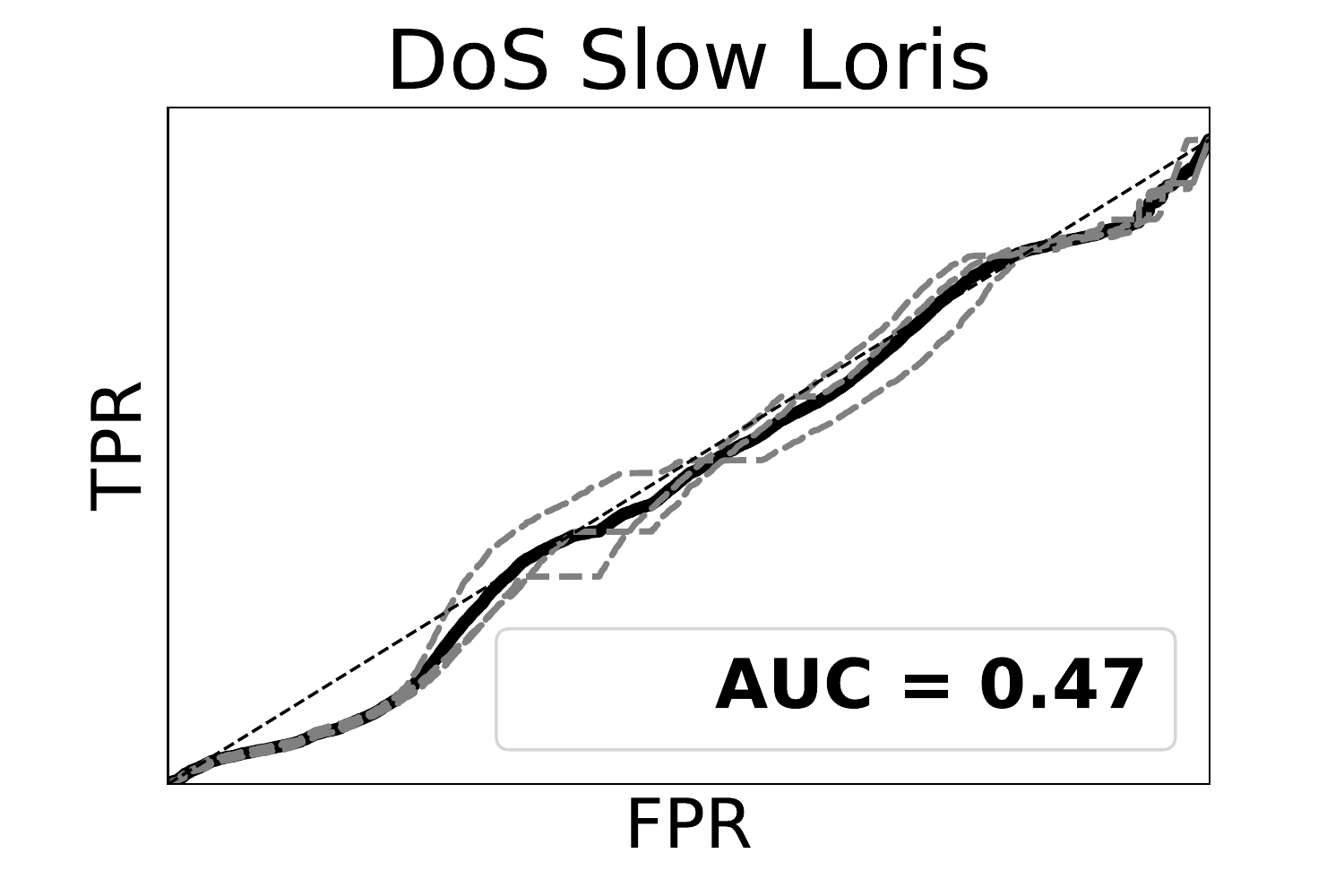} &
      \includegraphics[width=.19\textwidth]{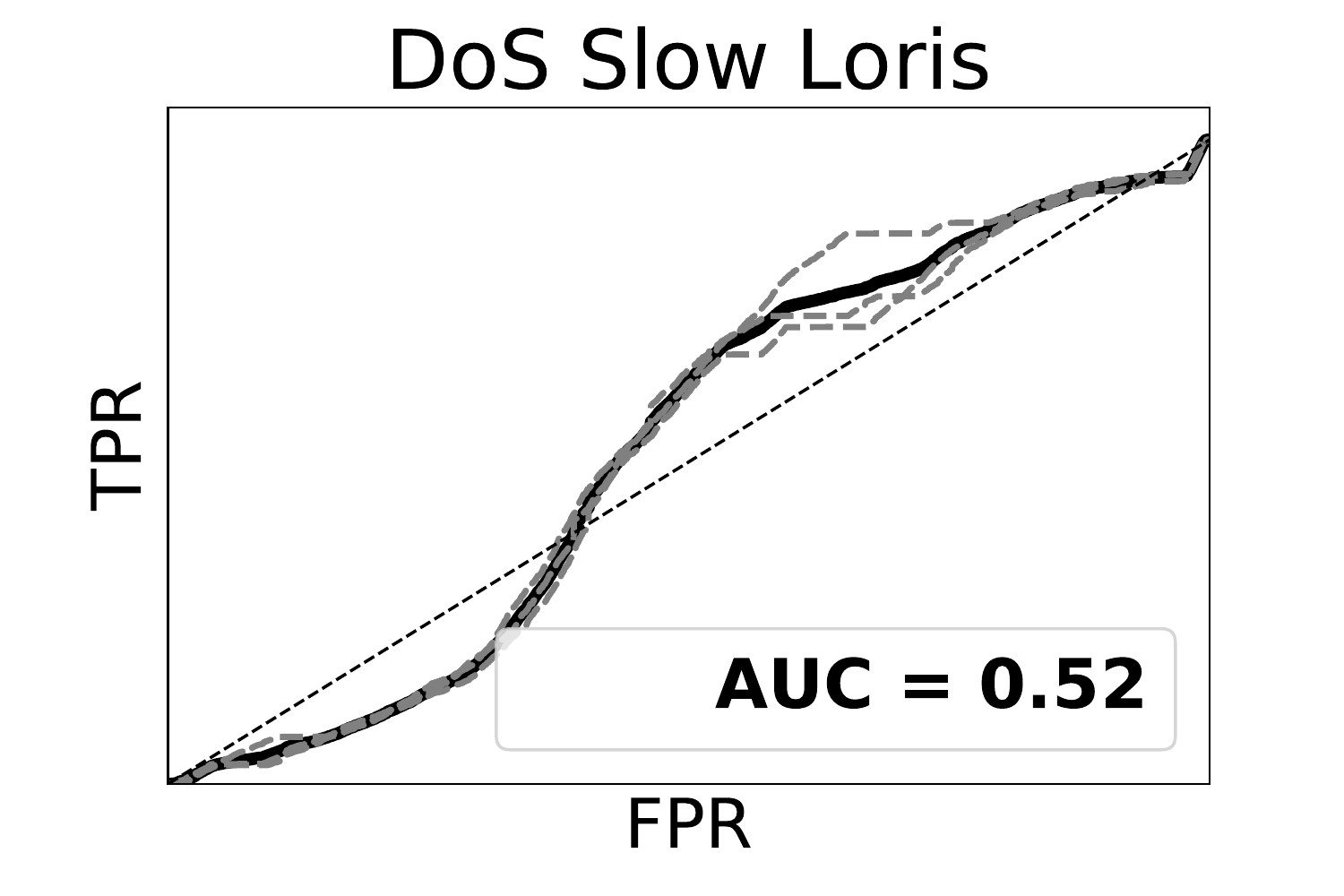}
      \\
      \includegraphics[width=.19\textwidth]{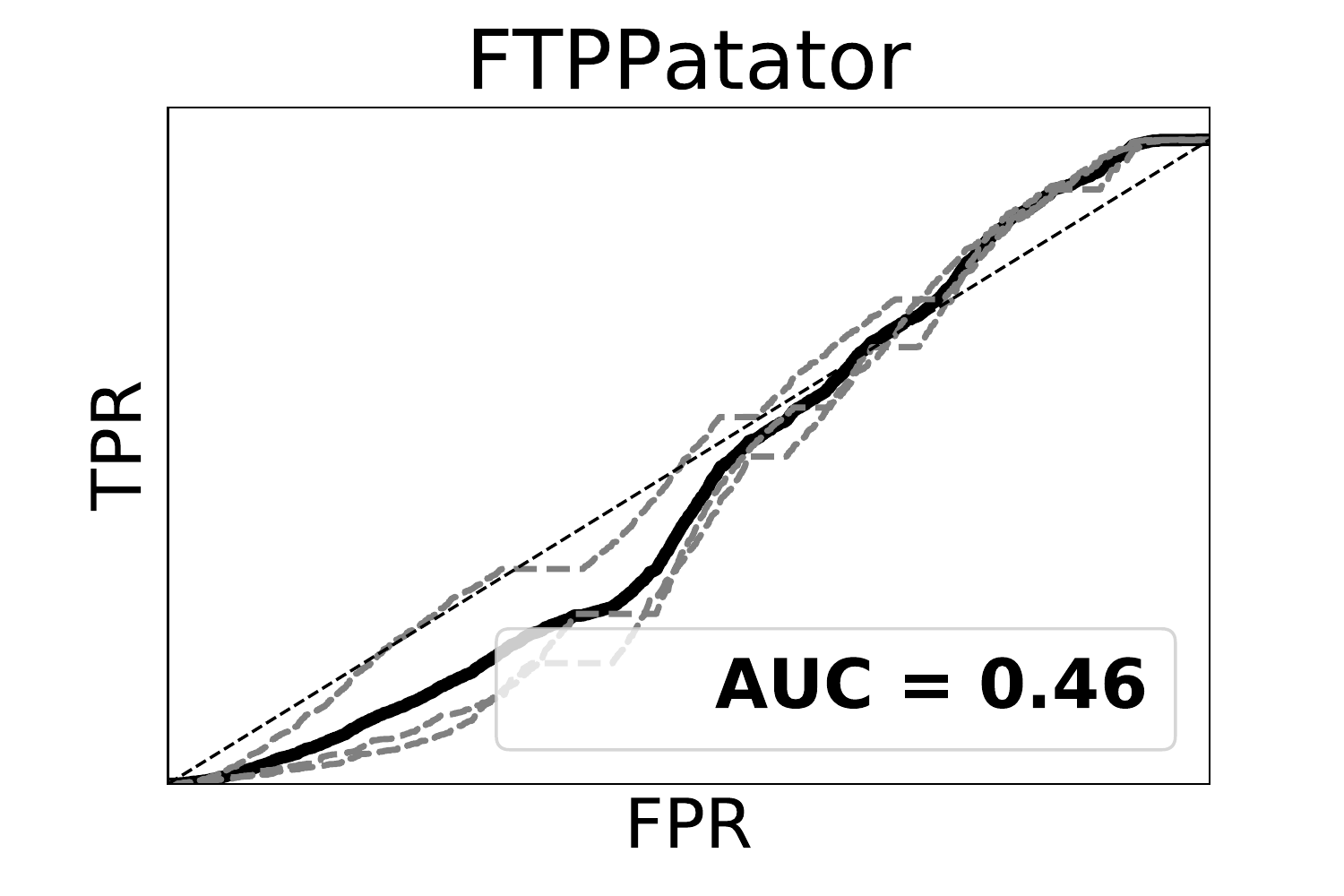} &
      \includegraphics[width=.19\textwidth]{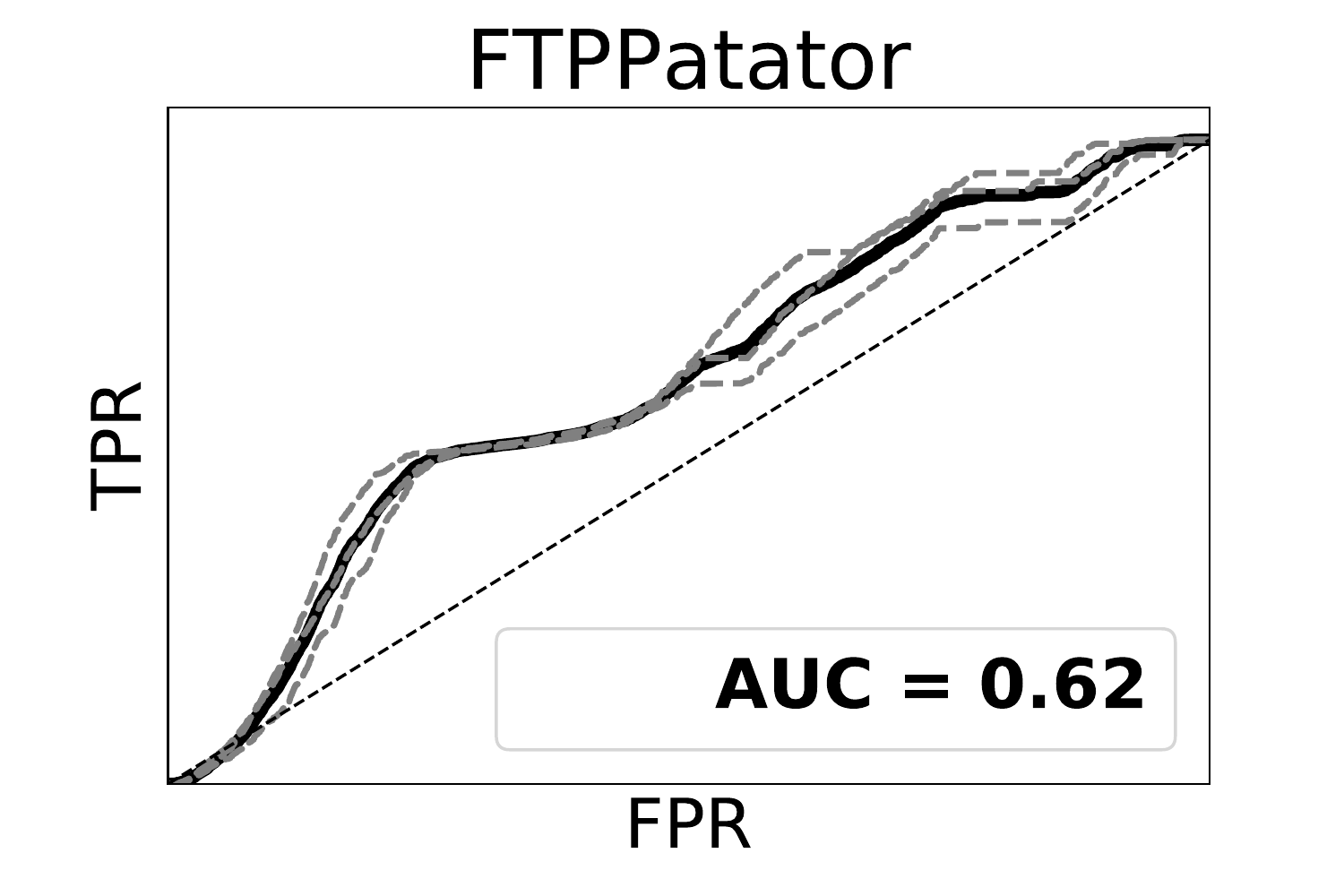} &
      \includegraphics[width=.19\textwidth]{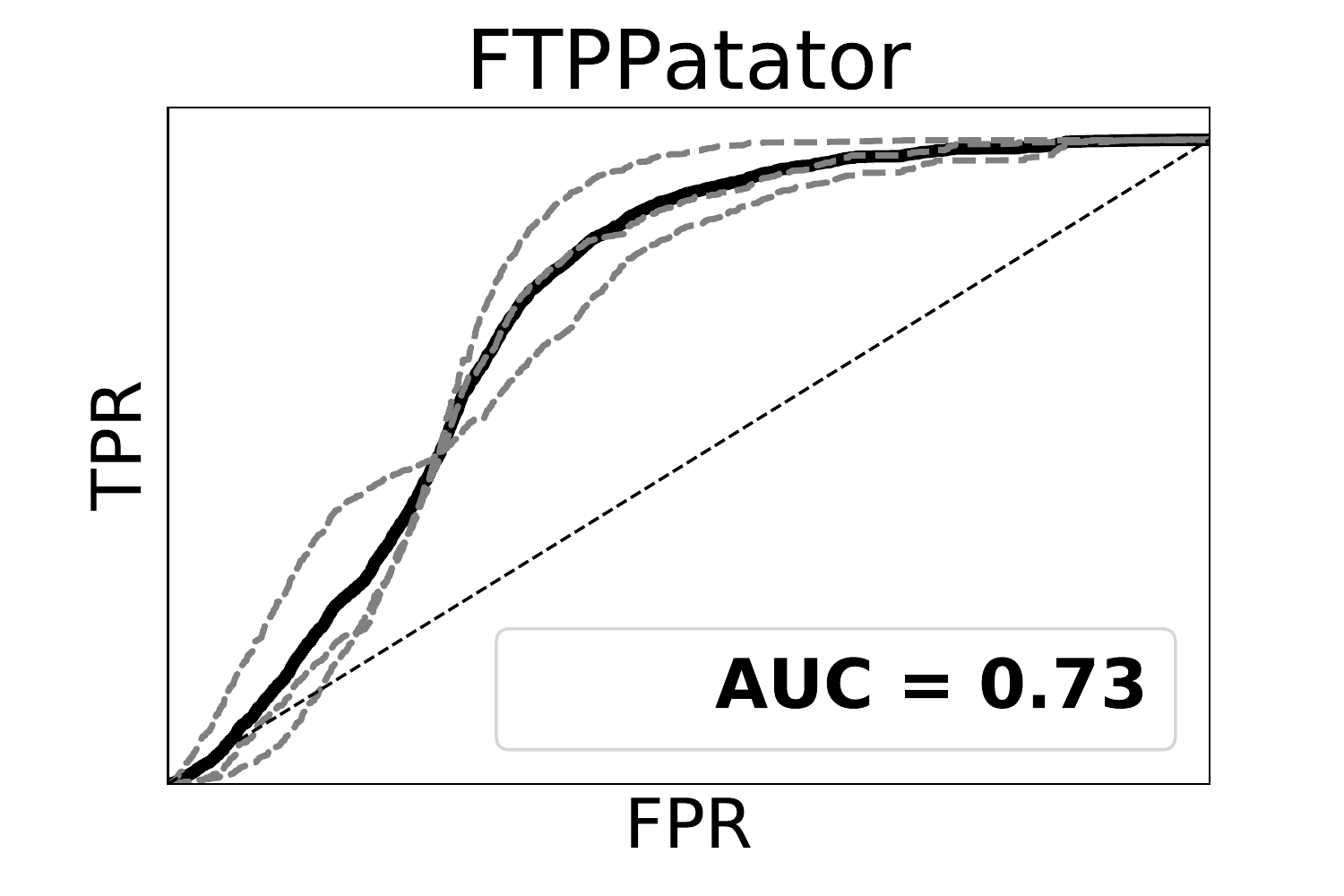} &
      \includegraphics[width=.19\textwidth]{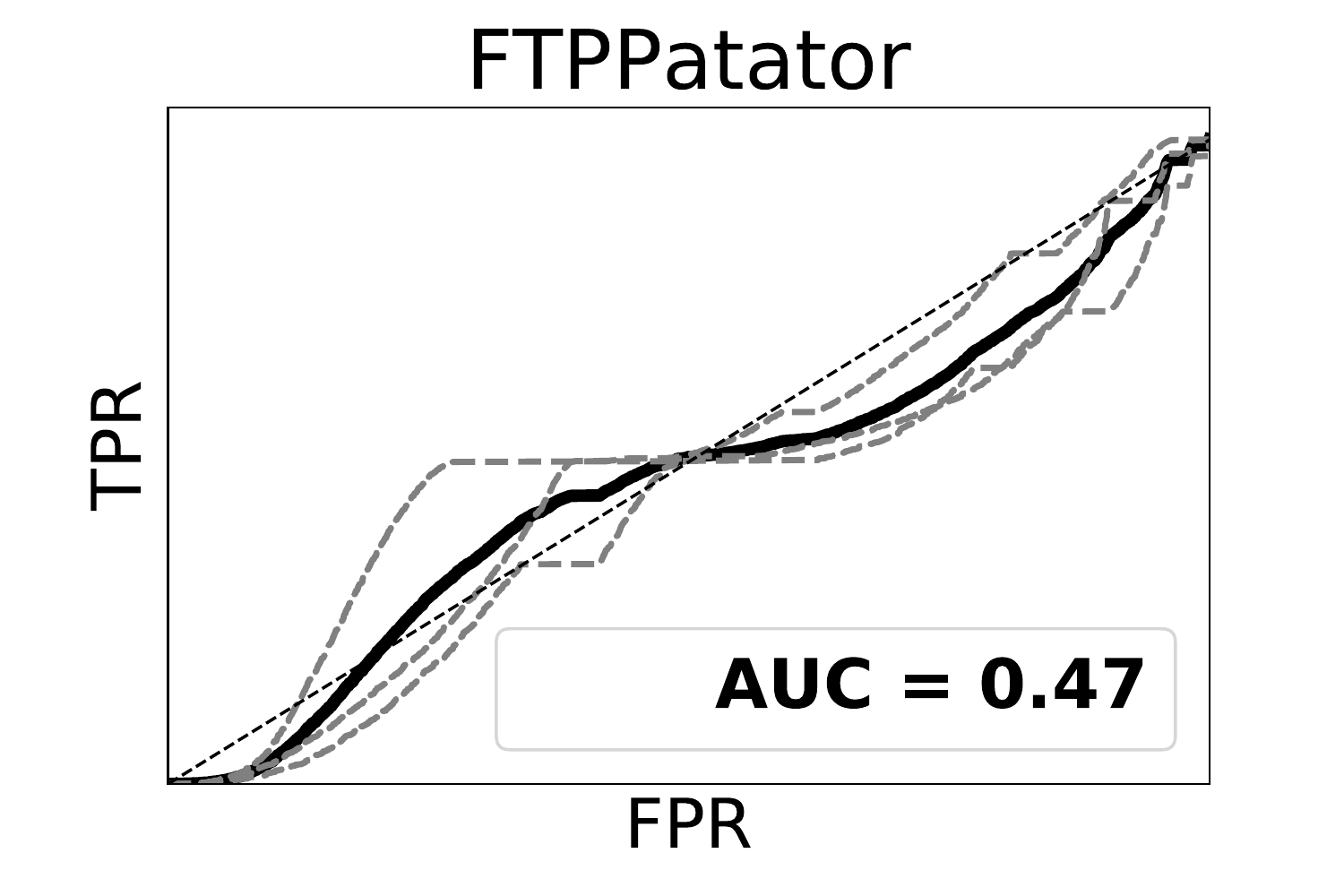} &
      \includegraphics[width=.19\textwidth]{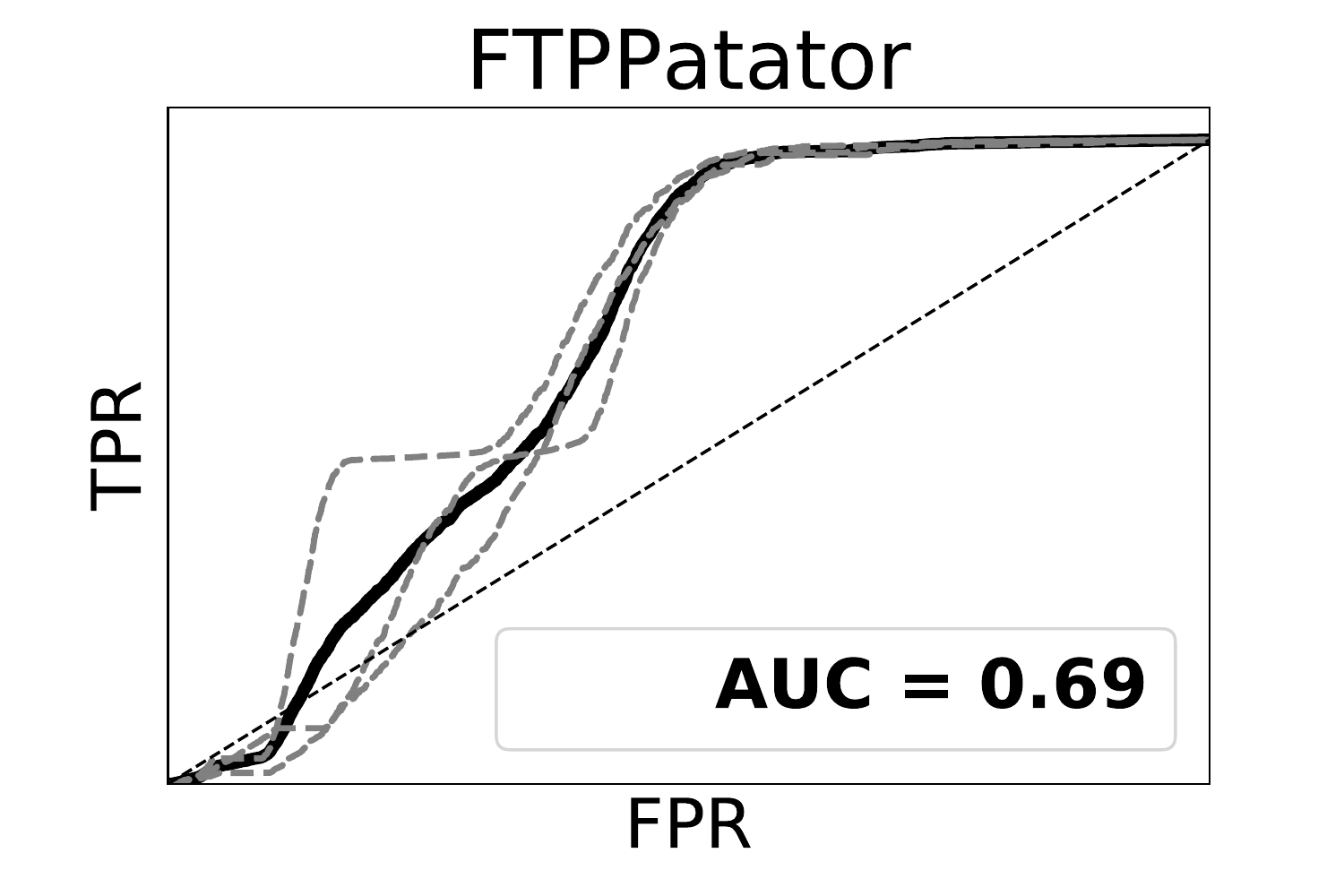}
      \\
  \end{tabular}
\end{table*}      
      
 \begin{table*}
  [ht] \caption{Protobyte Sequences B} \label{tab:protobytesb}
  \begin{tabular}{ccccc} 
  \hline 
  Source & Destination & Dyad & Internal & External \\
      \hline      
      \includegraphics[width=.19\textwidth]{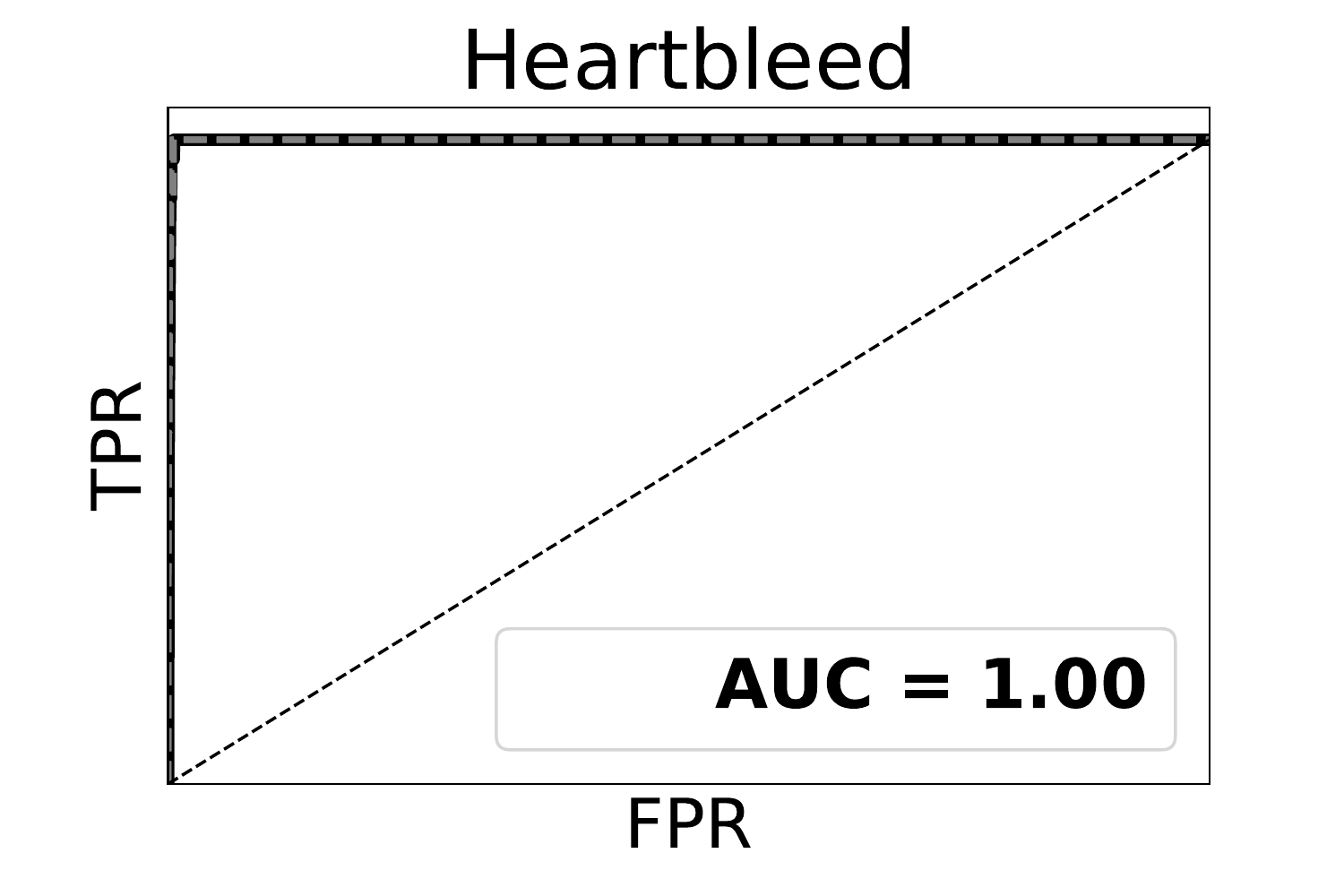} &
      \includegraphics[width=.19\textwidth]{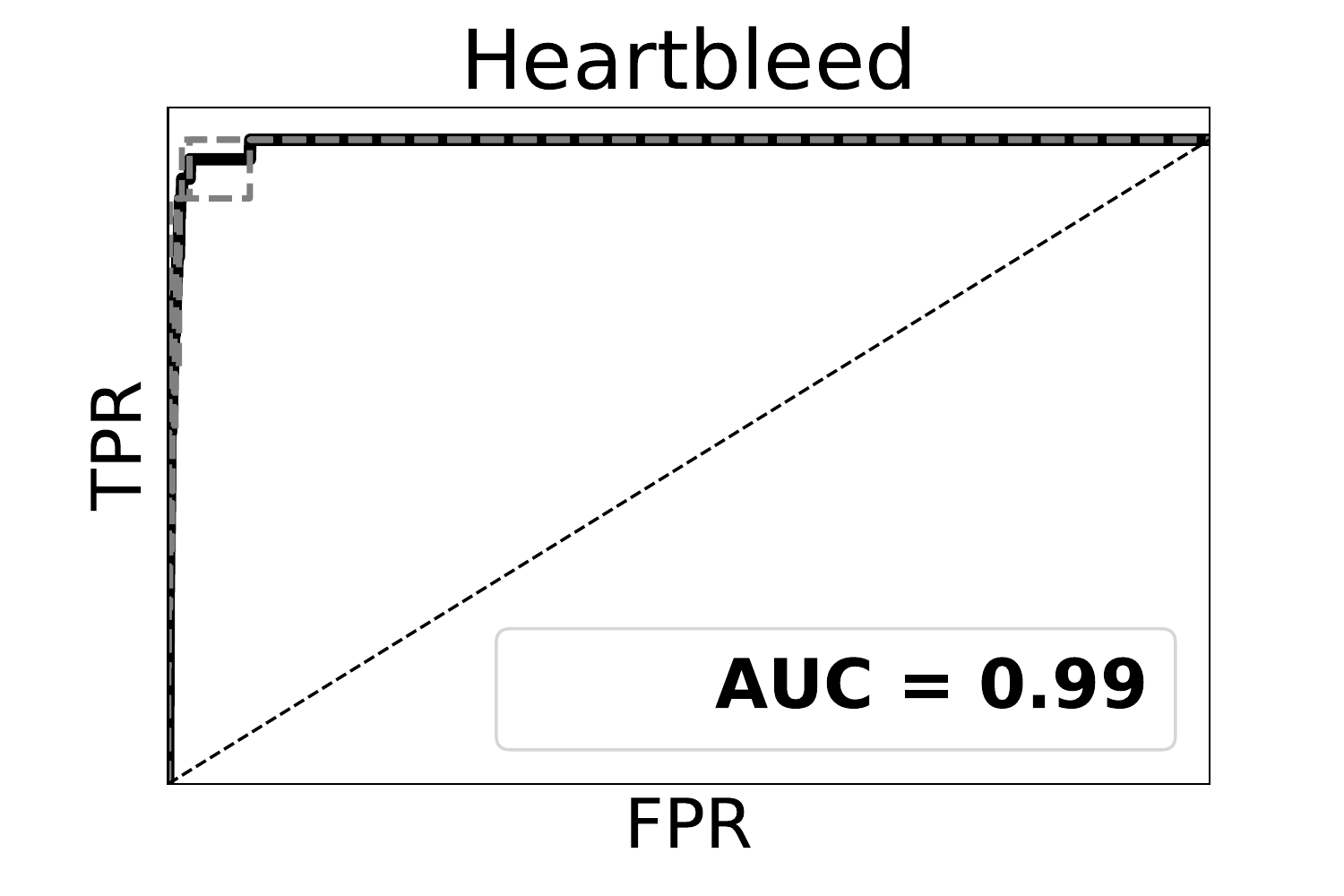} &
      \includegraphics[width=.19\textwidth]{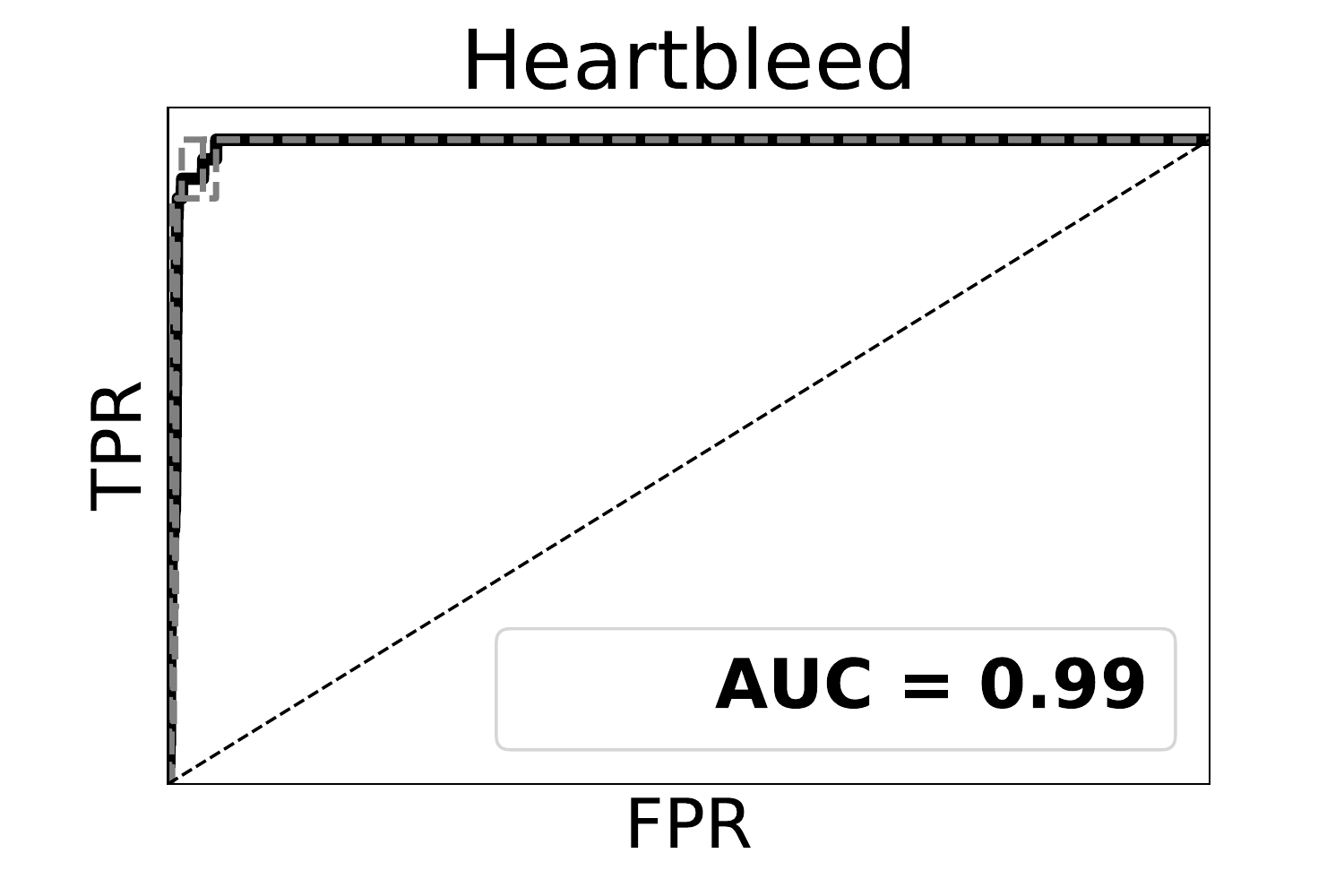} &
      \includegraphics[width=.19\textwidth]{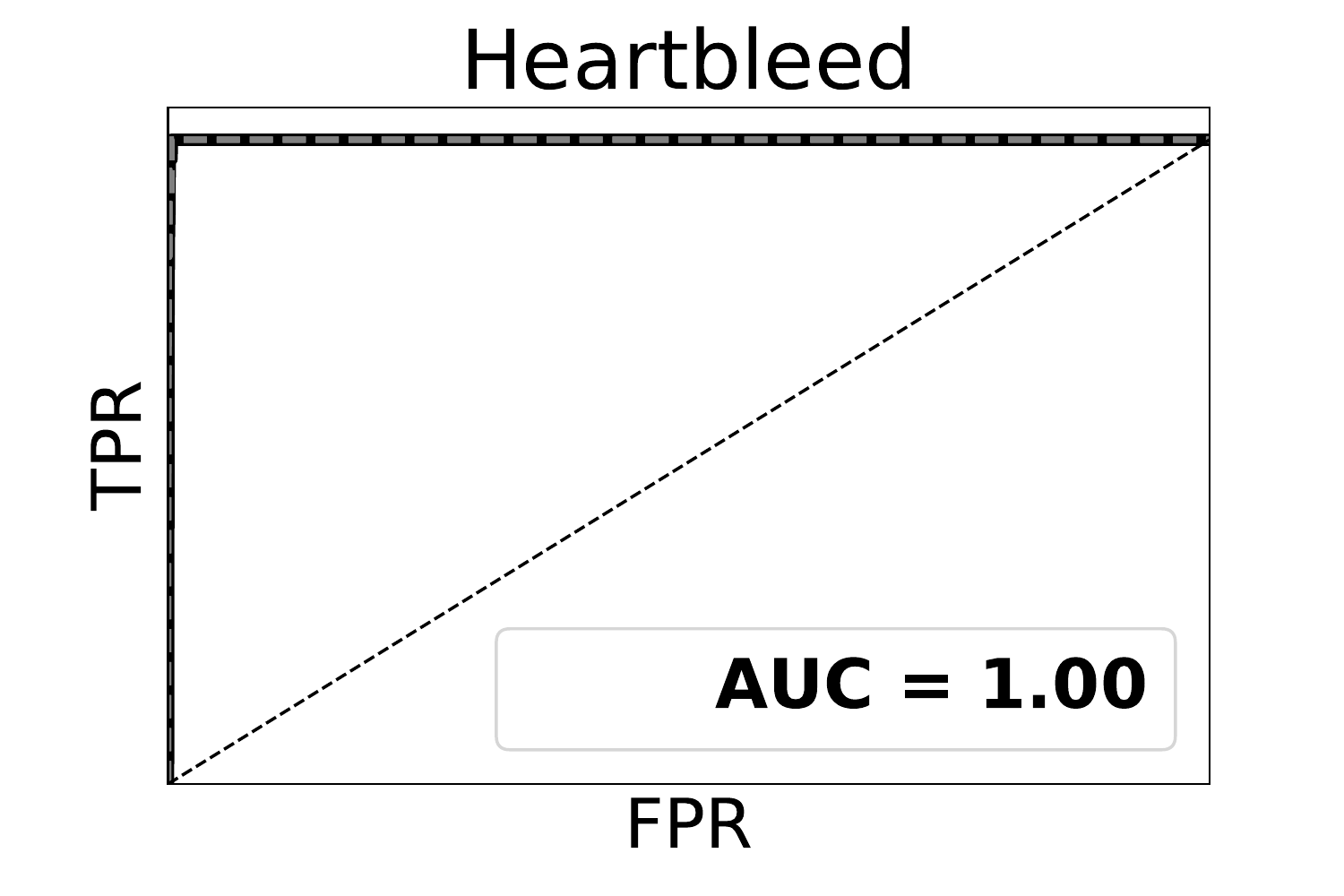} &
      \includegraphics[width=.19\textwidth]{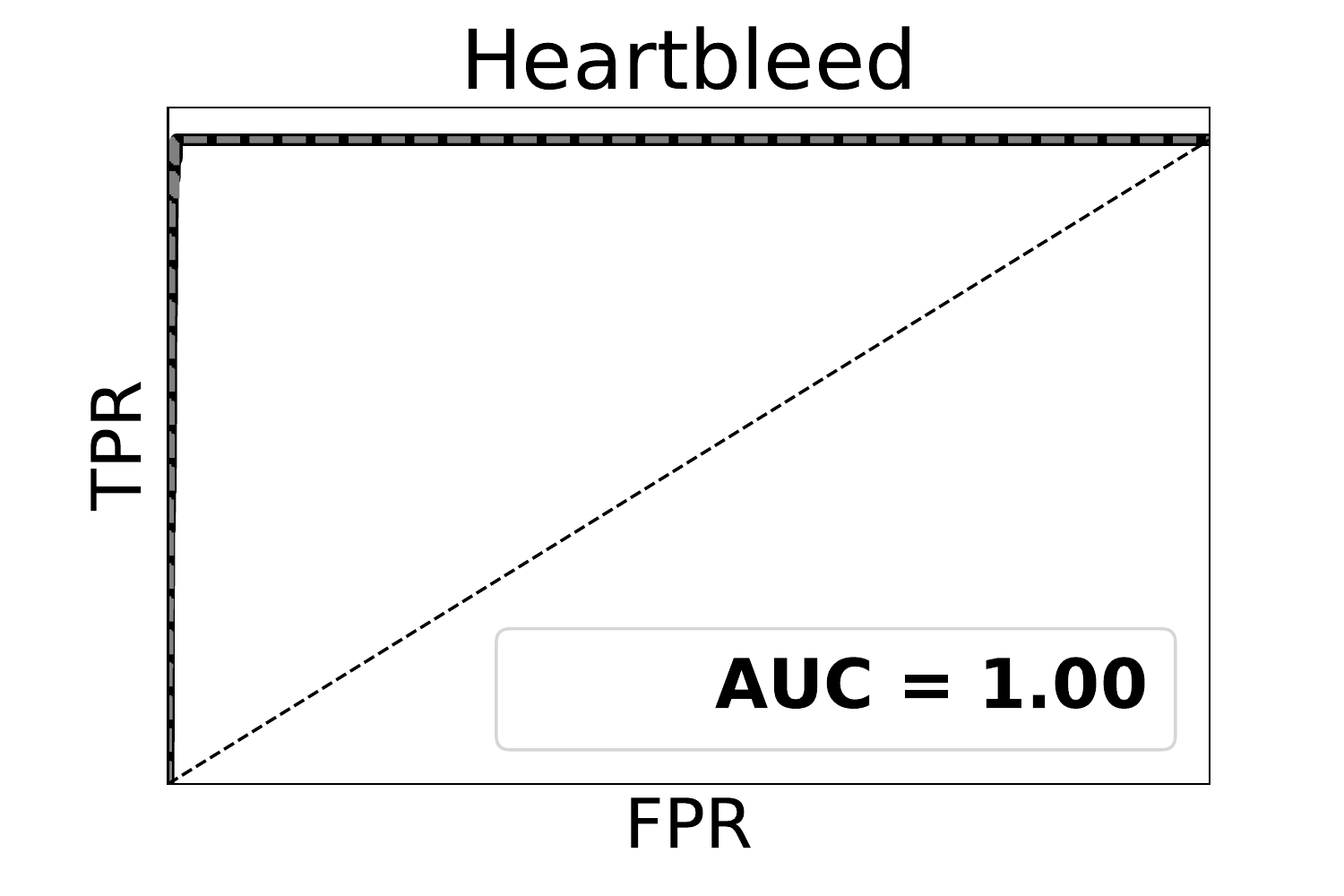}
      \\
      \includegraphics[width=.19\textwidth]{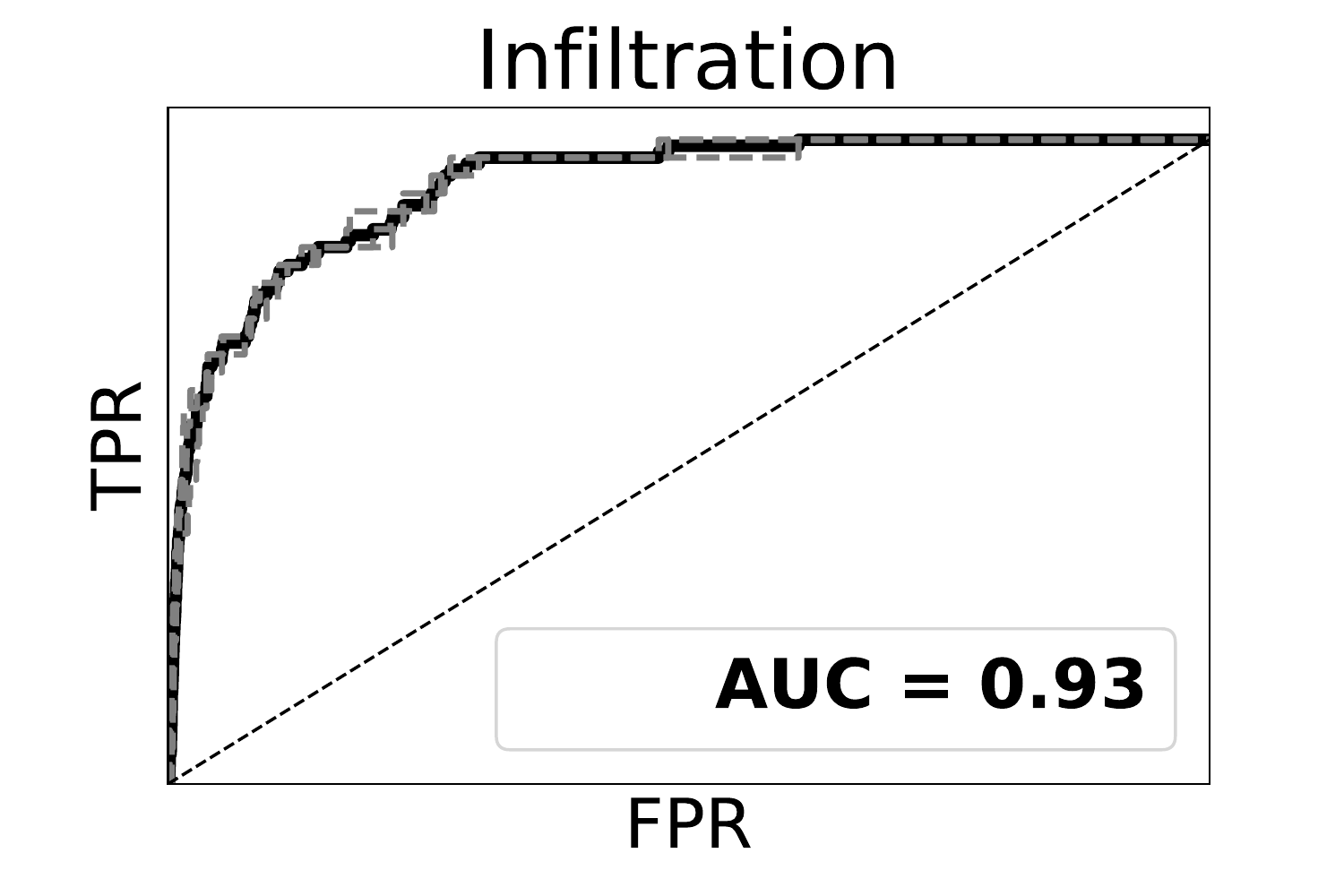} &
      \includegraphics[width=.19\textwidth]{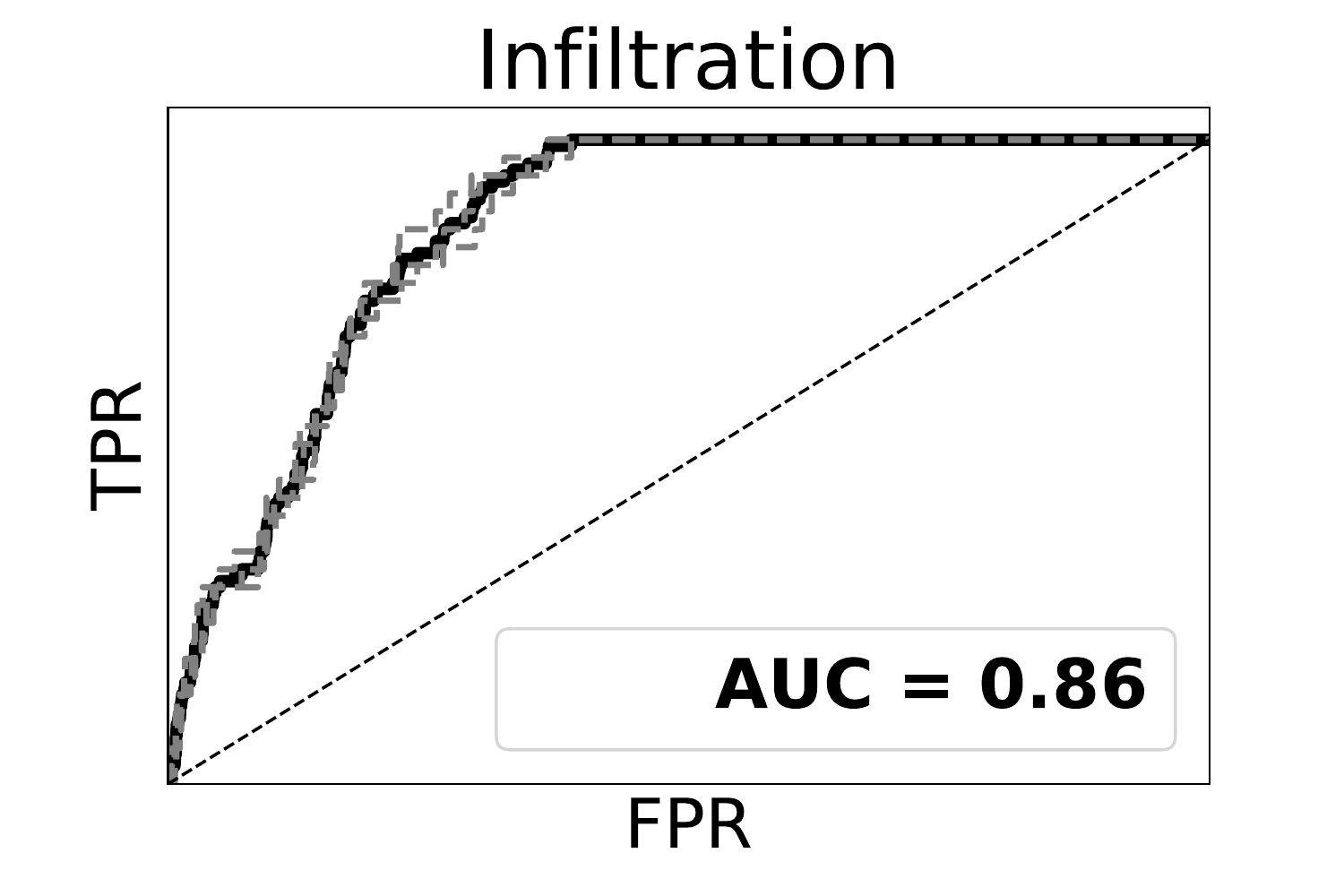} &
      \includegraphics[width=.19\textwidth]{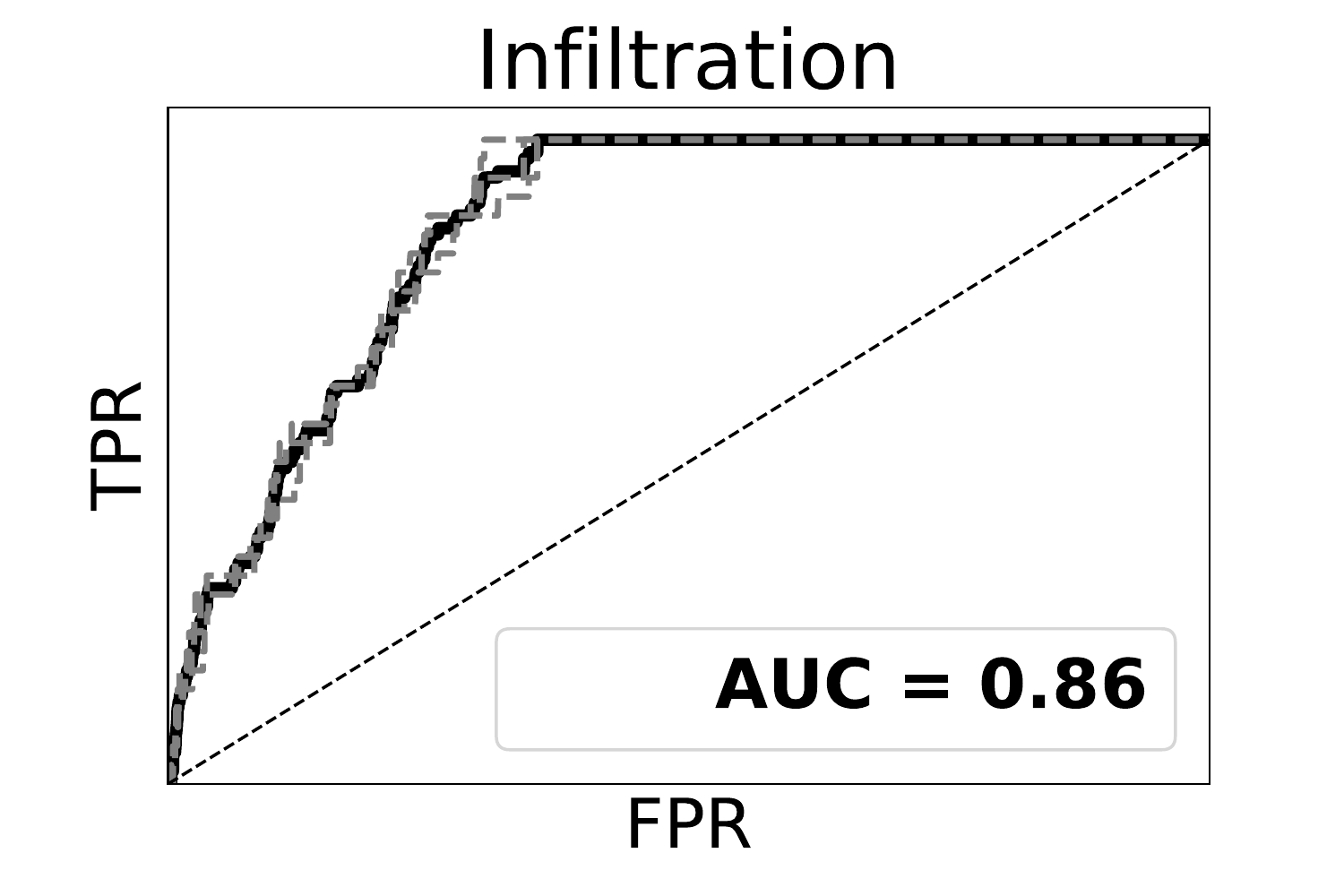} &
      \includegraphics[width=.19\textwidth]{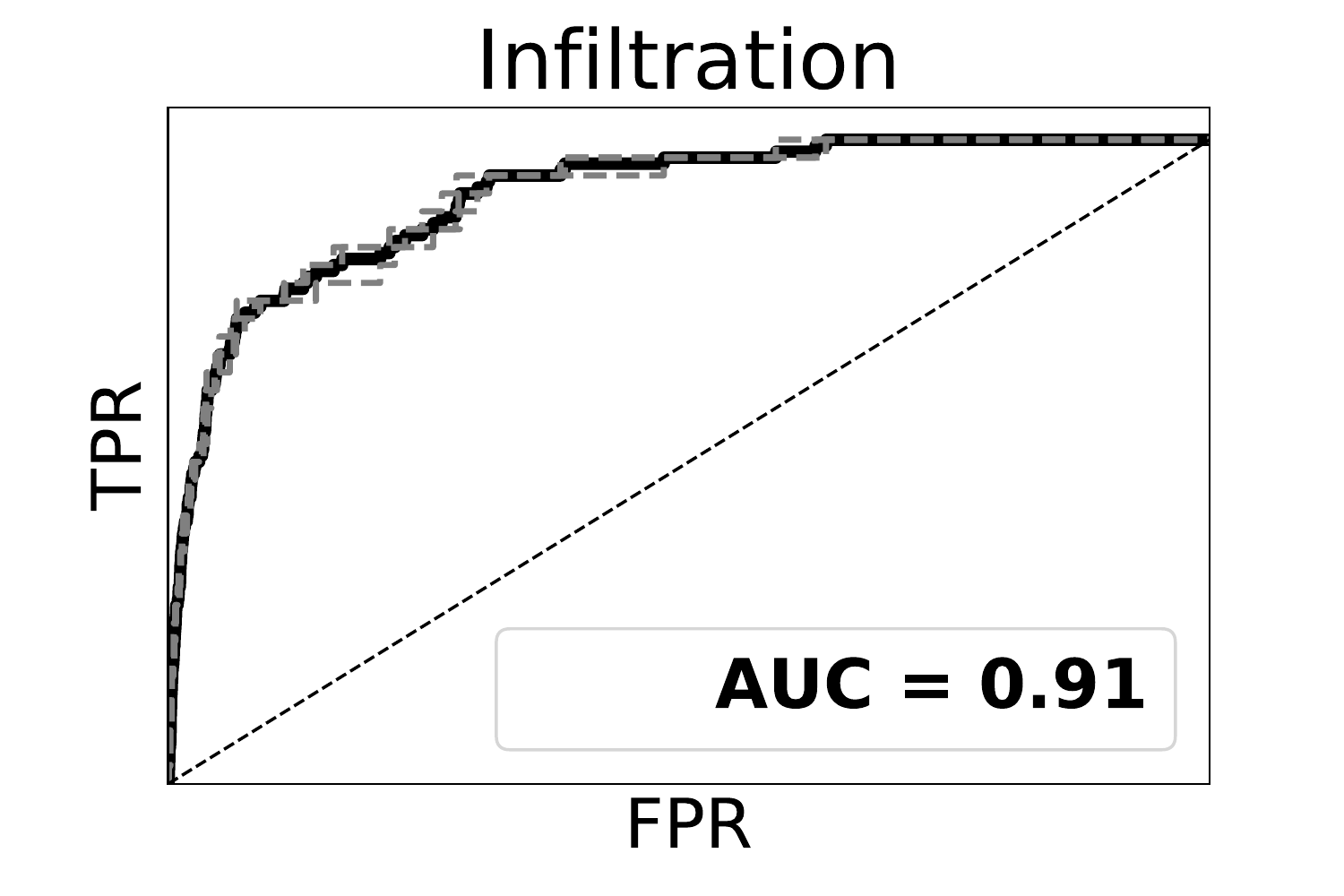} &
      \includegraphics[width=.19\textwidth]{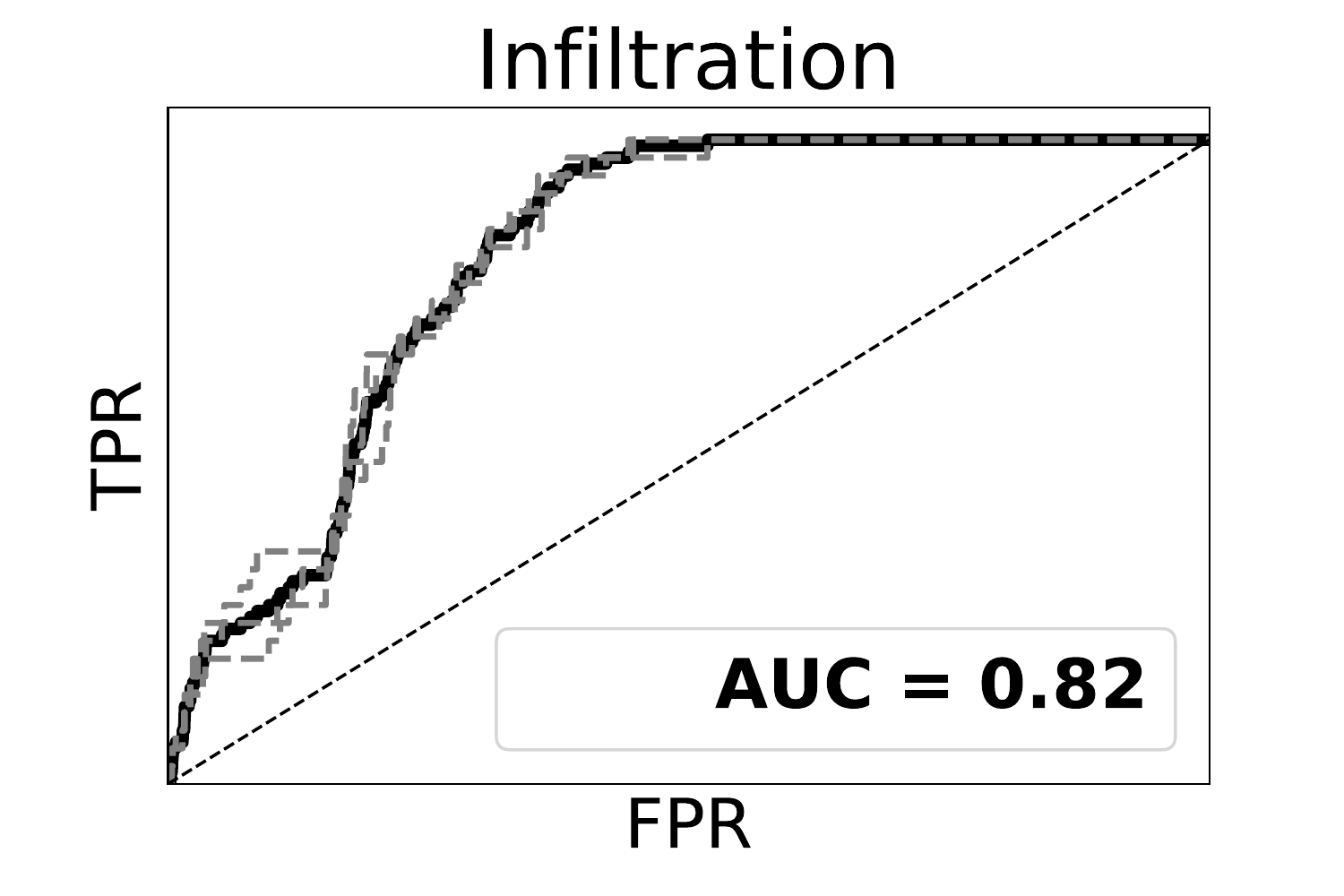}
      \\
      \includegraphics[width=.19\textwidth]{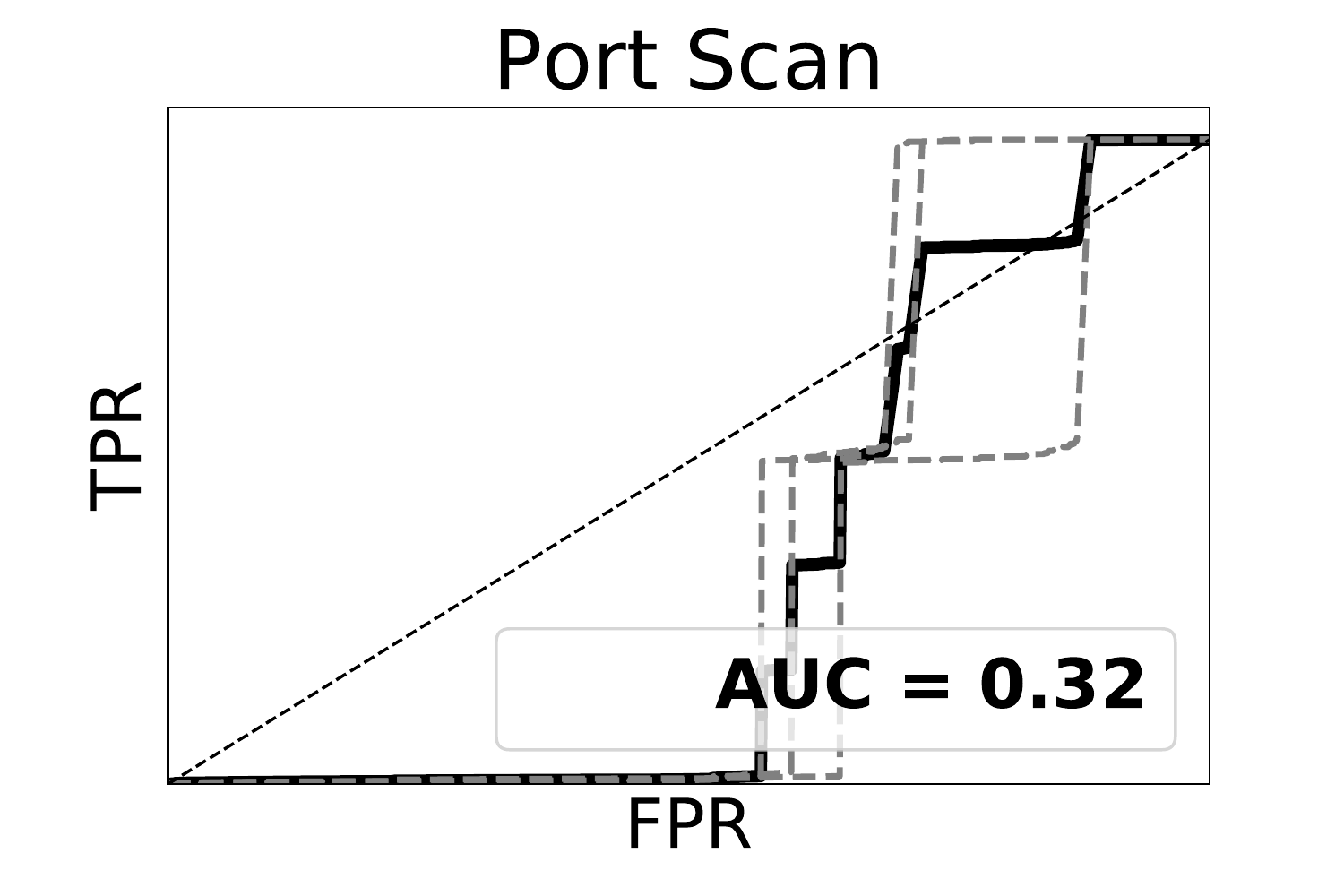} &
      \includegraphics[width=.19\textwidth]{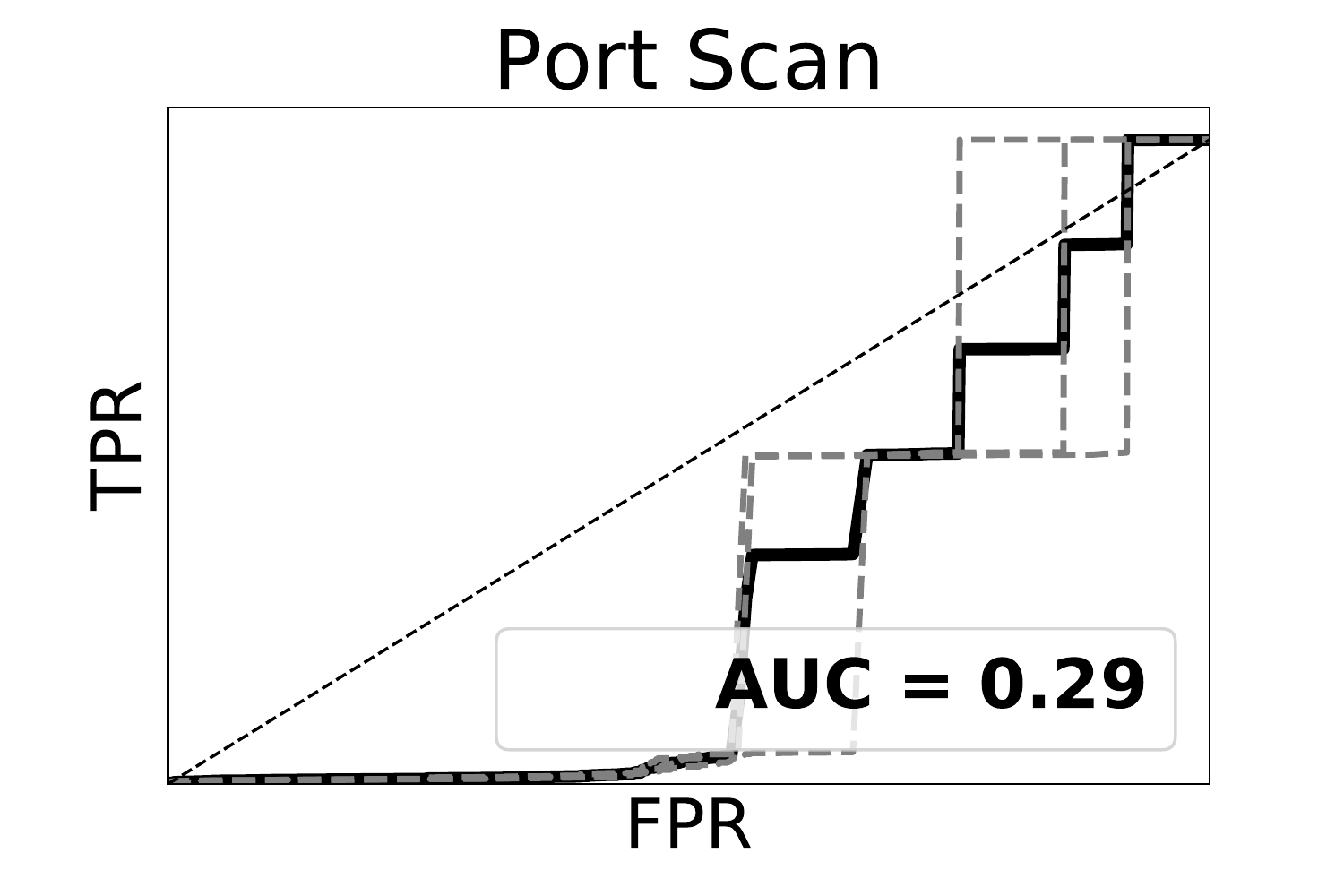} &
      \includegraphics[width=.19\textwidth]{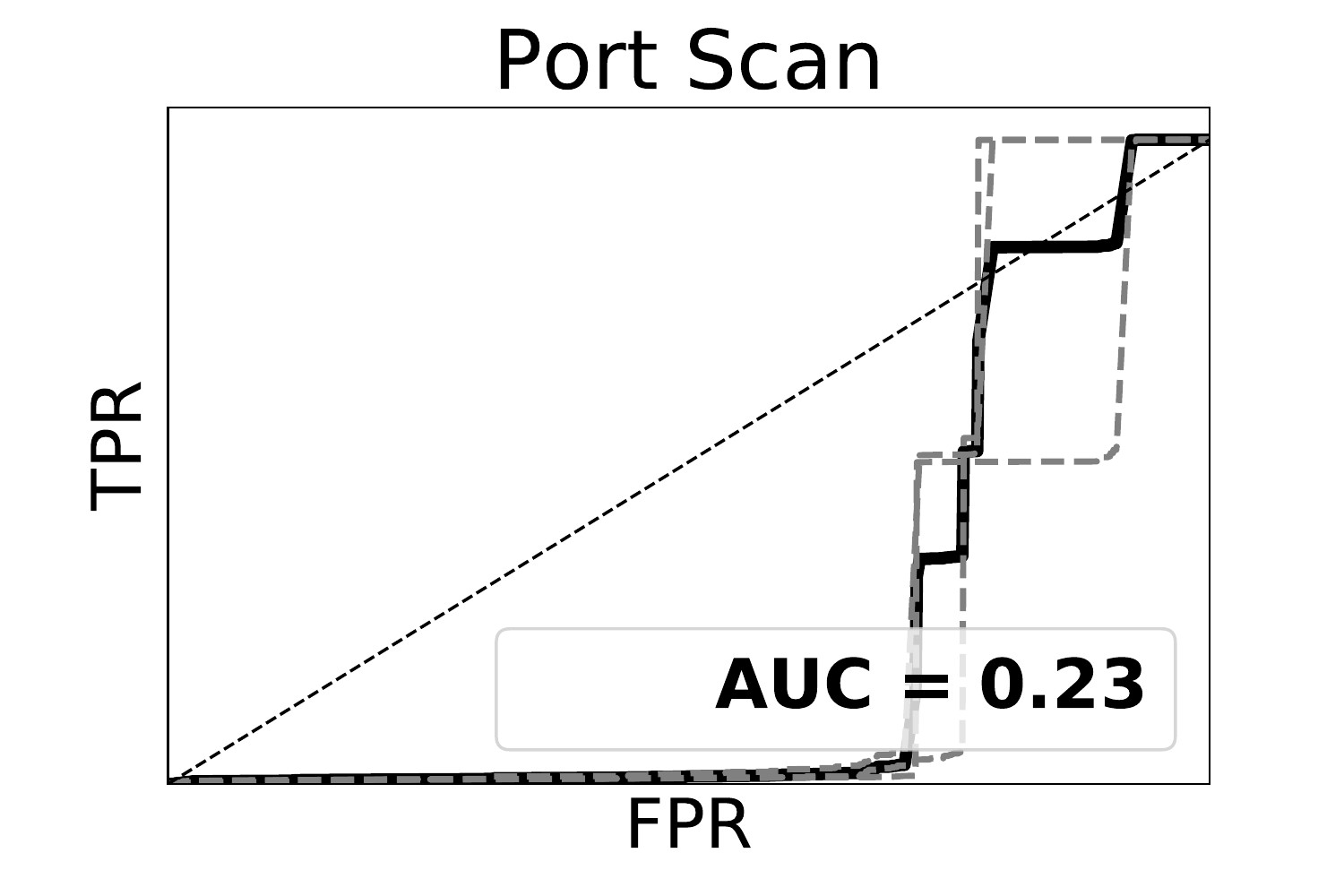} &
      \includegraphics[width=.19\textwidth]{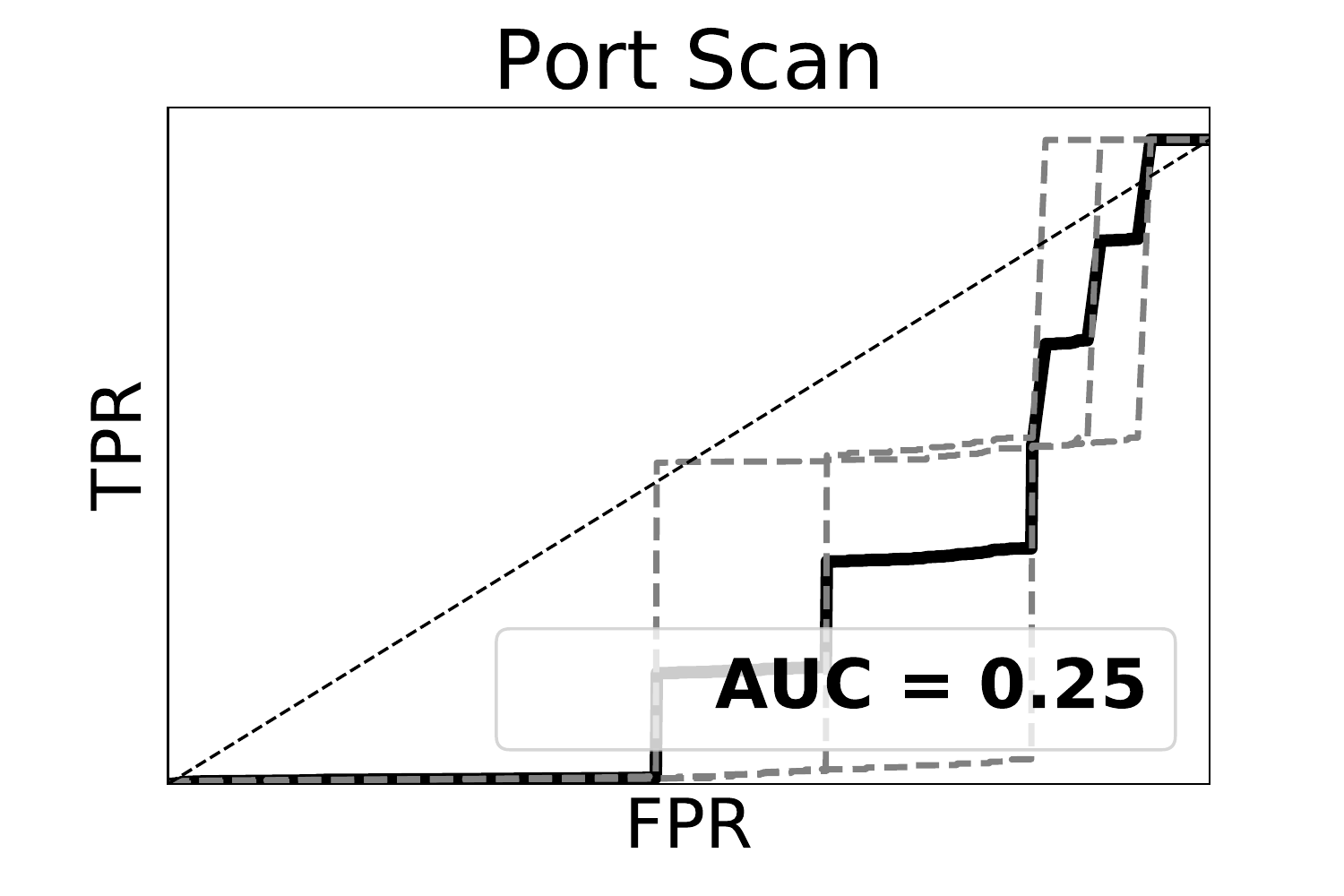} &
      \includegraphics[width=.19\textwidth]{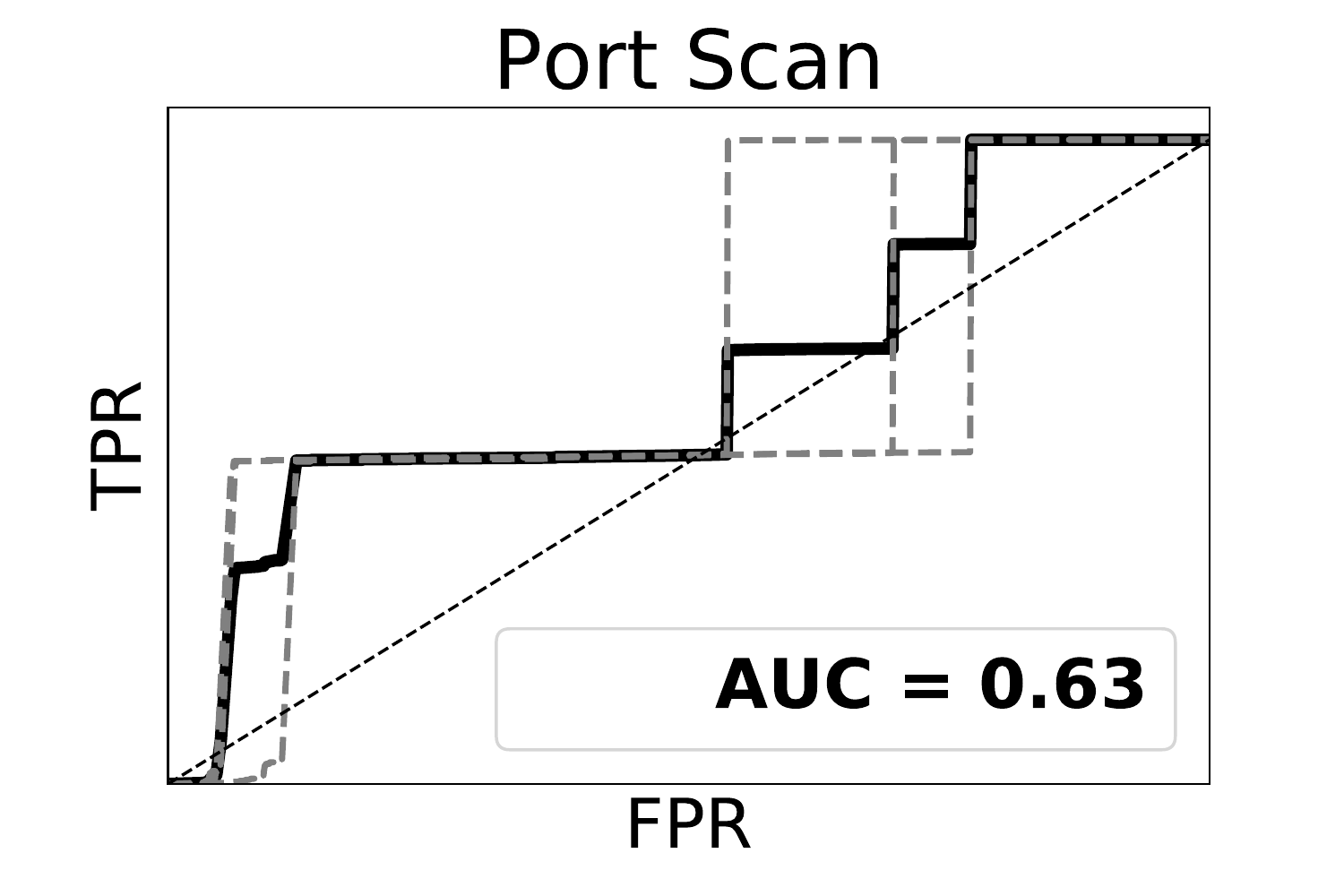}
      \\
      \includegraphics[width=.19\textwidth]{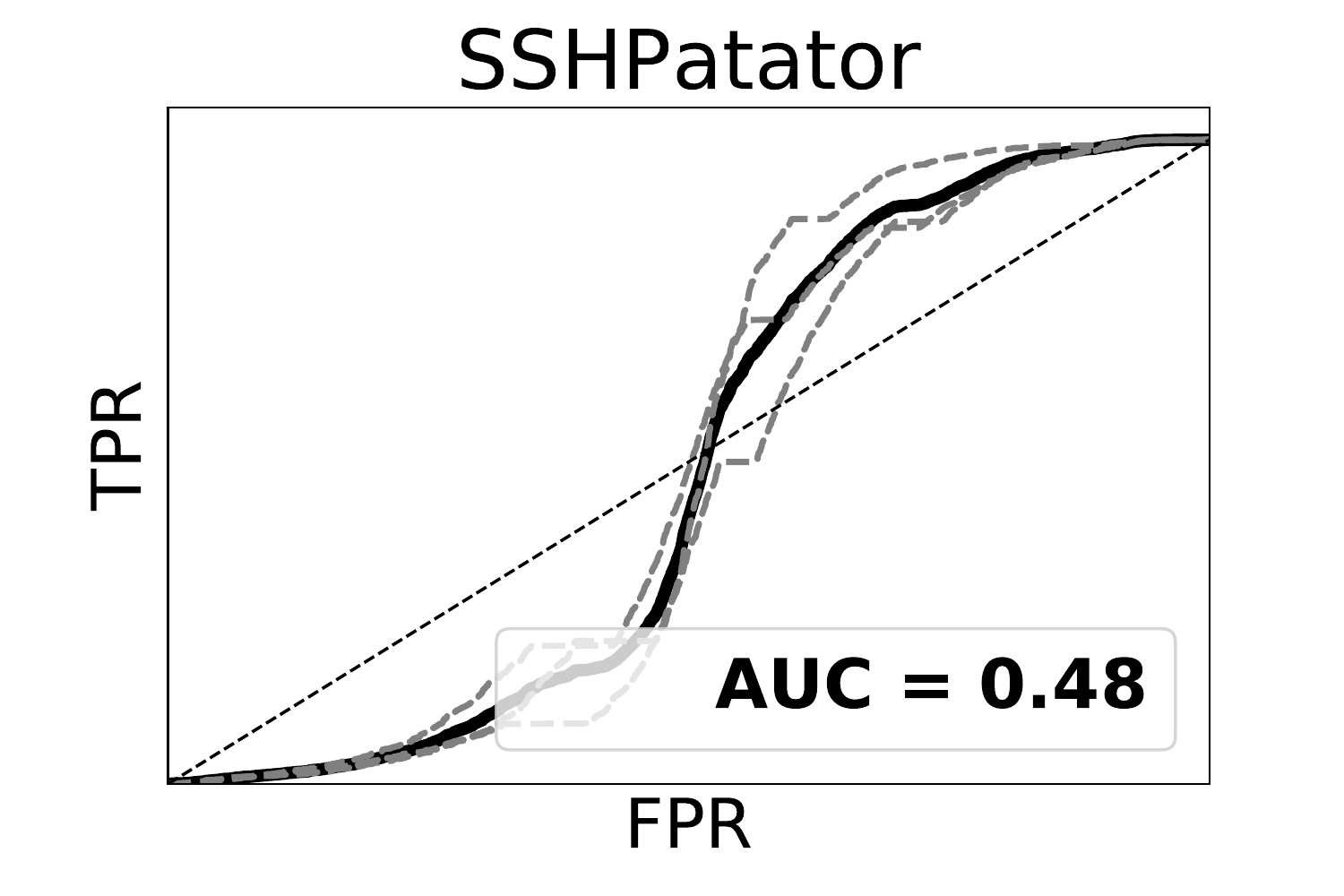} &
      \includegraphics[width=.19\textwidth]{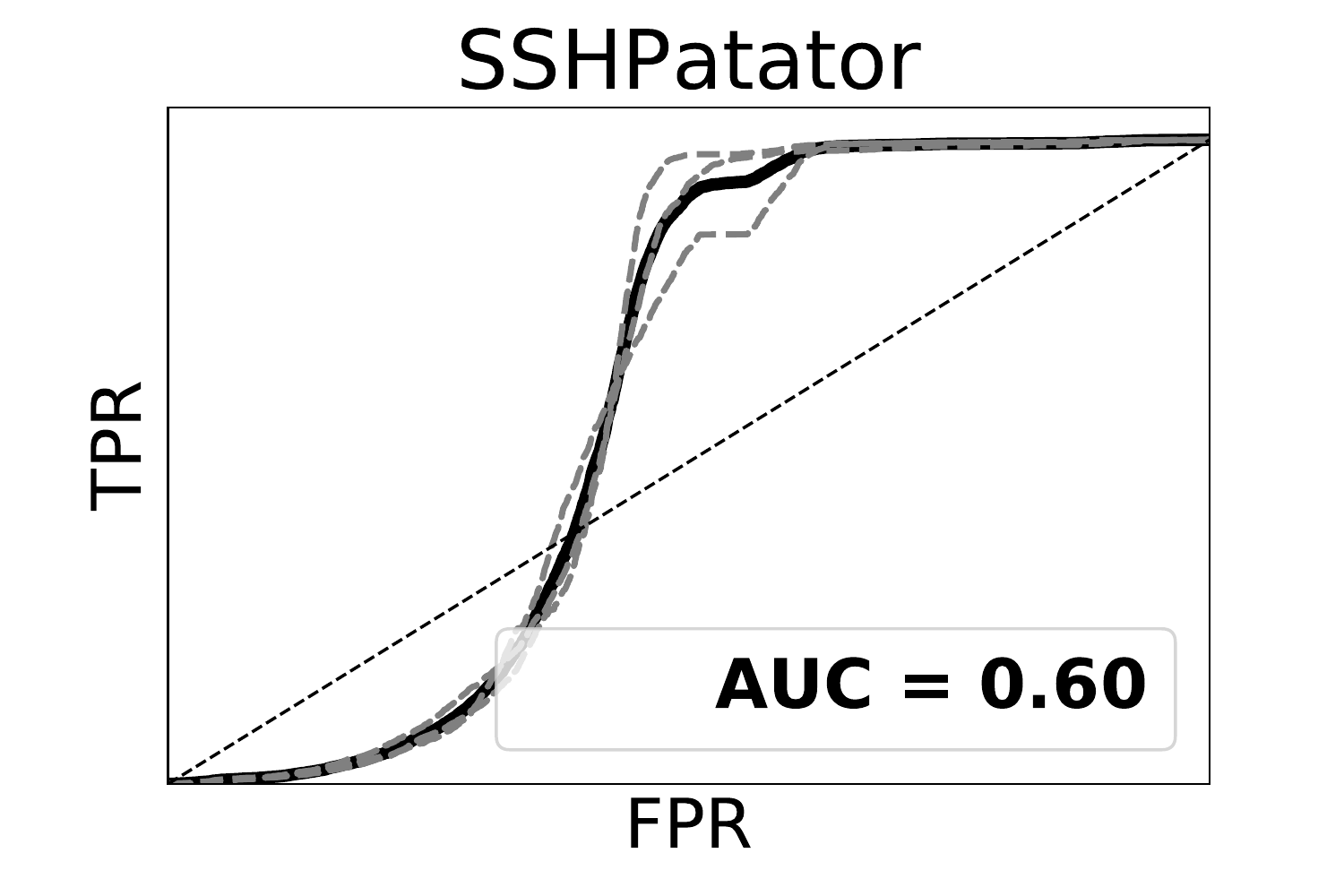} &
      \includegraphics[width=.19\textwidth]{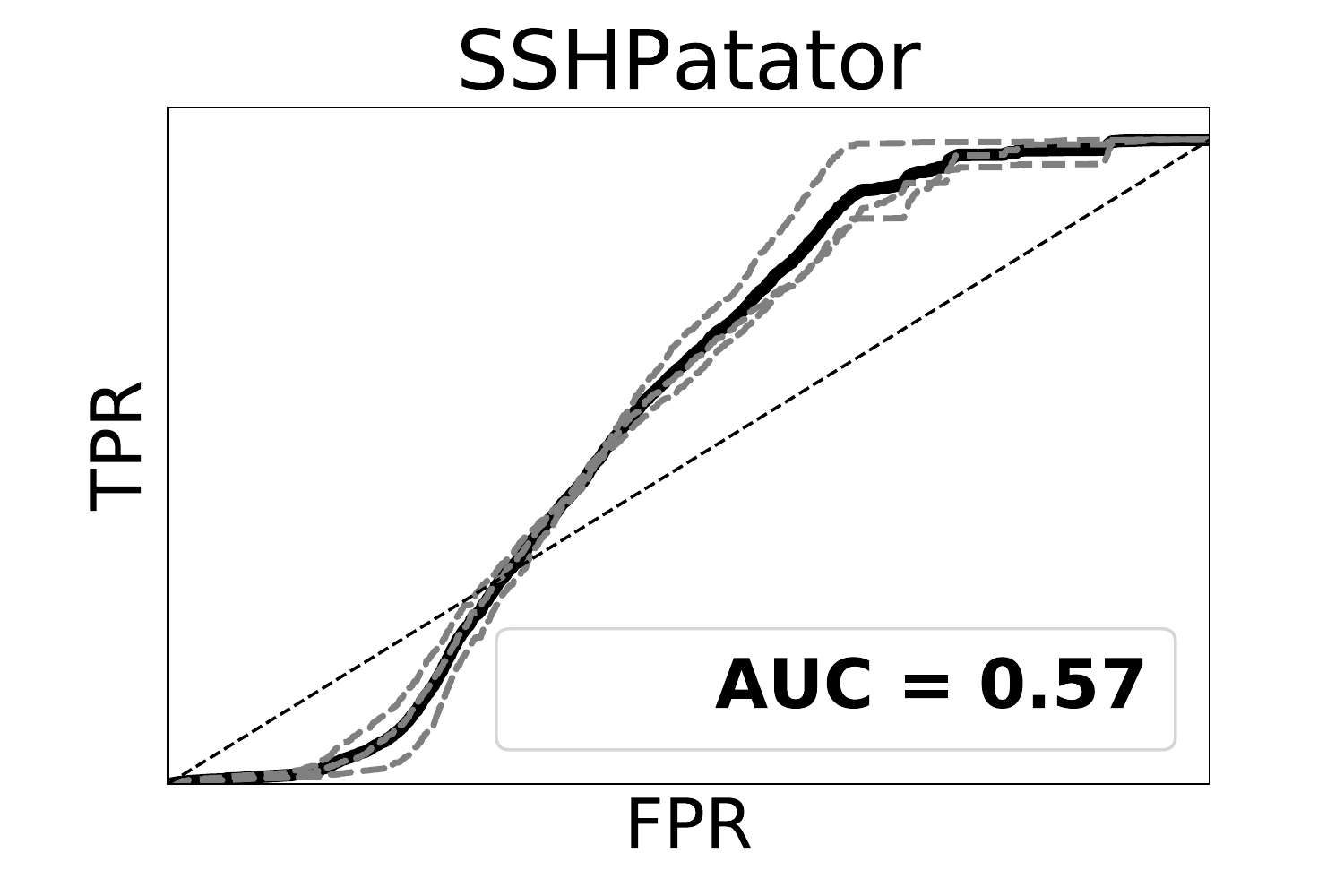} &
      \includegraphics[width=.19\textwidth]{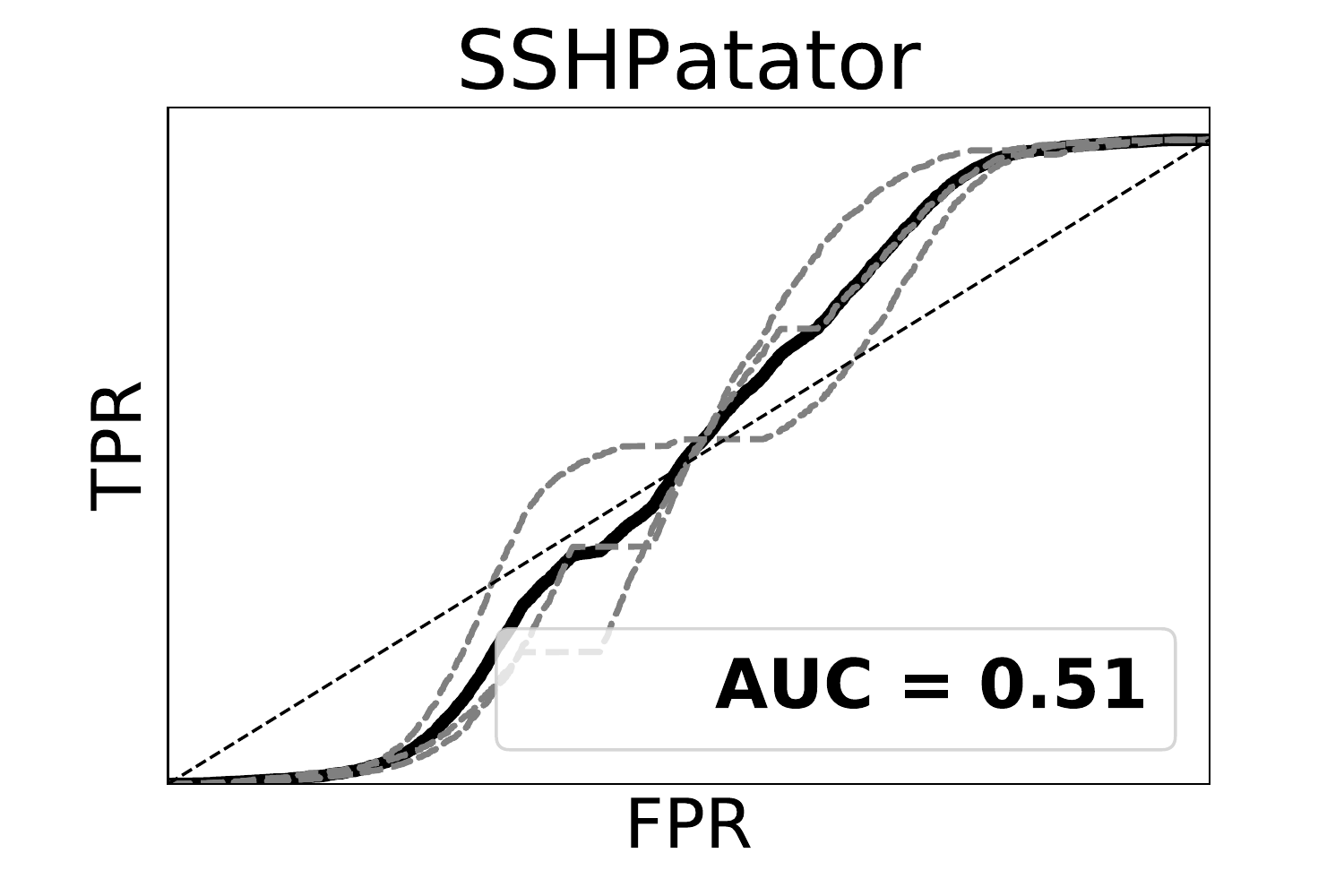} &
      \includegraphics[width=.19\textwidth]{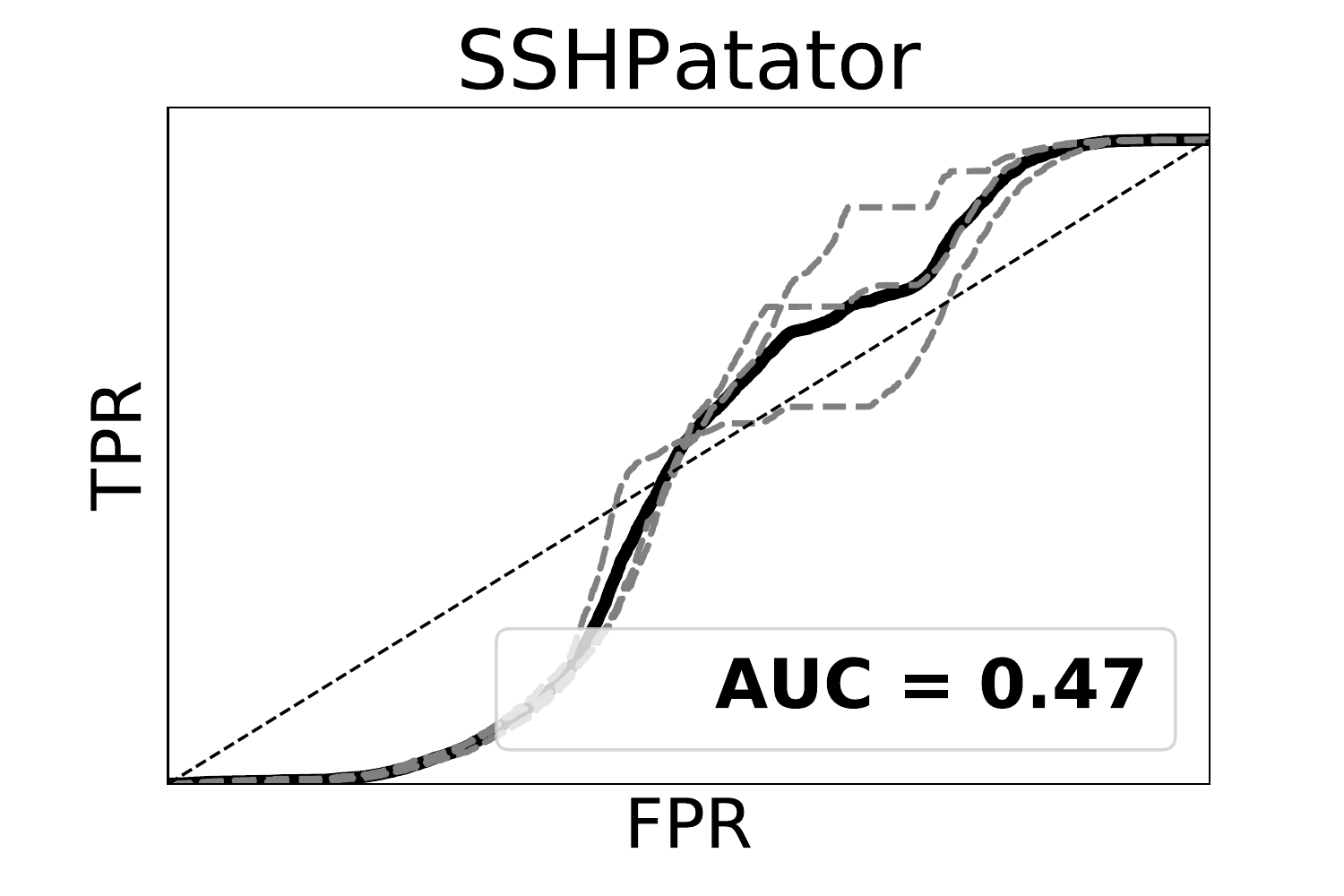}
      \\
      \includegraphics[width=.19\textwidth]{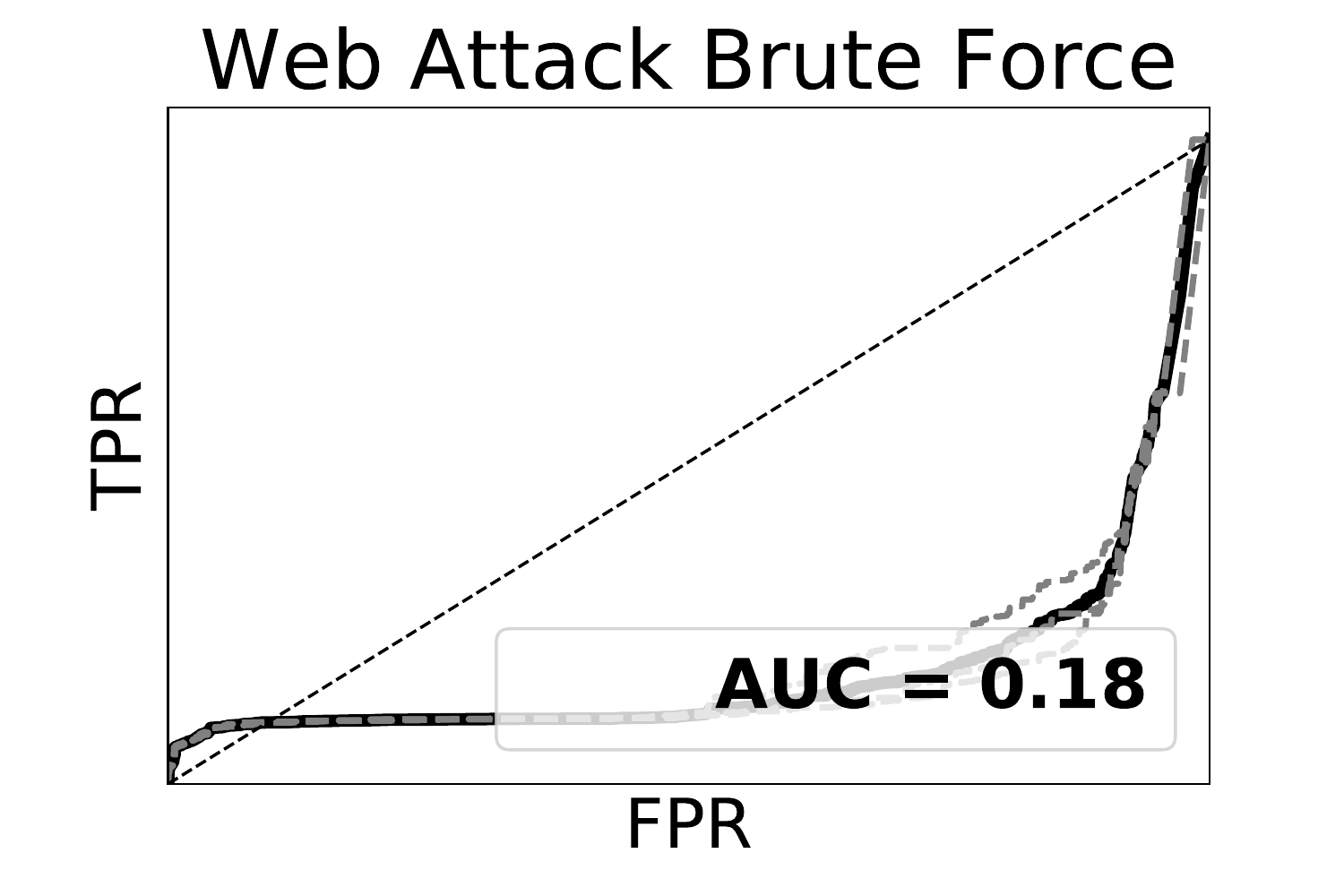} &
      \includegraphics[width=.19\textwidth]{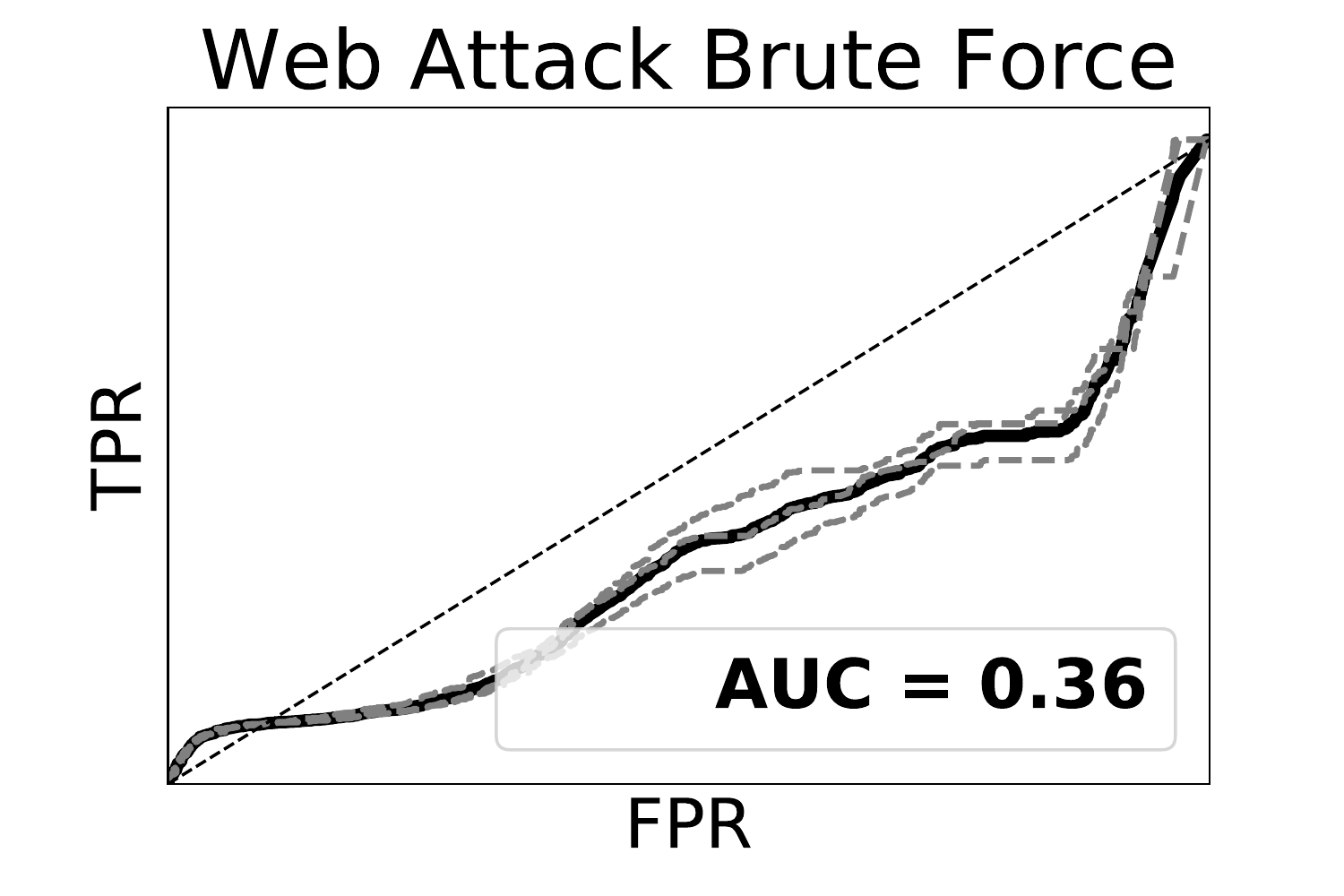} &
      \includegraphics[width=.19\textwidth]{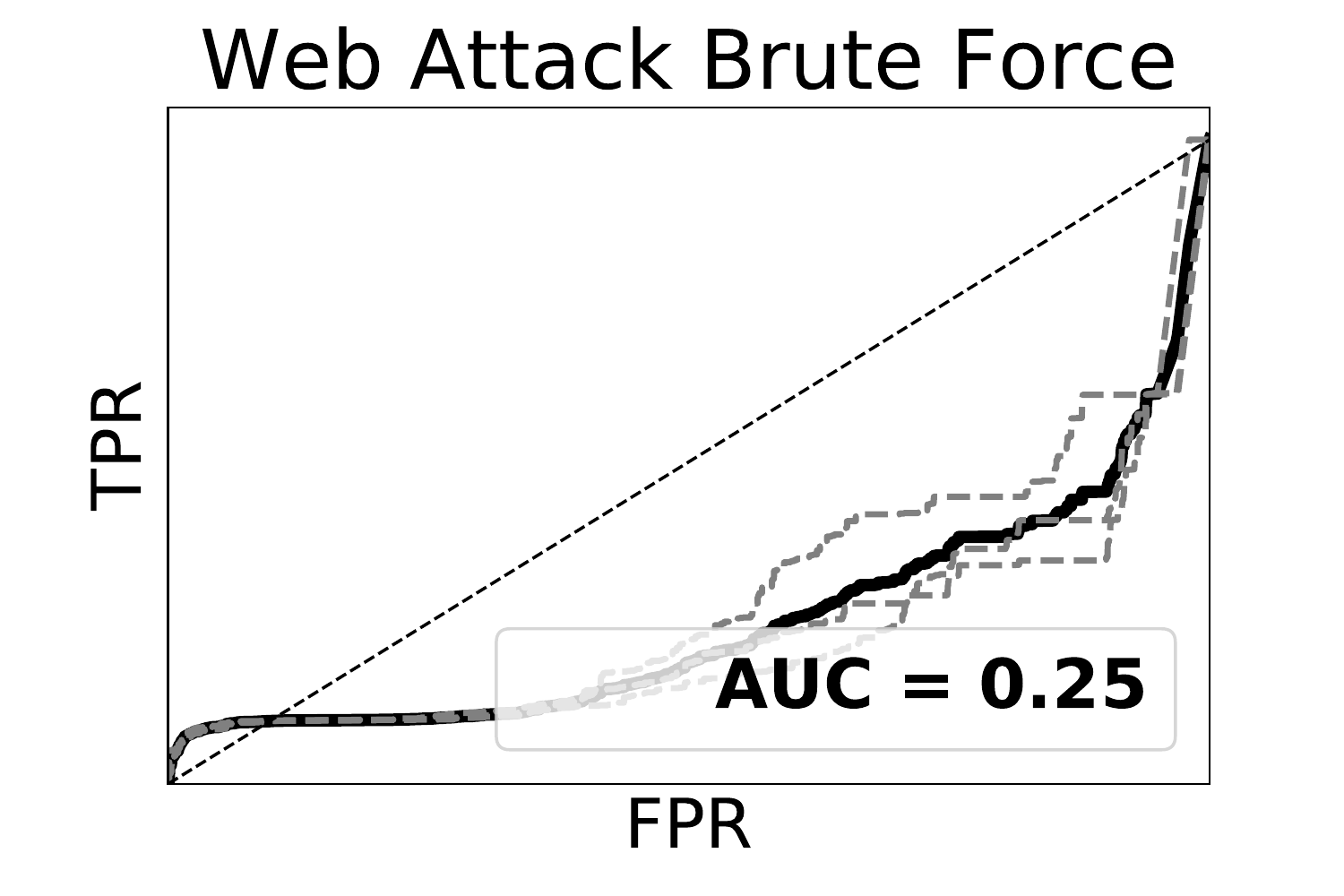} &
      \includegraphics[width=.19\textwidth]{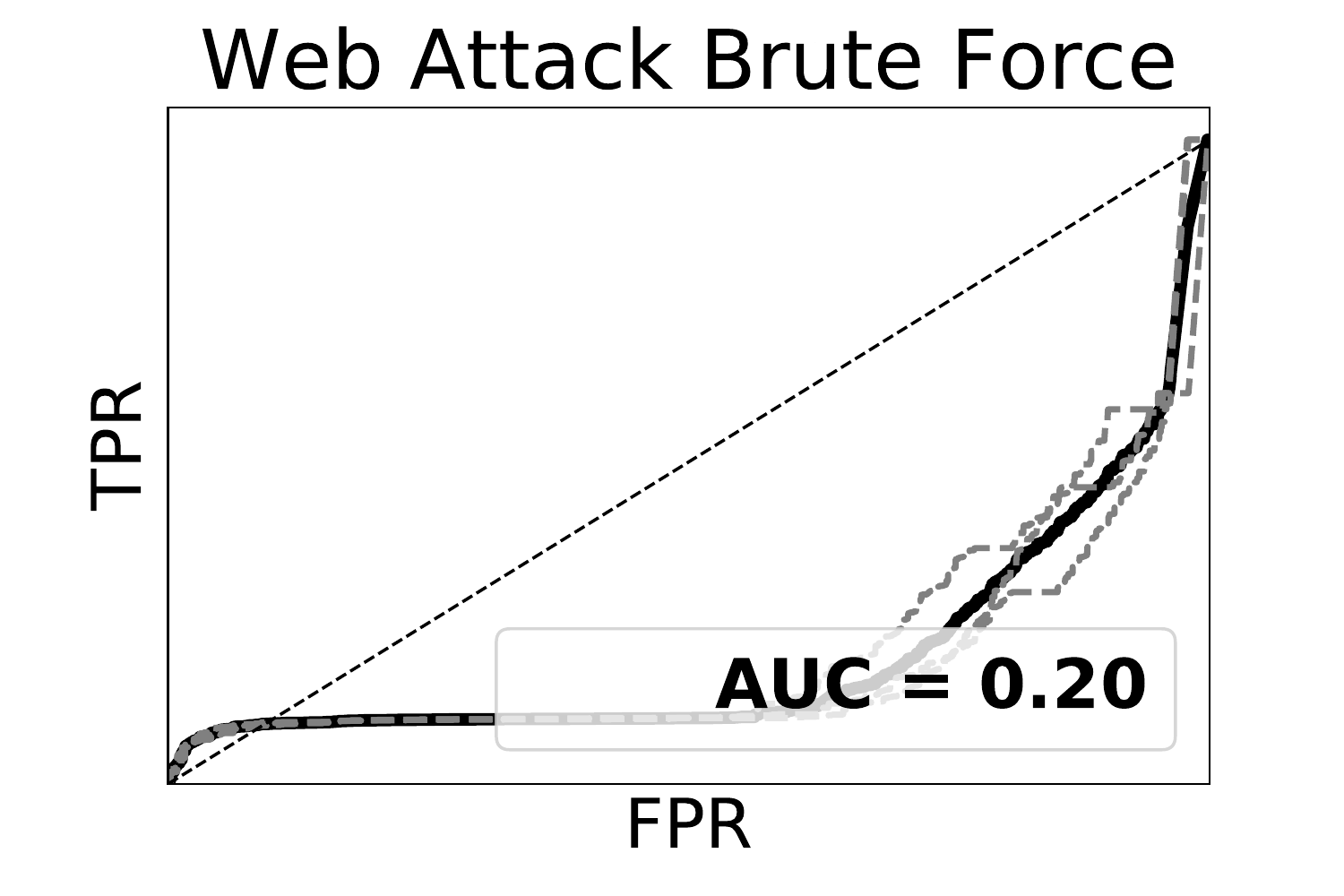} &
      \includegraphics[width=.19\textwidth]{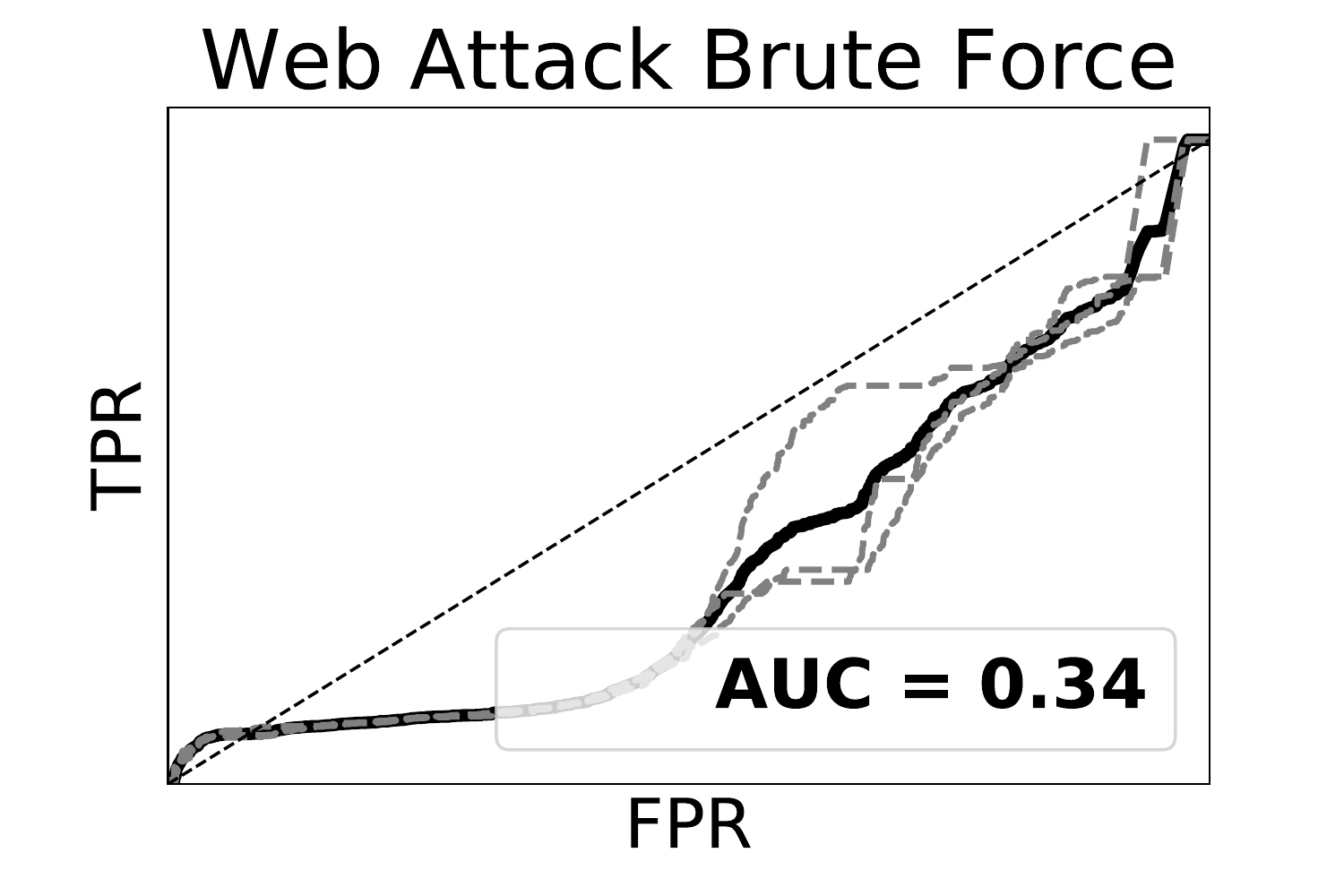}
      \\
      \includegraphics[width=.19\textwidth]{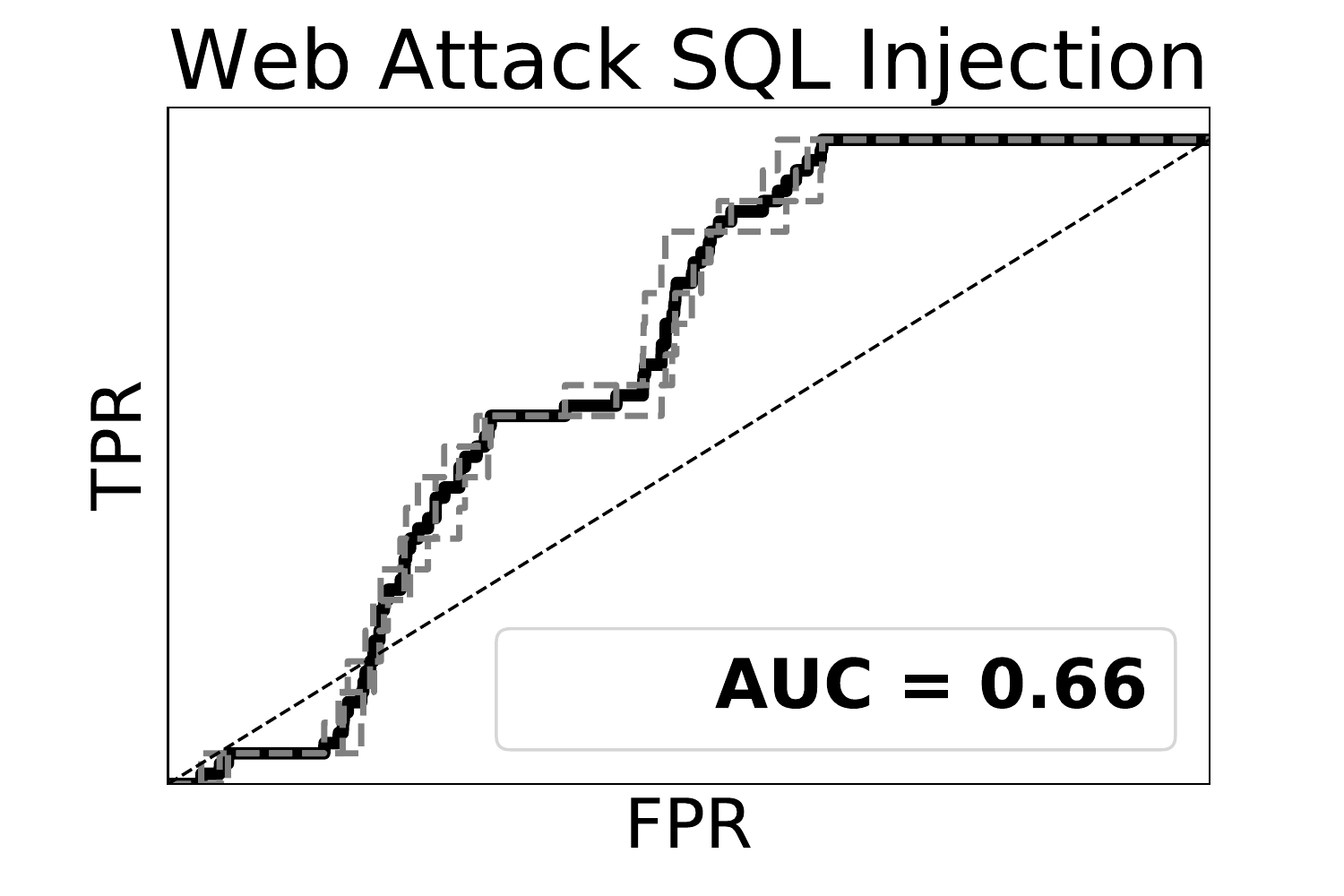} &
      \includegraphics[width=.19\textwidth]{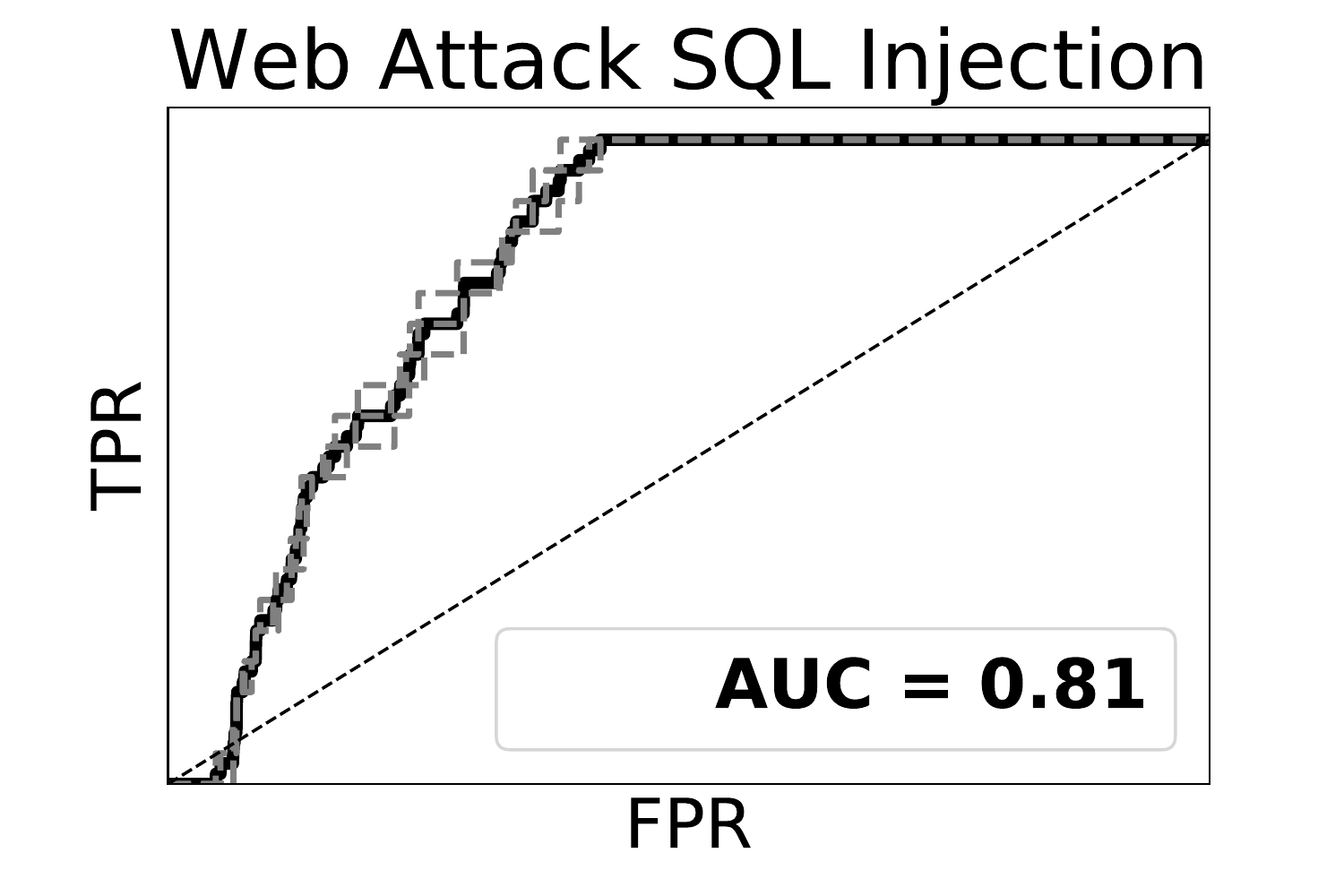} &
      \includegraphics[width=.19\textwidth]{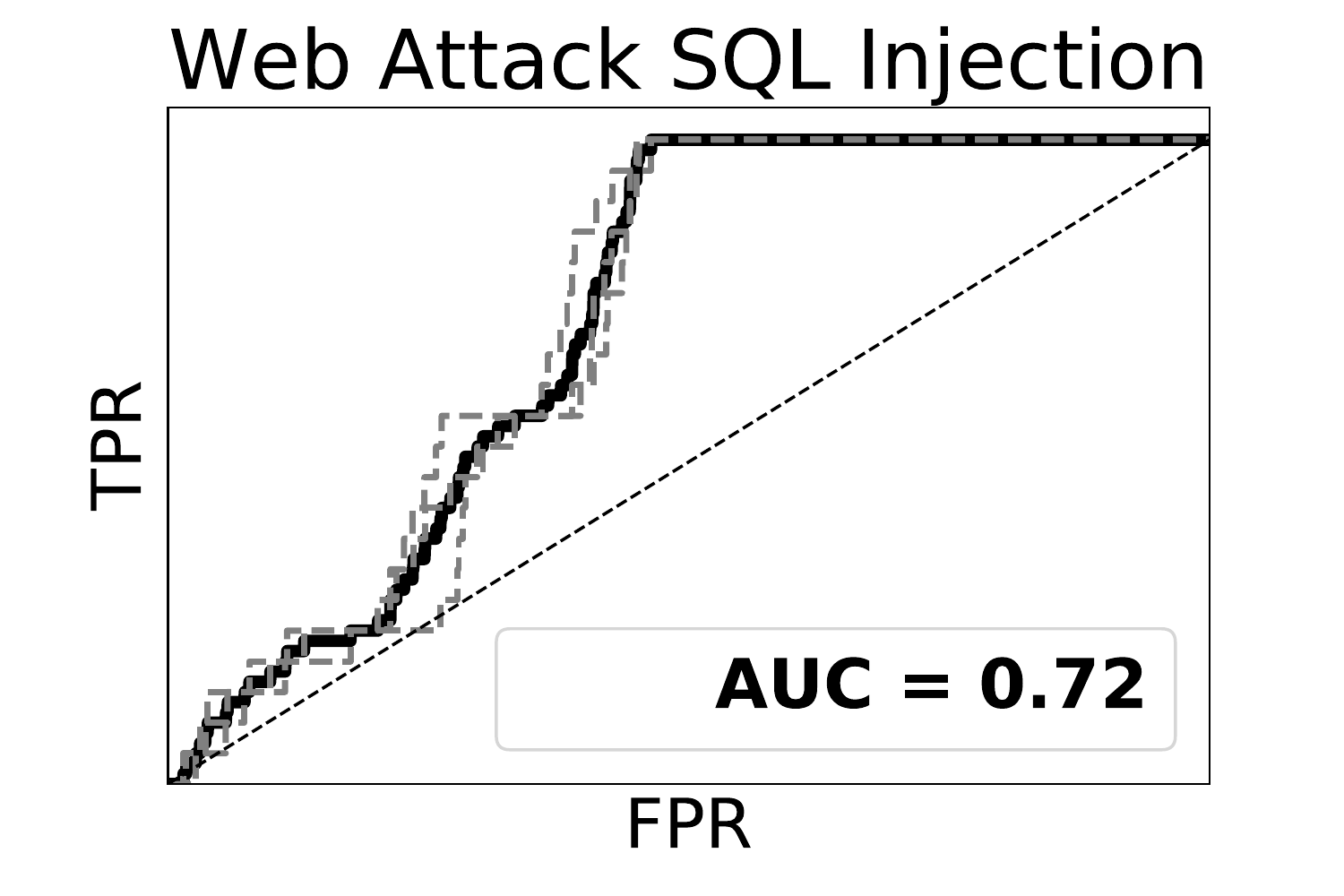} &
      \includegraphics[width=.19\textwidth]{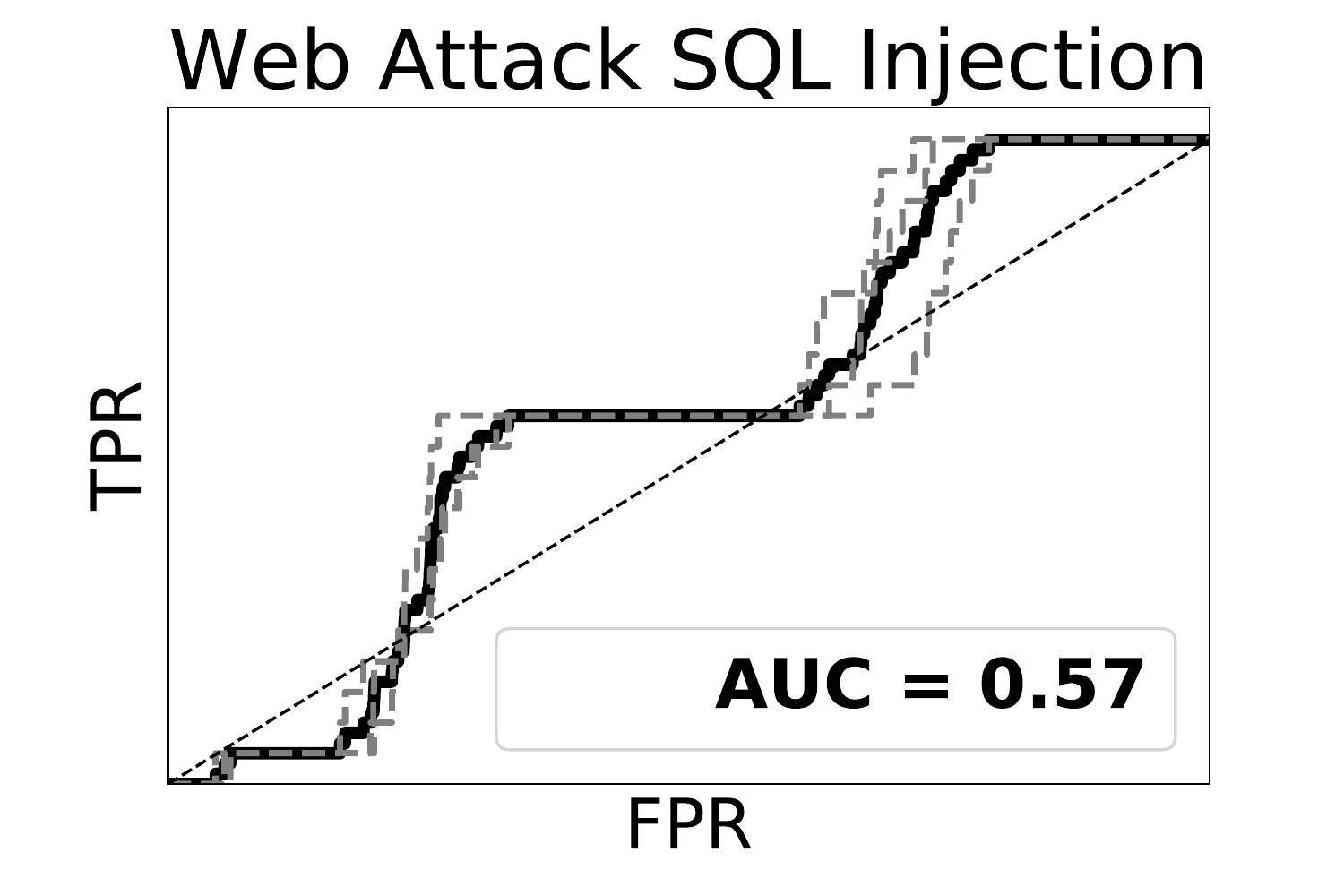} &
      \includegraphics[width=.19\textwidth]{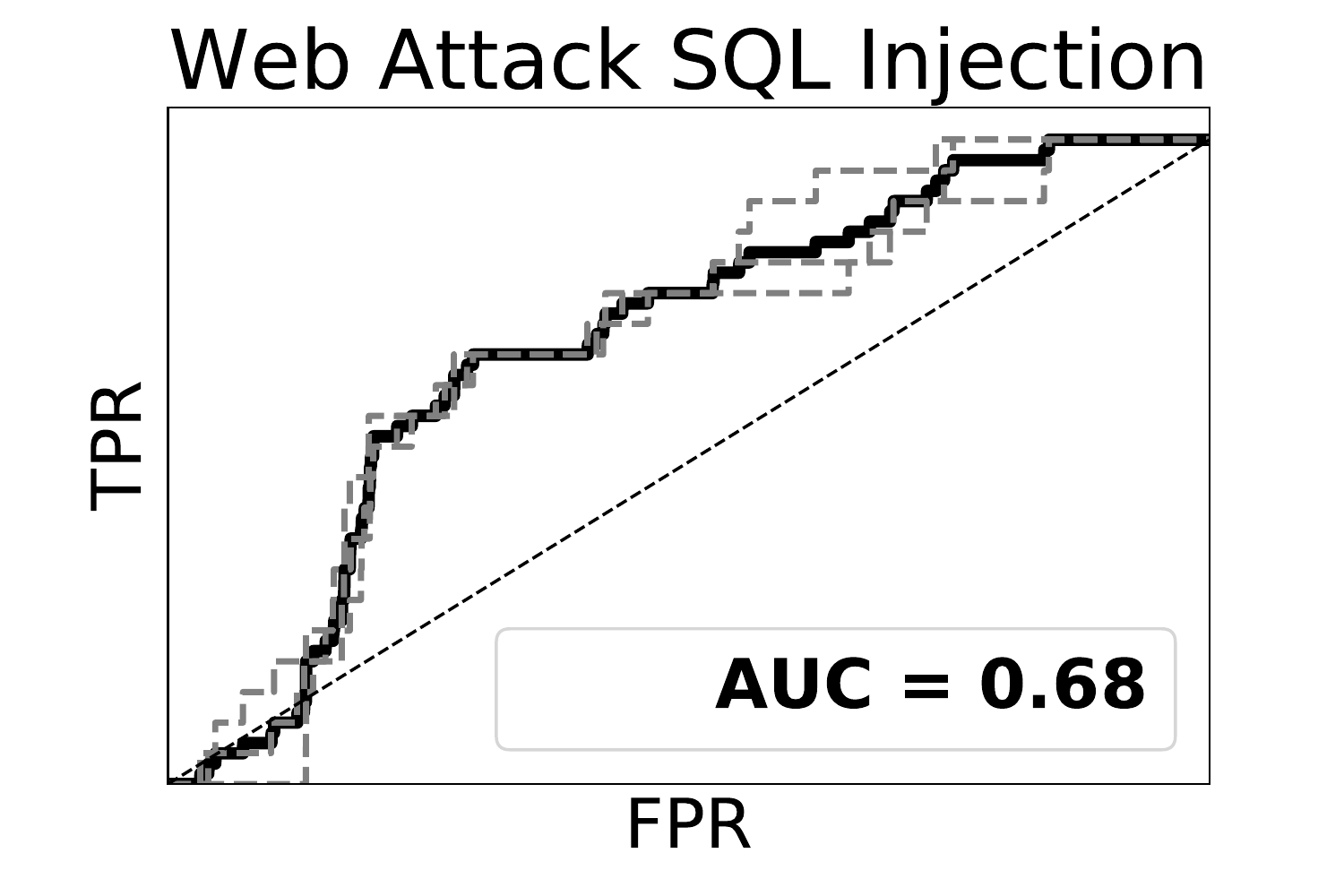}
      \\
      \includegraphics[width=.19\textwidth]{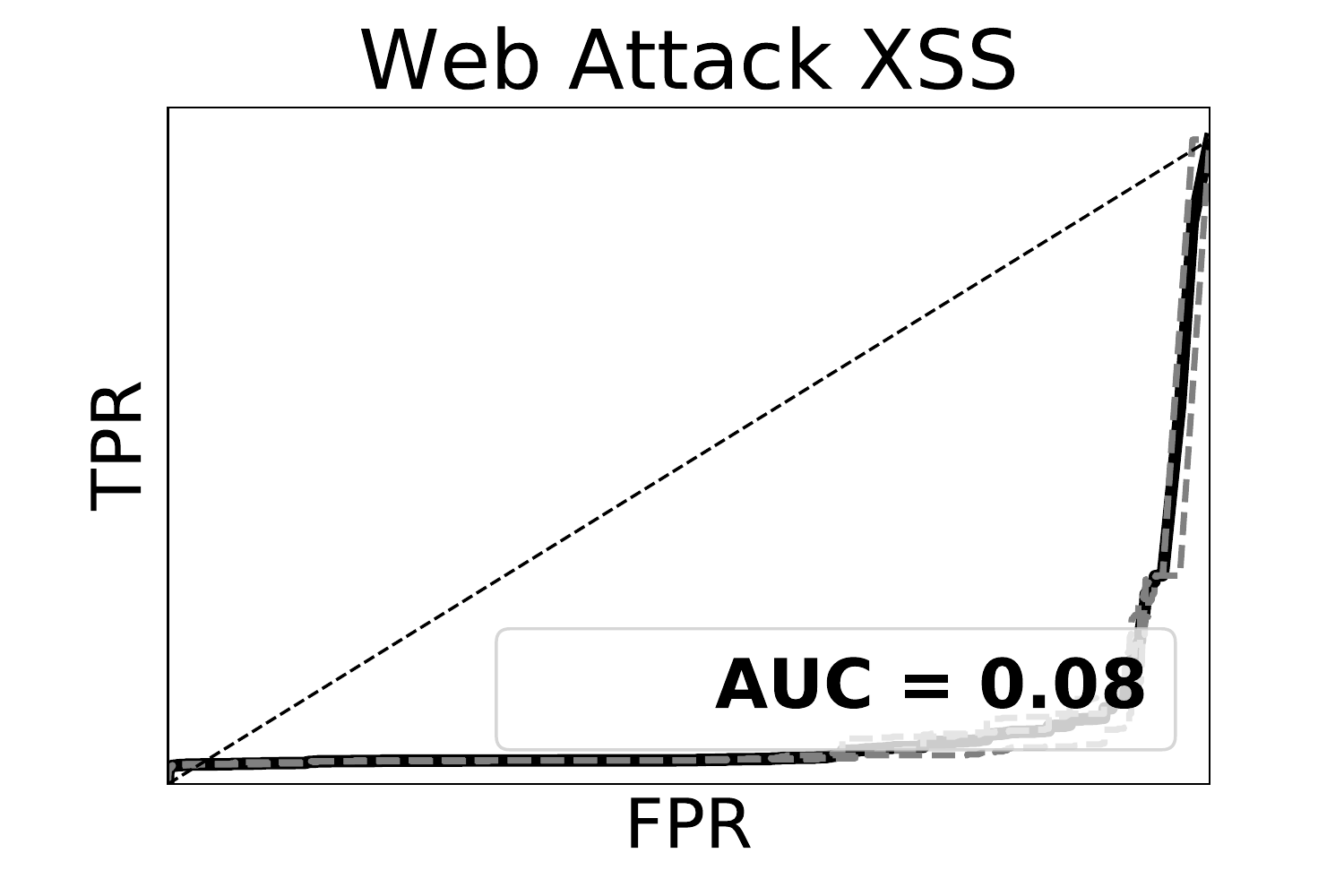} &
      \includegraphics[width=.19\textwidth]{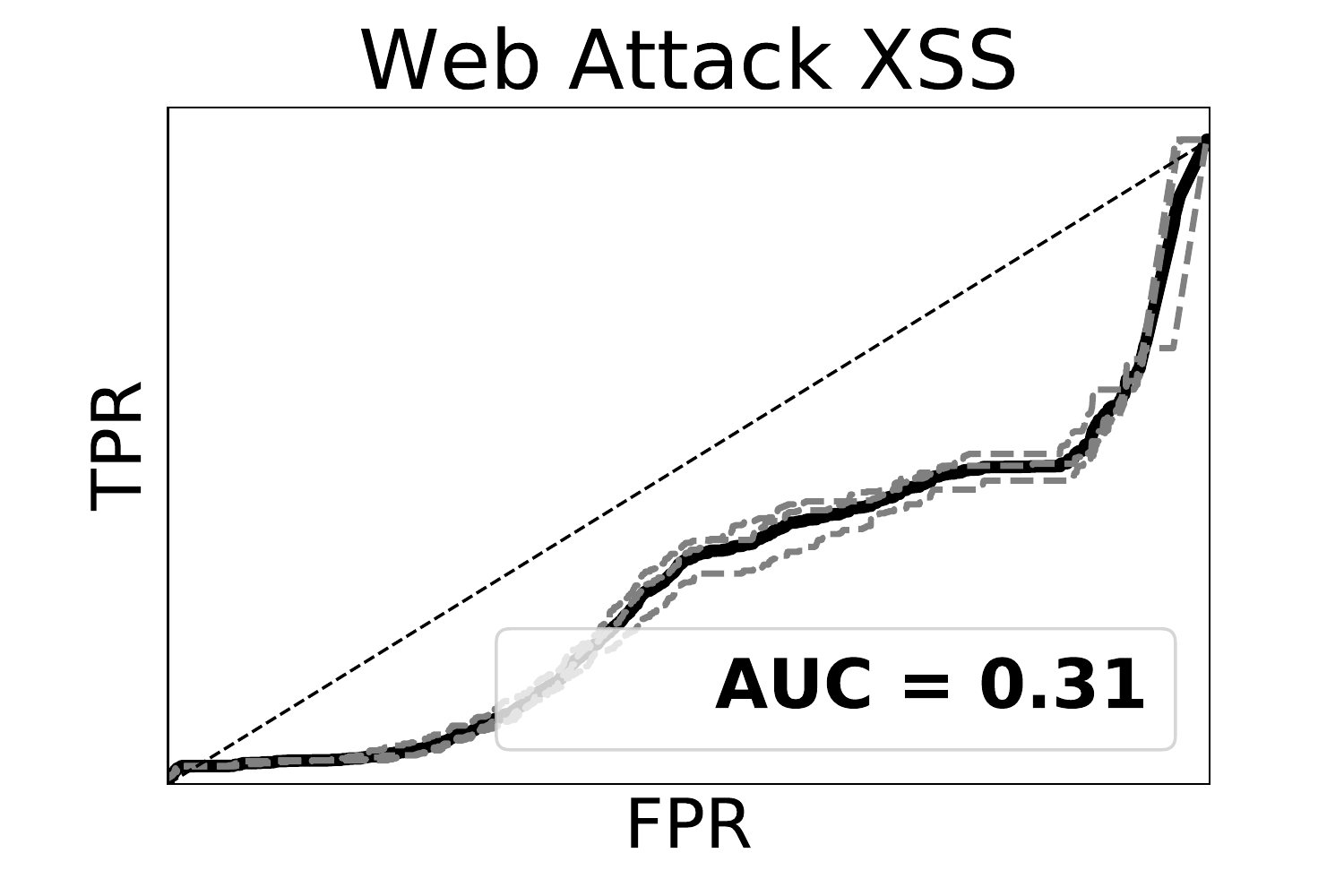} &
      \includegraphics[width=.19\textwidth]{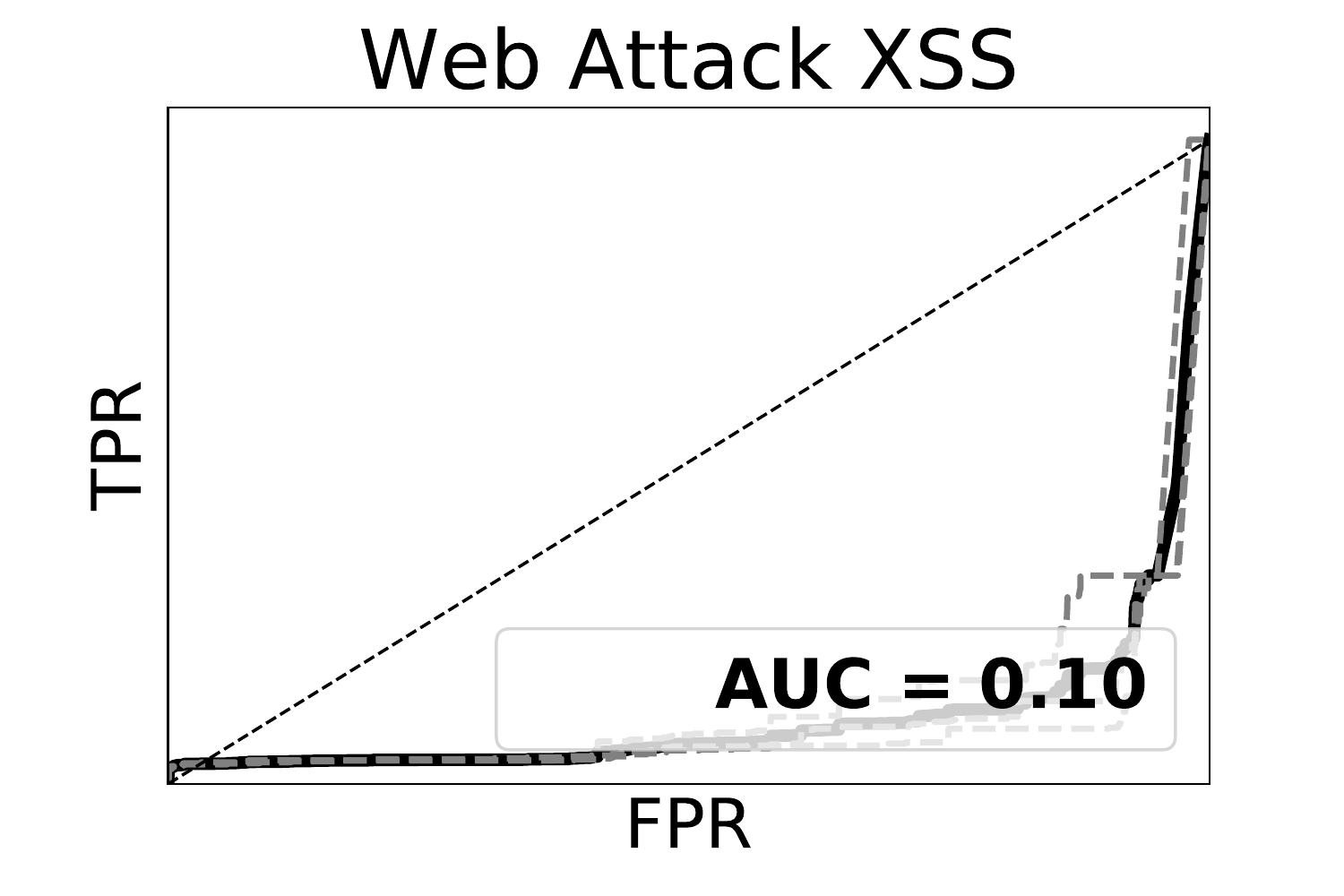} &
      \includegraphics[width=.19\textwidth]{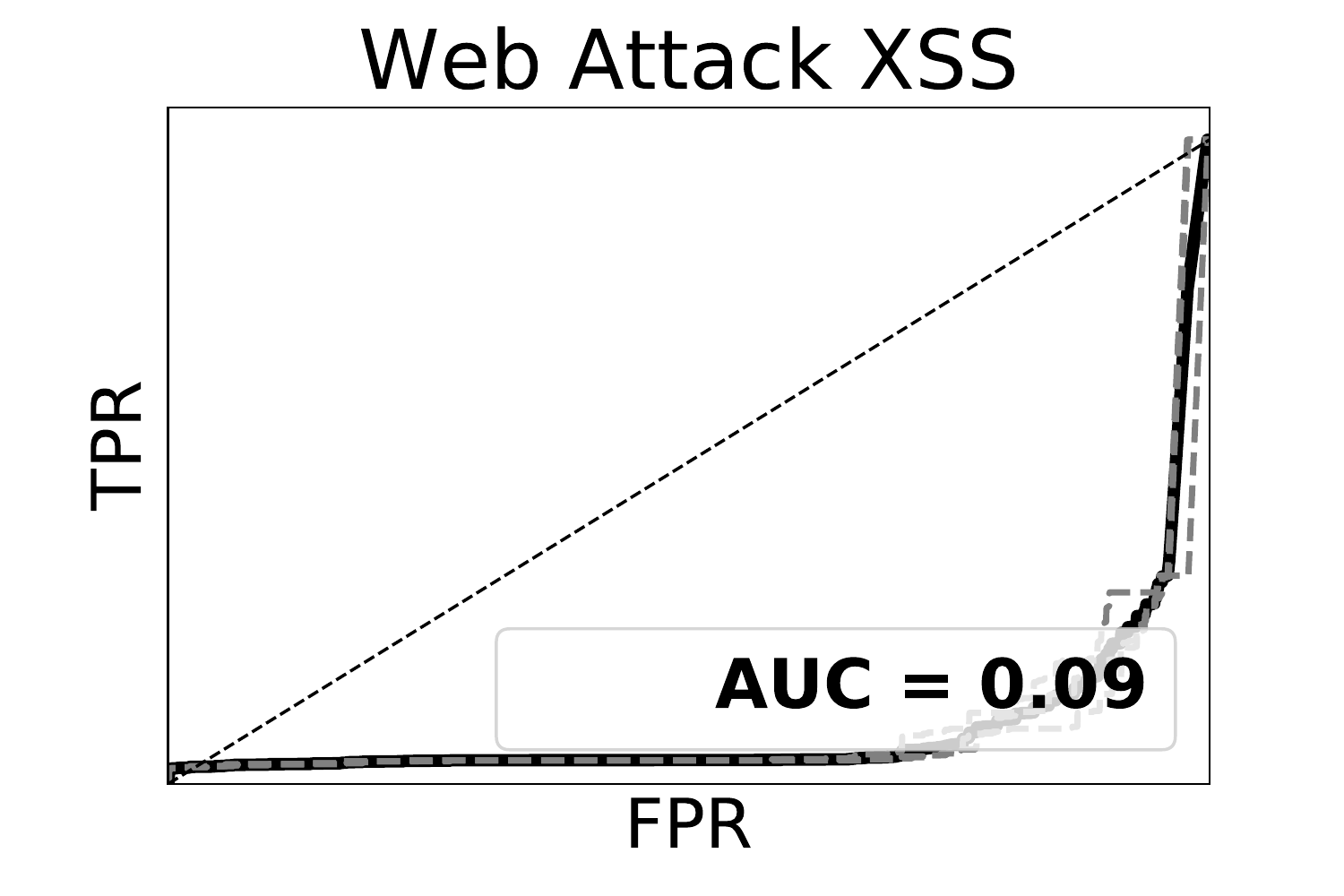} &
      \includegraphics[width=.19\textwidth]{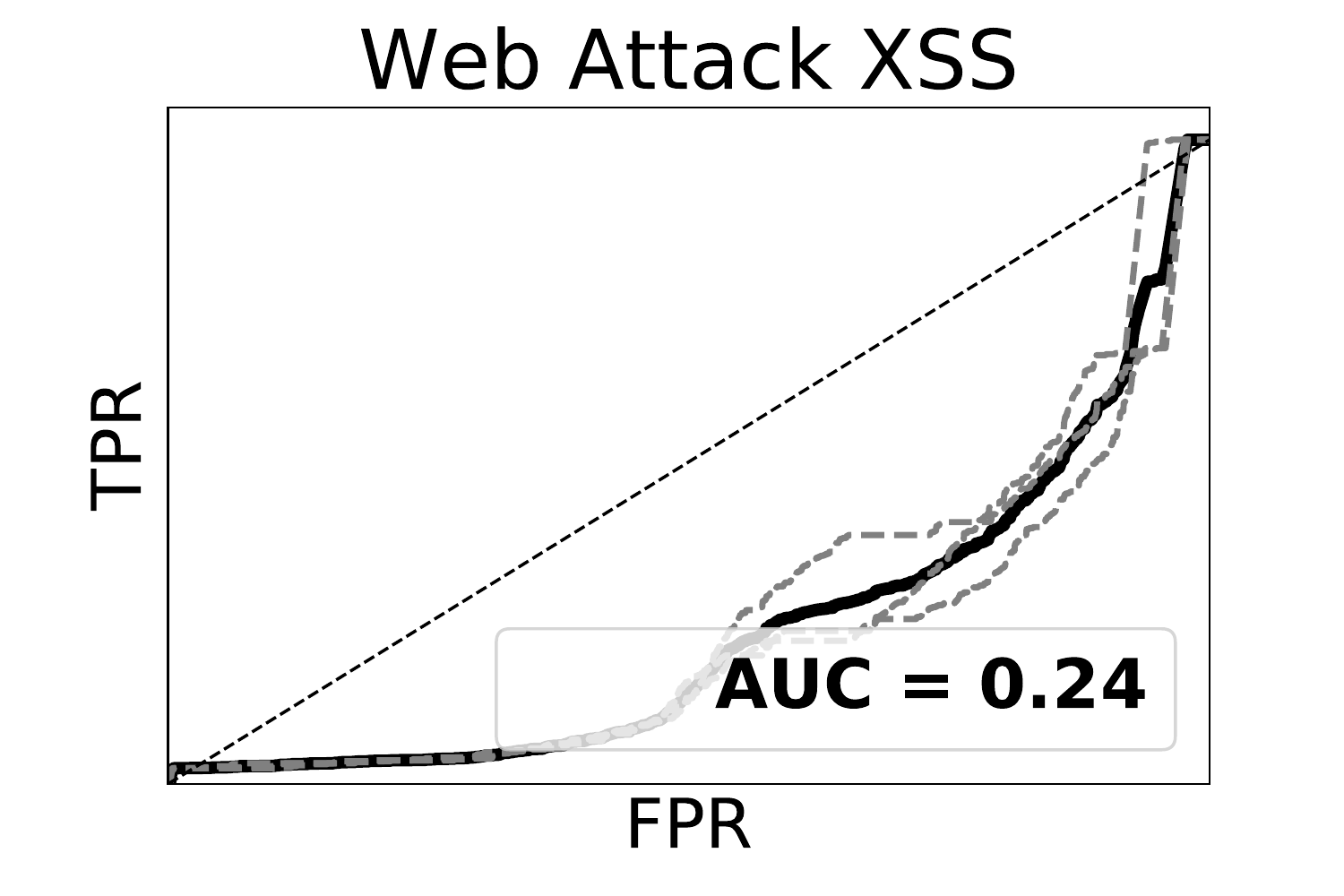}
      \\
  \end{tabular}
\end{table*}

\begin{table*}
  [ht] \caption{Service Port Sequences A} \label{tab:portsa}
  \begin{tabular}{ccccc} 
  \hline 
  Source & Destination & Dyad & Internal & External \\
      \hline 
      \includegraphics[width=.19\textwidth]{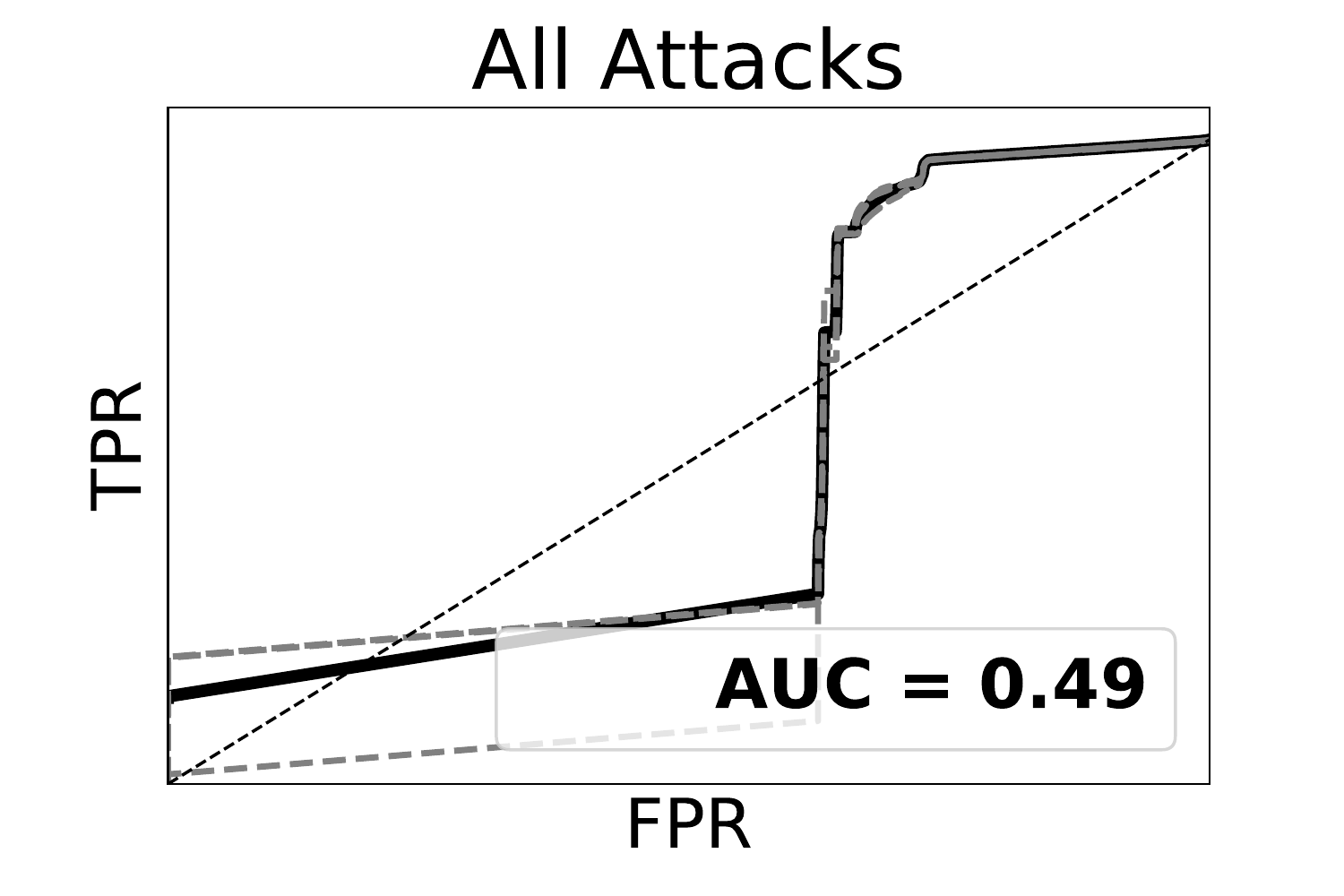} &
      \includegraphics[width=.19\textwidth]{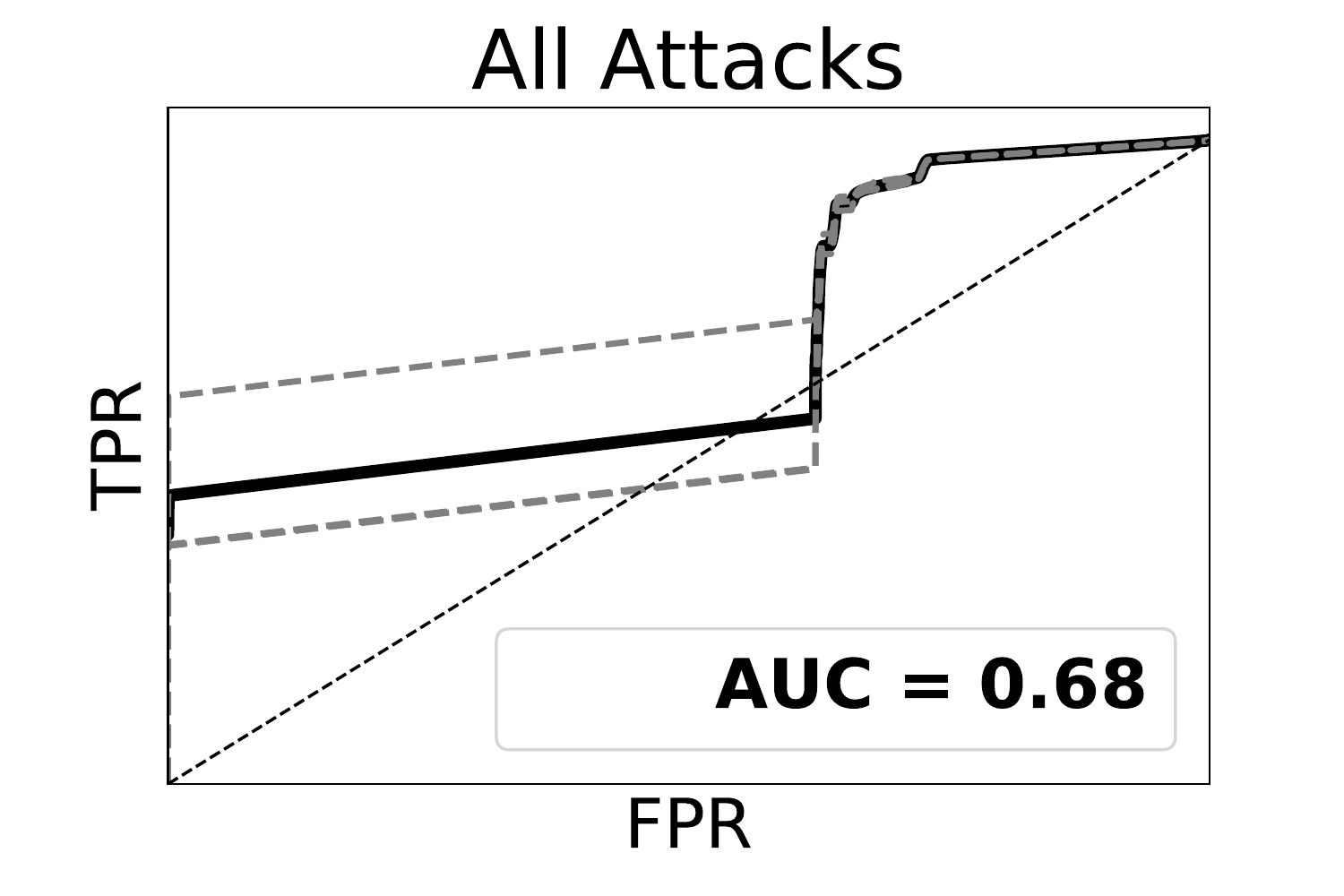} &
      \includegraphics[width=.19\textwidth]{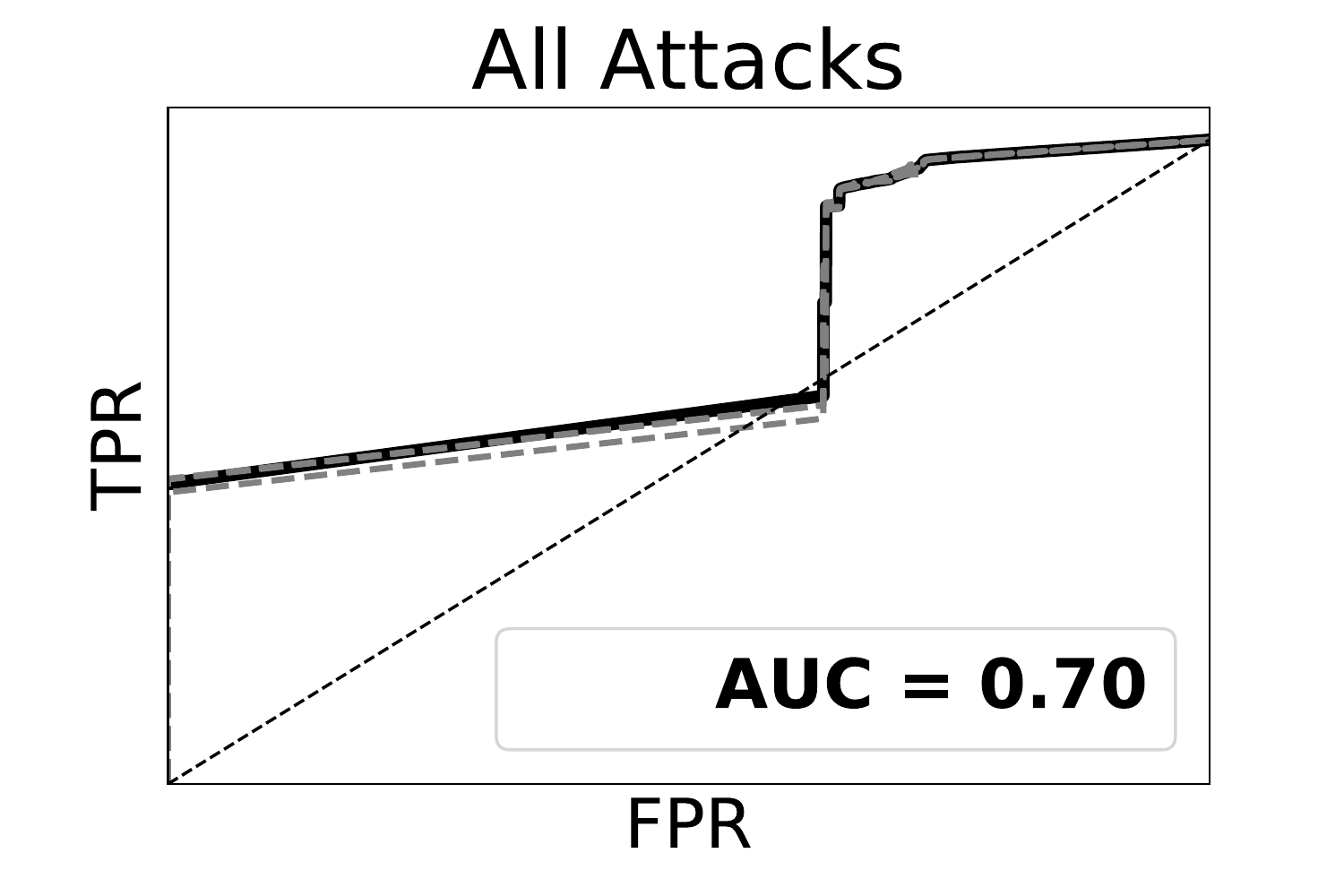} &
      \includegraphics[width=.19\textwidth]{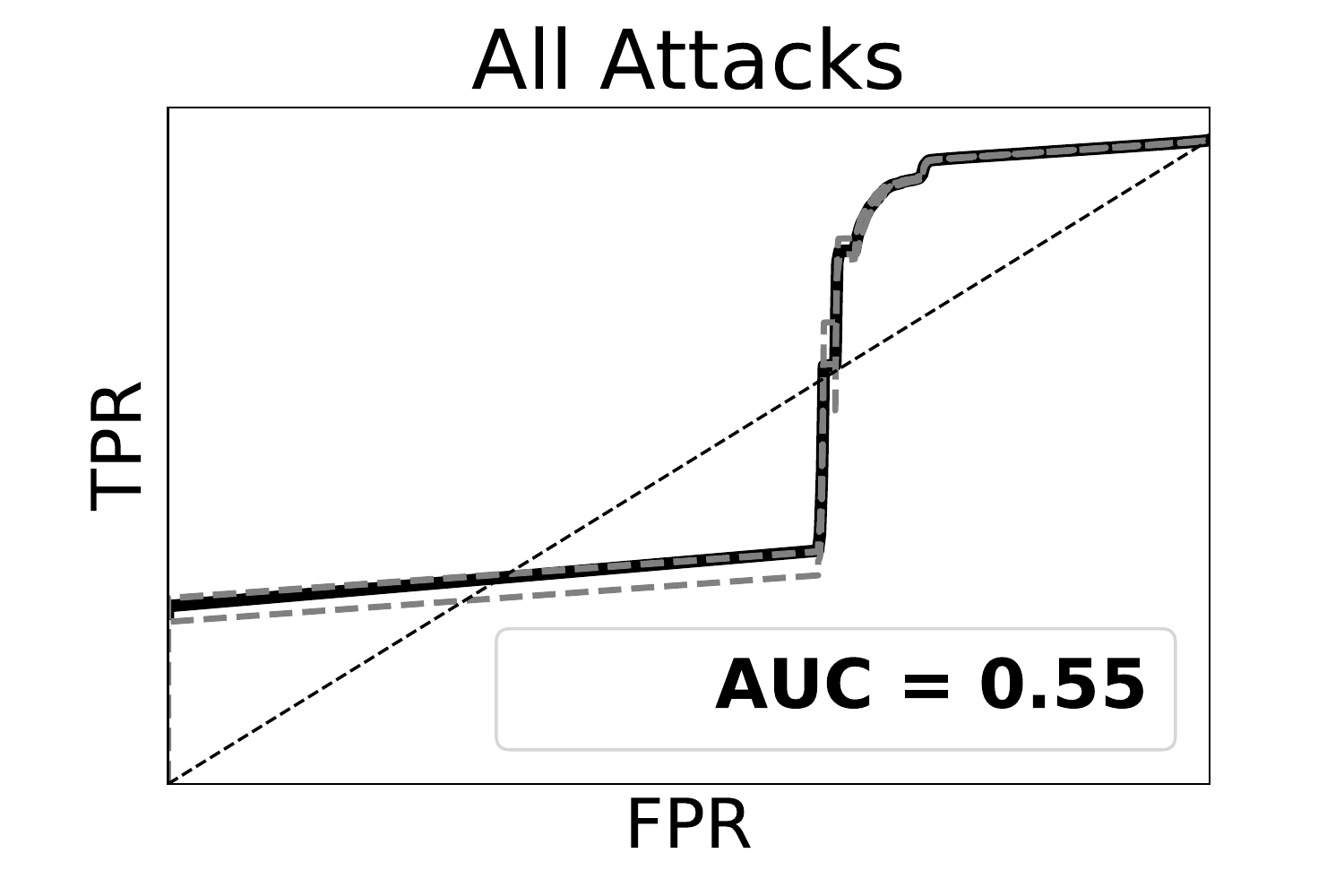} &
      \includegraphics[width=.19\textwidth]{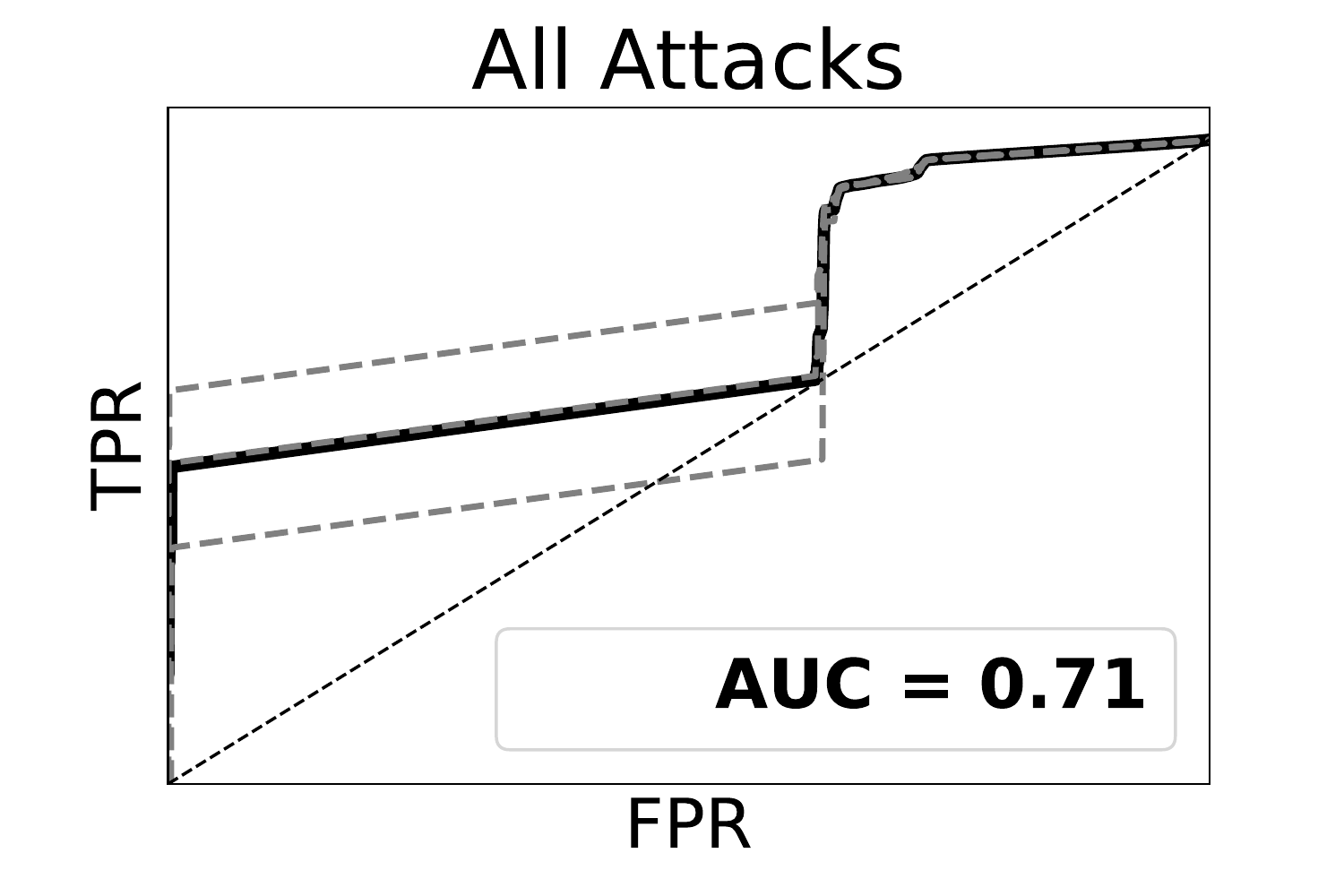}
      \\
            \includegraphics[width=.19\textwidth]{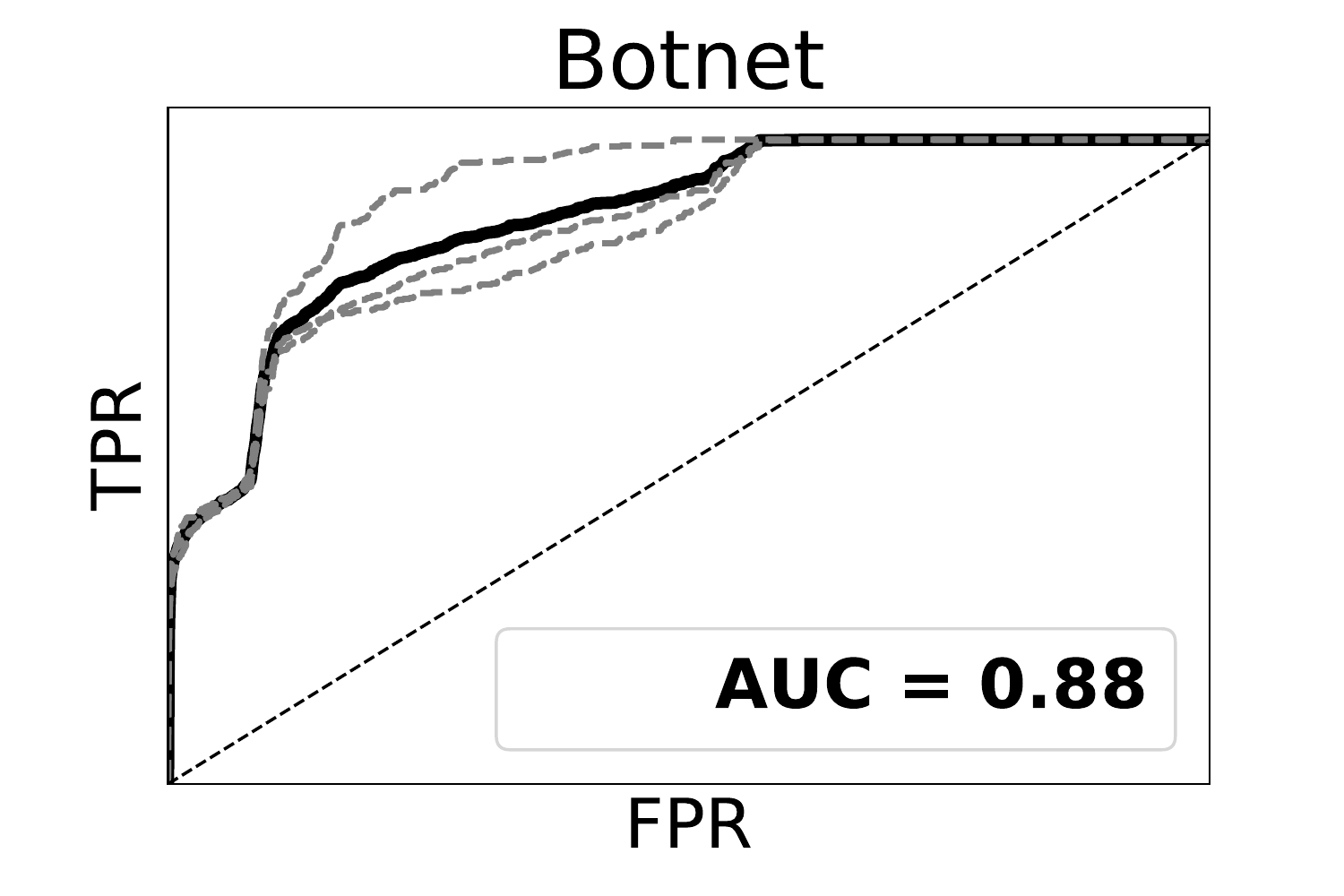} &
      \includegraphics[width=.19\textwidth]{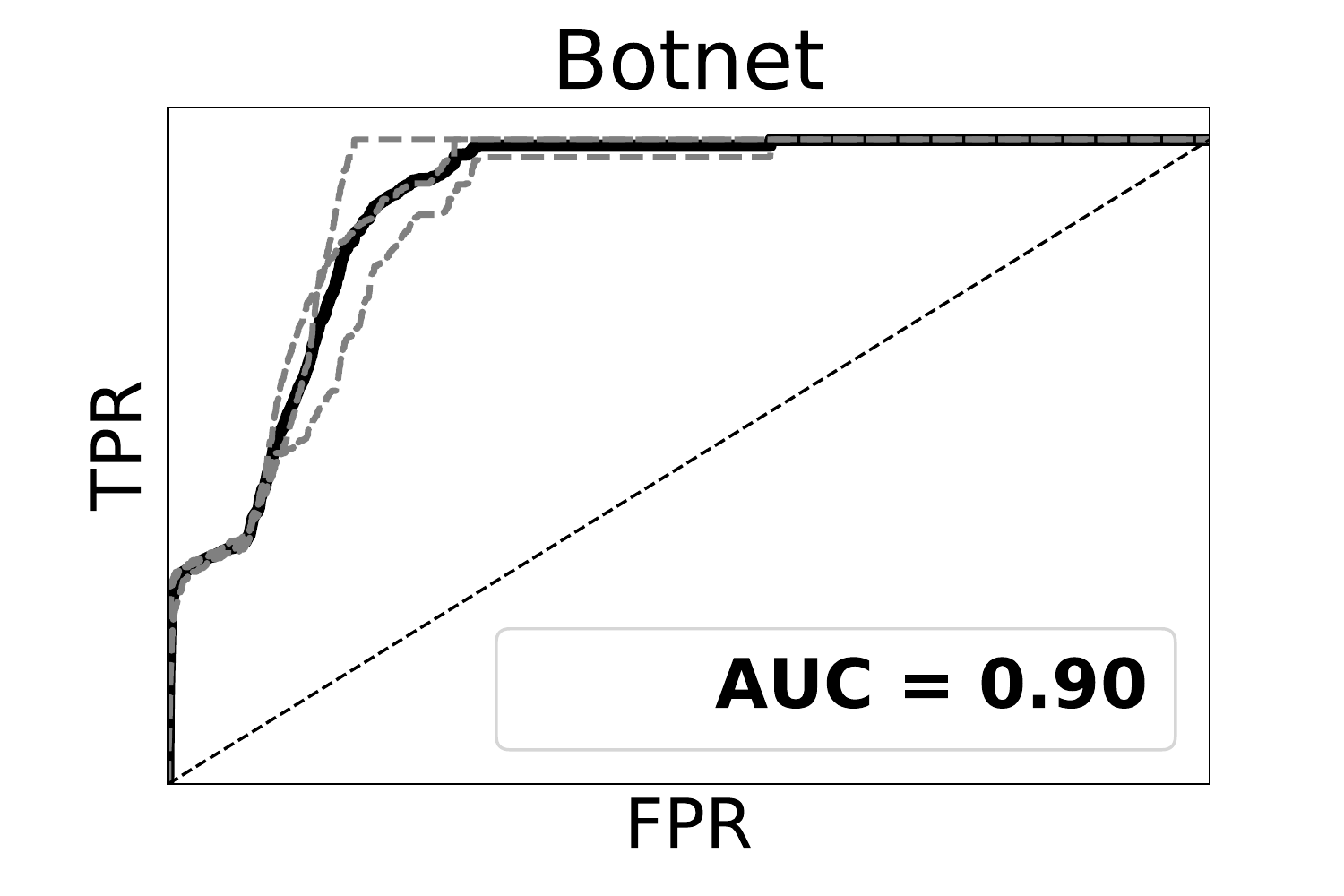} &
      \includegraphics[width=.19\textwidth]{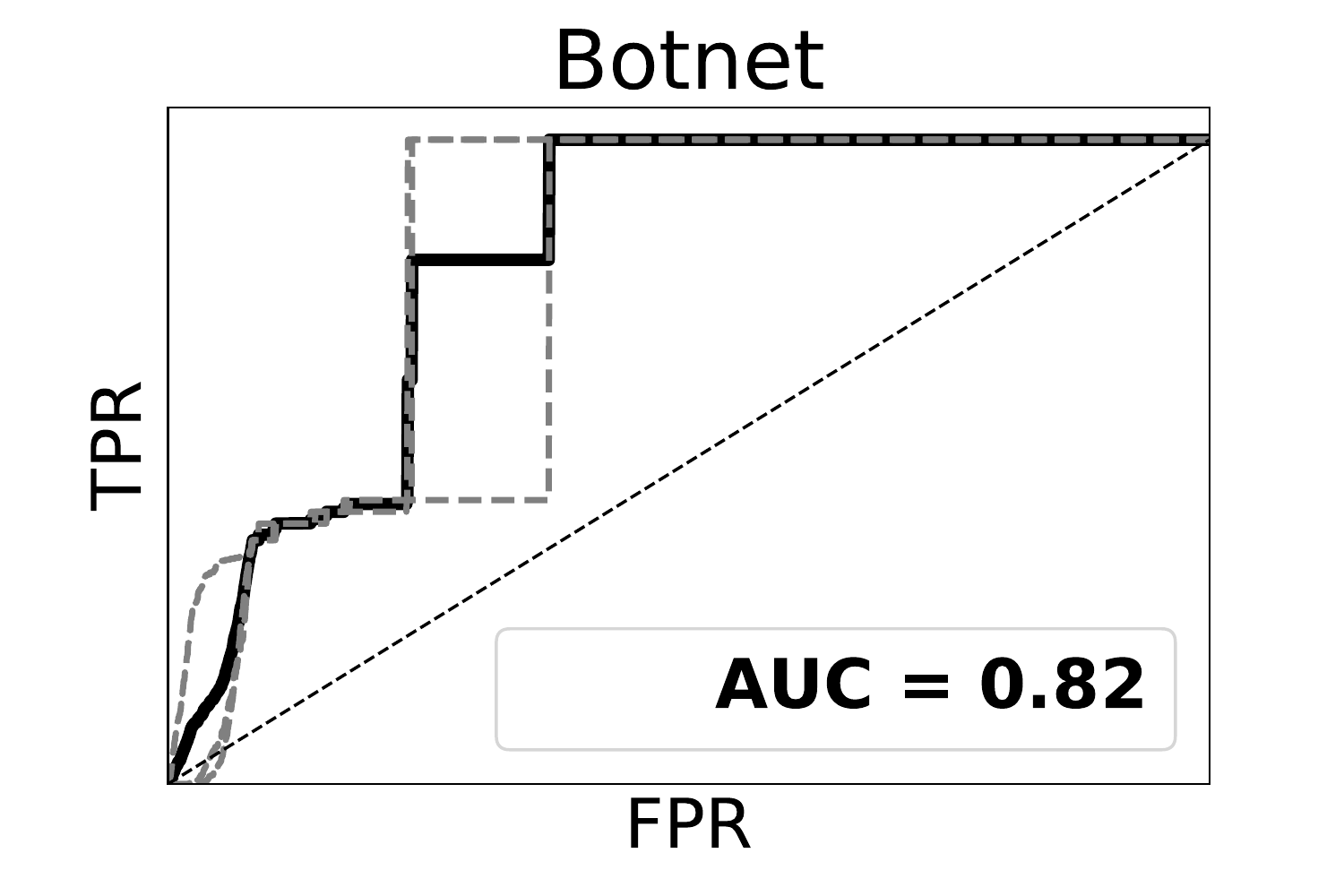} &
      \includegraphics[width=.19\textwidth]{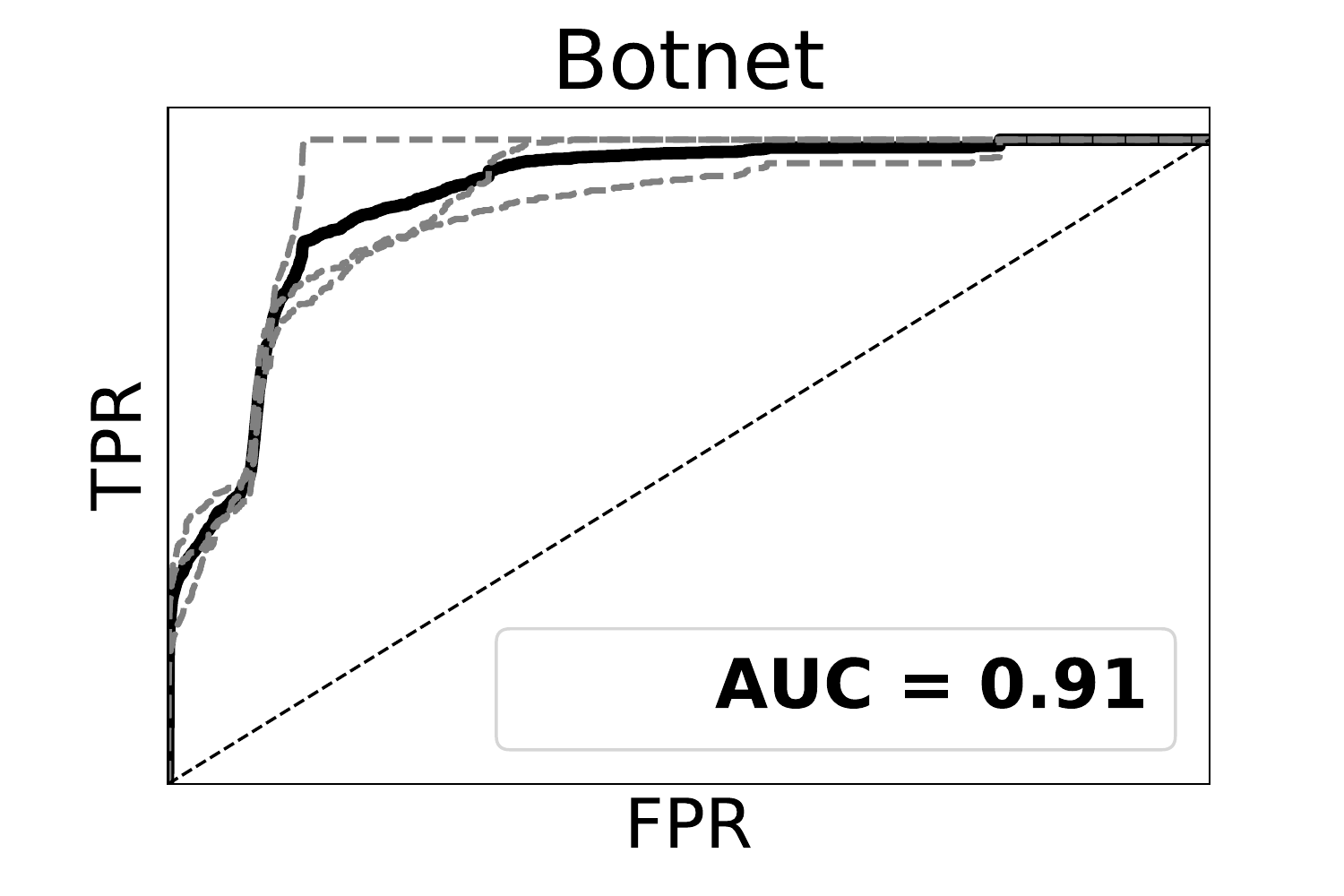} &
      \includegraphics[width=.19\textwidth]{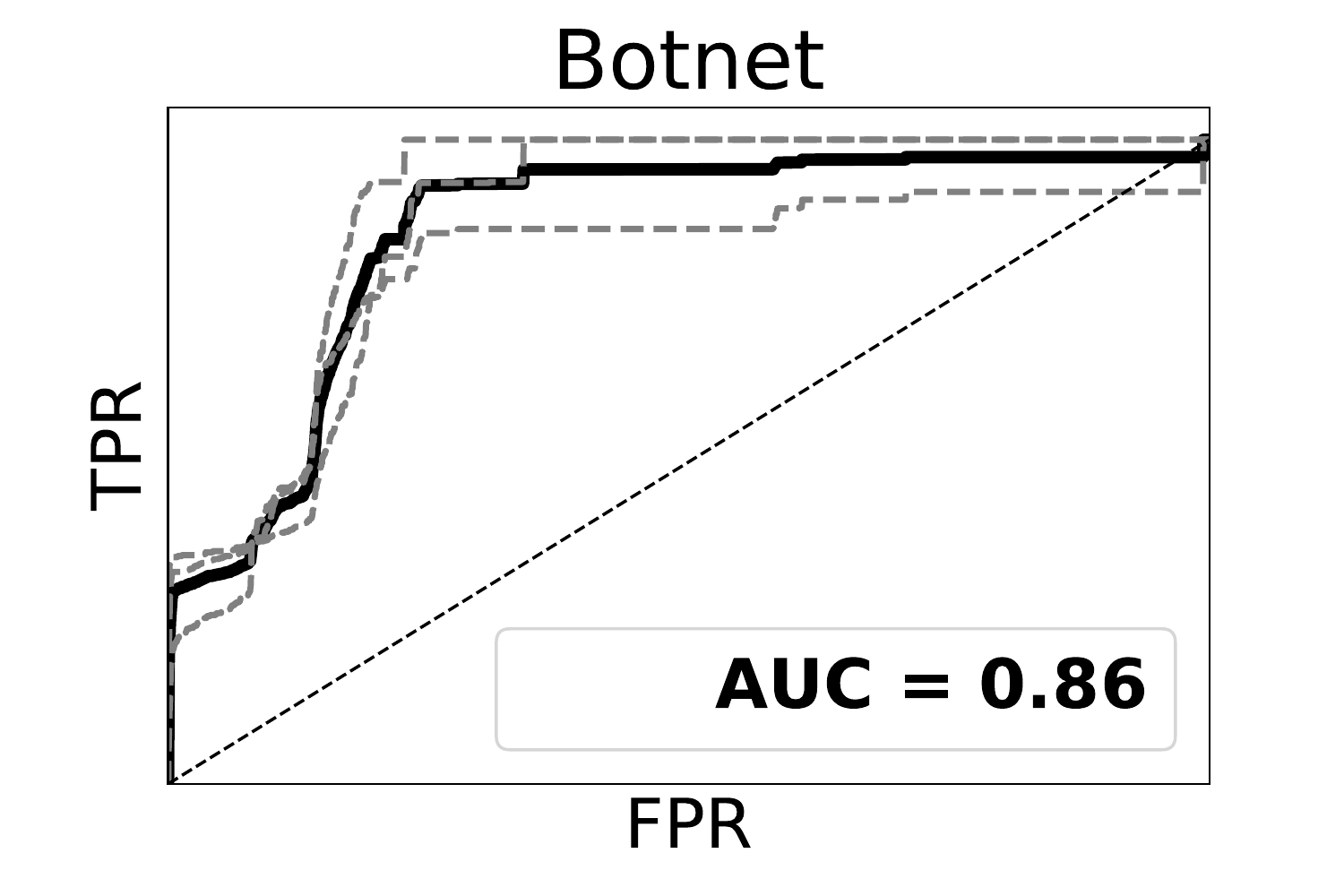}
      \\
            \includegraphics[width=.19\textwidth]{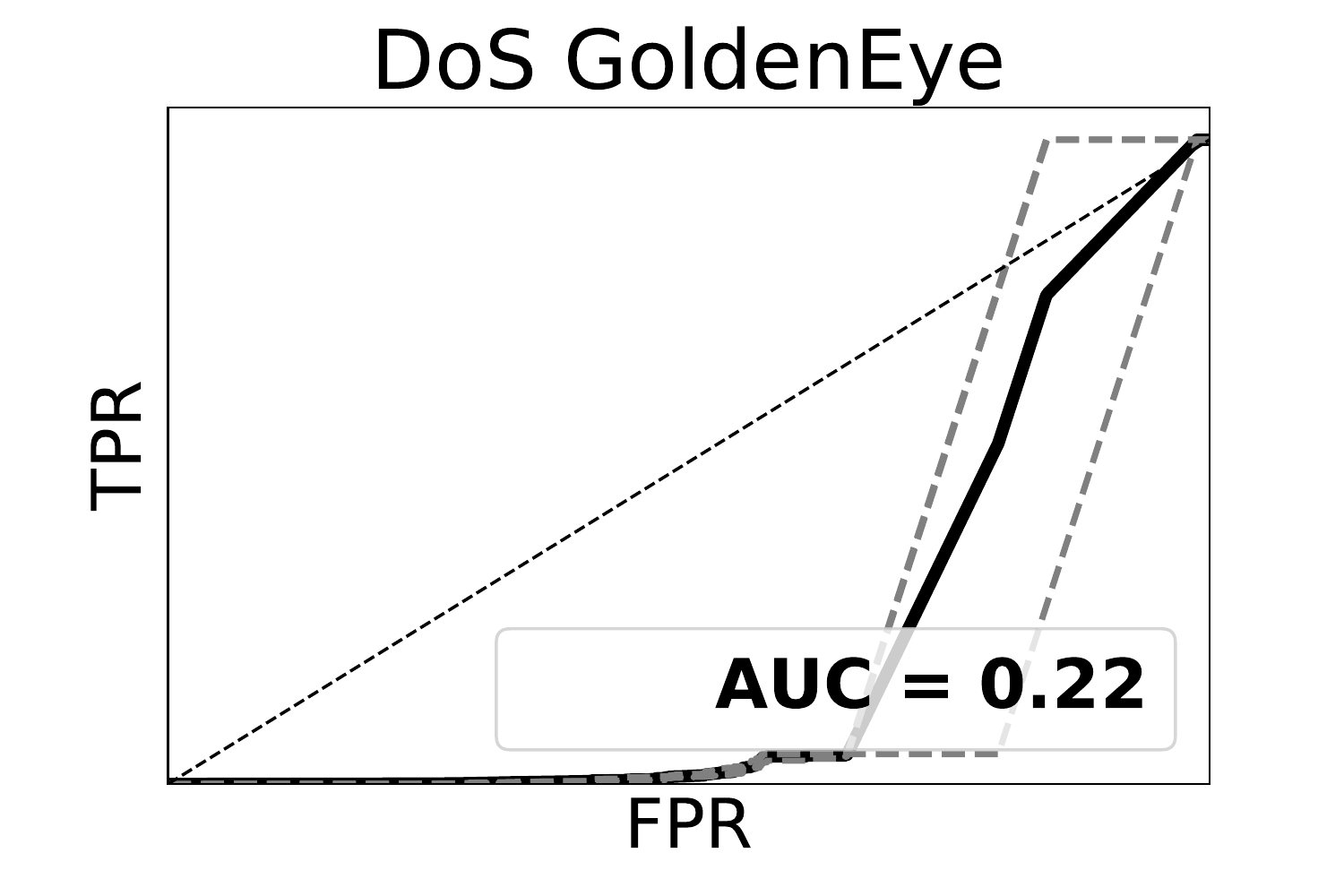} &
      \includegraphics[width=.19\textwidth]{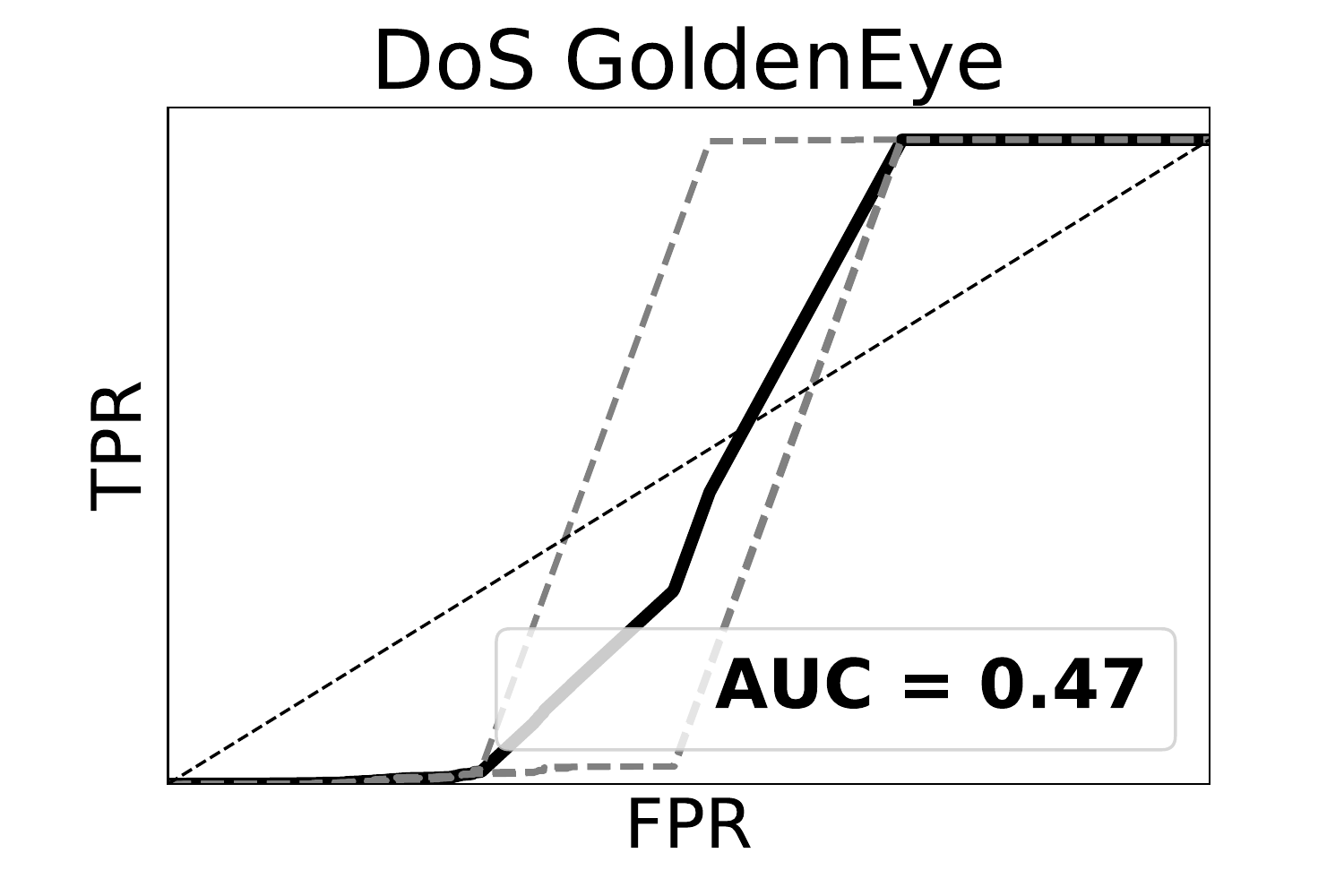} &
      \includegraphics[width=.19\textwidth]{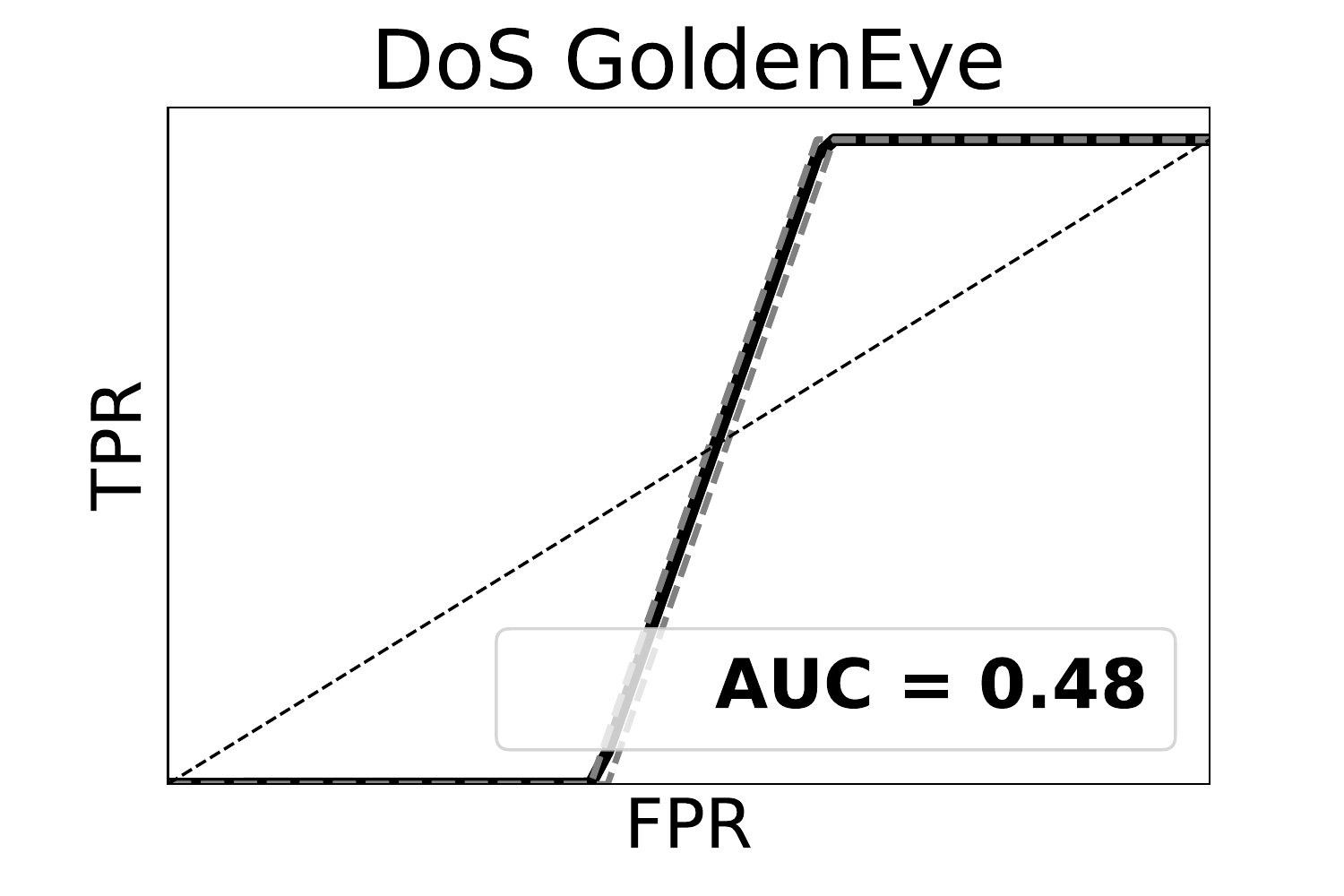} &
      \includegraphics[width=.19\textwidth]{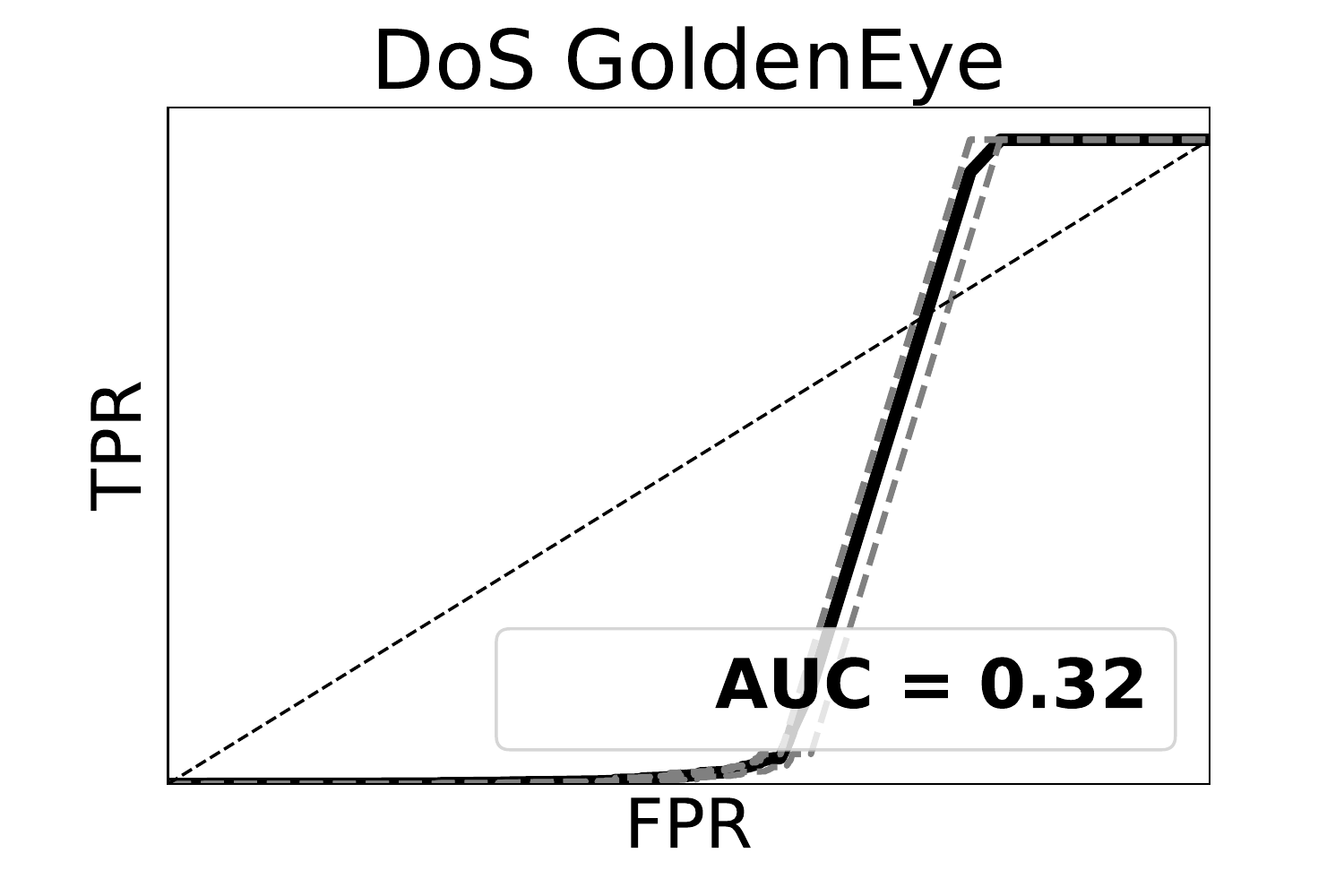} &
      \includegraphics[width=.19\textwidth]{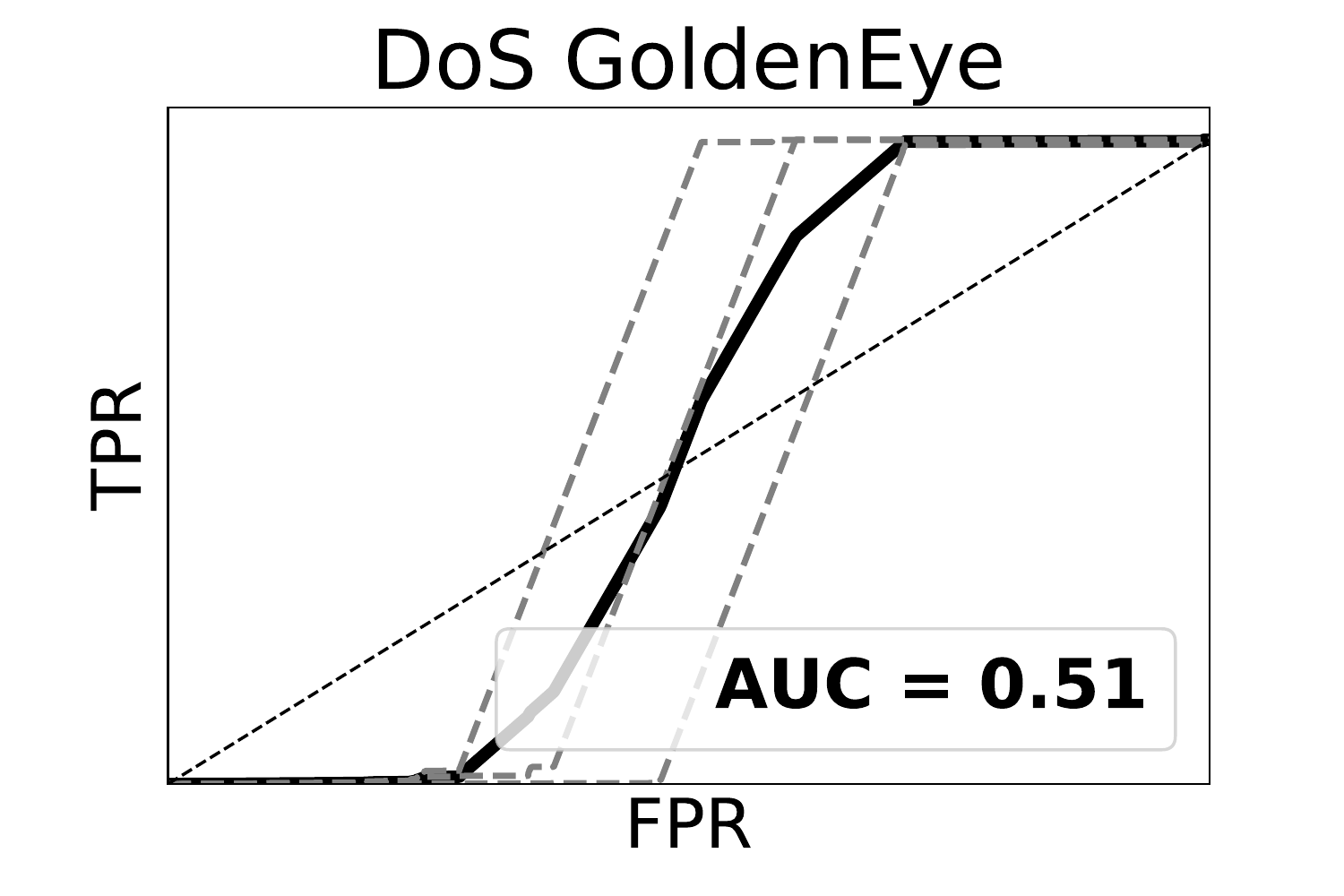}
      \\
      \includegraphics[width=.19\textwidth]{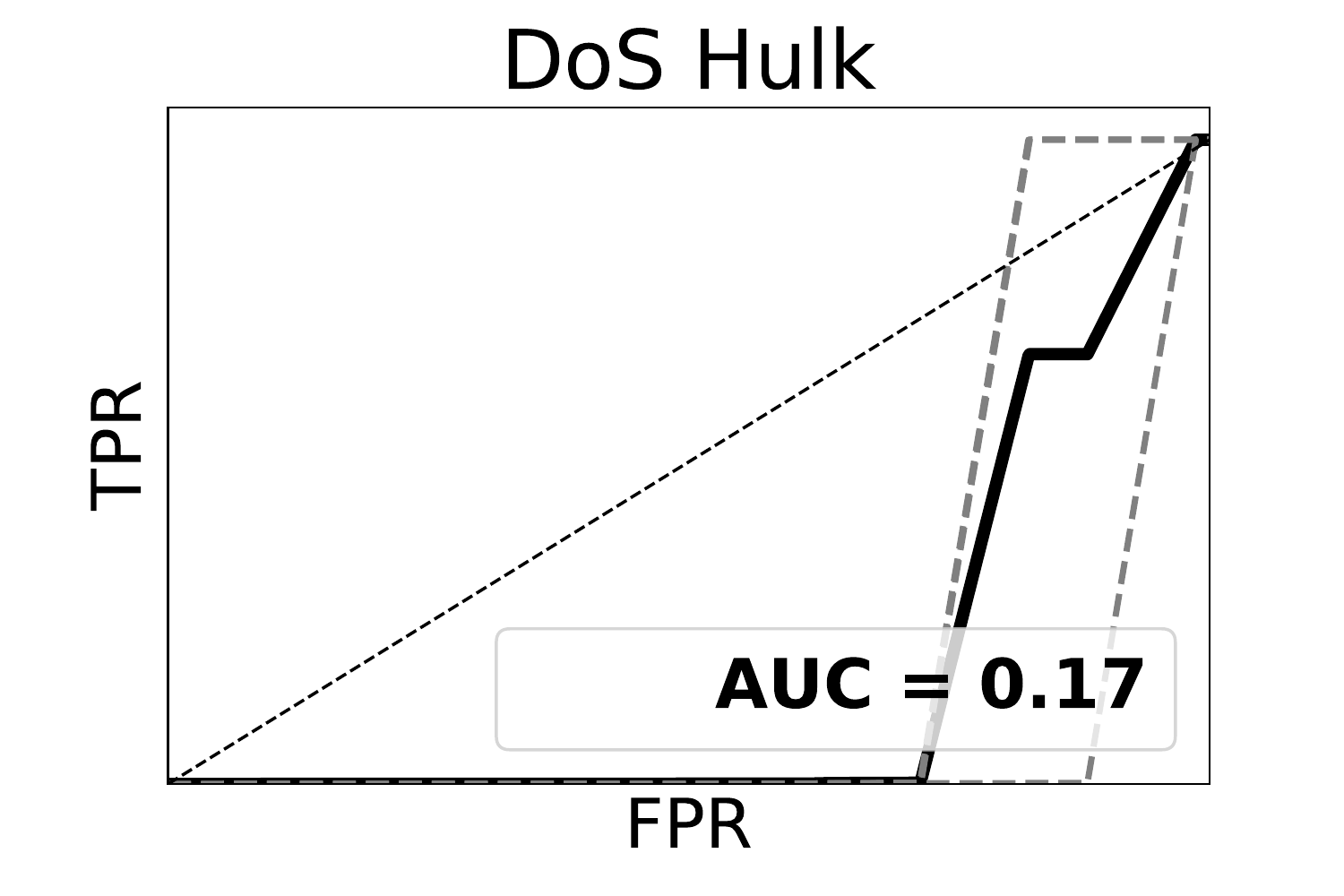} &
      \includegraphics[width=.19\textwidth]{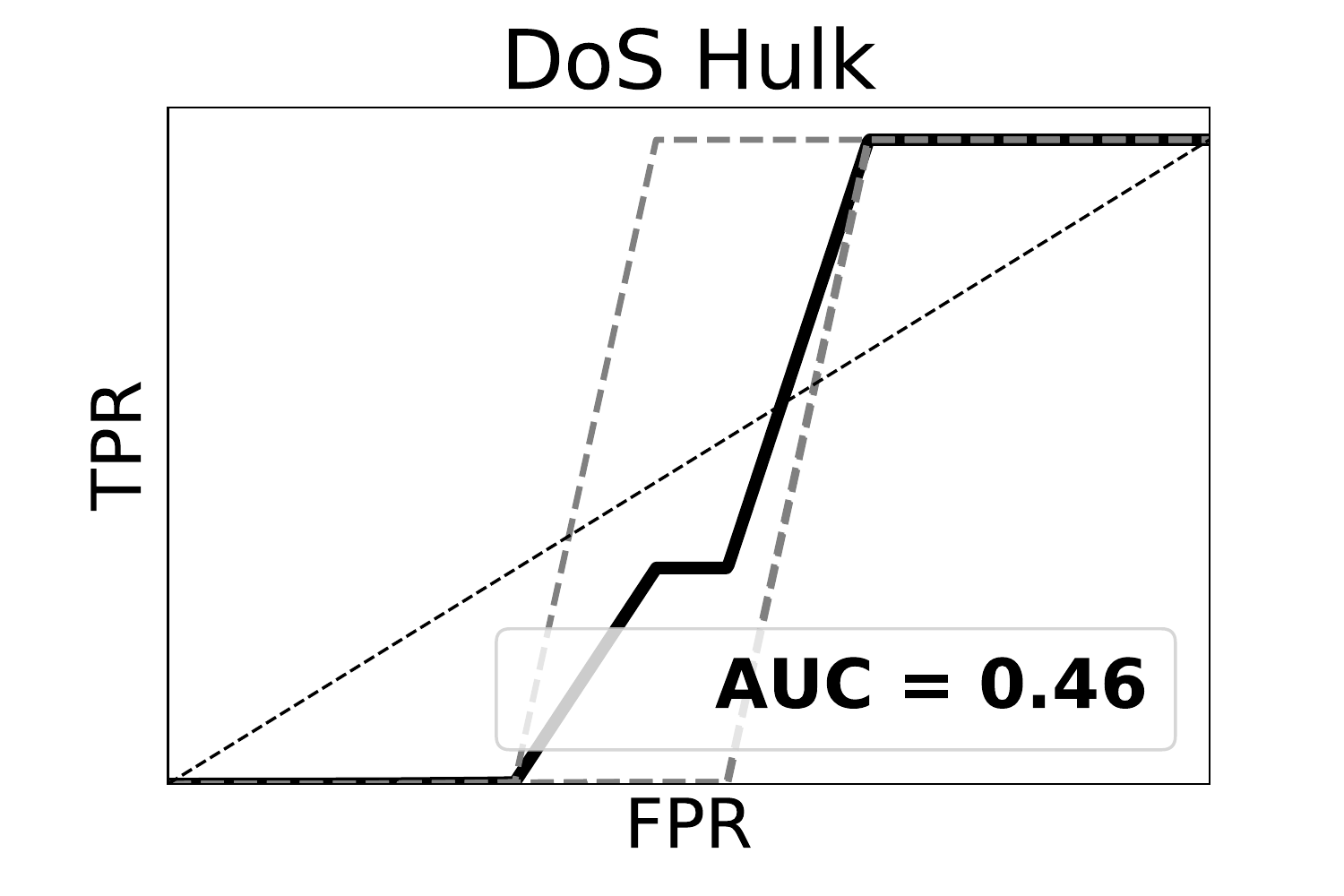} &
      \includegraphics[width=.19\textwidth]{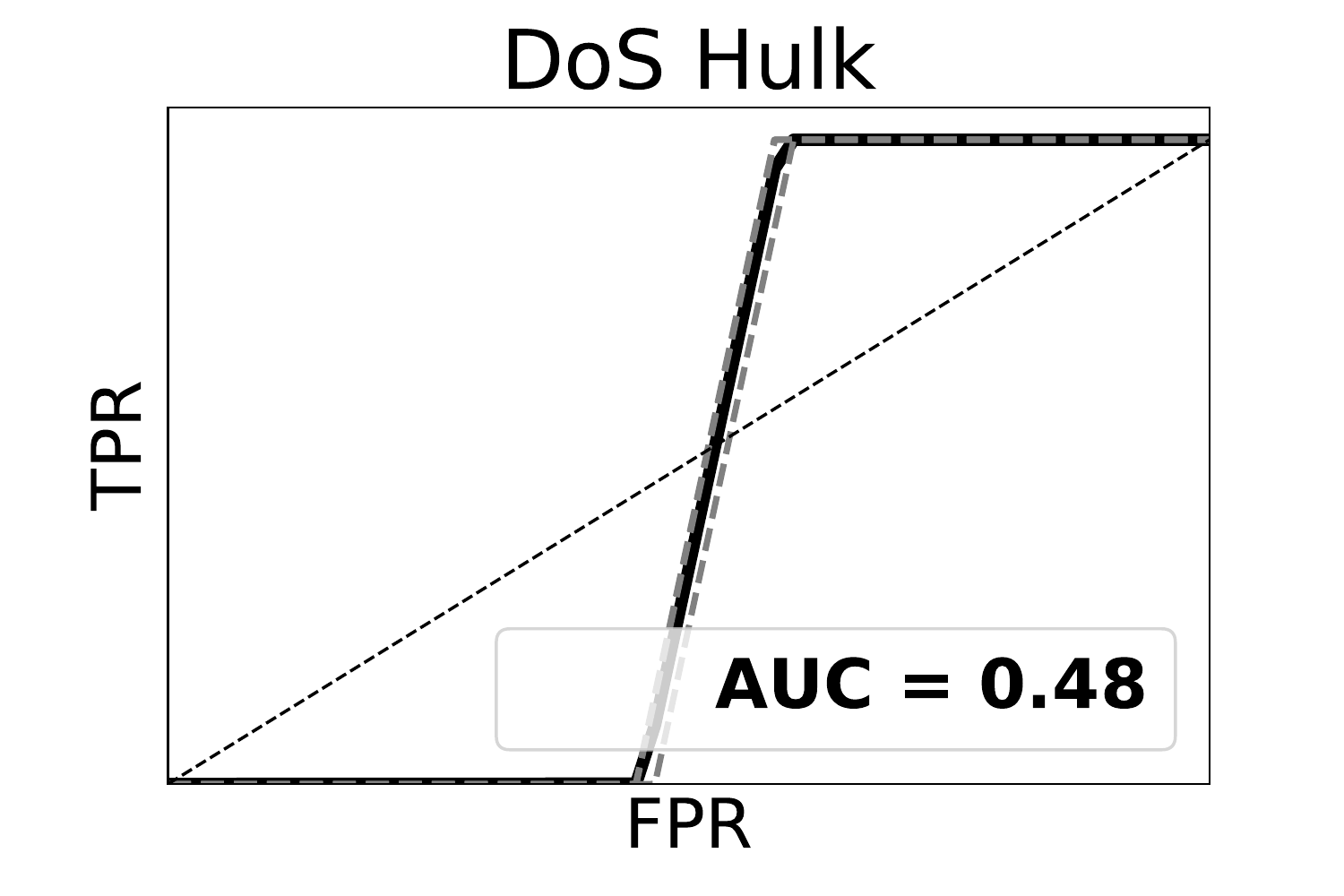} &
      \includegraphics[width=.19\textwidth]{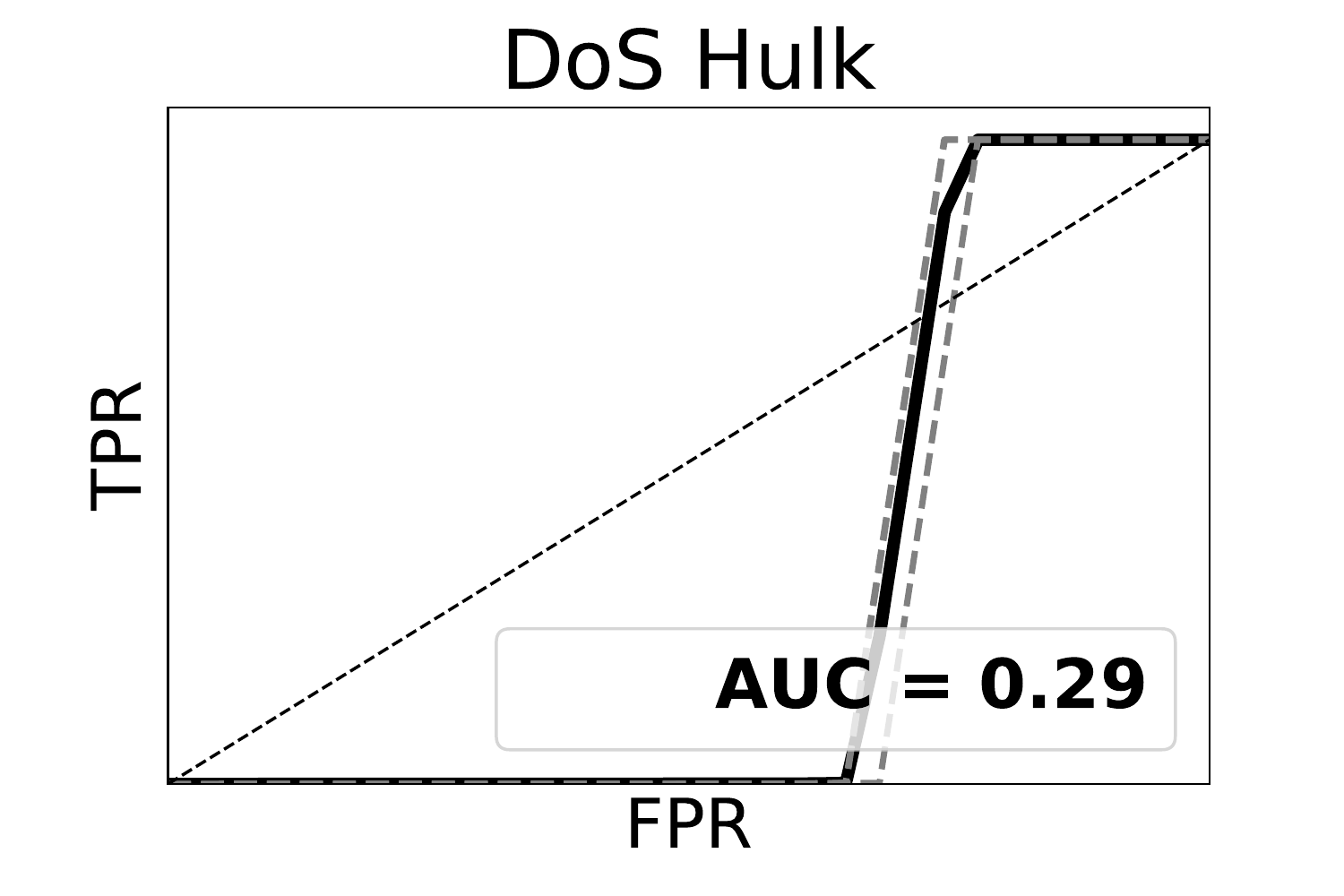} &
      \includegraphics[width=.19\textwidth]{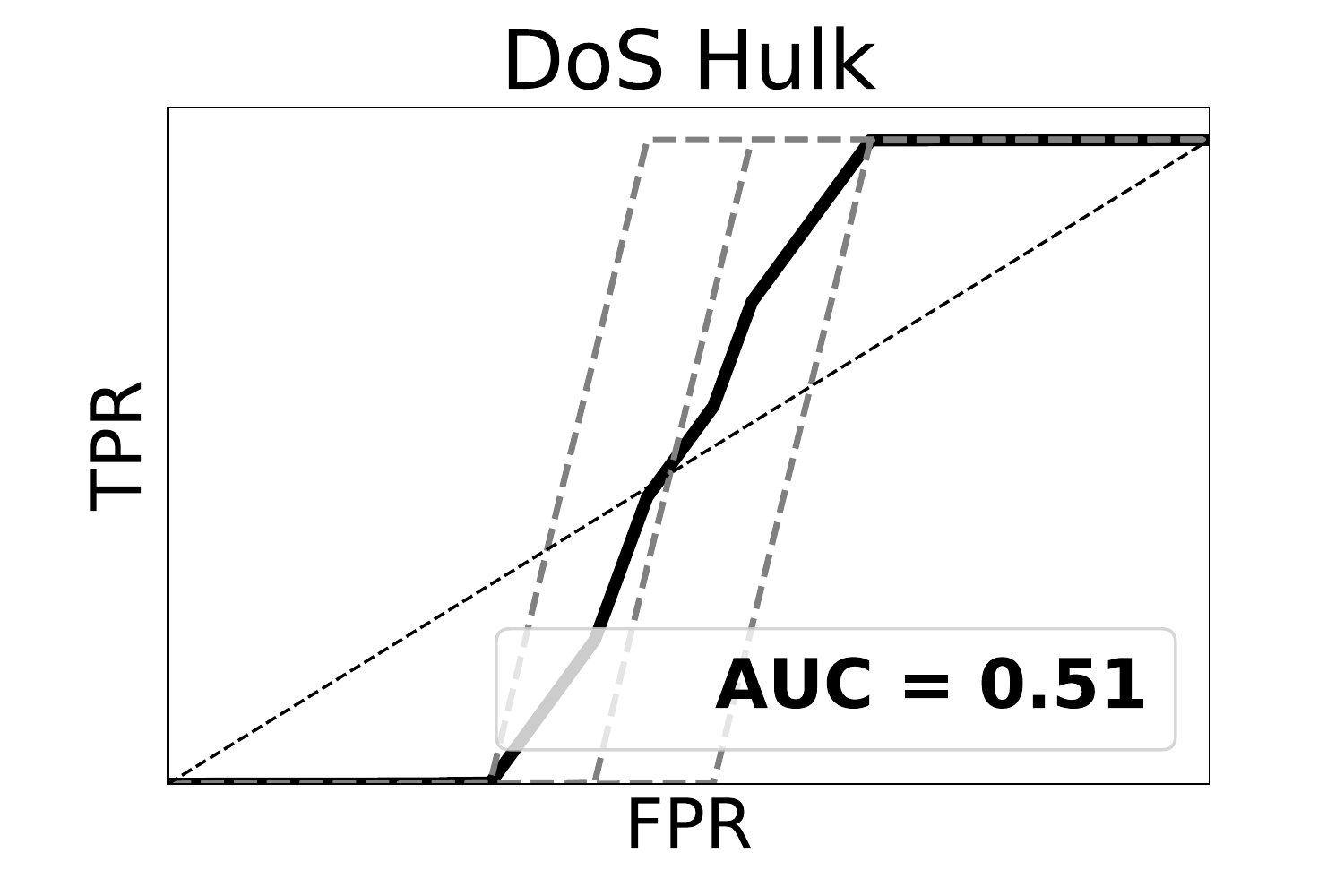}
      \\
            \includegraphics[width=.19\textwidth]{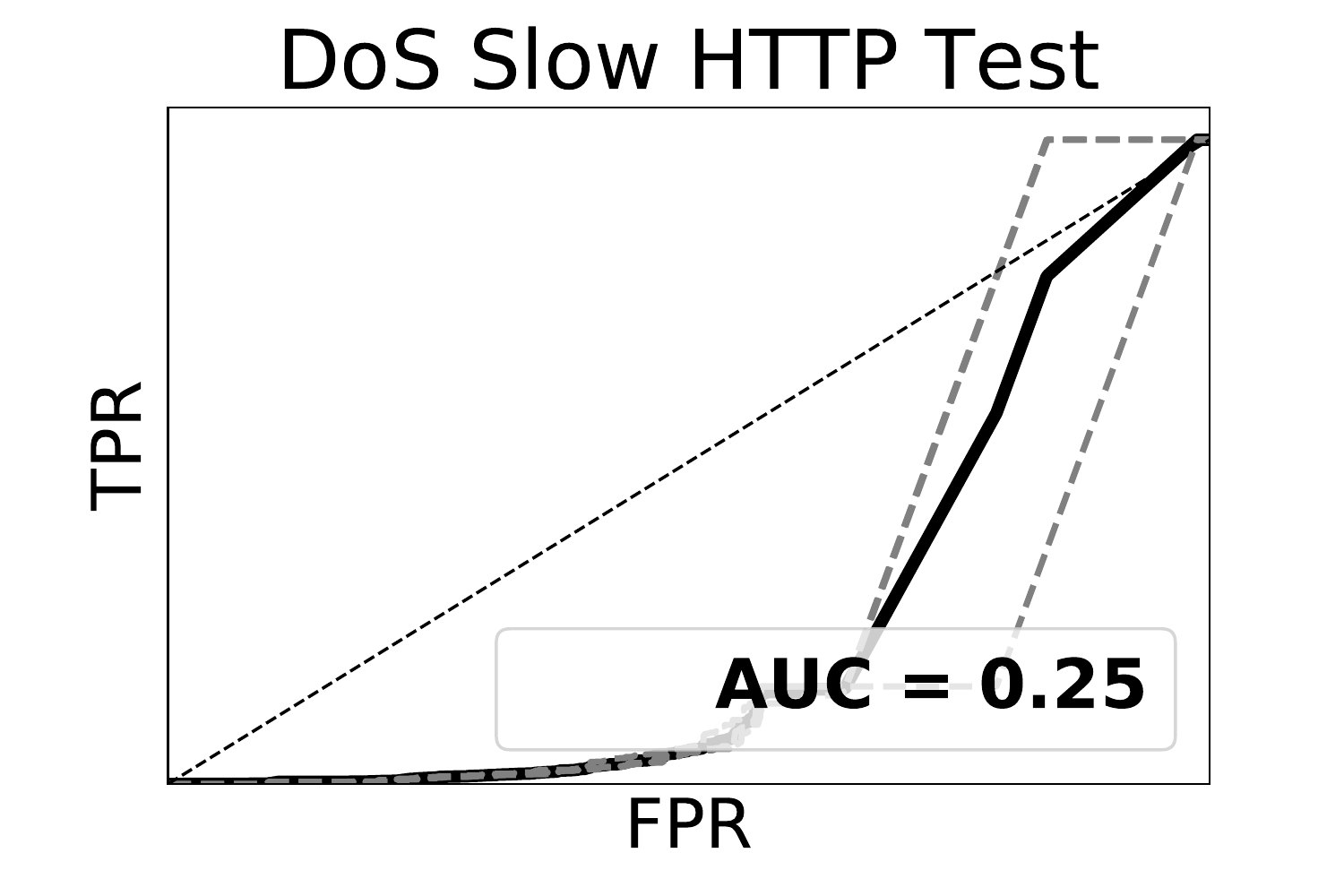} &
      \includegraphics[width=.19\textwidth]{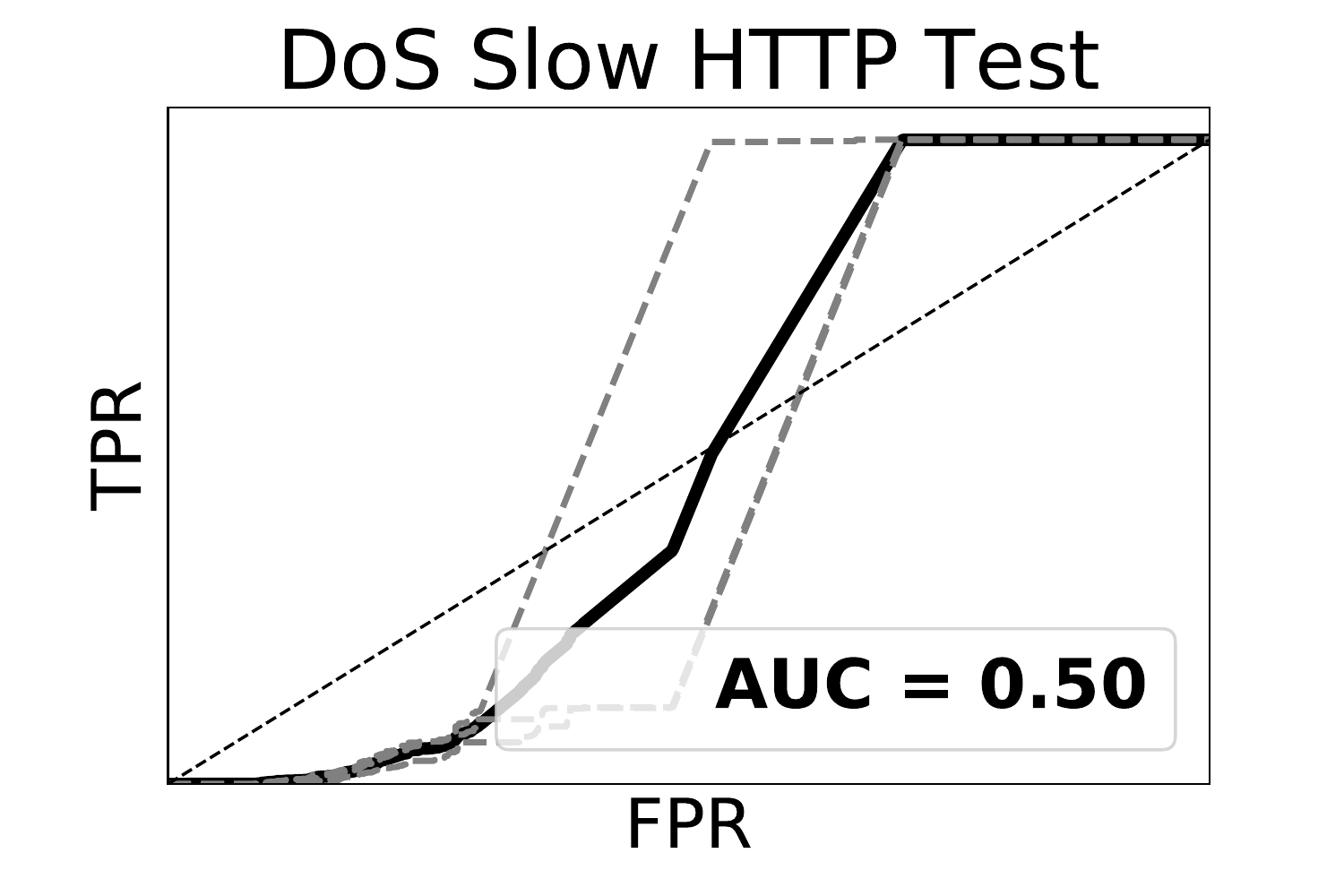} &
      \includegraphics[width=.19\textwidth]{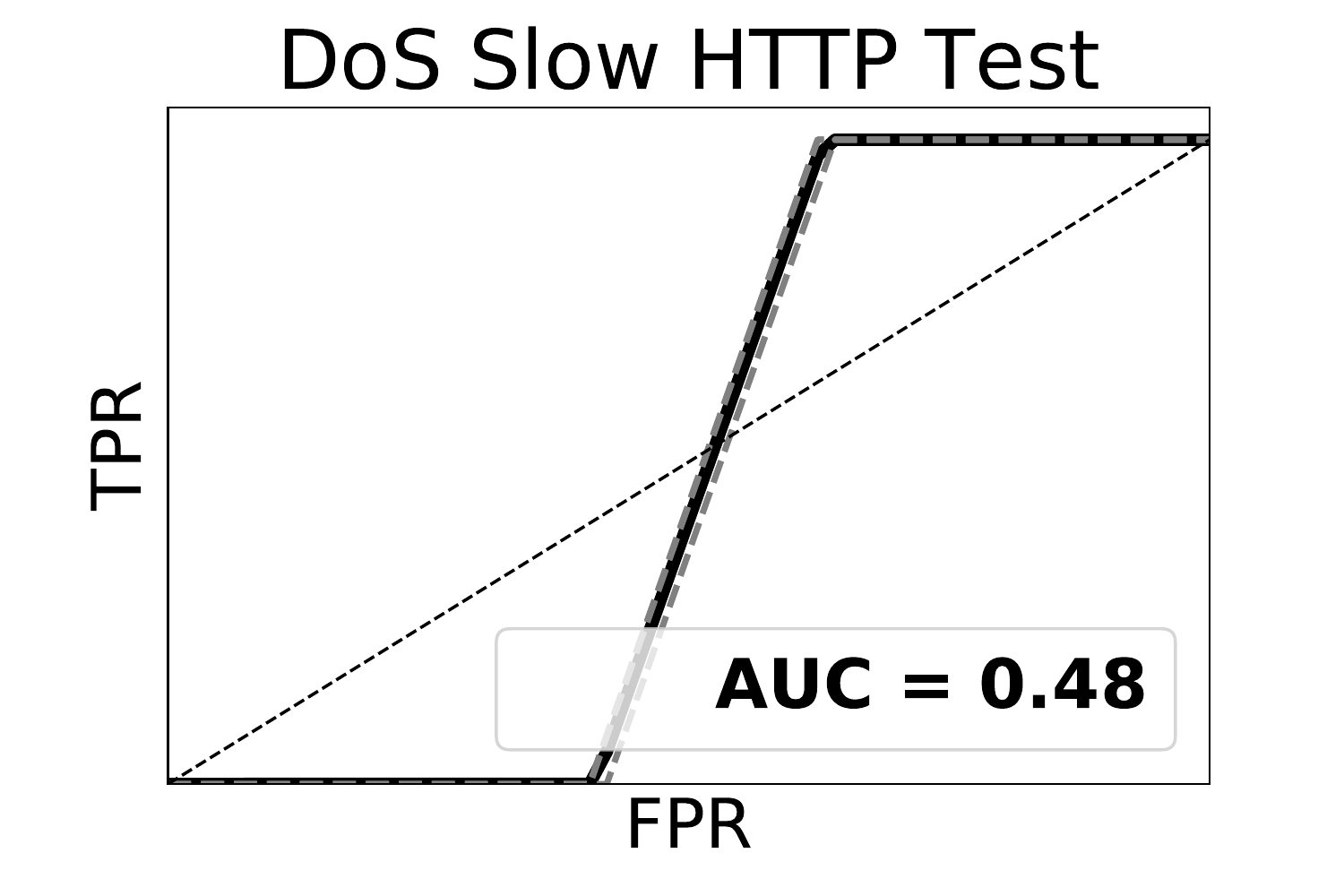} &
      \includegraphics[width=.19\textwidth]{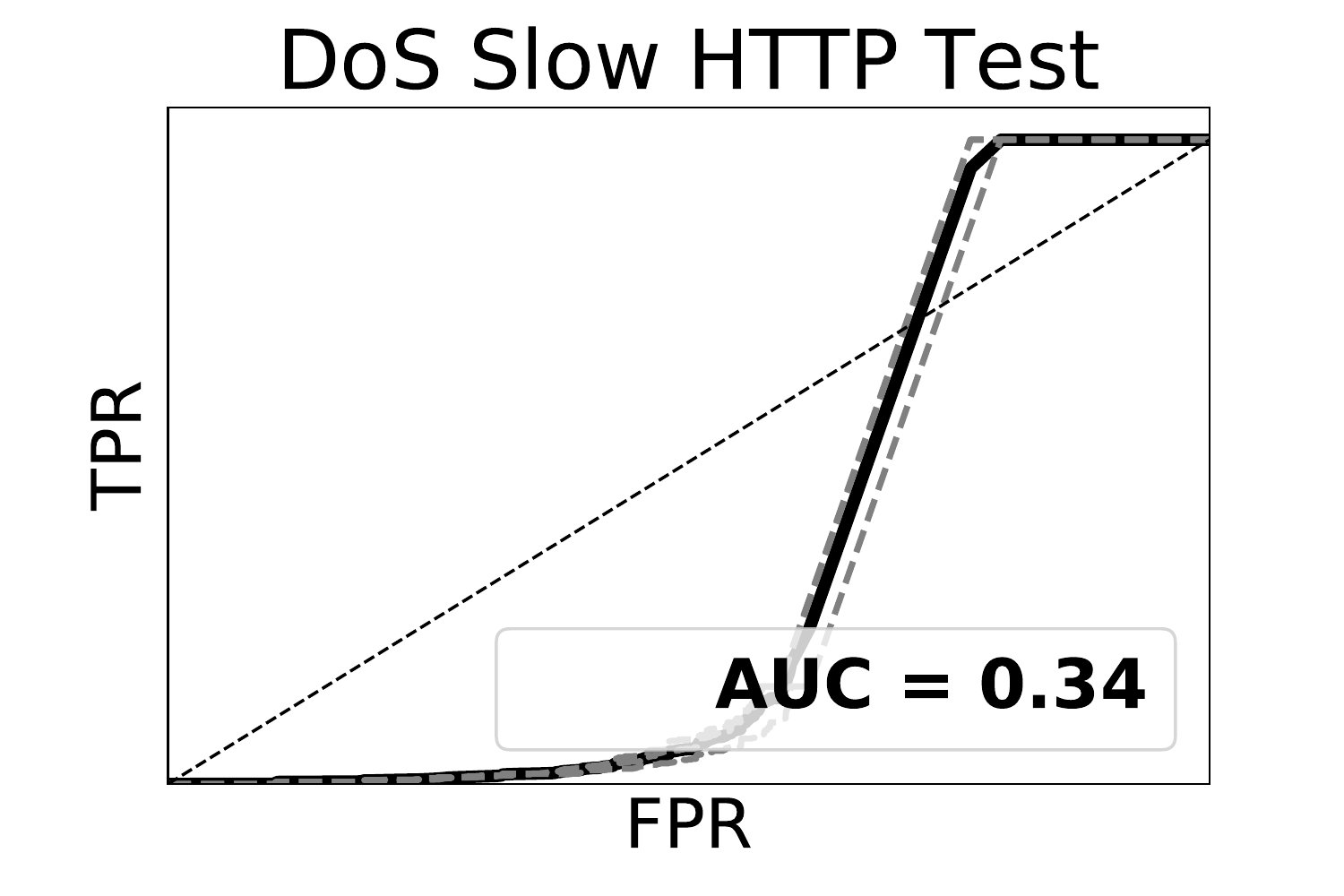} &
      \includegraphics[width=.19\textwidth]{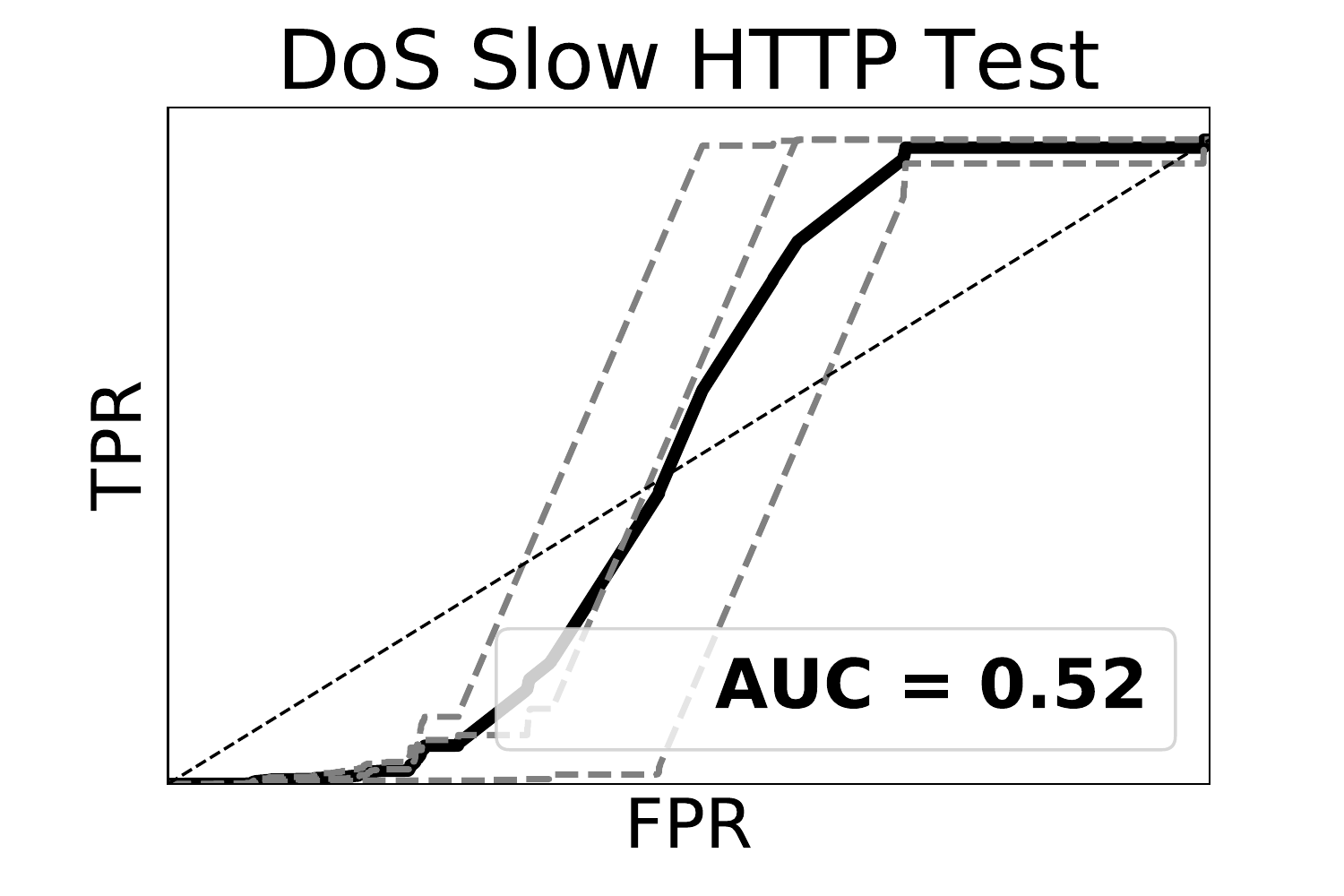}
      \\
            \includegraphics[width=.19\textwidth]{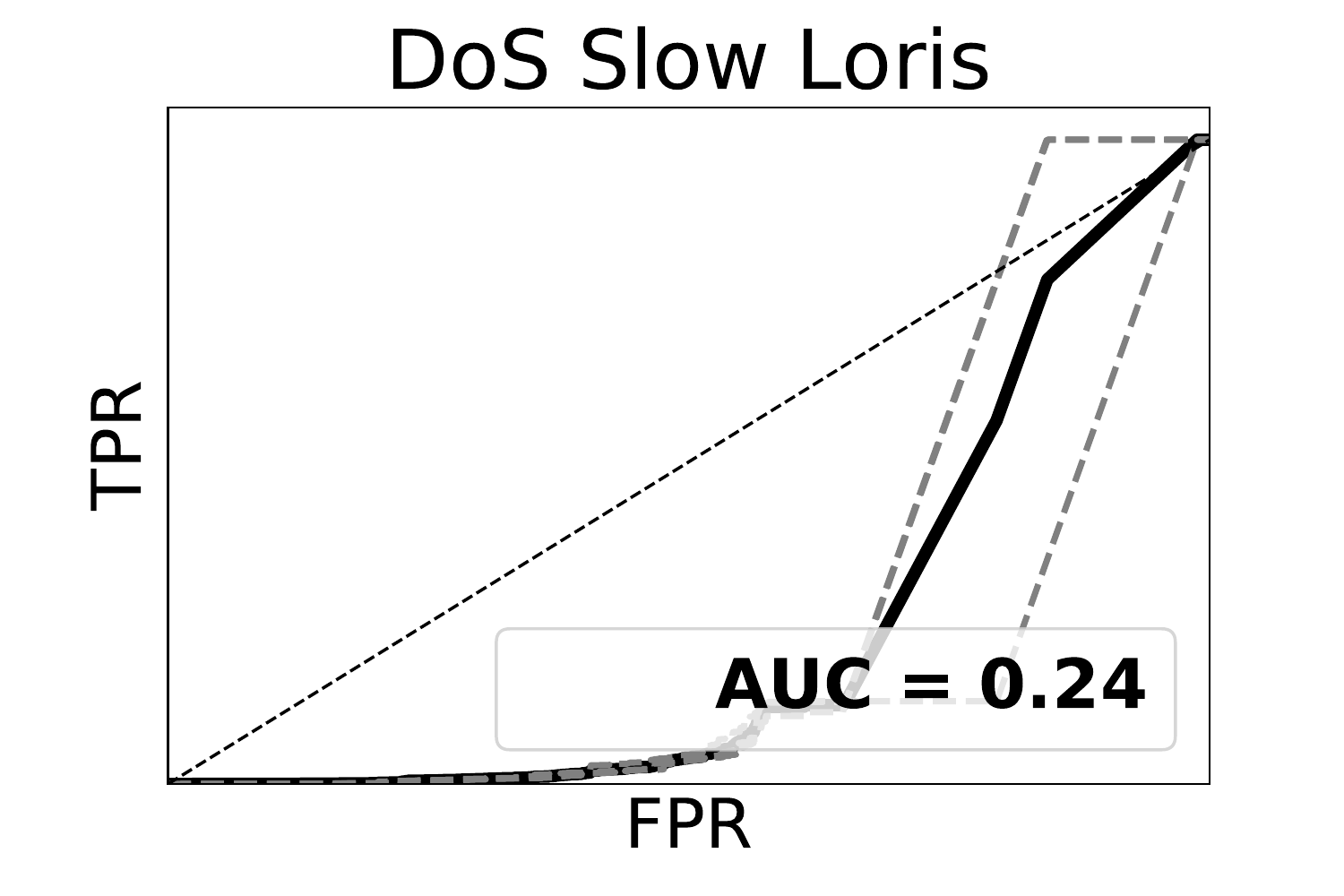} &
      \includegraphics[width=.19\textwidth]{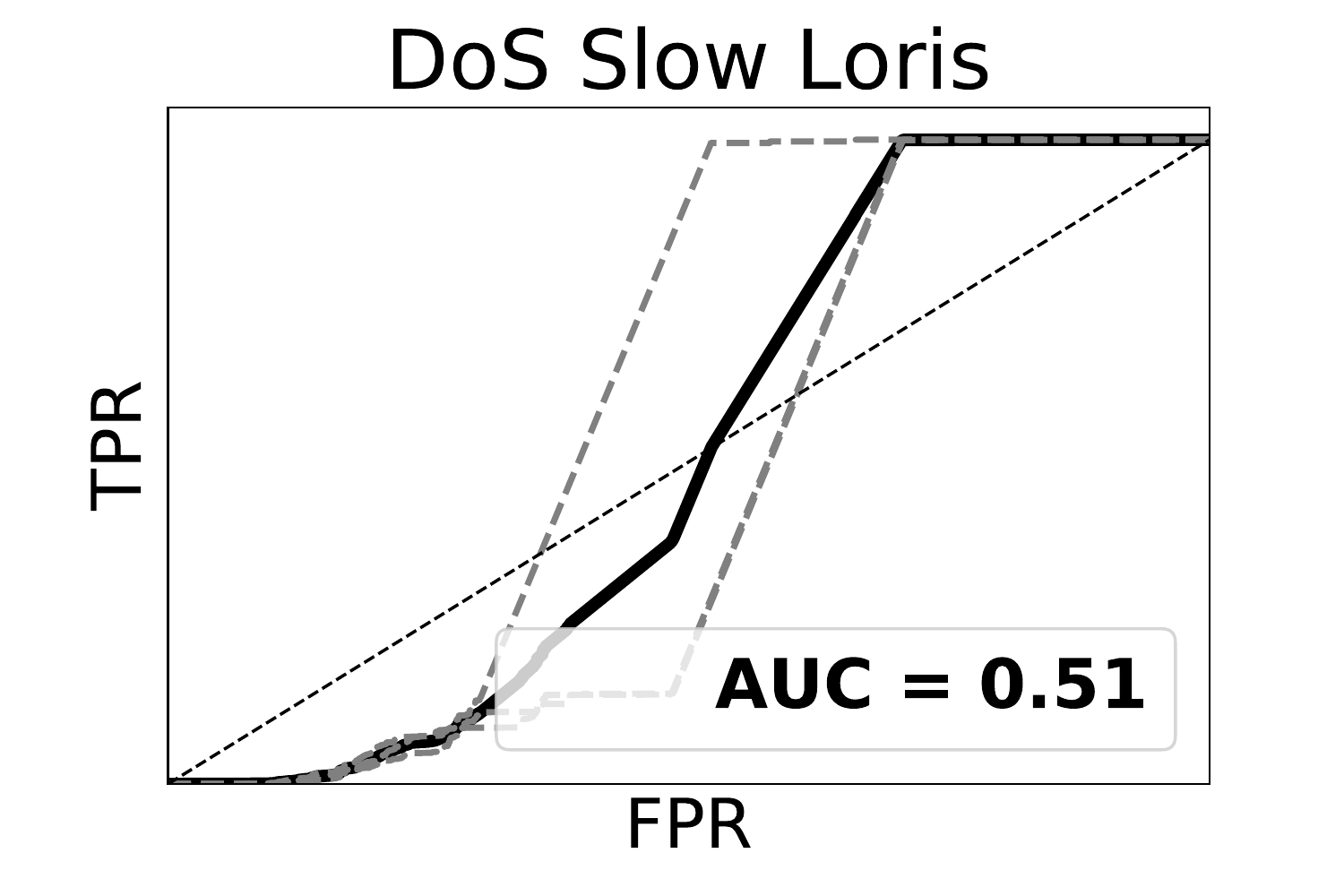} &
      \includegraphics[width=.19\textwidth]{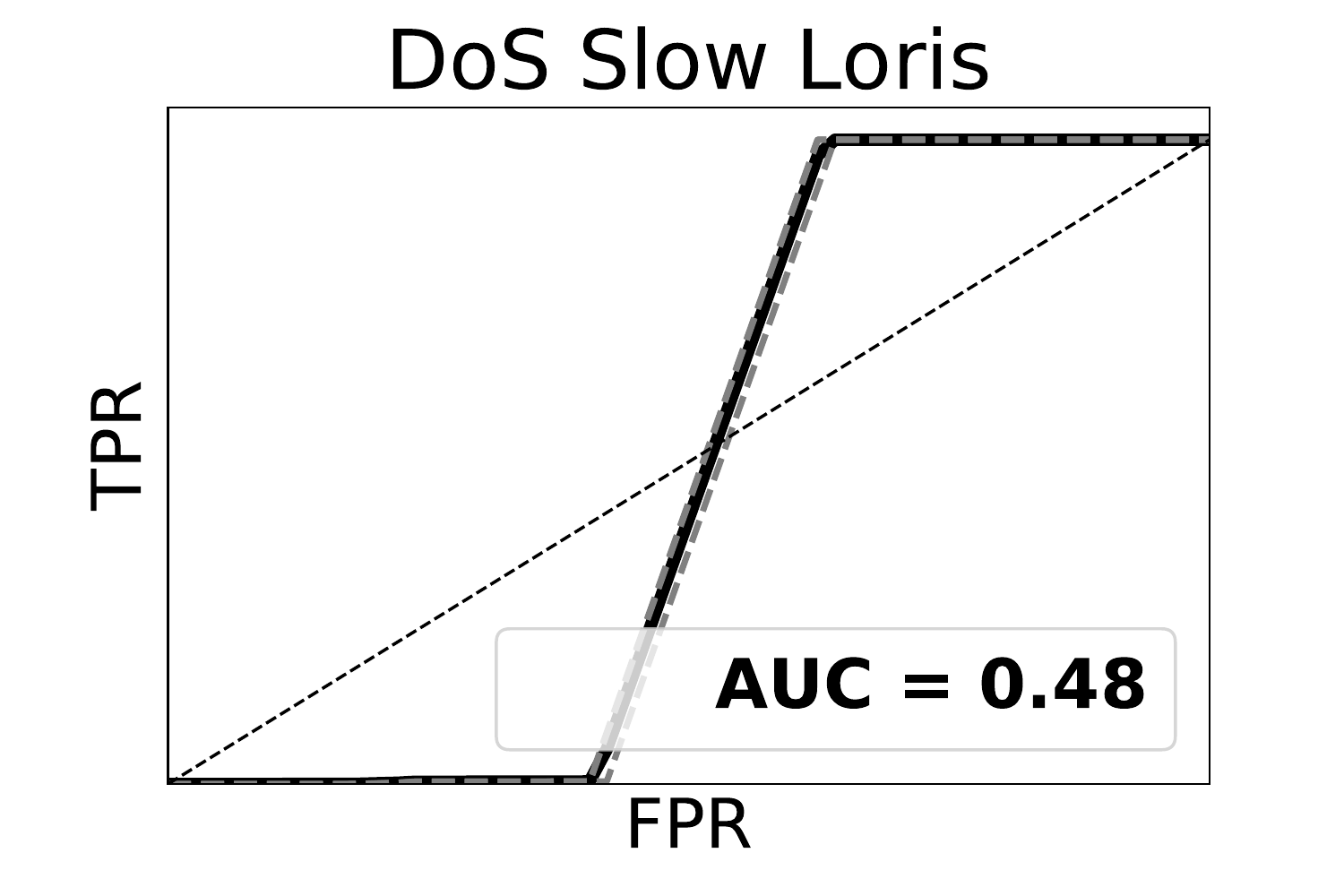} &
      \includegraphics[width=.19\textwidth]{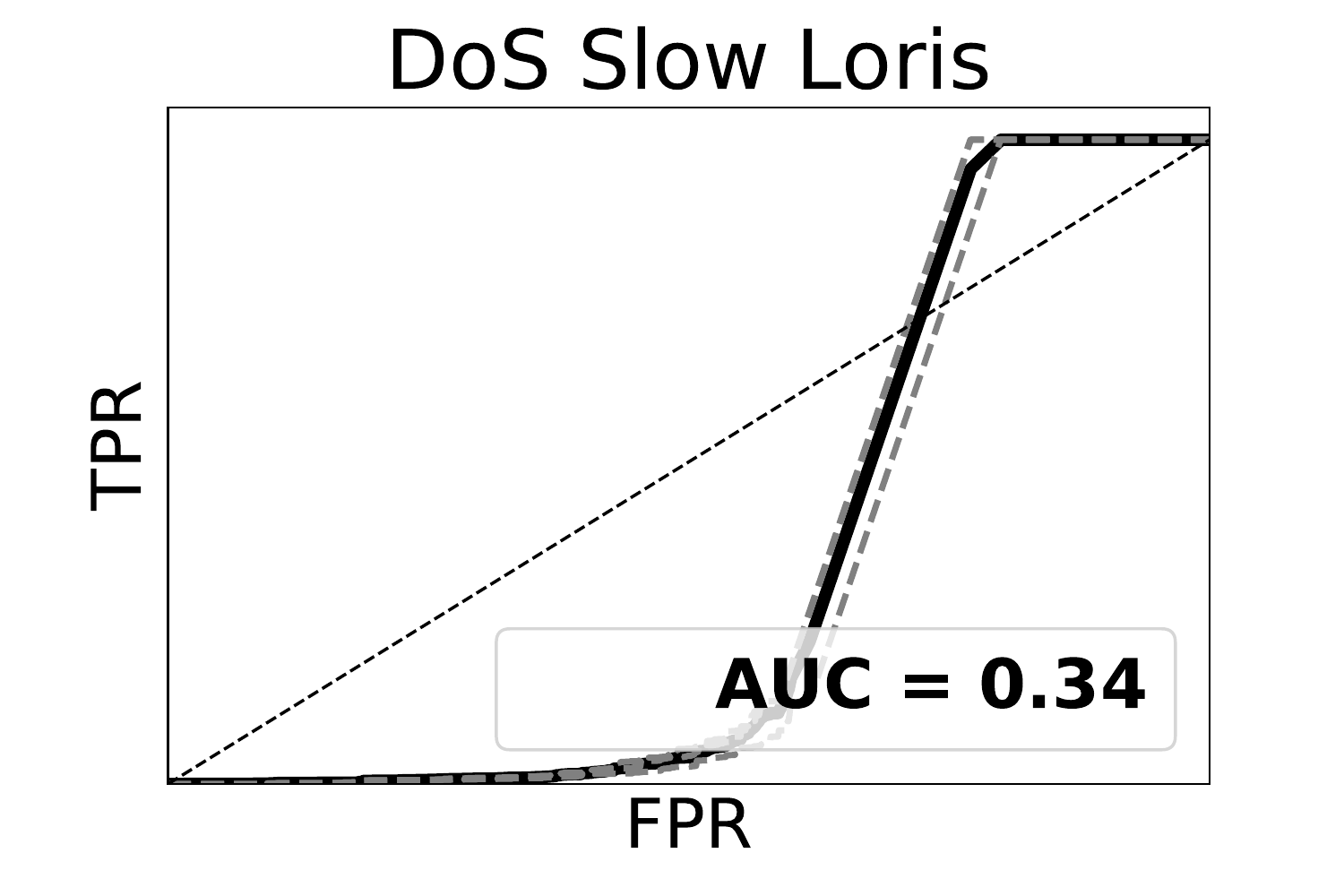} &
      \includegraphics[width=.19\textwidth]{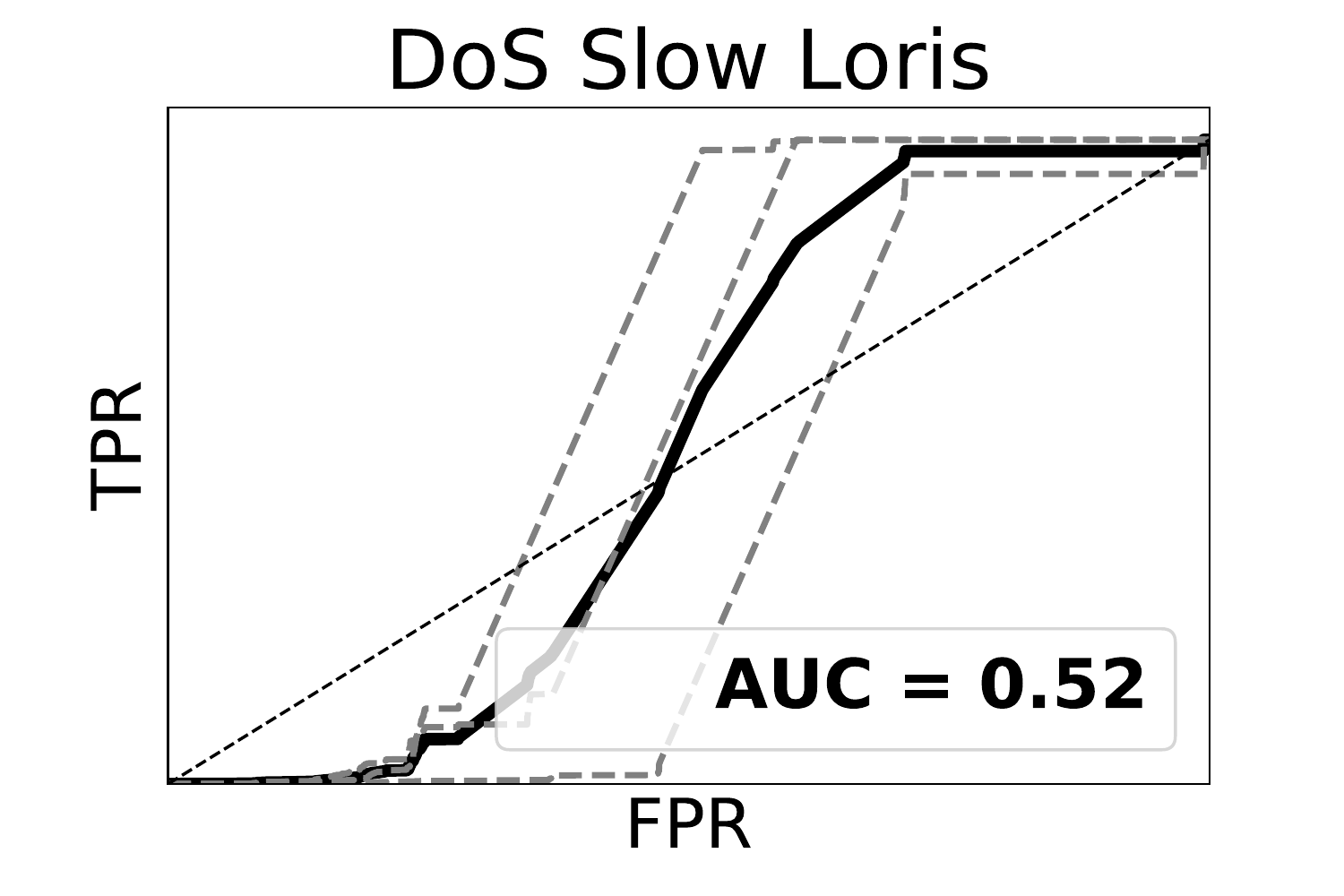}
      \\
      \includegraphics[width=.19\textwidth]{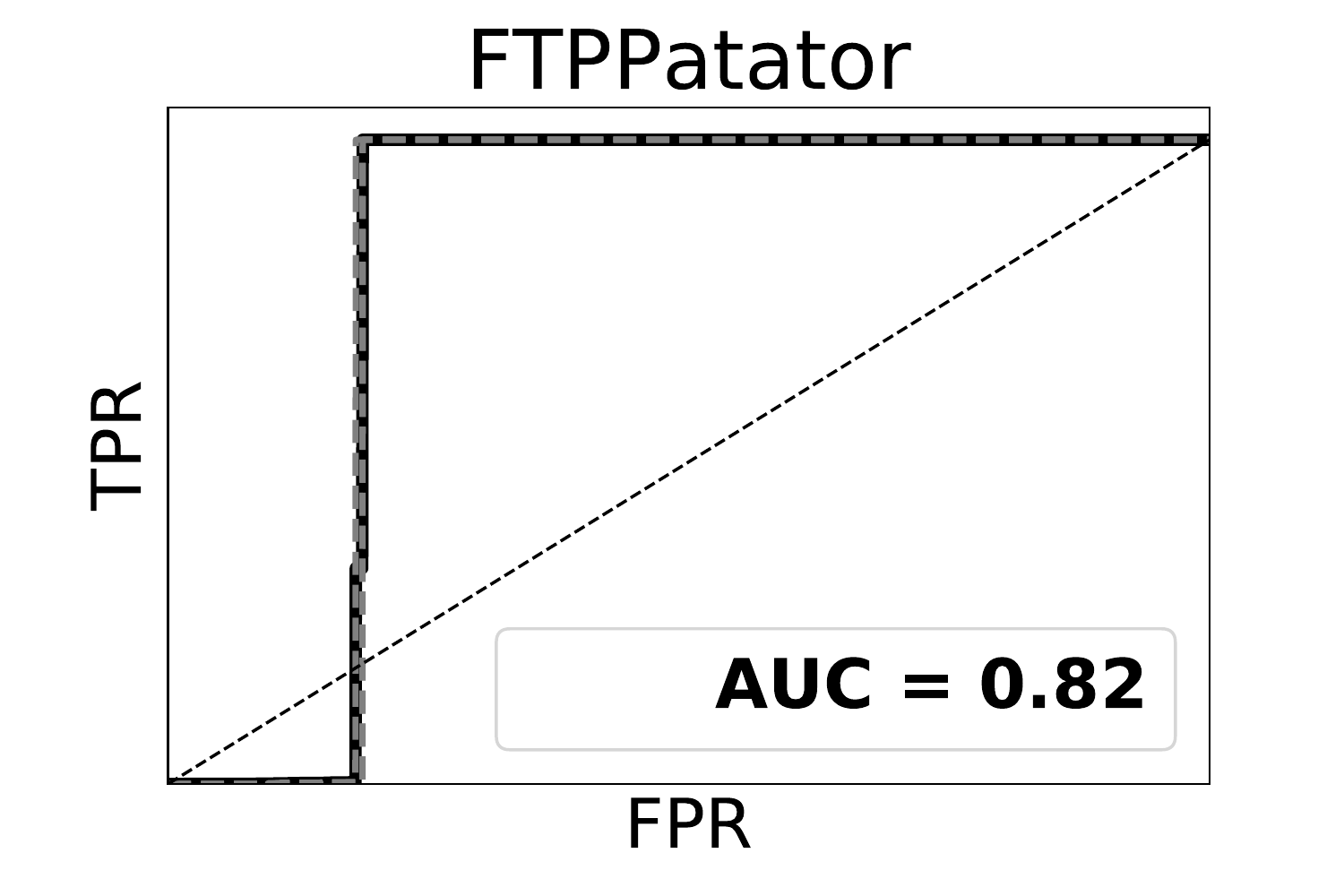} &
      \includegraphics[width=.19\textwidth]{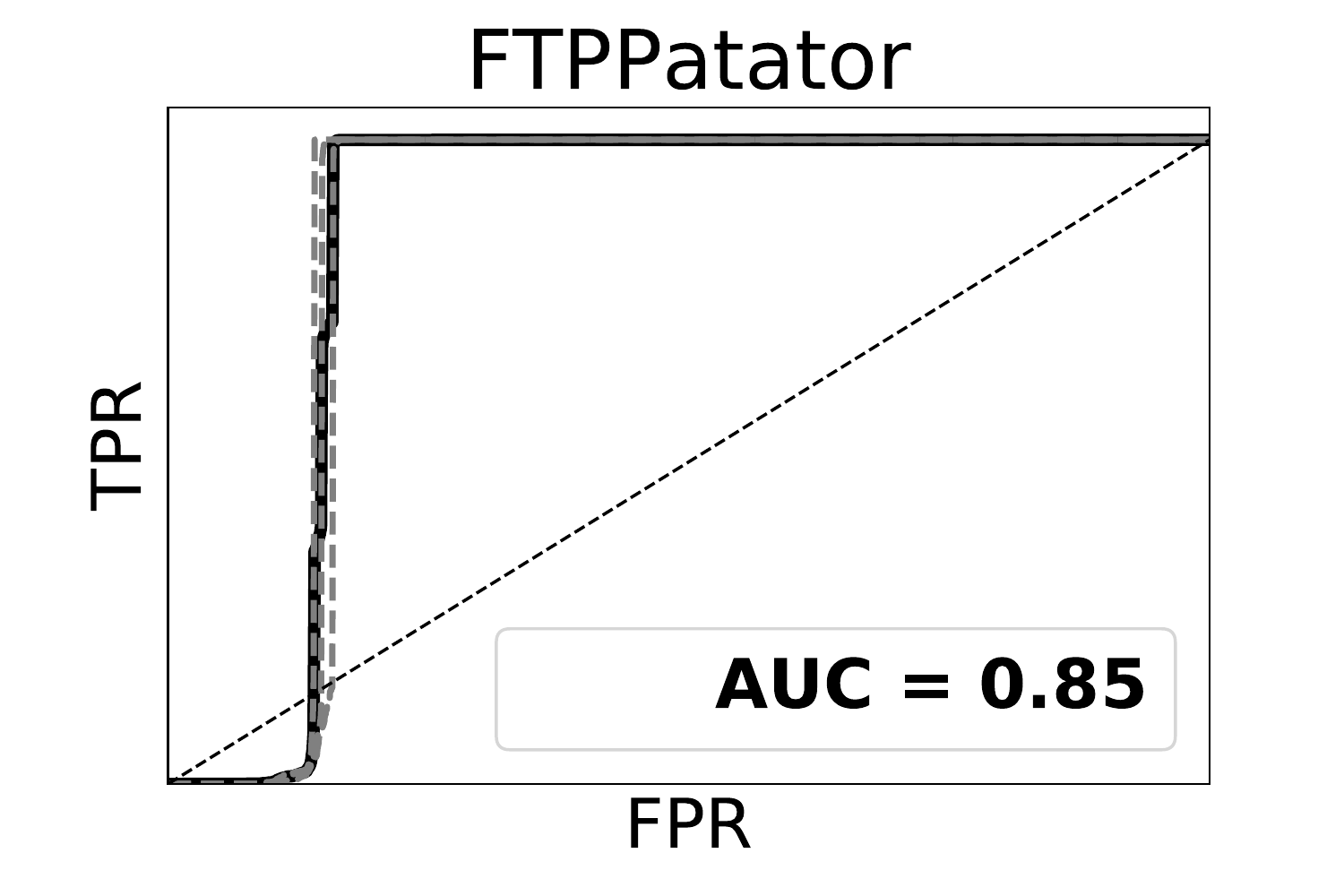} &
      \includegraphics[width=.19\textwidth]{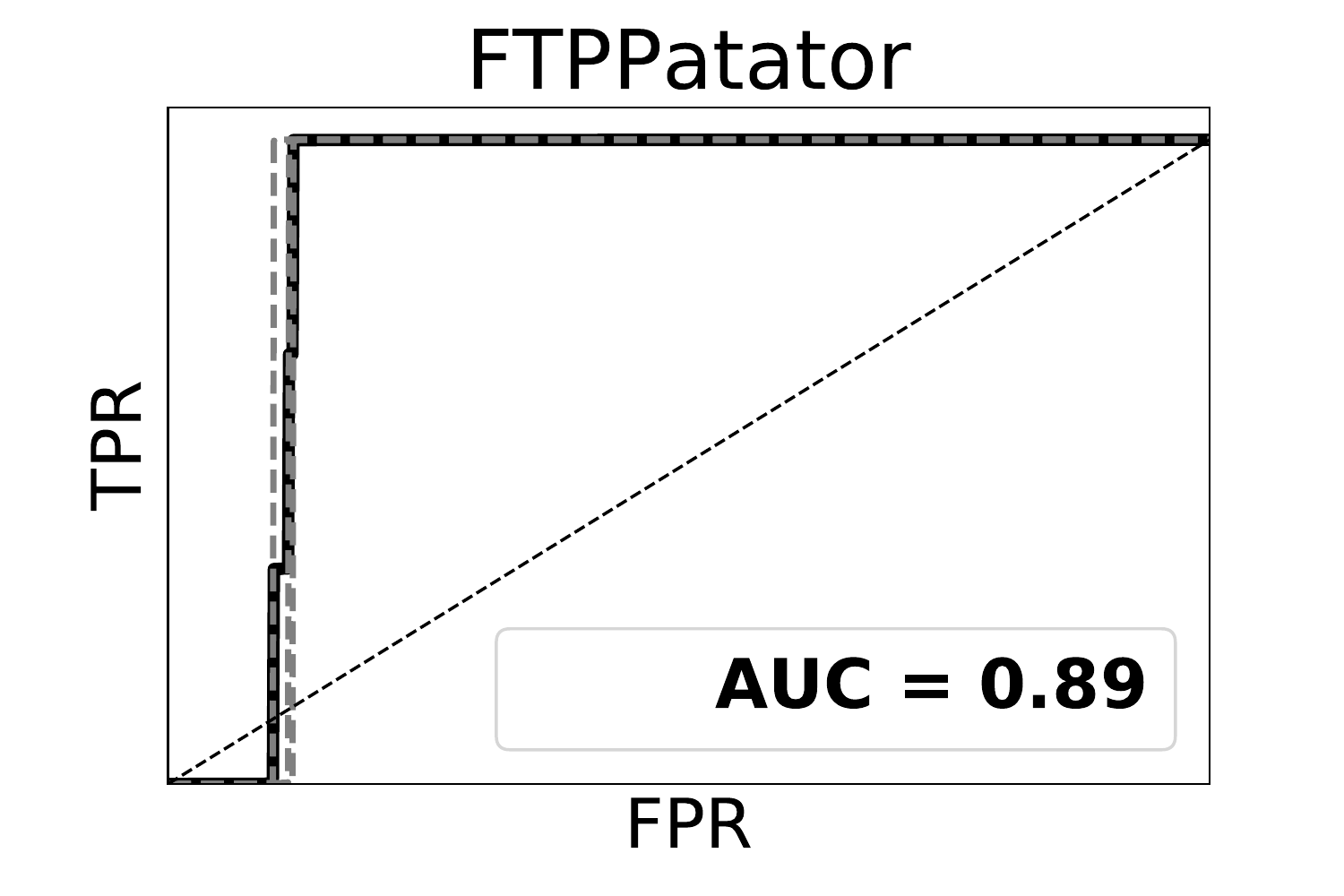} &
      \includegraphics[width=.19\textwidth]{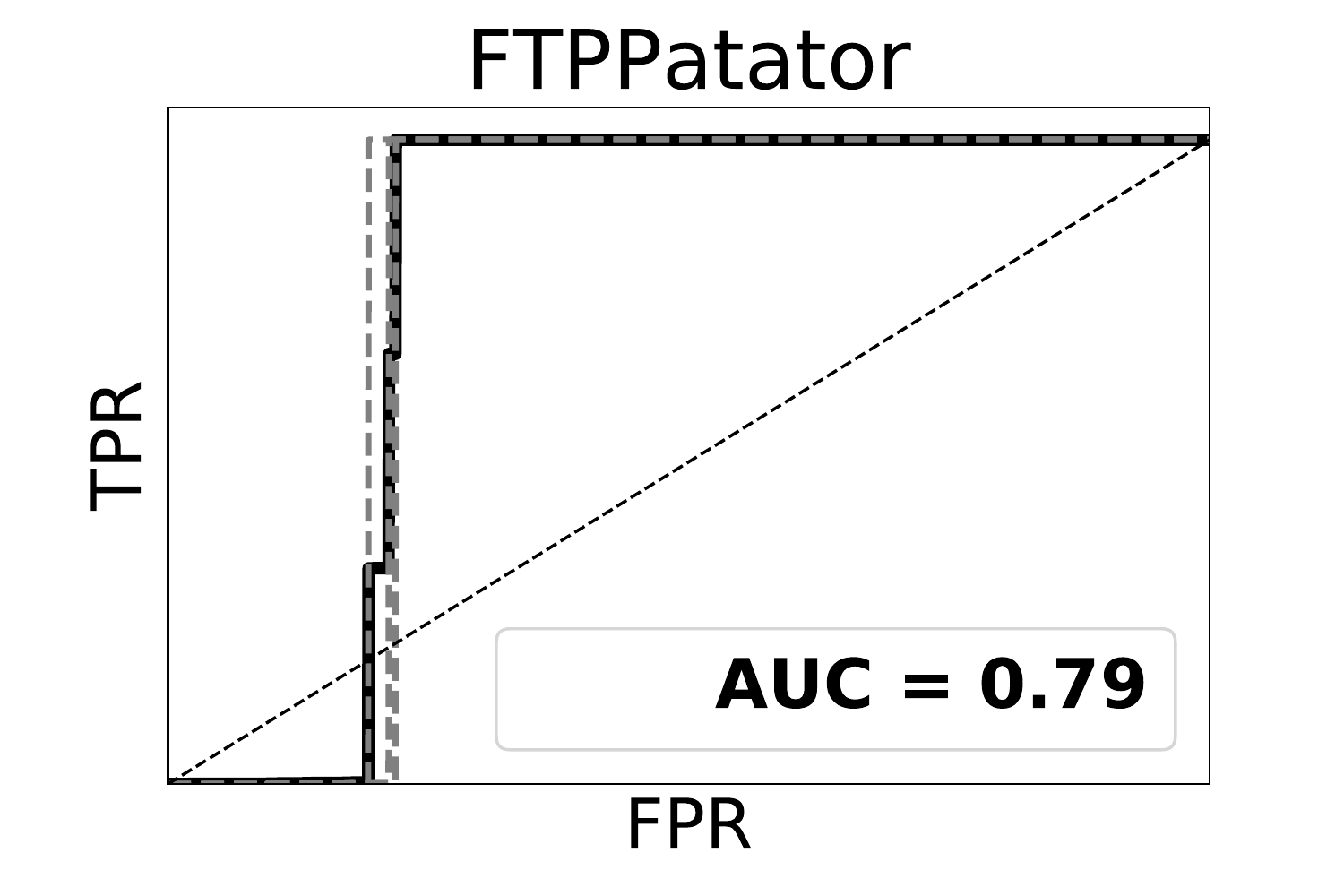} &
      \includegraphics[width=.19\textwidth]{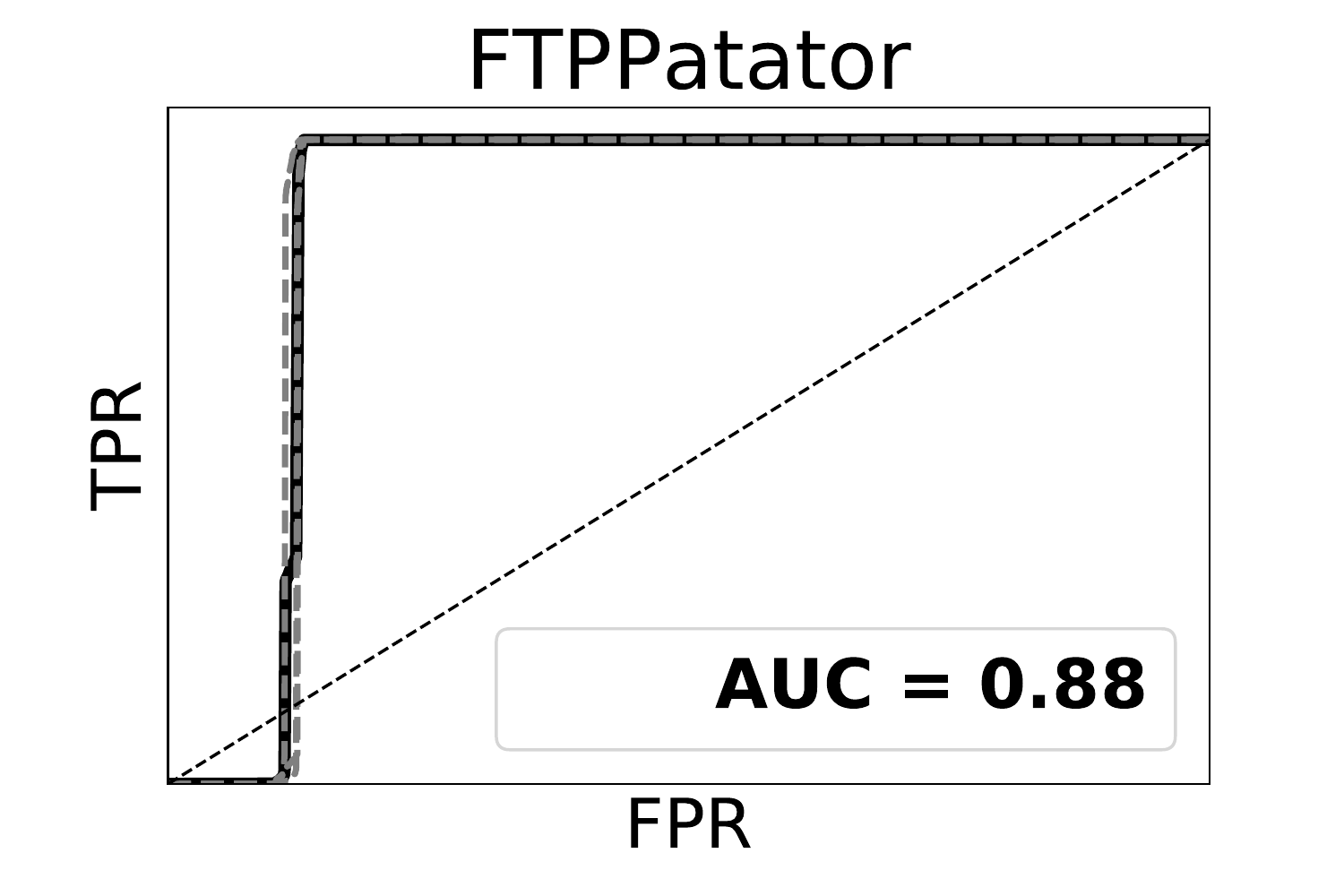}
      \\
  \end{tabular}
\end{table*}      
      
 \begin{table*}
  [ht] \caption{Service Port Sequences B} \label{tab:portsb}
  \begin{tabular}{ccccc} 
  \hline 
  Source & Destination & Dyad & Internal & External \\
      \hline      
      \includegraphics[width=.19\textwidth]{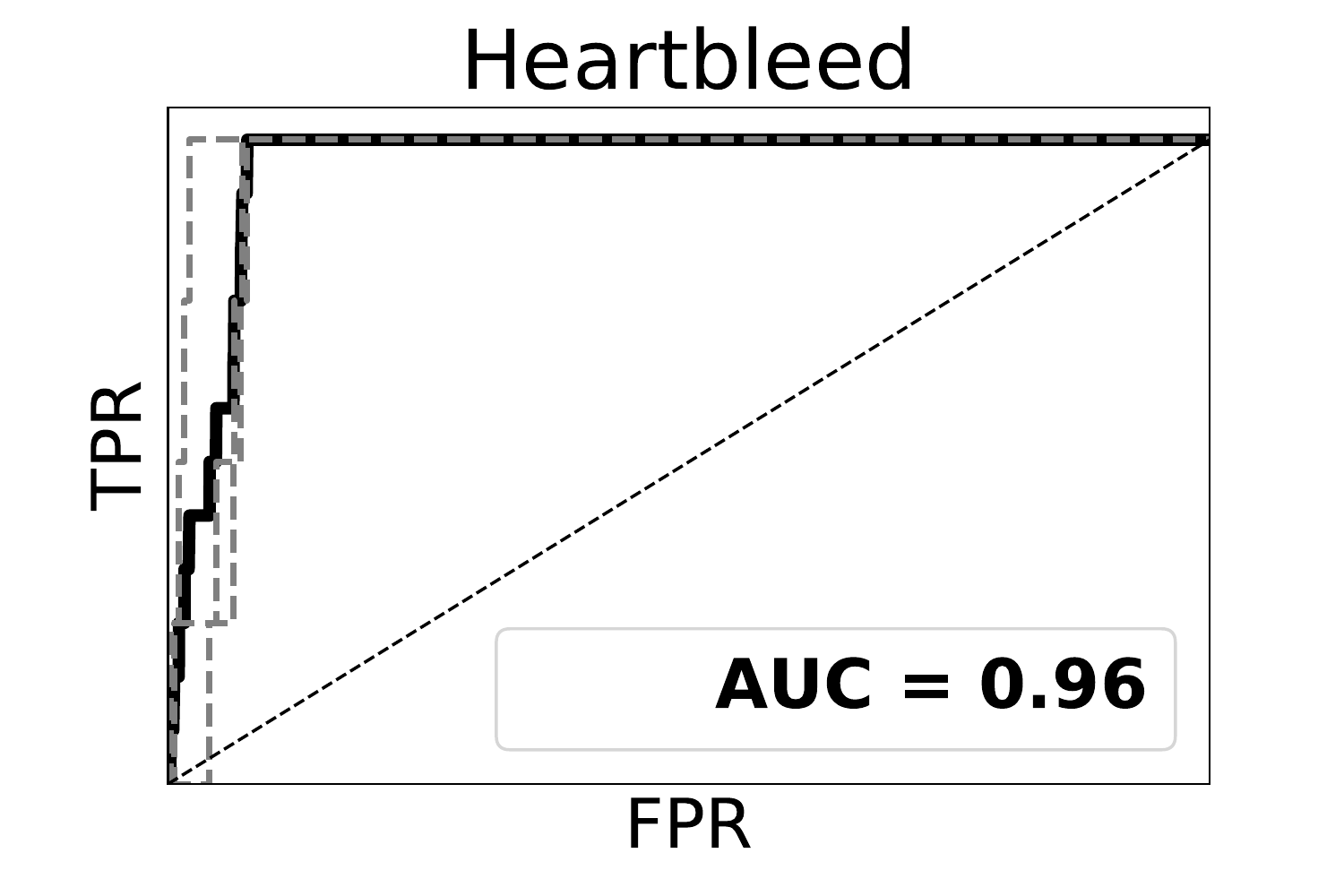} &
      \includegraphics[width=.19\textwidth]{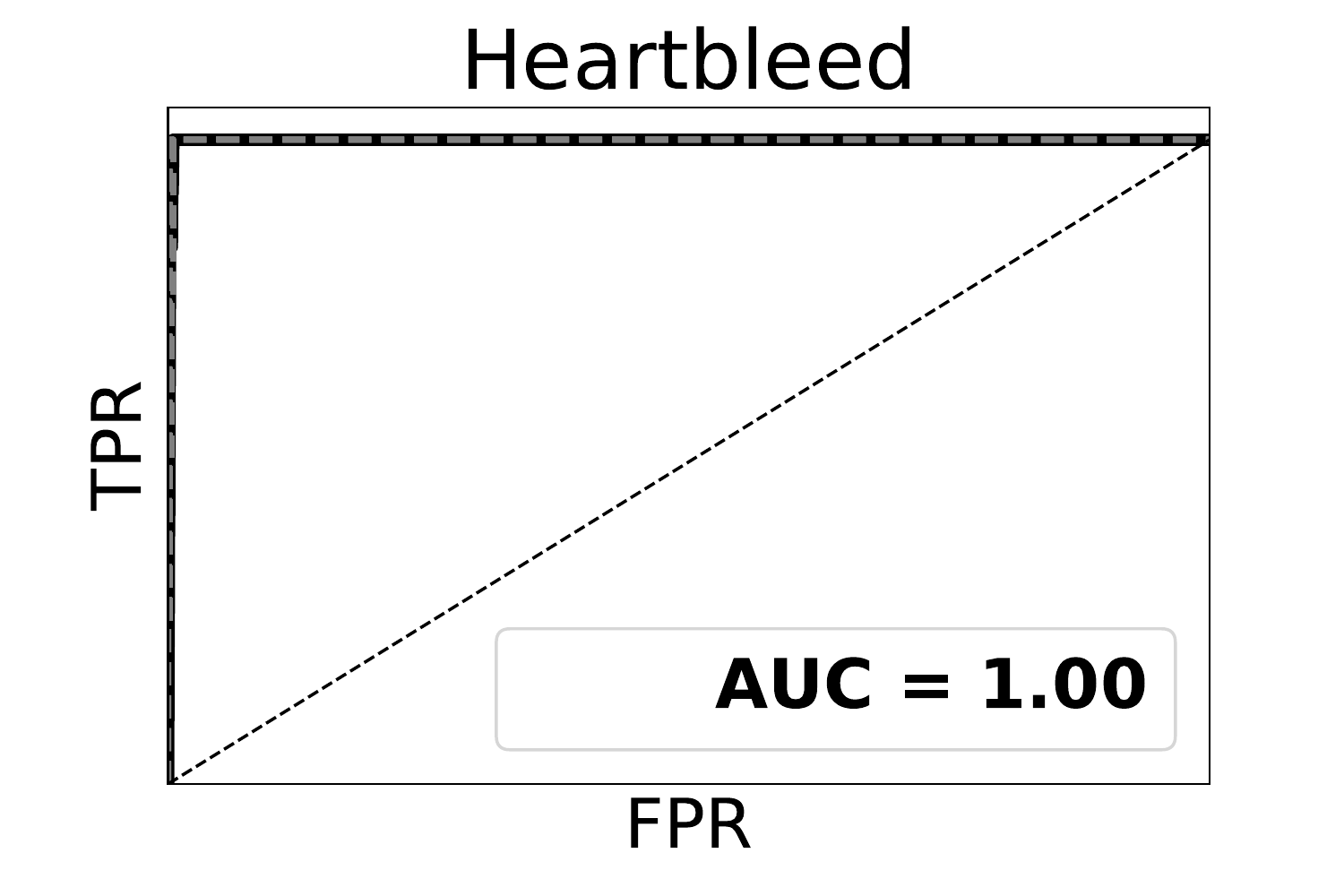} &
      \includegraphics[width=.19\textwidth]{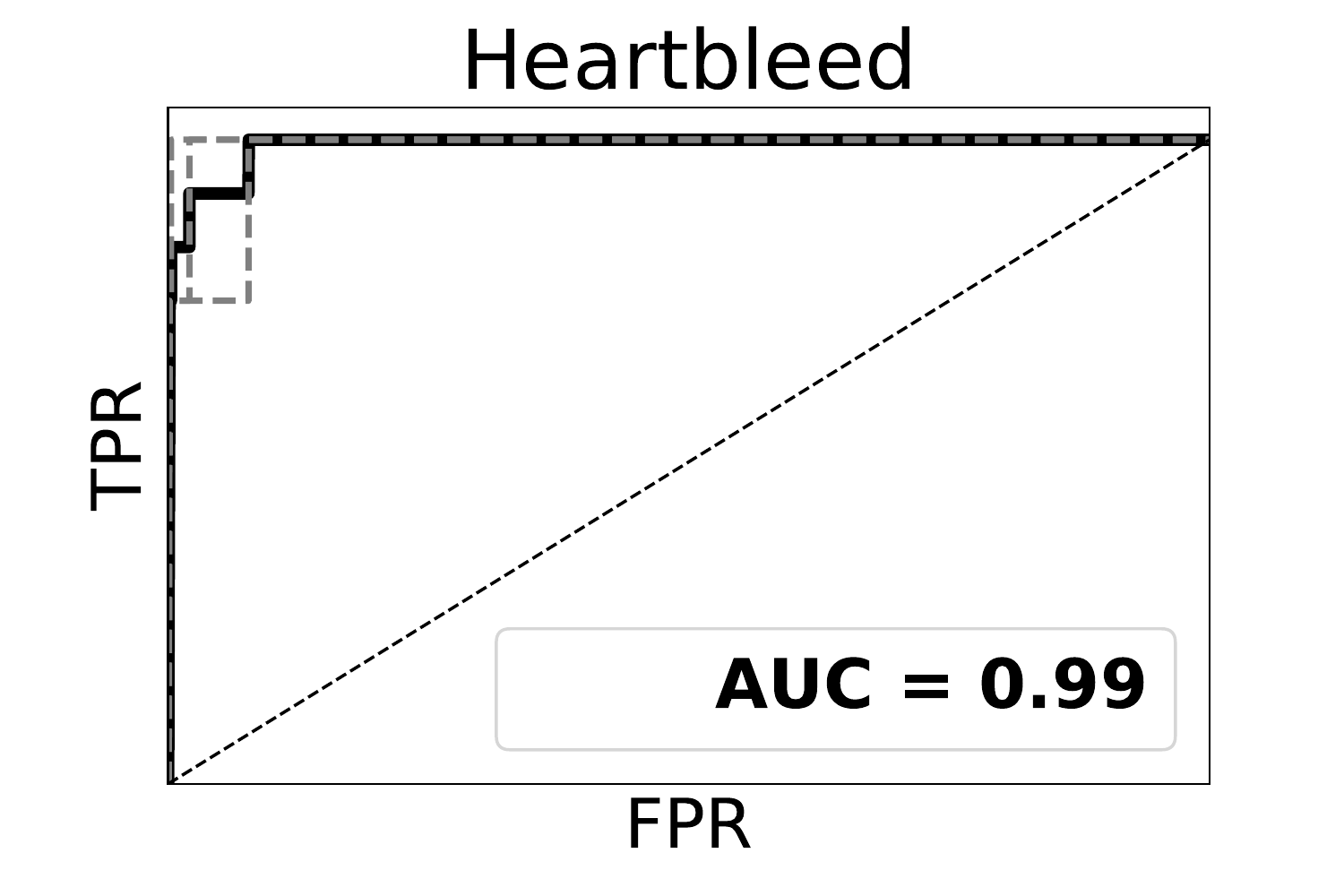} &
      \includegraphics[width=.19\textwidth]{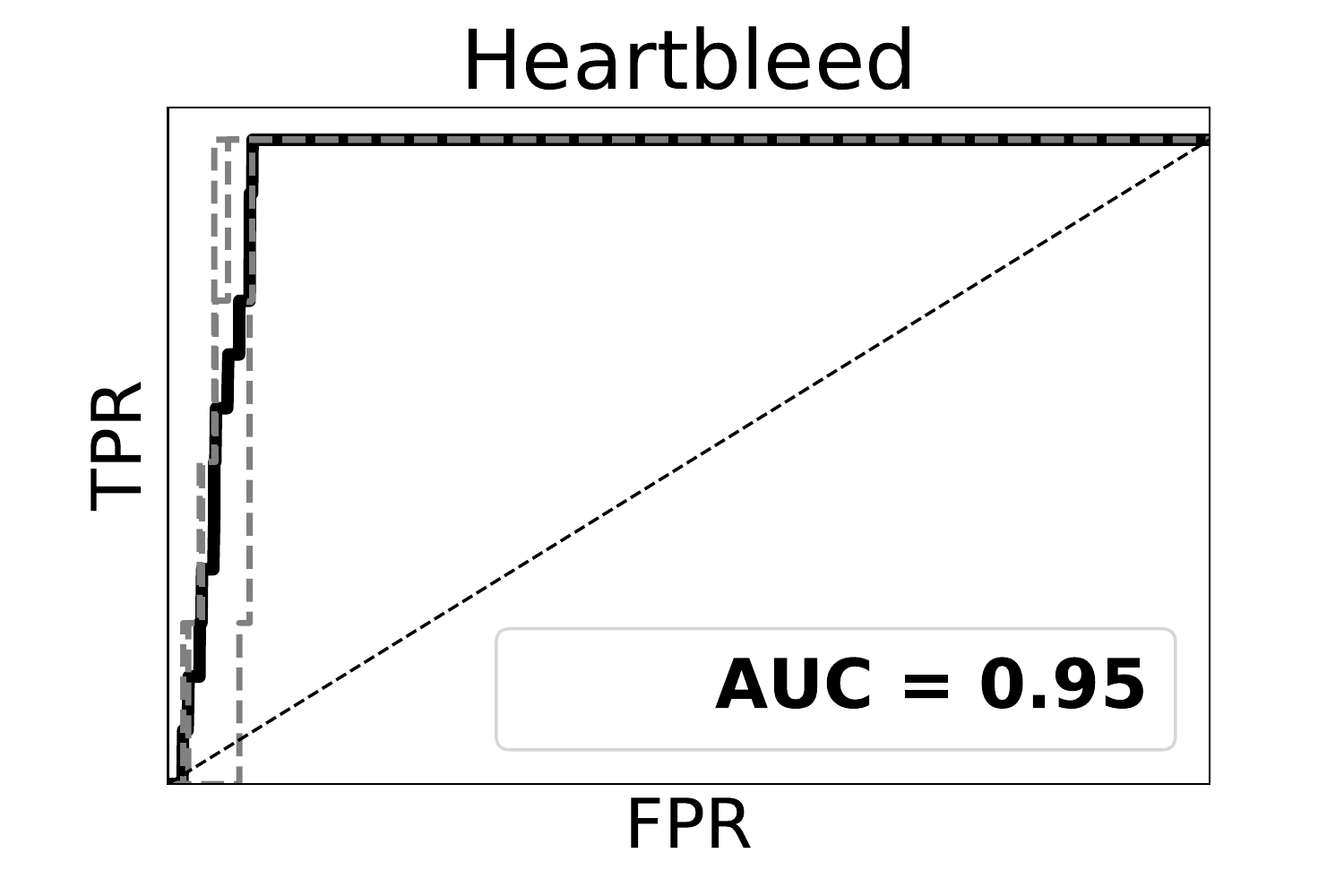} &
      \includegraphics[width=.19\textwidth]{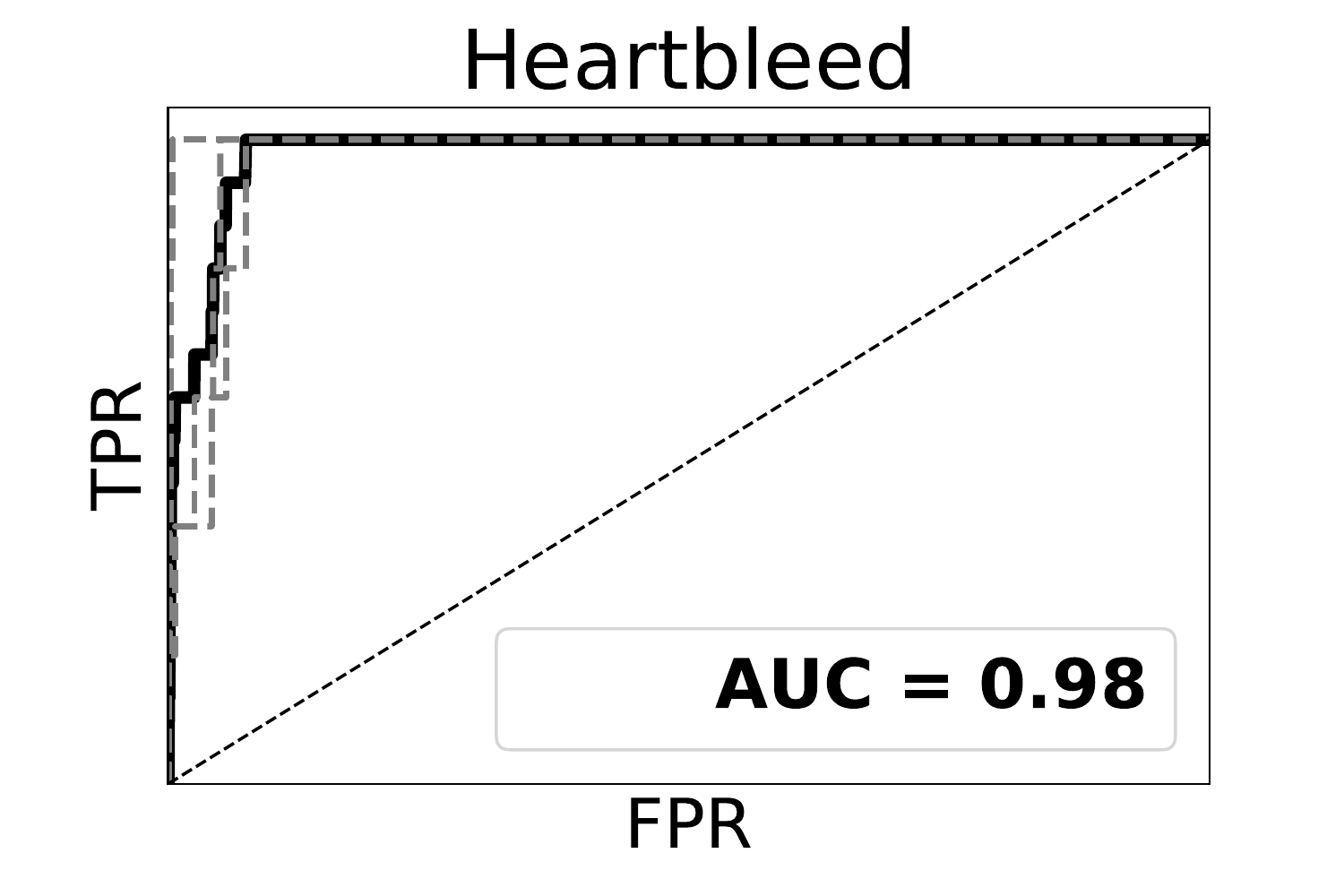}
      \\
      \includegraphics[width=.19\textwidth]{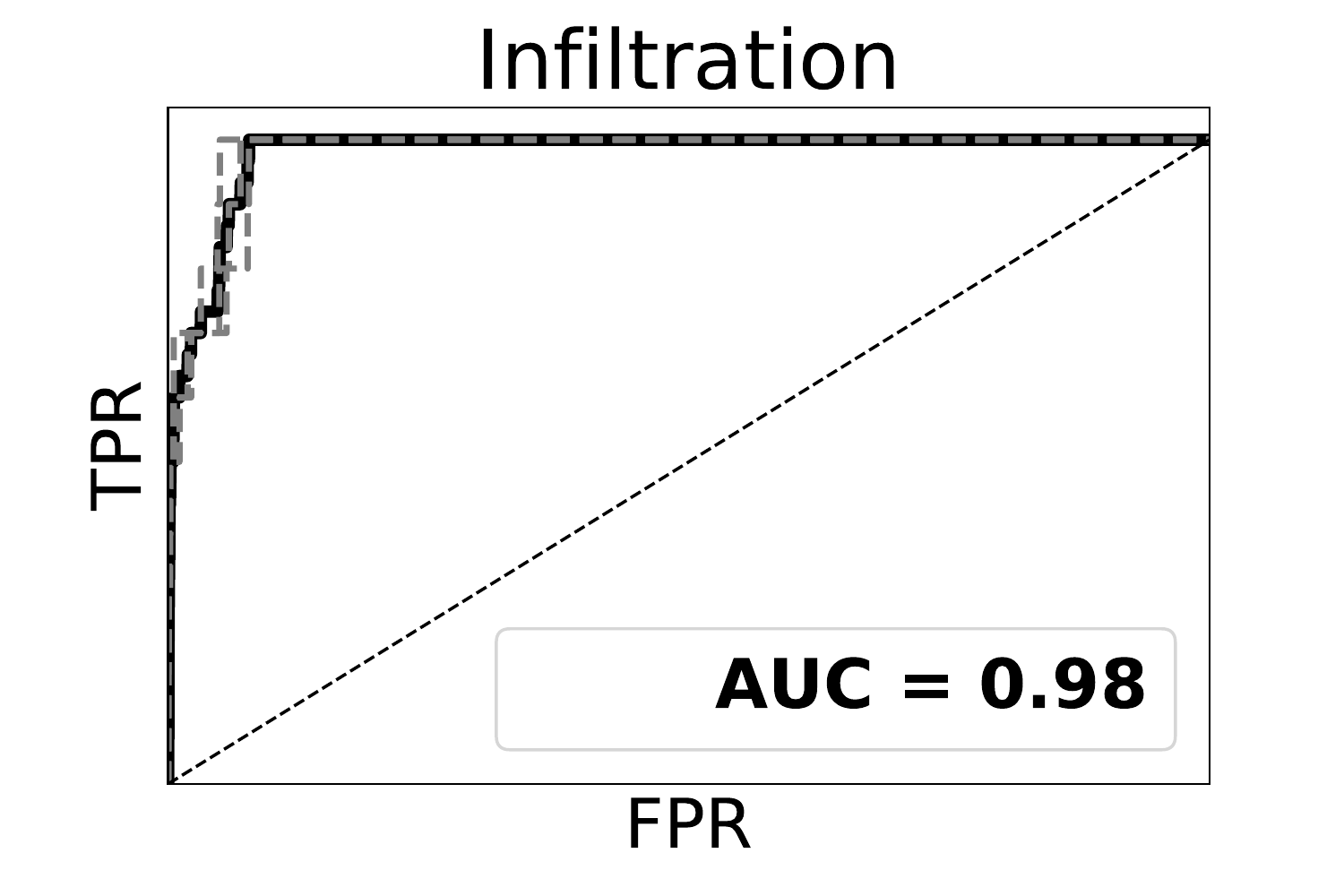} &
      \includegraphics[width=.19\textwidth]{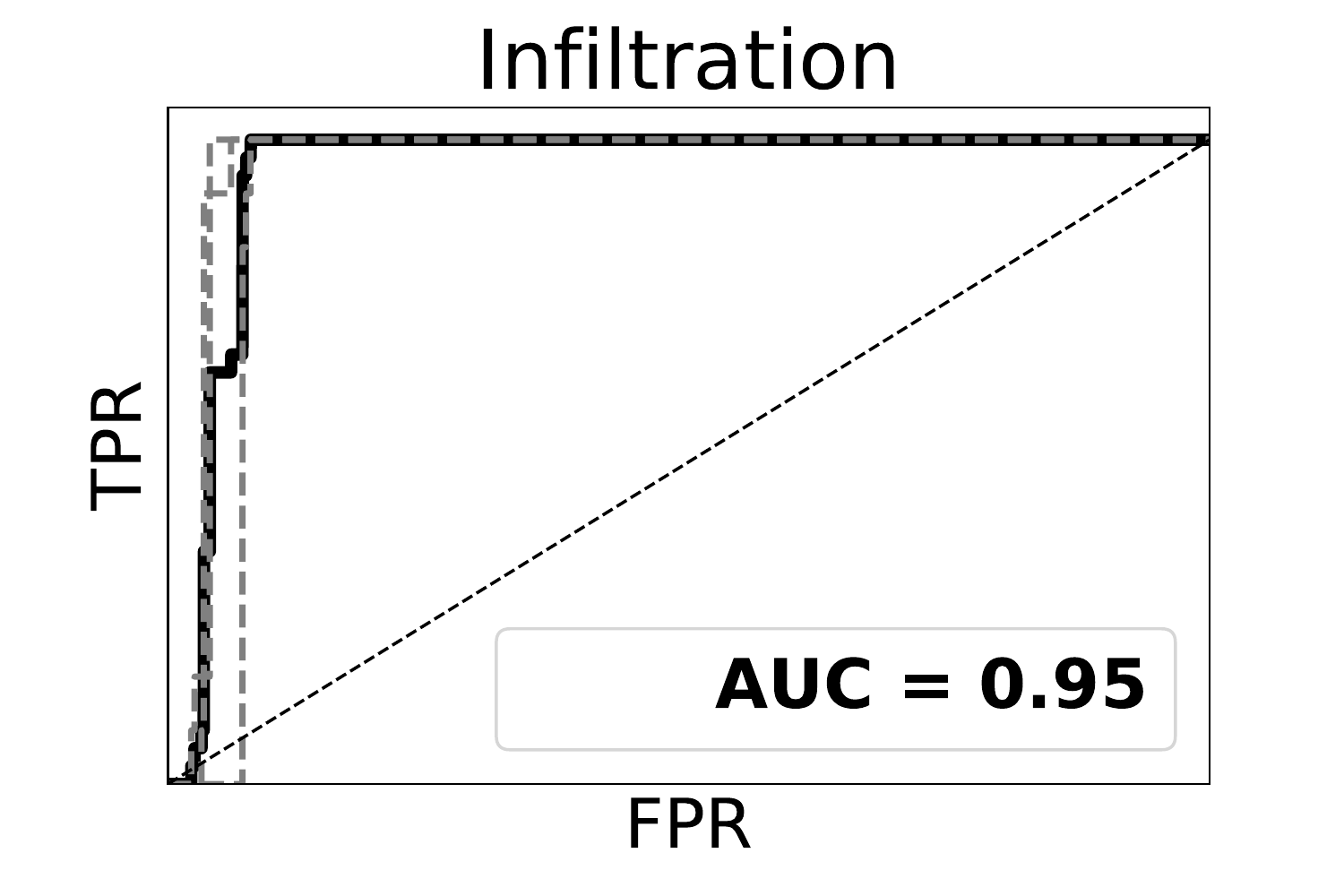} &
      \includegraphics[width=.19\textwidth]{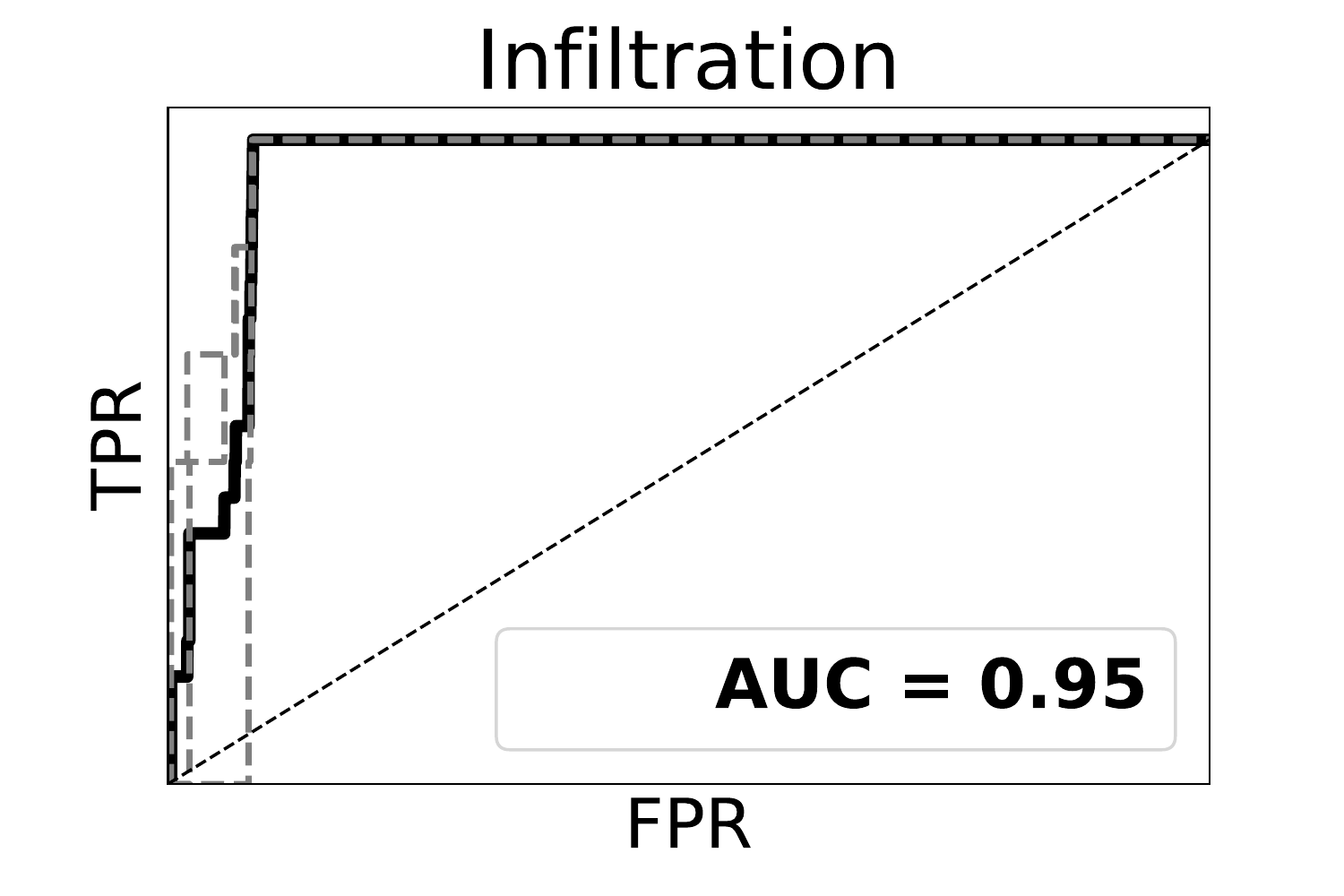} &
      \includegraphics[width=.19\textwidth]{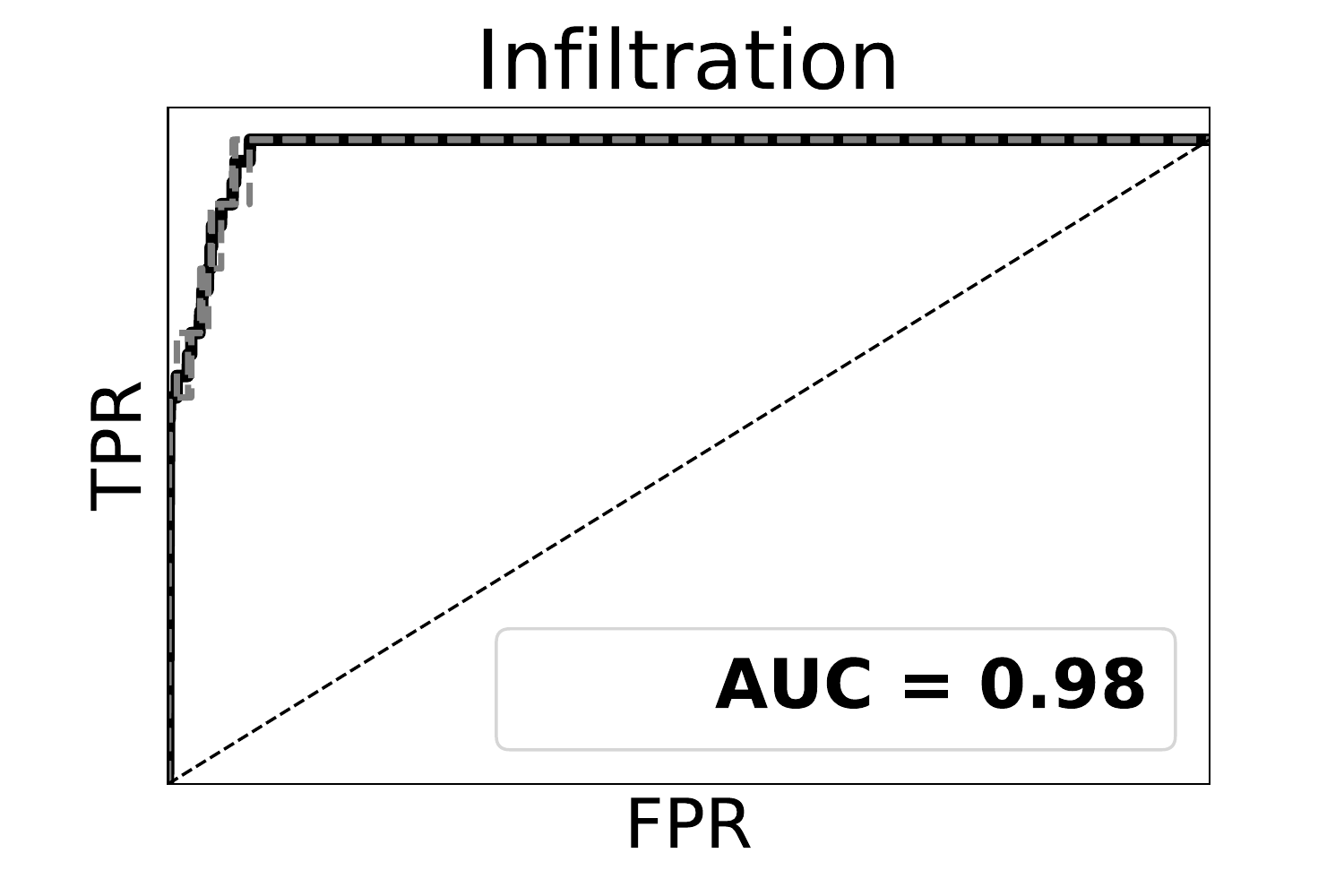} &
      \includegraphics[width=.19\textwidth]{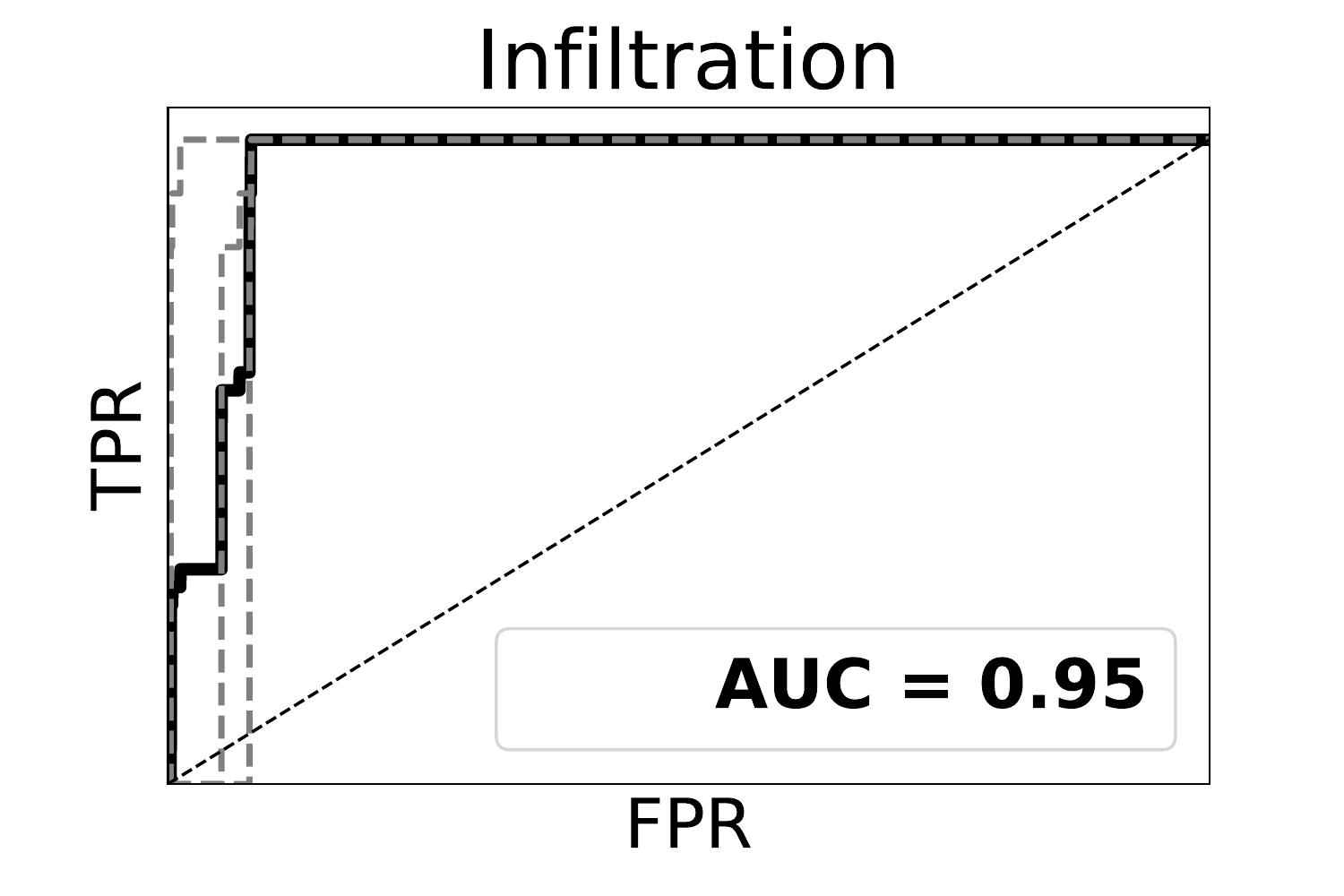}
      \\
      \includegraphics[width=.19\textwidth]{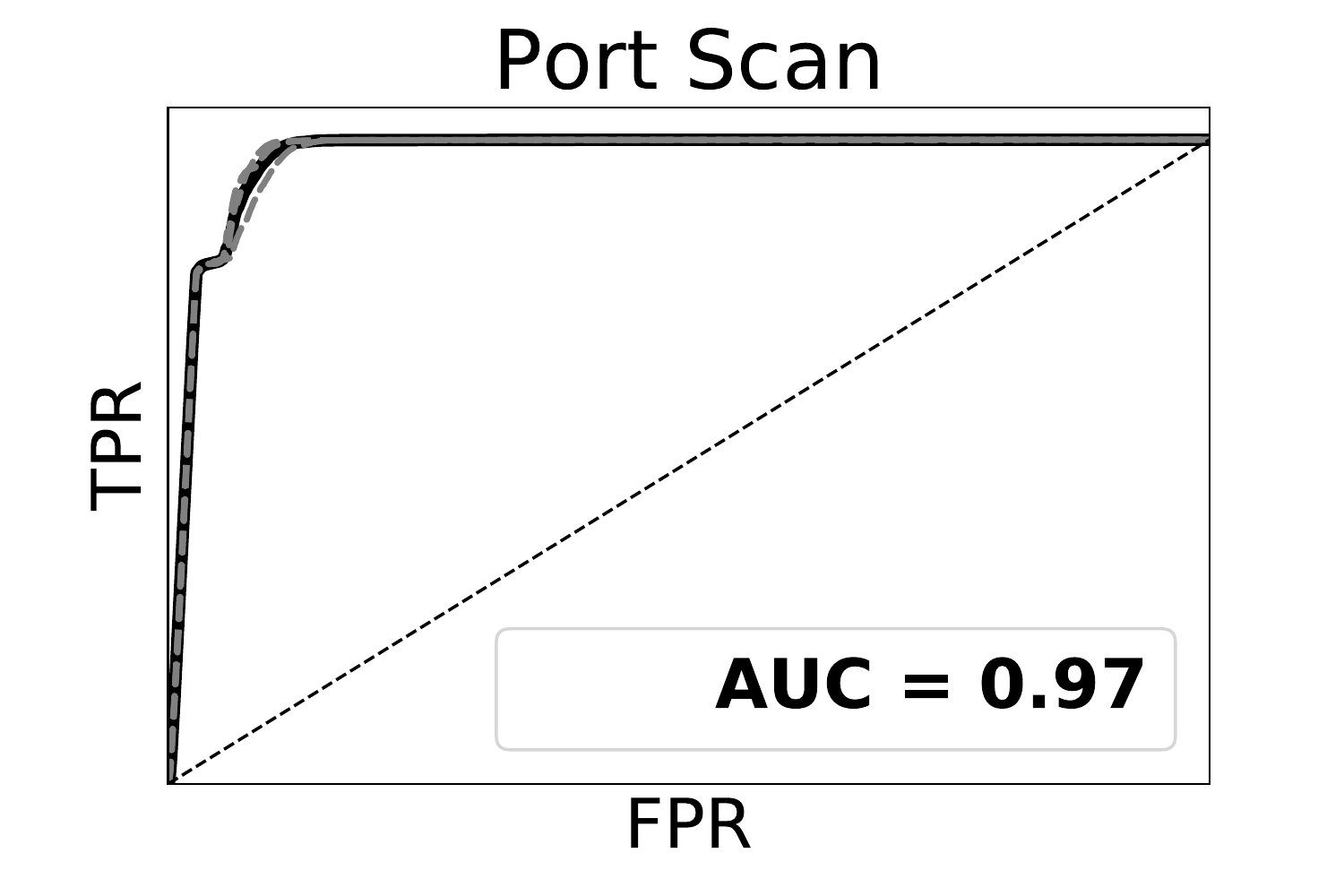} &
      \includegraphics[width=.19\textwidth]{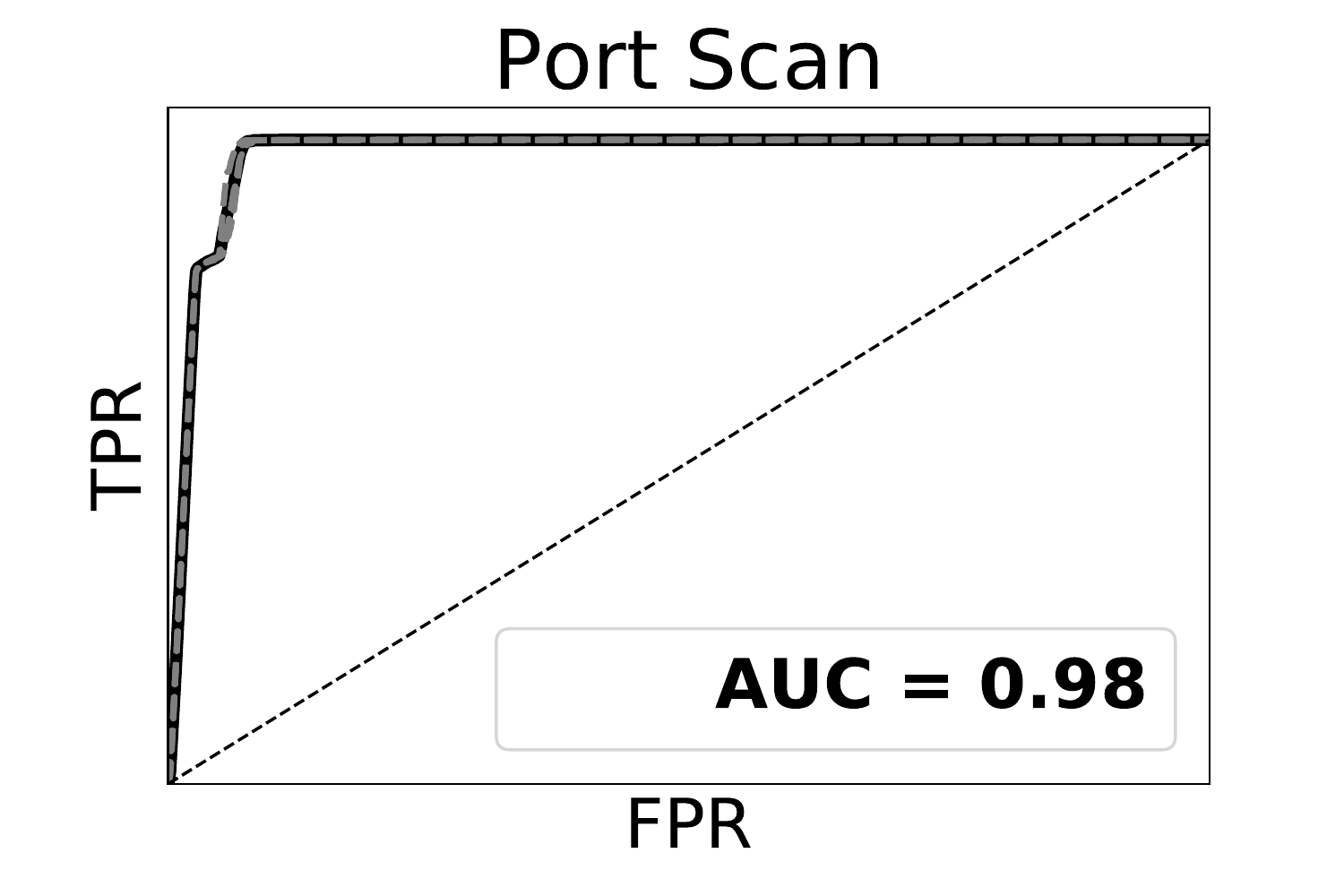} &
      \includegraphics[width=.19\textwidth]{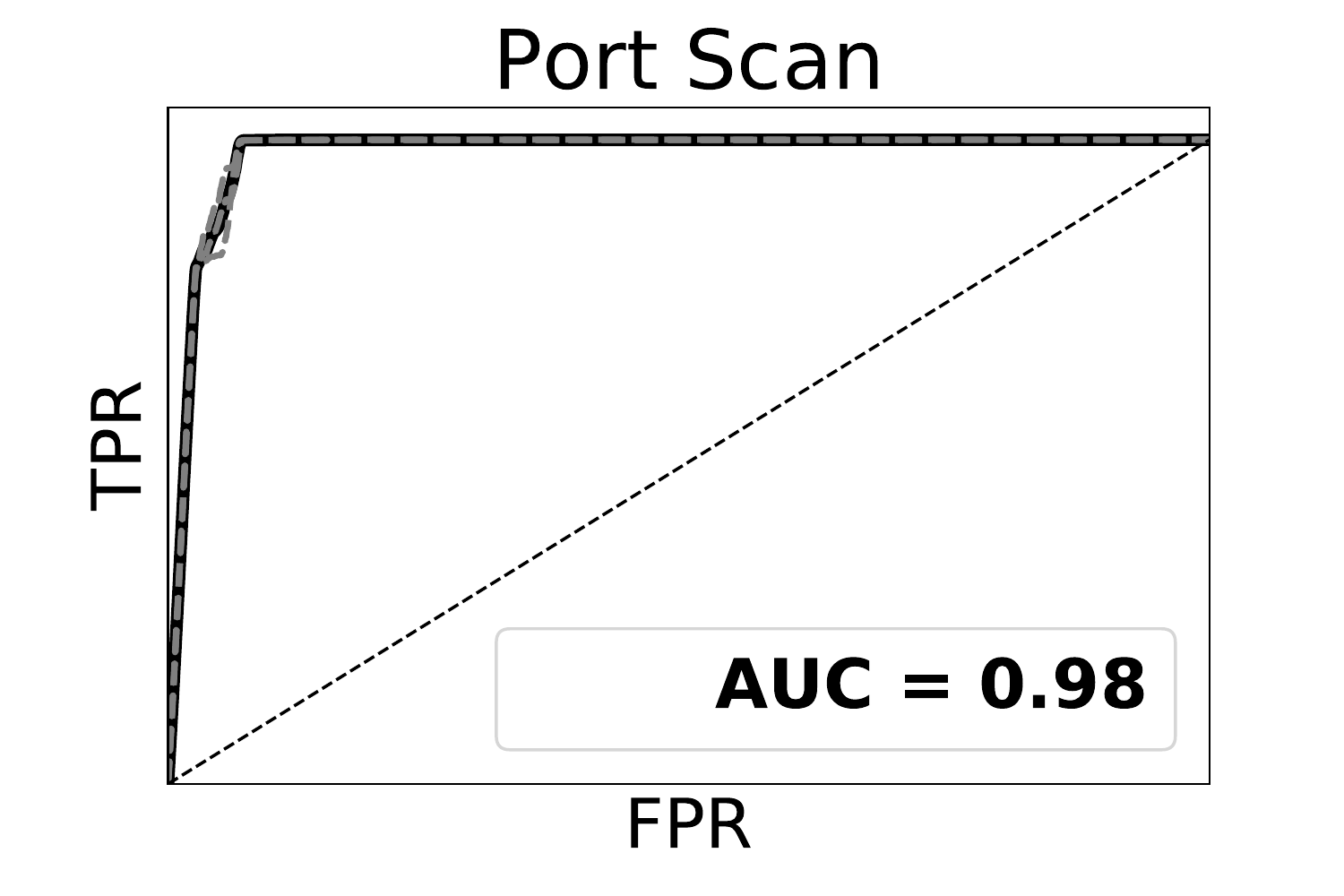} &
      \includegraphics[width=.19\textwidth]{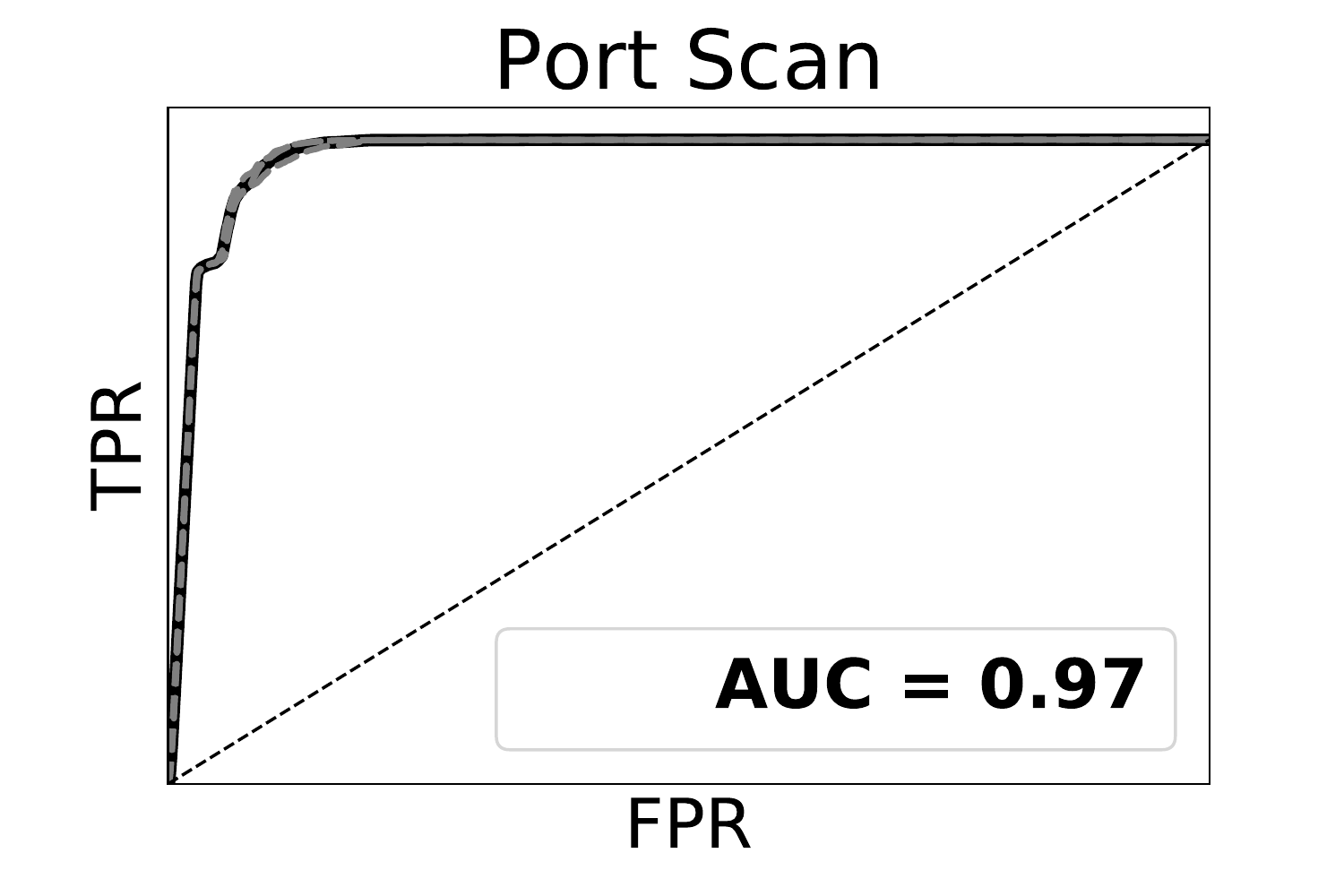} &
      \includegraphics[width=.19\textwidth]{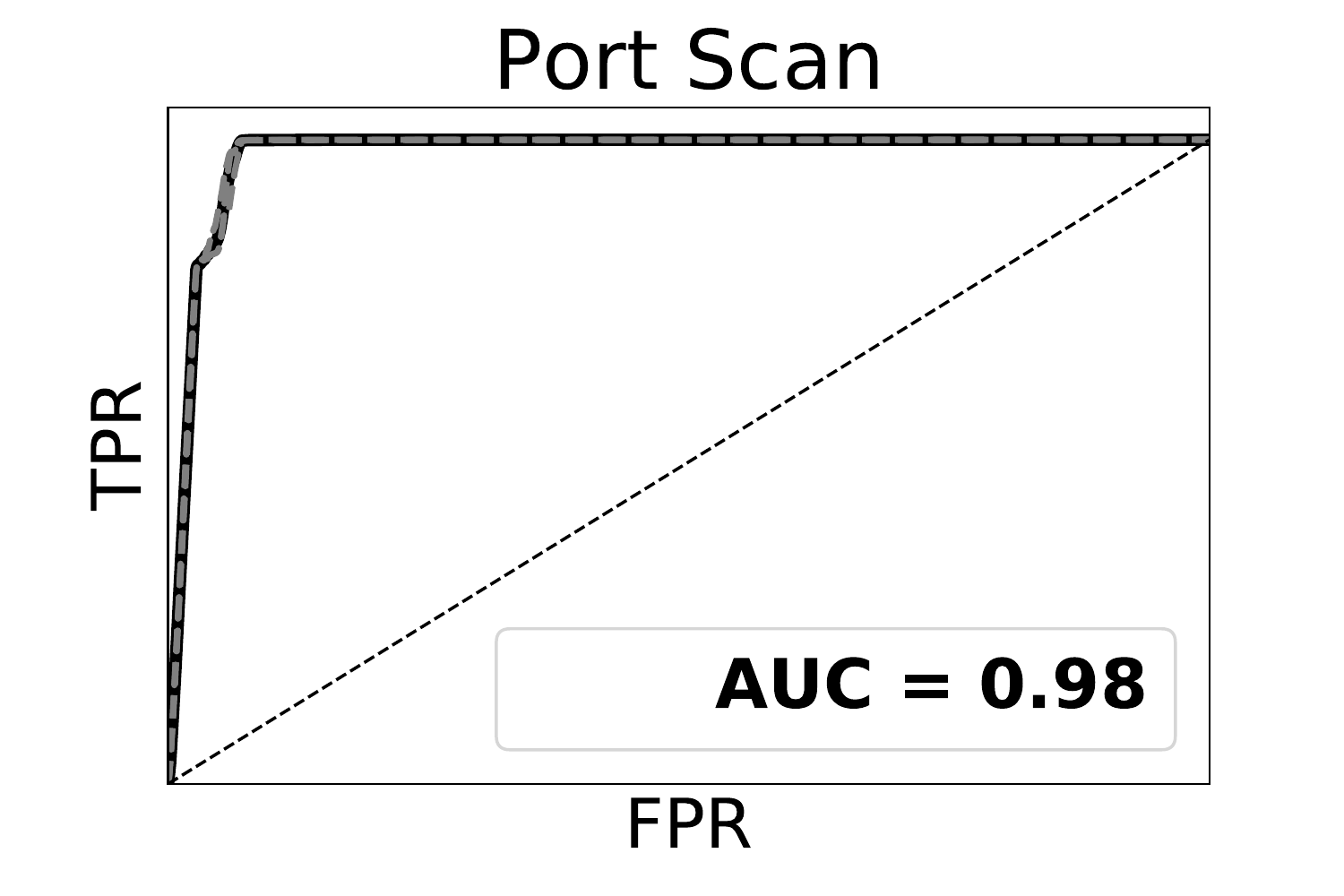}
      \\
      \includegraphics[width=.19\textwidth]{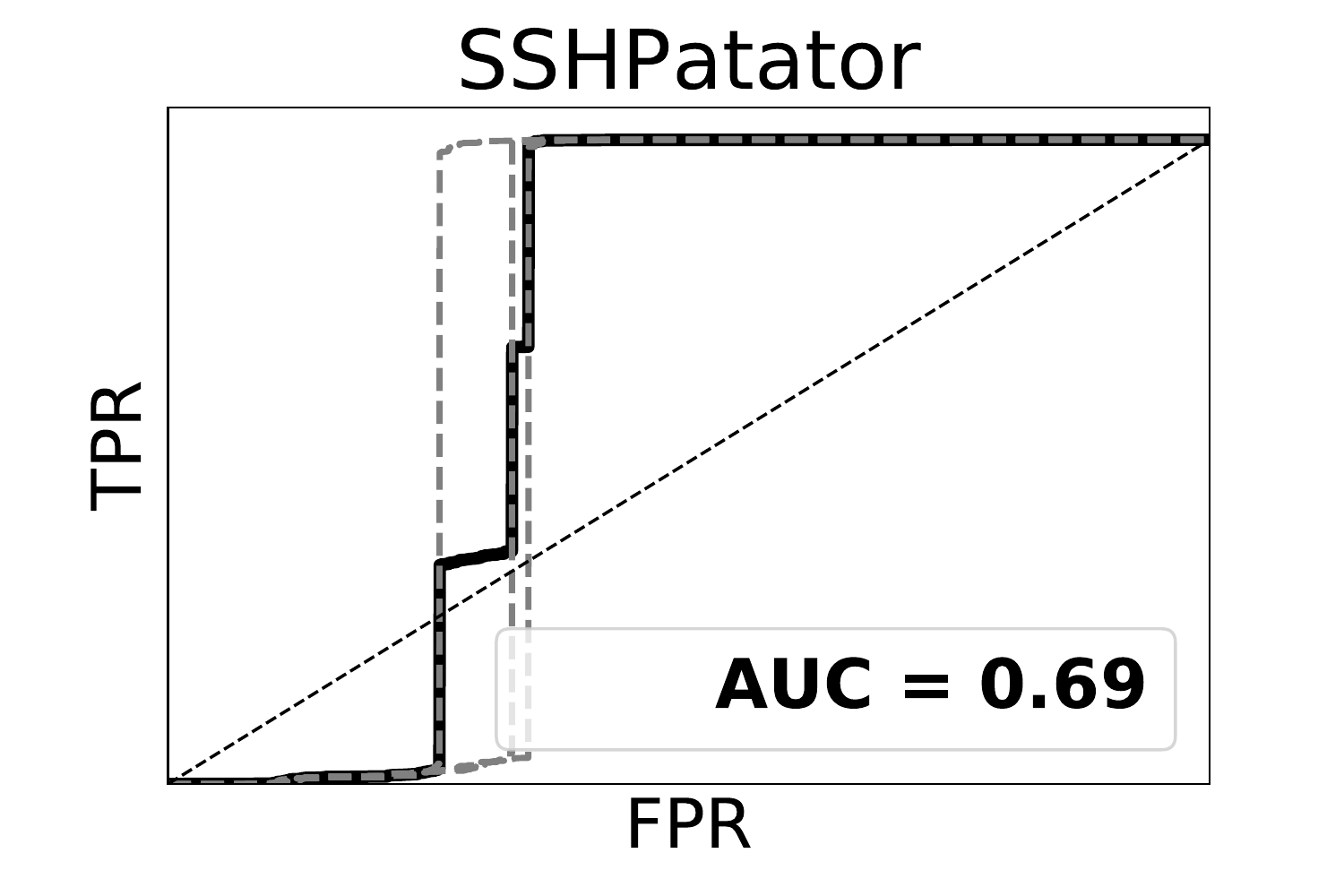} &
      \includegraphics[width=.19\textwidth]{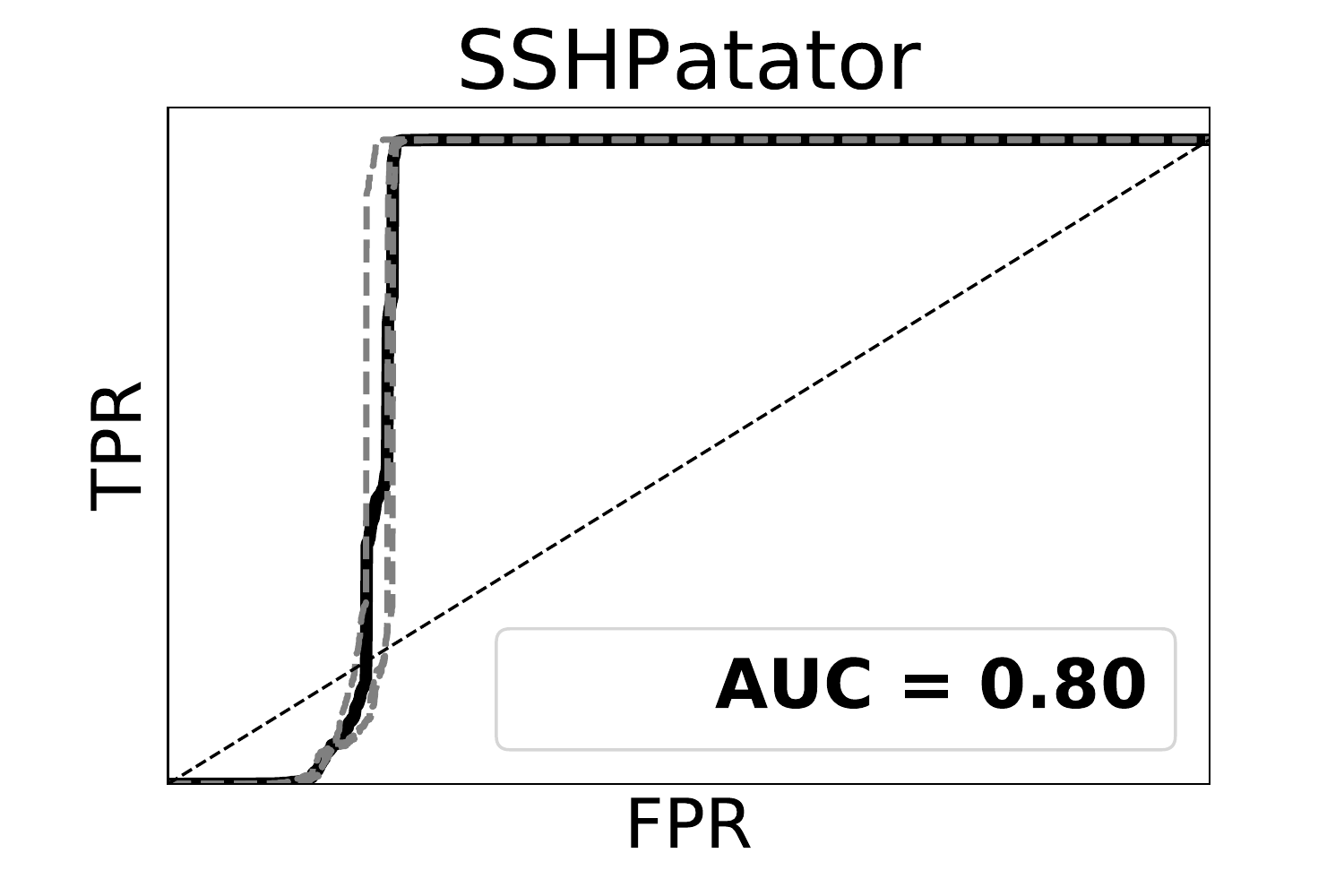} &
      \includegraphics[width=.19\textwidth]{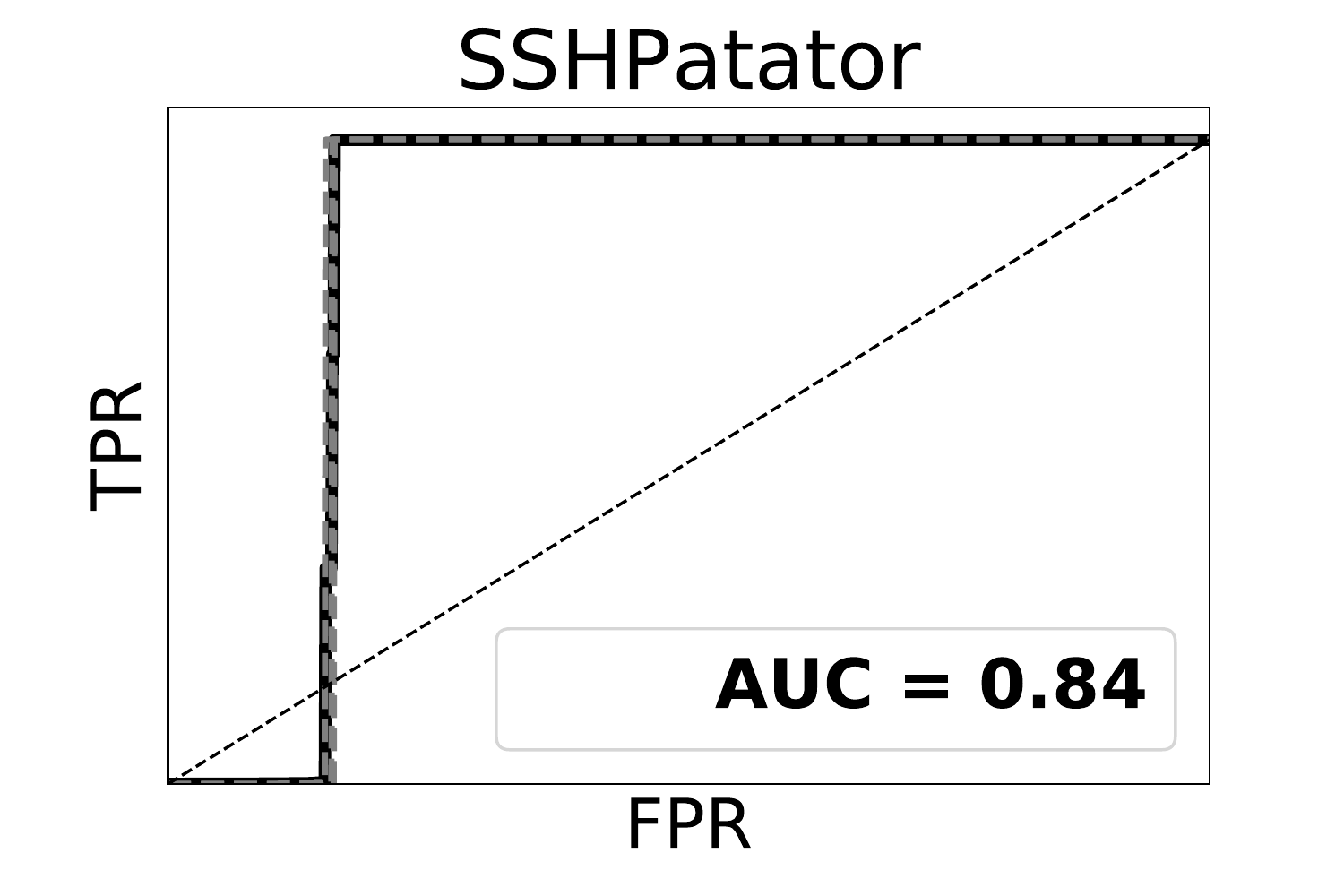} &
      \includegraphics[width=.19\textwidth]{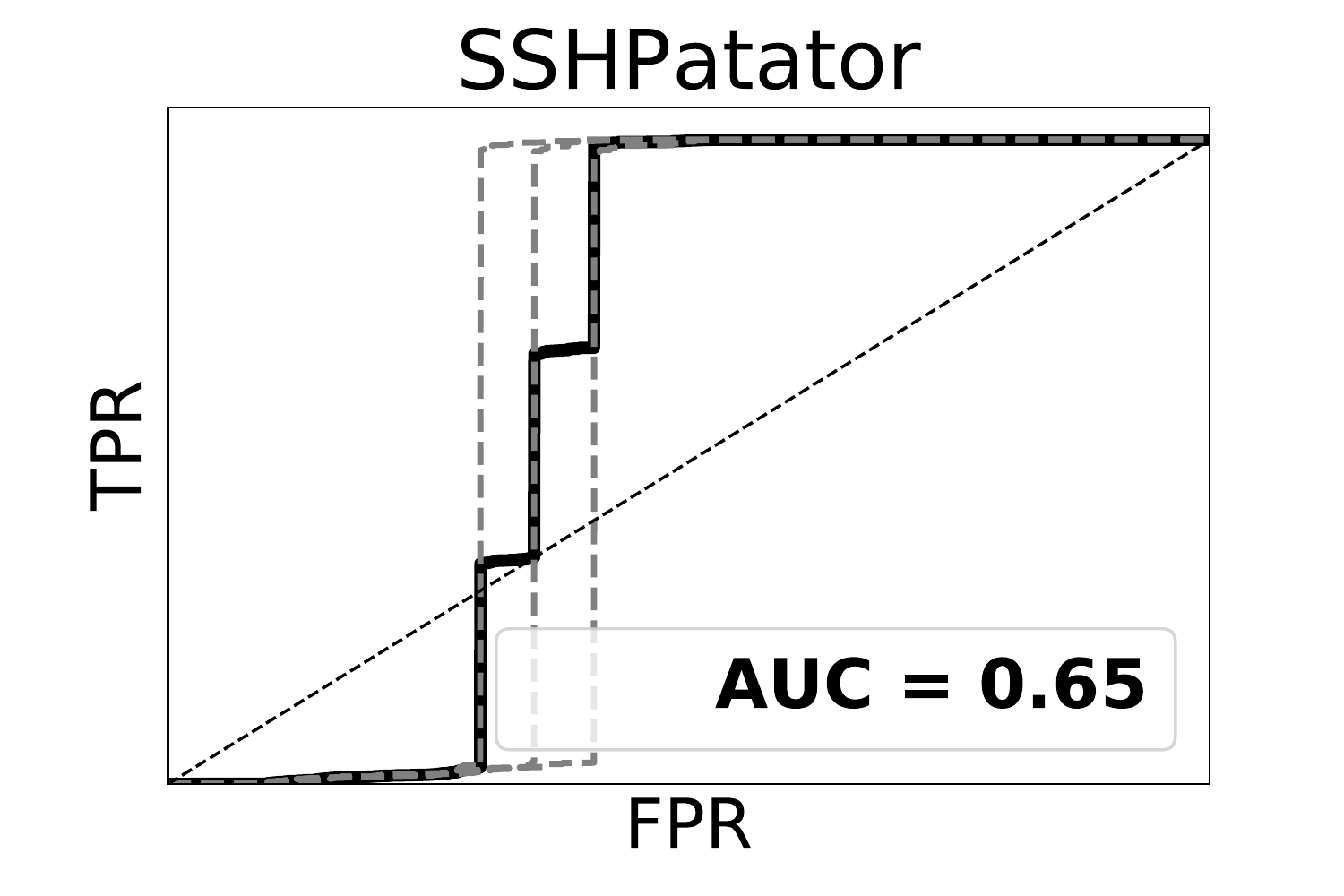} &
      \includegraphics[width=.19\textwidth]{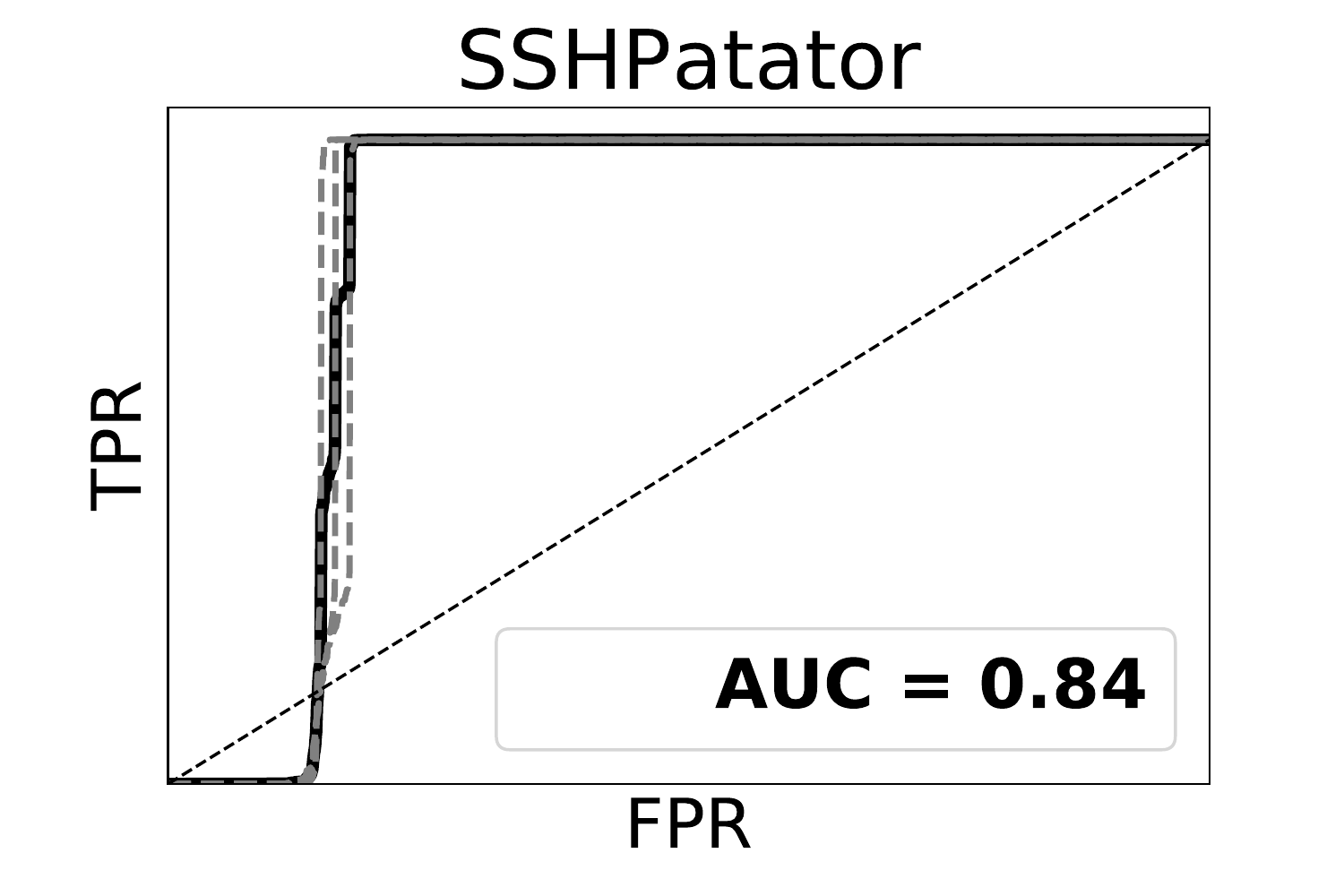}
      \\
      \includegraphics[width=.19\textwidth]{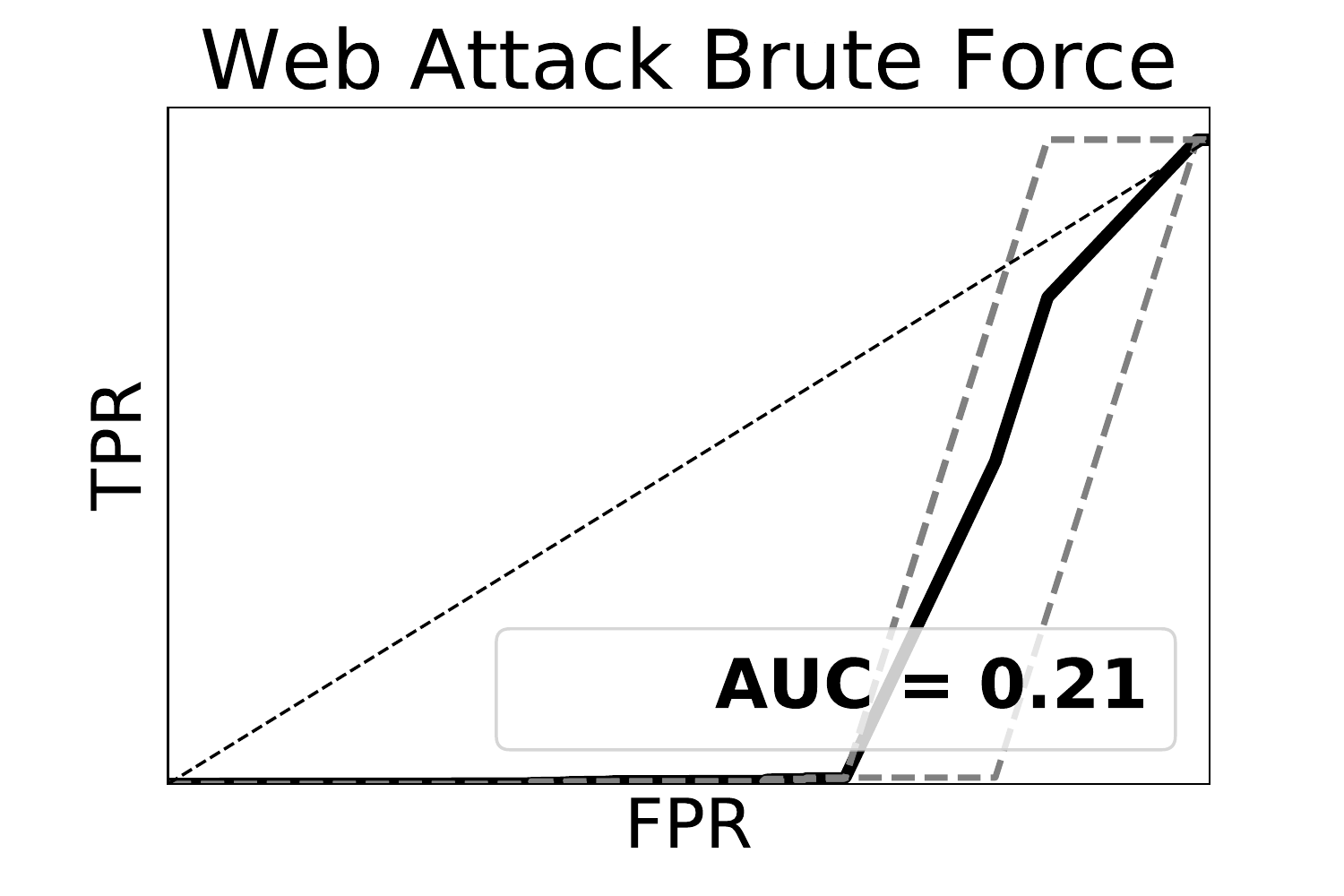} &
      \includegraphics[width=.19\textwidth]{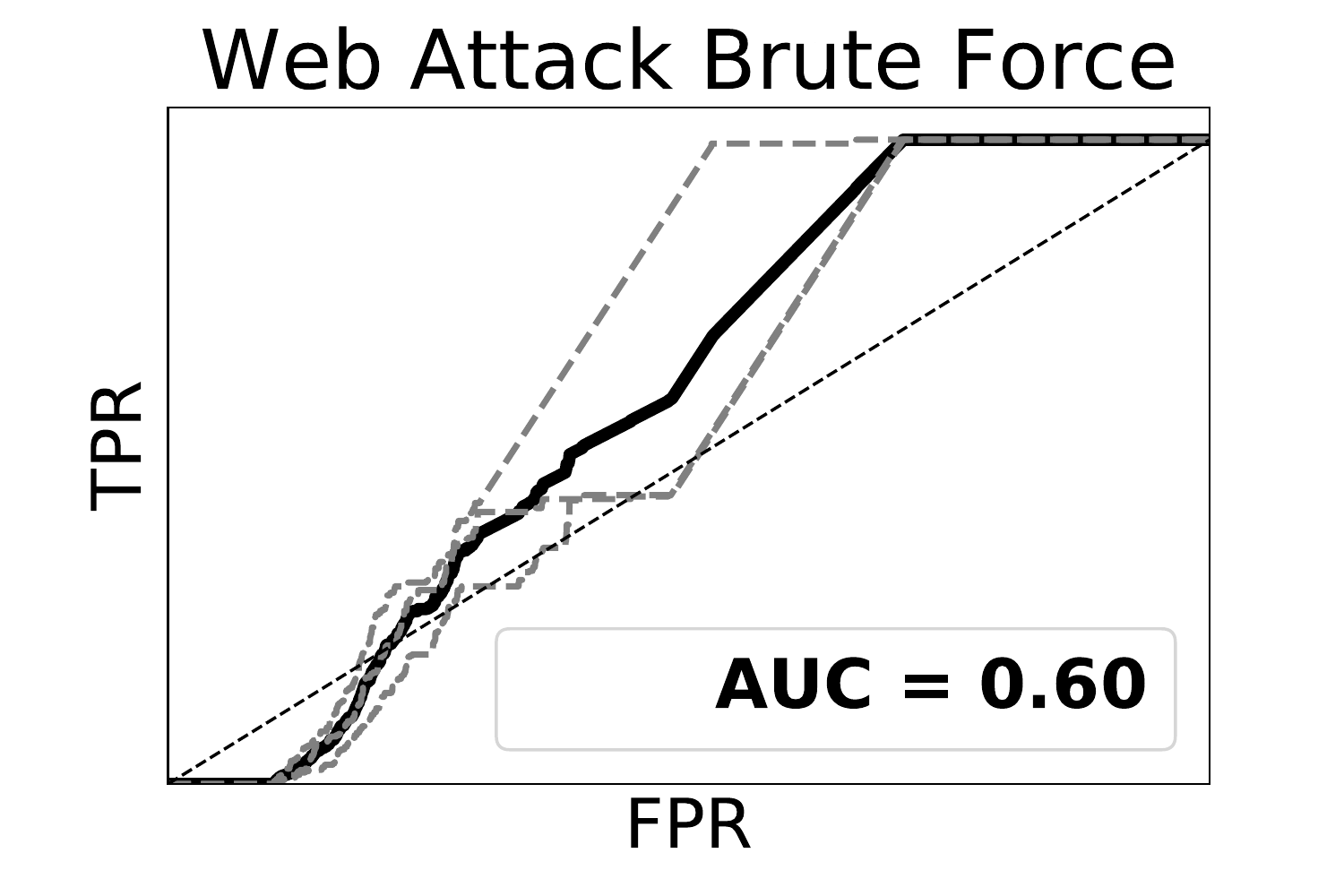} &
      \includegraphics[width=.19\textwidth]{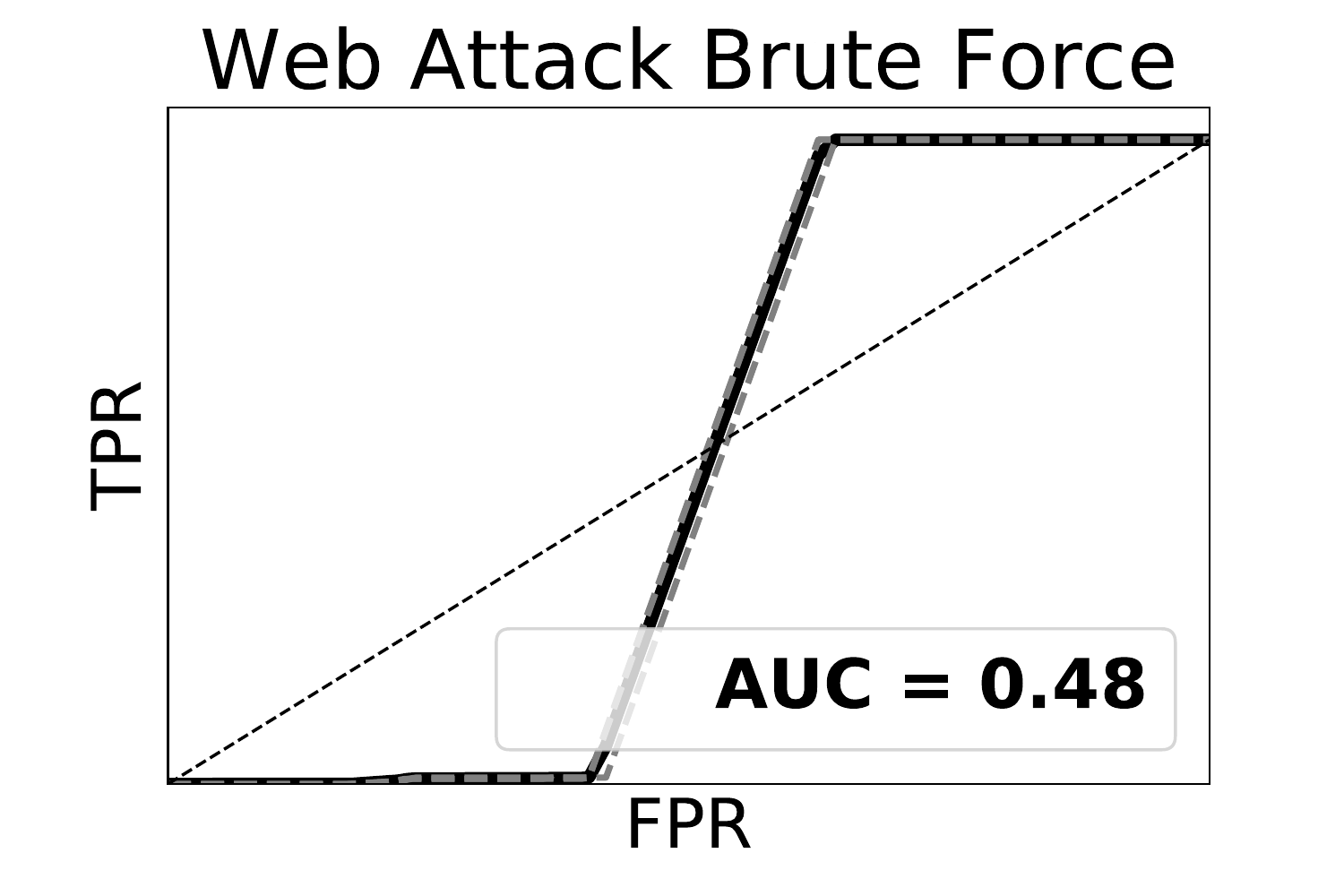} &
      \includegraphics[width=.19\textwidth]{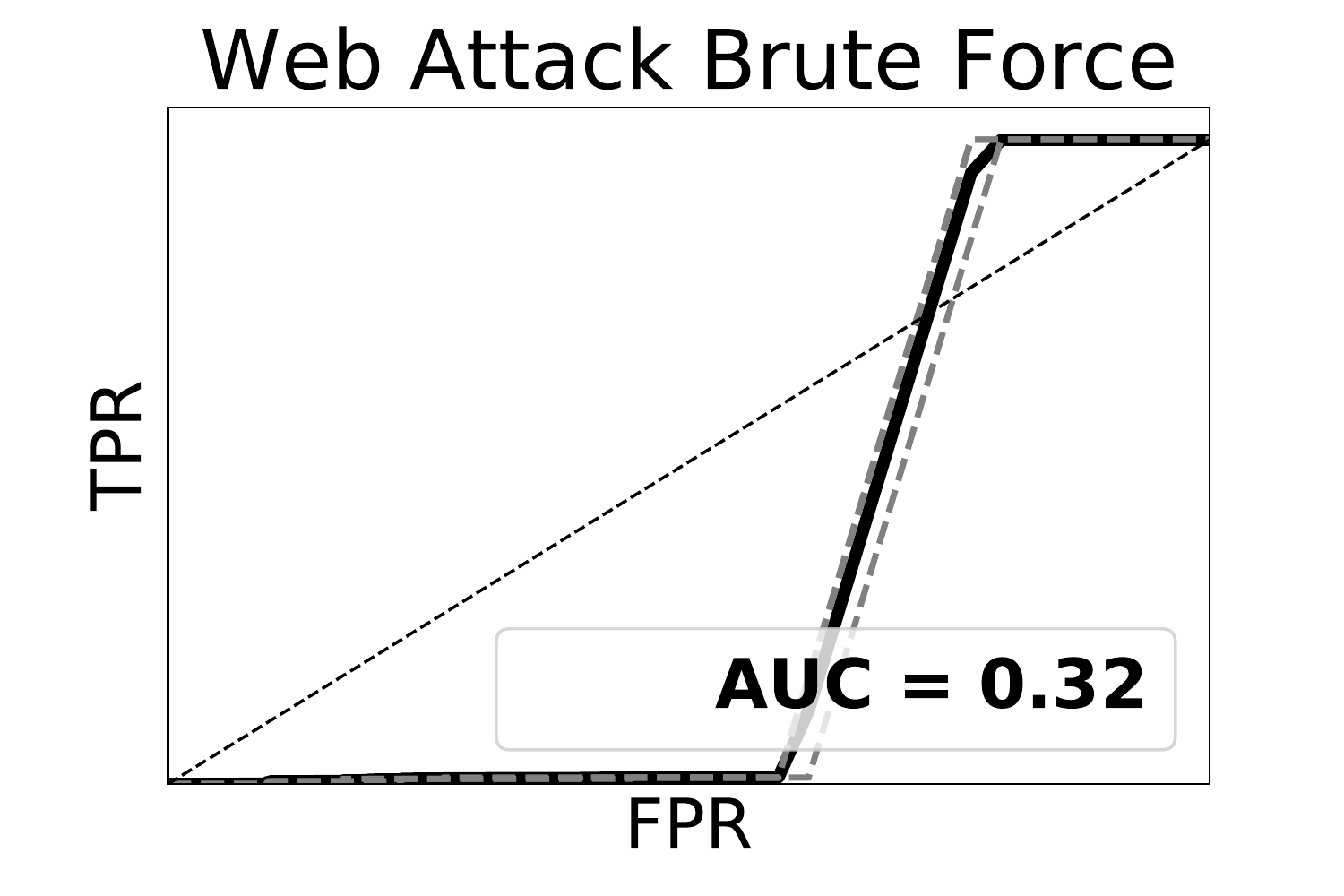} &
      \includegraphics[width=.19\textwidth]{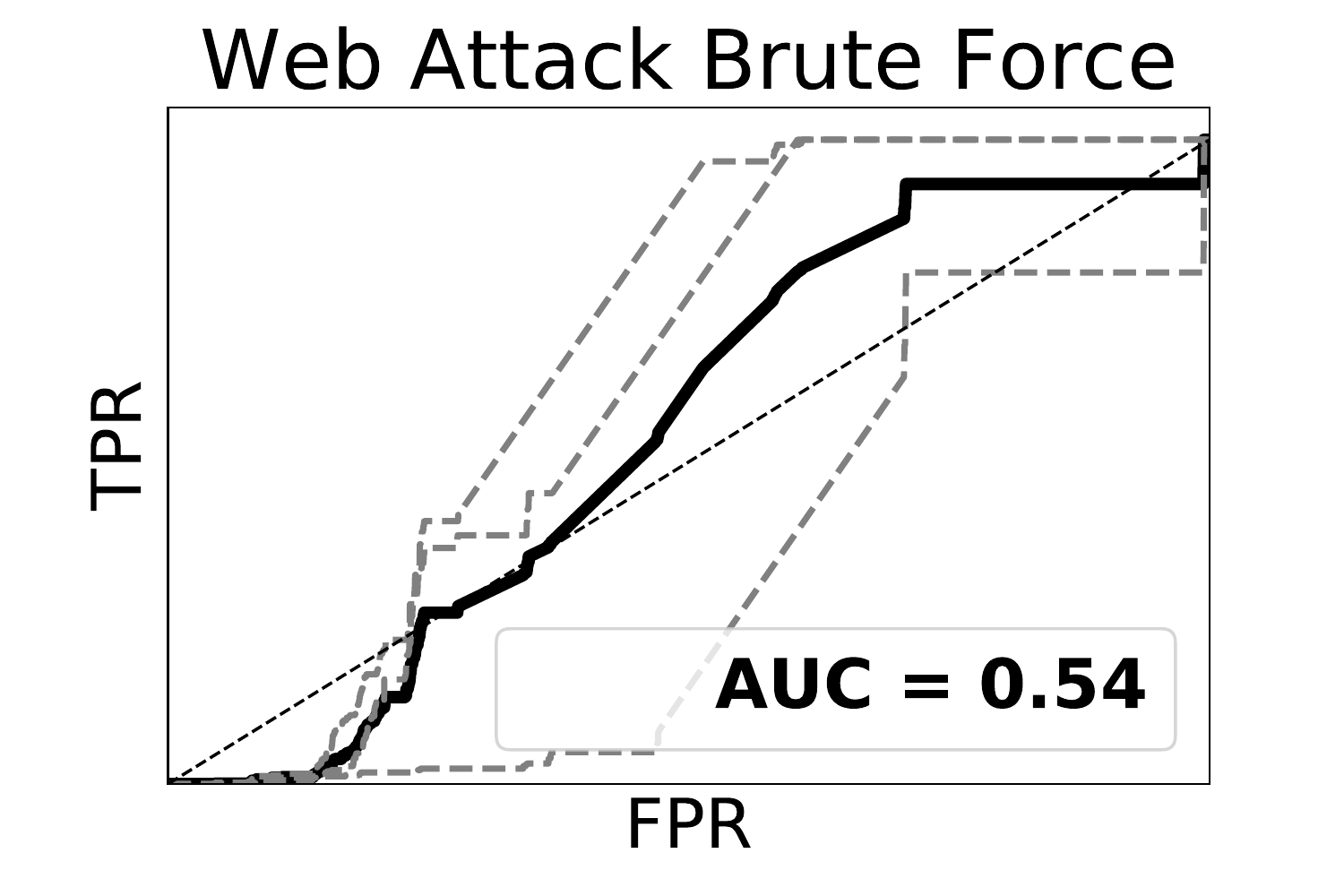}
      \\
      \includegraphics[width=.19\textwidth]{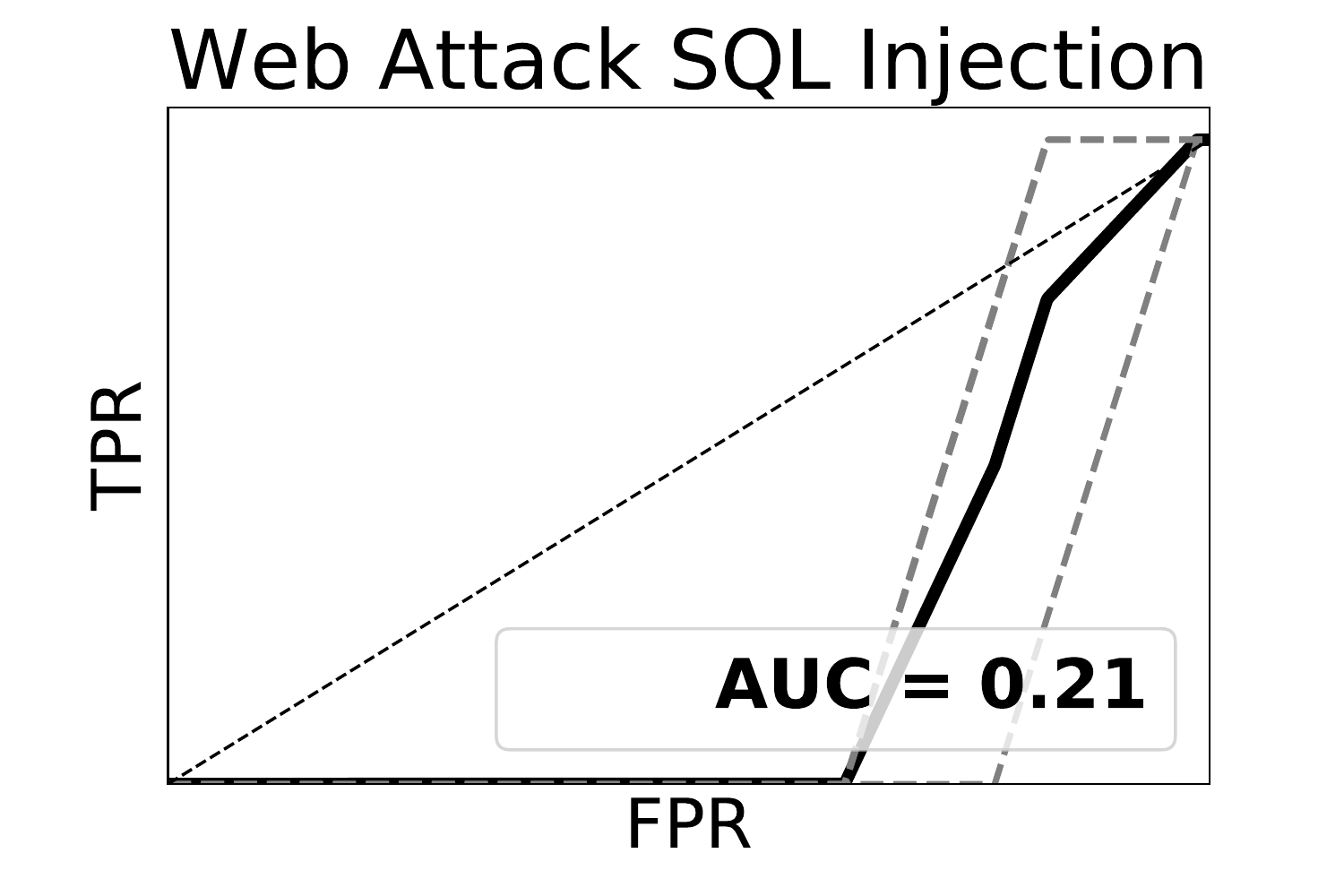} &
      \includegraphics[width=.19\textwidth]{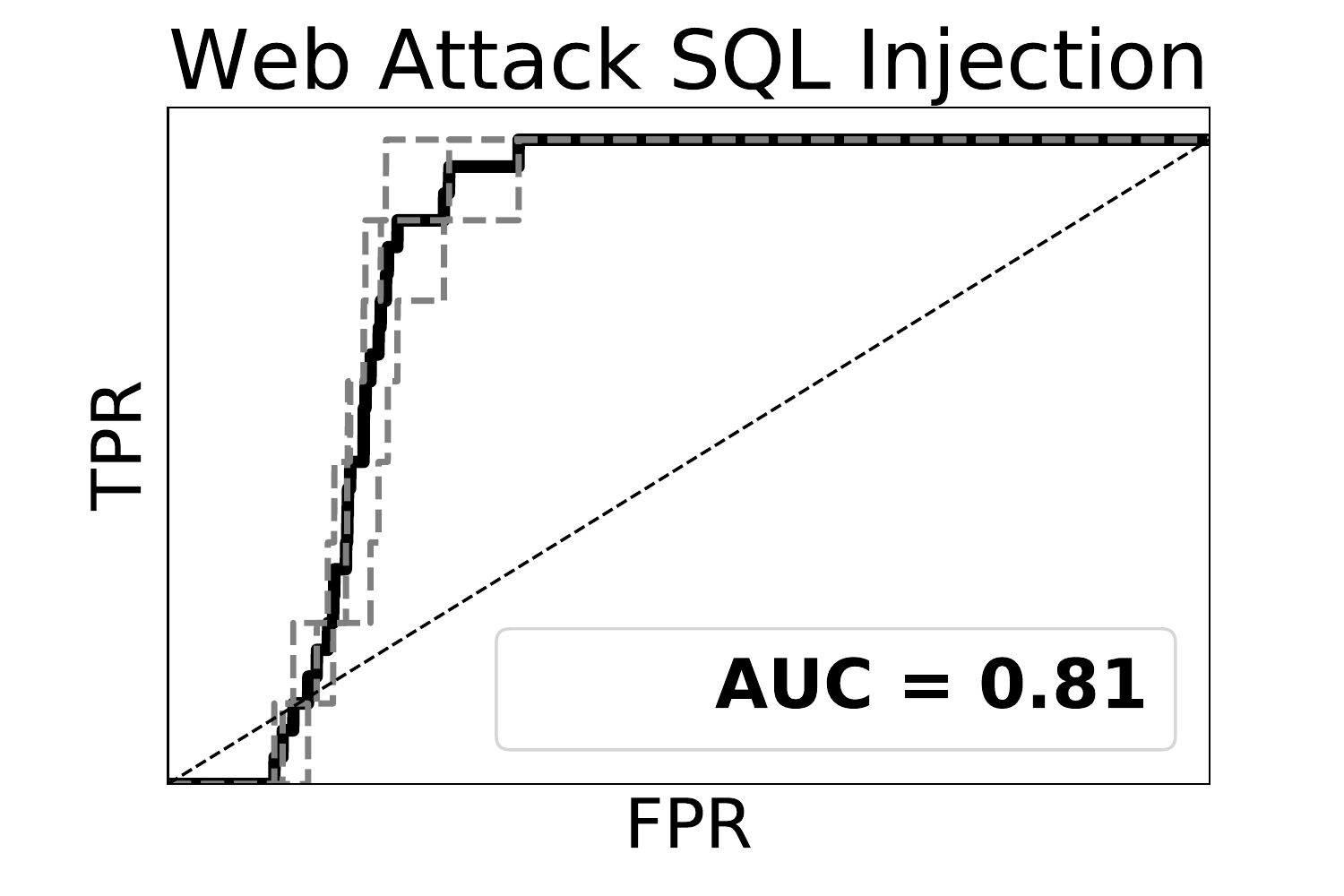} &
      \includegraphics[width=.19\textwidth]{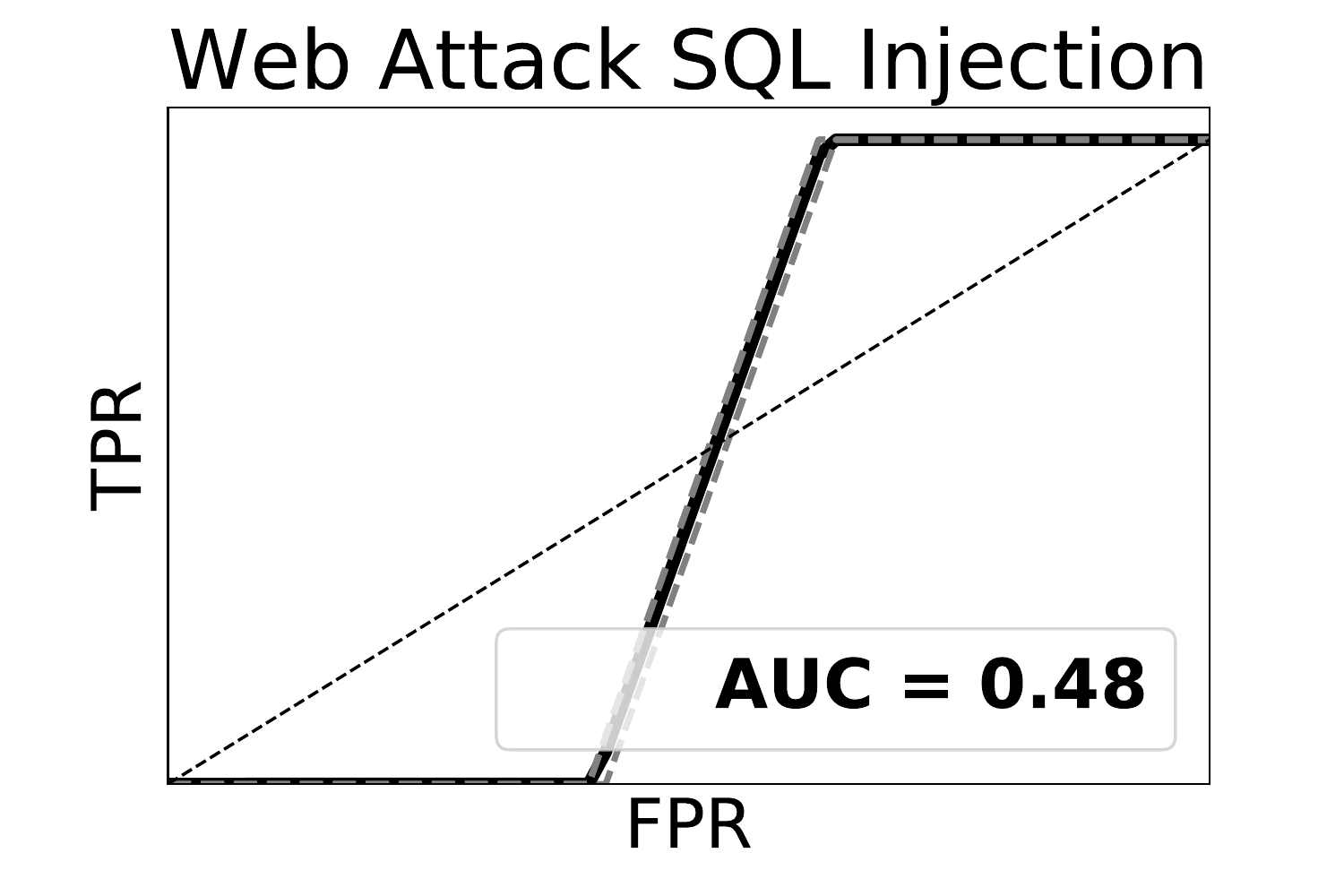} &
      \includegraphics[width=.19\textwidth]{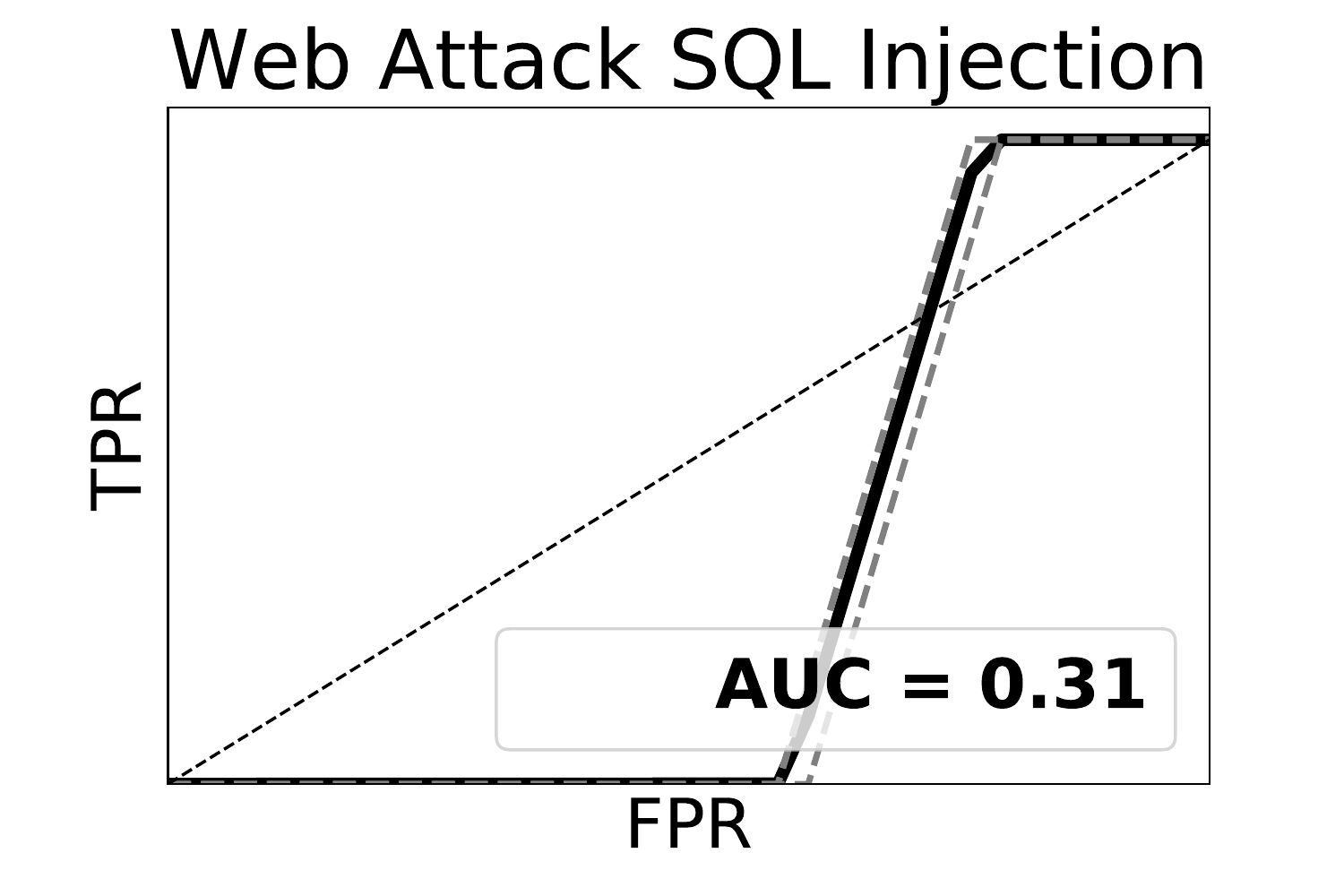} &
      \includegraphics[width=.19\textwidth]{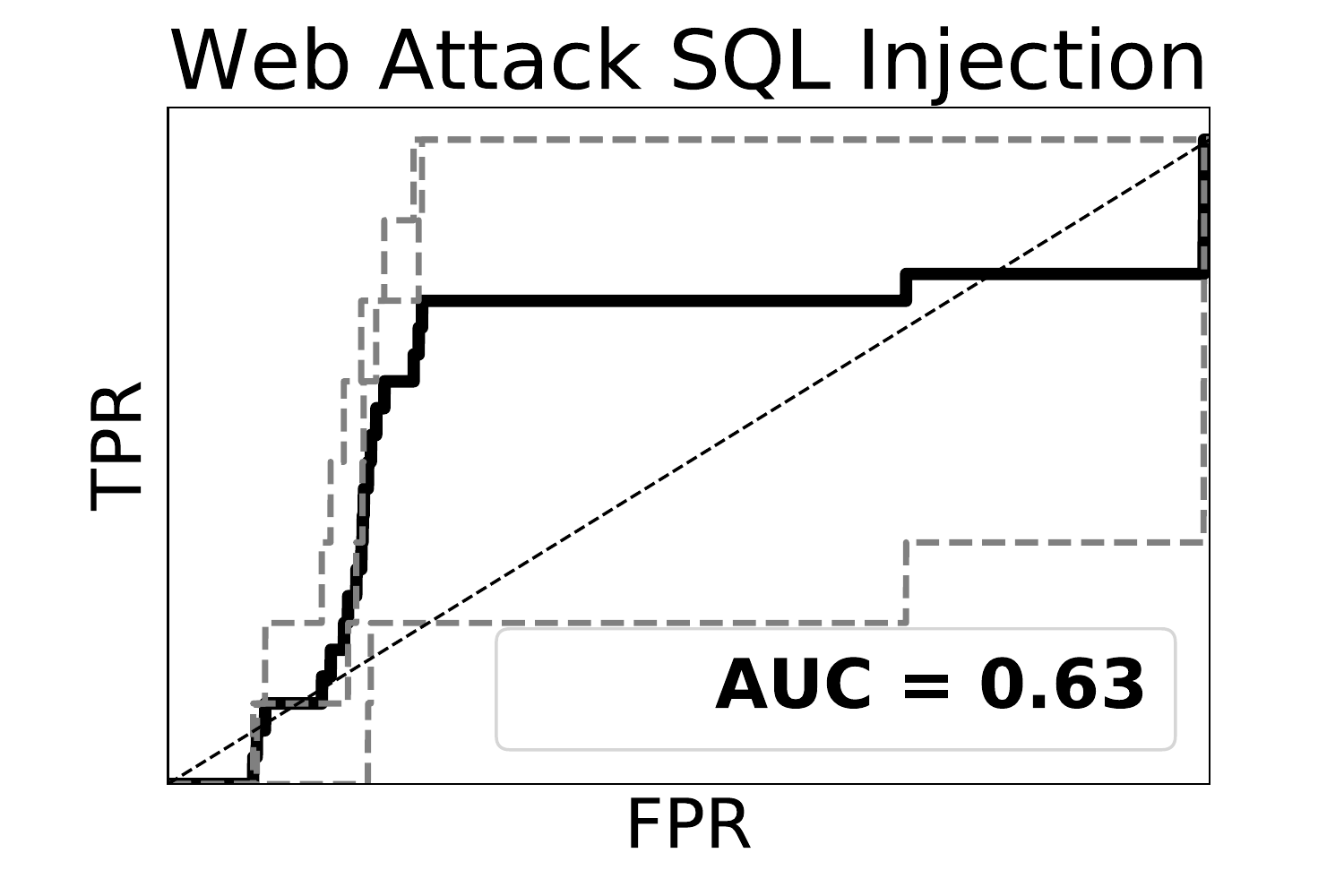}
      \\
      \includegraphics[width=.19\textwidth]{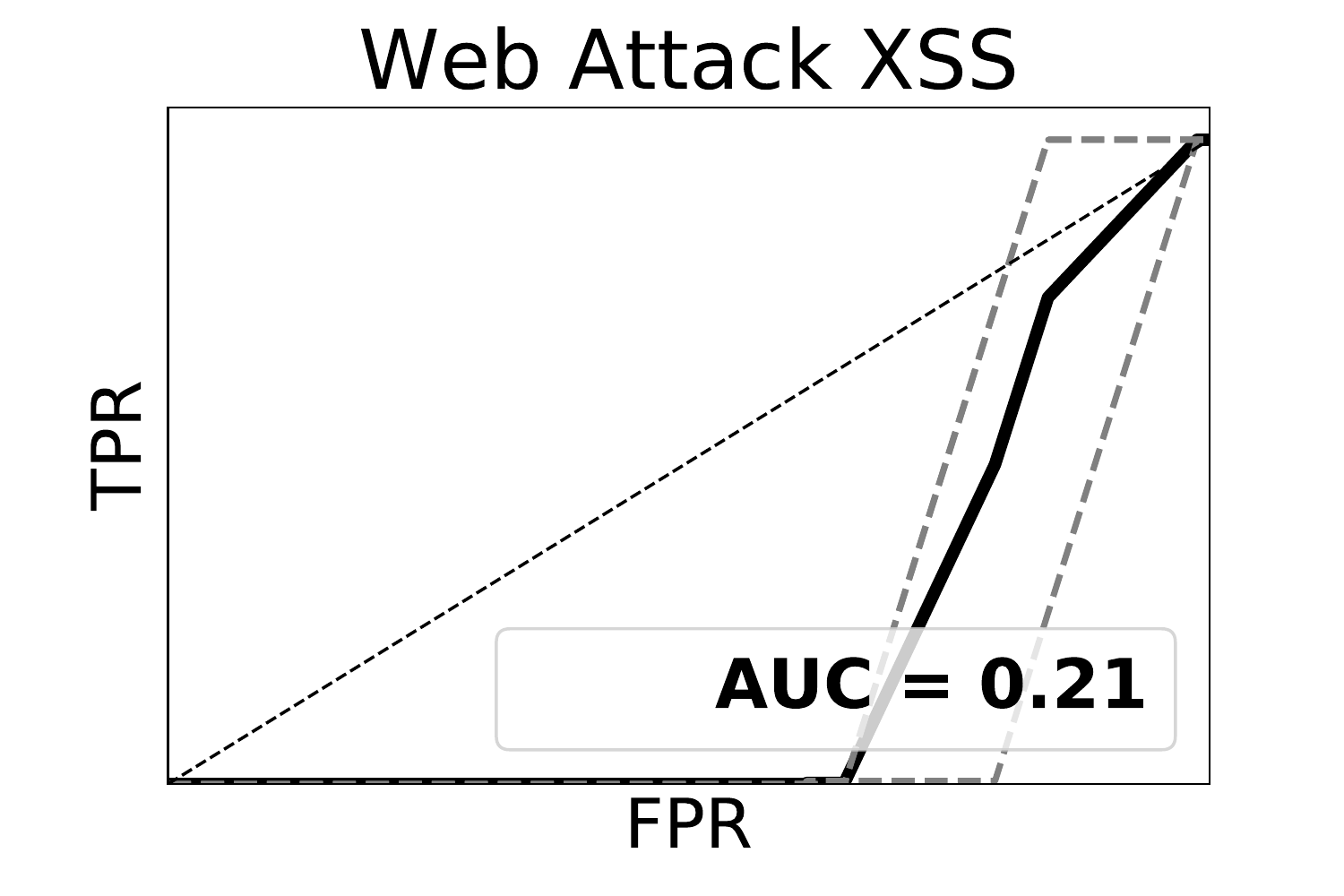} &
      \includegraphics[width=.19\textwidth]{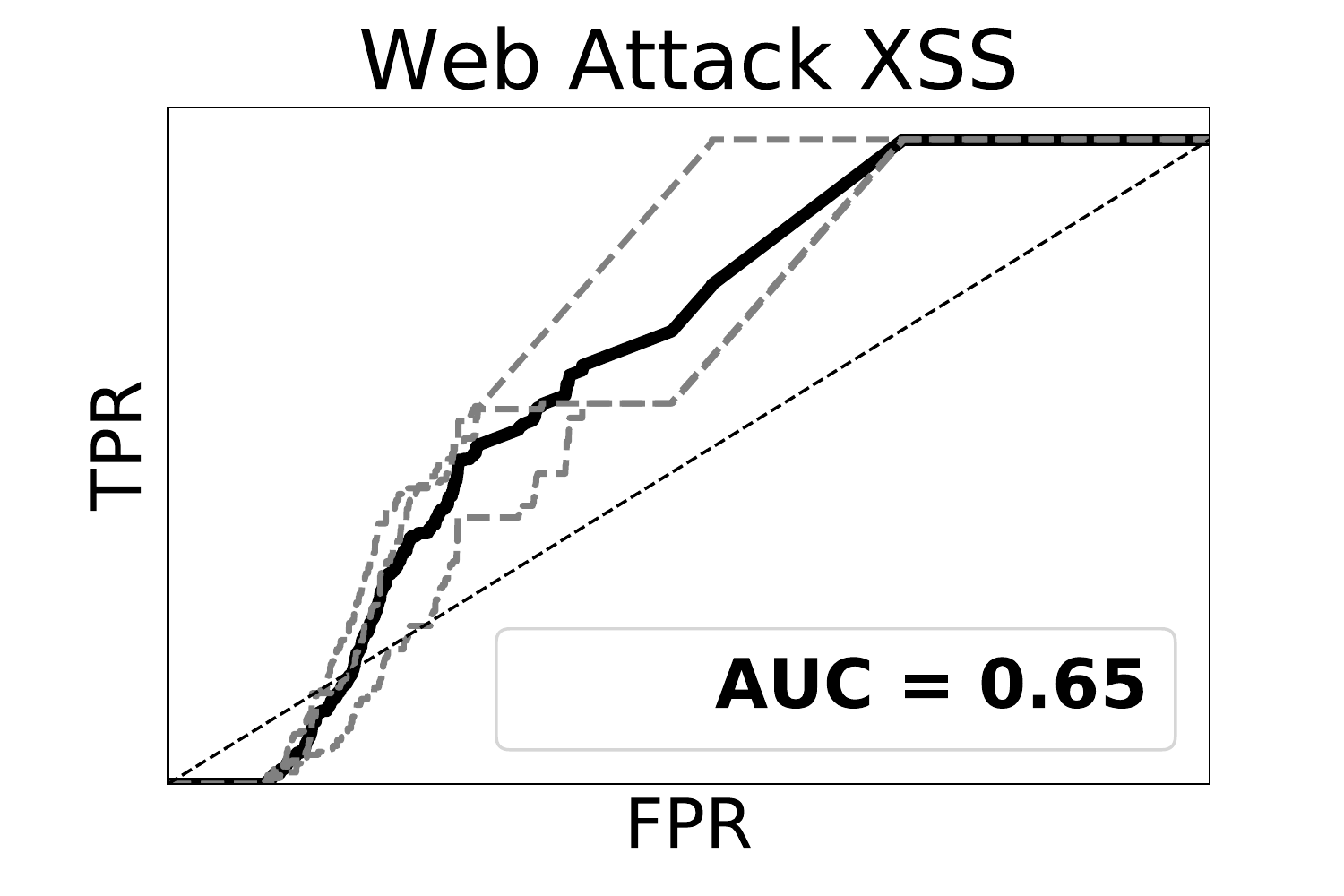} &
      \includegraphics[width=.19\textwidth]{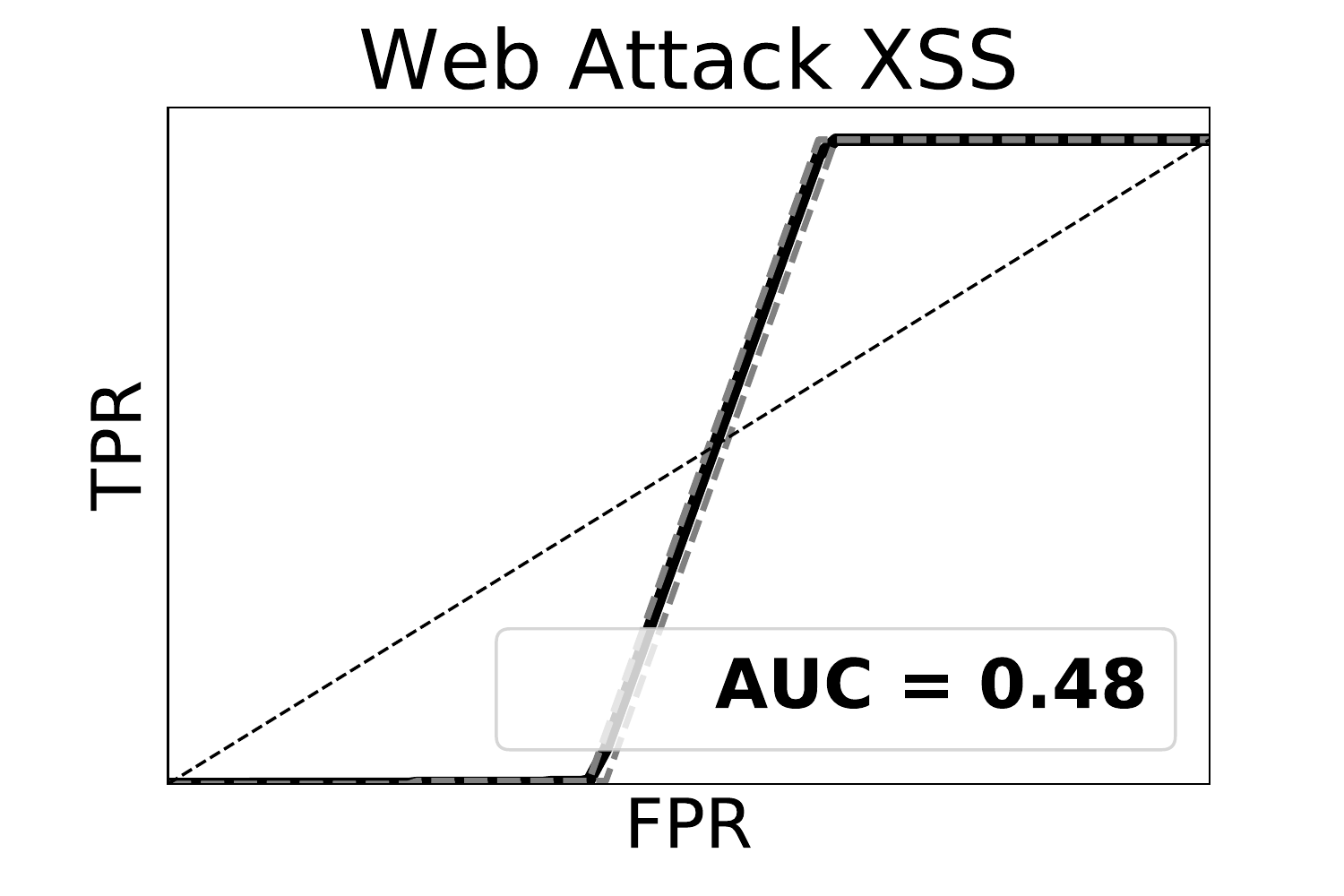} &
      \includegraphics[width=.19\textwidth]{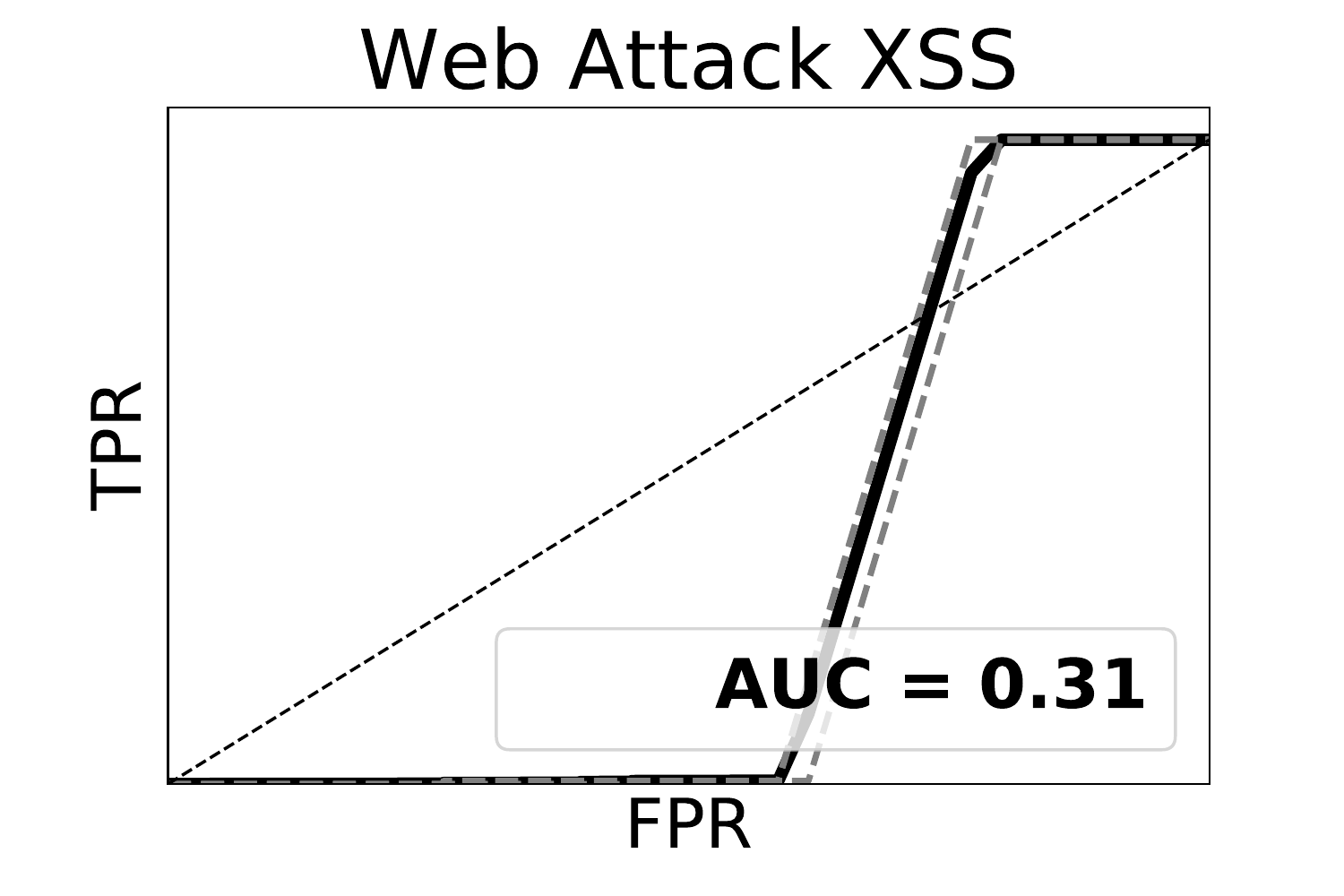} &
      \includegraphics[width=.19\textwidth]{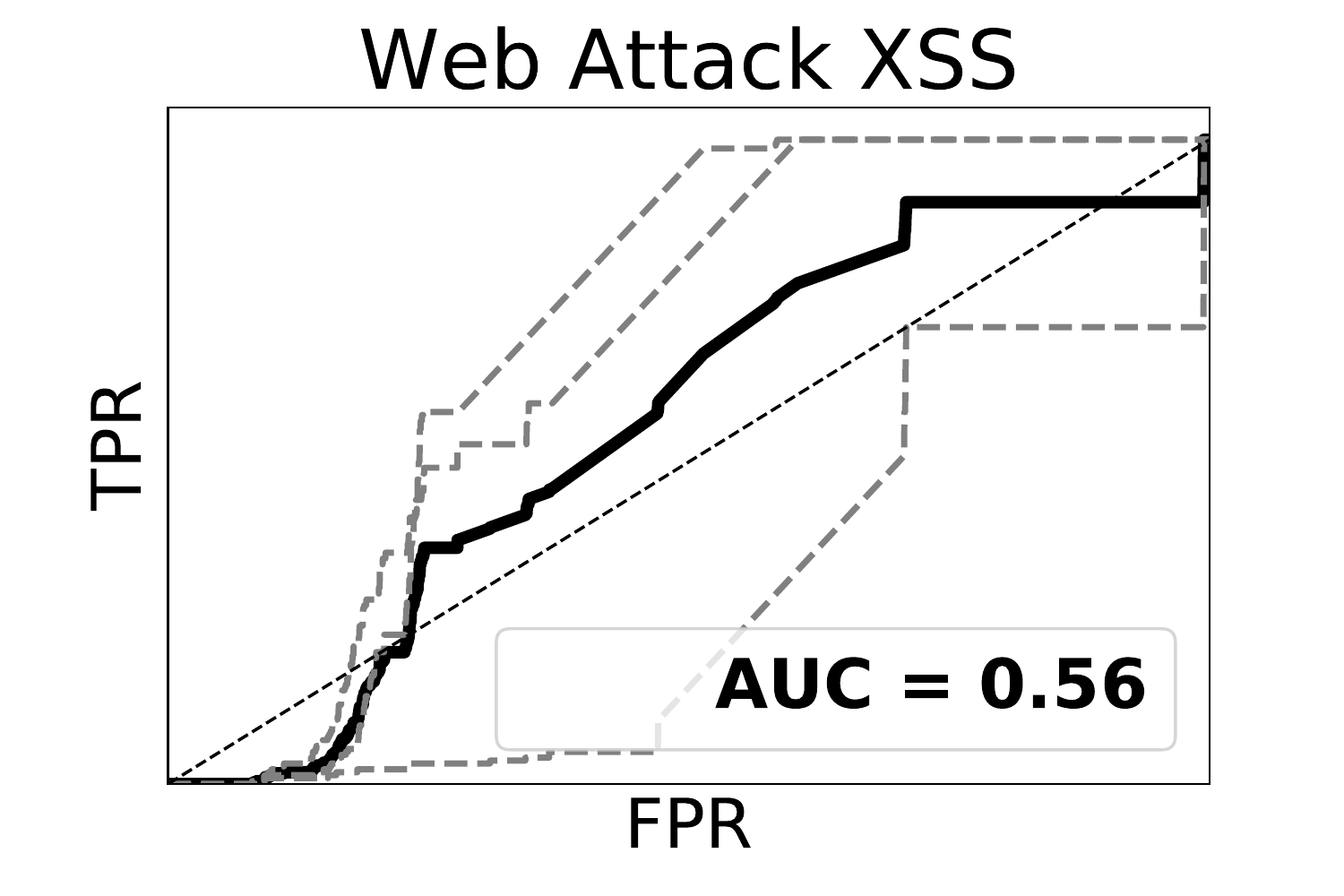}
      \\
  \end{tabular}
\end{table*}

\end{document}